\documentclass[a4paper,11pt]{article}
\usepackage{jheppub}
\usepackage{lineno}
\usepackage{array}
\usepackage{booktabs}
\usepackage{pgfplots}
\usepackage{verbatim}
\usepackage[english]{babel}
\usepackage{iftex}
\ifPDFTeX
  \usepackage[utf8x]{inputenc}
\fi
\usepackage[T1]{fontenc}
\usepackage{amsmath, amssymb, graphicx, setspace}
\usepackage[mathscr]{euscript}
\usepackage{enumitem}
\usepackage{bbold}
\usepackage[compat=1.1.0]{tikz-feynman}
\usepackage{contour}
\usepackage{tikz}
\usepackage{float}
\usepackage{physics}
\usepackage{breqn}
\usepackage{subfigure}
\usepackage{placeins}
\usepackage{xstring}
\usepackage{listings}
\tikzfeynmanset{warn luatex=false}
\definecolor{lightgray}{gray}{0.95}
\usepackage{comment}
\usepackage{caption}
\usepackage{tikz}
\usetikzlibrary{calc,decorations.markings}
\usetikzlibrary{arrows}
\usetikzlibrary{decorations.markings}
\usetikzlibrary{decorations.pathreplacing,decorations.markings} 

\usepackage{mathtools,slashed}
\usepackage{xcolor}
\definecolor{brown}{rgb}{1.0, 0.8, 0.0}
\definecolor{airforceblue}{rgb}{0.36, 0.54, 0.66}\definecolor{amethyst}{rgb}{0.6, 0.4, 0.8}
\definecolor{darkgreen}{RGB}{0,100,0}
\newcommand{\eqn}{&=&}
\newcommand{\non}{\notag \\}

\newcommand{\jc}[4]{\left\langle #1|#2+#3|#4\right]}
\newcommand{\kb}[3]{%
	\ifx\\#3\\
	{#1}_{#2,\perp}%
	\else
	\bar{#1}_{#2,\perp}%
	\fi
}
\usepackage{shuffle}

\newcommand{\cC}{\mathcal {C}}
\newcommand{\cP}{\mathcal {P}}
\newcommand{\cA}{\mathcal{A}}

\newcommand{\ab}[2]{\left\langle #1 #2\right\rangle }
\newcommand{\sbb}[2]{\left[ #1 #2\right] }
\definecolor{egcolor}{rgb}{0, 0, 1} 

\setlist[itemize]{leftmargin=*, nosep}

\title{\boldmath Regge factorization of tree-level QCD amplitudes using a minimal set of lightcone variables}

\author[a,b]{Emmet P. Byrne,}
\author[c]{Vittorio Del Duca,}
\author[a]{Einan Gardi,}
\author[a]{Yuyu Mo,}
\author[a]{Jennifer M. Smillie}
\affiliation[a]{Higgs Centre for Theoretical Physics, School of Physics and Astronomy, \\
The University of Edinburgh, Edinburgh EH9 3FD, Scotland, UK}
\affiliation[b]{Department of Physics and Astronomy,\\
The University of Manchester, Manchester M13 9PL, UK}
\affiliation[c]{INFN, Laboratori Nazionali di Frascati, 00044 Frascati (RM), Italy}

\emailAdd{Emmet.Byrne@manchester.ac.uk, Vittorio.DelDuca@lnf.infn.it,\\Einan.Gardi@ed.ac.uk, Y.Y.Mo@sms.ed.ac.uk, J.M.Smillie@ed.ac.uk}

\abstract{We represent the multi-leg tree-level amplitudes of quarks and gluons using a minimal set of lightcone variables, which incorporate all on-shell and momentum conservation conditions and naturally captures the separate longitudinal and transverse momentum components. These variables make it easy to eliminate spurious poles and consider multi-Regge kinematic limits.  
In this framework we examine the factorization of tree-level amplitudes in rapidity and extract 
all two, three and four parton Multi-Regge Emission Vertices (MREVs), both central and peripheral, and summarise them in a Mathematica library, \texttt{MREV}. We investigate in detail how relations between amplitudes translate into relations between MREVs. 
These relations, along with factorization properties in further kinematic limits, provide robust consistency checks of the results.  
}

\begin{document}
\maketitle
\flushbottom
\section{Introduction}

The study of kinematic limits has a special significance in our understanding of scattering amplitudes. The interest in the Regge limit~\cite{Regge:1959mz,Collins:1977jy}, in particular, predates the birth of QCD as the theory of the strong interactions.
Within QCD the Regge limit and its generalizations open up a 
plethora of physics phenomena, which can be investigated perturbatively owing to factorization and 
rapidity evolution, BFKL~\cite{Lipatov:1976zz,Kuraev:1976ge,Kuraev:1977fs,Balitsky:1978ic,Fadin:1998py,Ciafaloni:1998gs,Kotikov:2000pm,Kotikov:2002ab} and its generalizations.

The factorization property is already visible at tree level: in line with general expectations~\cite{Regge:1959mz,Collins:1977jy}, the dominant interactions in the high-energy limit, where the centre-of-mass energy ($s$) is much larger than the momentum transfer ($-t$), are exclusively those where the particle with the highest spin -- the gluon -- is exchanged in the $t$ channel, and where the helicity of the target and the projectile are conserved. All other contributions are power suppressed by $-t/s$, and are therefore irrelevant in the limit considered.
Thus, the dominant $t$-channel interactions trivially factorize in rapidity. This factorization property extends in a non-trivial way to loop level, where dependence on $s$ enters exclusively through logarithms, which are governed by the phenomenon of Reggeization~\cite{Lipatov:1976zz,Fadin:1995km,Fadin:1996tb,Fadin:1995xg,Blumlein:1998ib,DelDuca:2001gu,Fadin:2015zea,Fadin:2006bj,Falcioni:2021buo,DelDuca:2021vjq,Caola:2021izf,Falcioni:2021dgr} and said evolution equations. 
Furthermore, factorization in rapidity also extends to multi-leg amplitudes where all final state particles -- or subsets thereof -- are widely separated in rapidity. Such multi-Regge kinematic limits are the central topic of the present paper. 

Our main goal is to study the factorization of multi-leg tree-level amplitudes in rapidity, and systematically extract the universal building blocks, dubbed \emph{multi-Regge emission vertices} (MREVs), which feature at leading power. This includes Peripheral-Emission Vertices (PEVs)~\cite{Kuraev:1976ge,Fadin:1989kf,DelDuca:1995ki,Fadin:1996nw,DelDuca:1996nom,DelDuca:1996km,Duhr:2009uxa,Fadin:1993wh,Fadin:1992zt,Fadin:1993qb,DelDuca:1998kx,Bern:1998sc,DelDuca:1999iql,Canay:2021yvm,Byrne:2023nqx,DelDuca:2014cya,Caron-Huot:2017fxr,DelDuca:2021vjq,Falcioni:2021dgr,Caola:2021izf}, describing emission at rapidities comparable to the target or the projectile, but which are widely separated in rapidity from the rest of the process, and Central-Emission Vertices (CEVs)~\cite{Lipatov:1976zz,Fadin:1989kf,DelDuca:1995ki,Fadin:1996nw,DelDuca:1996nom,DelDuca:1996km,Fadin:1993wh,Fadin:1994fj,Fadin:1996yv,DelDuca:1998cx,Bern:1998sc,DelDuca:1999iql,Antonov:2004hh,Duhr:2009uxa,Byrne:2022wzk,Buccioni:2024gzo,Abreu:2024xoh} describing emissions that are well separated in rapidity from both ends. In particular, CEVs are widely separated from both the target and the projectile, and are hence process independent. 

The computation of scattering amplitudes and our understanding of their properties has progressed a lot over the past couple of decades~\cite{Dixon:1996wi,Elvang:2015rqa,Badger:2023eqz,Weinzierl:2016bus}. Thanks to recursion-relation methods, such as BCFW~\cite{Britto:2004ap,Britto:2005fq}, tree-level scattering amplitudes in Yang-Mills theory are accessible to high multiplicity. One of the key elements underlying 
this progress is the understanding of the kinematic space of massless particles with spin, and in particular the use of spinor helicity variables~\cite{Dixon:1996wi}. While the redundancy of these variables facilitates writing relatively compact expressions for multi-leg amplitudes, it also obscures certain properties.
For example, spinor helicity variables make 
it challenging to eliminate spurious poles from the expressions. They also make it difficult to take certain kinematic limits, such as the multi-Regge limit, in which we are interested here.   

Lightcone variables provide a natural alternative to express amplitudes of massless particles.
Specifically, considering colour-ordered amplitudes of $n$ partons, and taking into account all on-shell and momentum conservation conditions, we introduce a minimal set of $3n-9$ (real) variables which can be used to express all spinor products, and hence any helicity amplitude. These variables are particularly well-suited to study the factorization of multi-leg amplitudes in a variety of high-energy limits. 
Such variables have been used for example in Ref.~\cite{Byrne:2022wzk}, to study the central next-to-multi-Regge kinematic limit of the six-point amplitude in ${\cal N}=4$ supersymmetric Yang-Mills theory (sYM) at tree level and at one loop, and hence to extract the two-gluon central emission vertex. Here we consider similar limits of multi-leg tree amplitudes with up to eight partons, from which we extract all MREVs of up to four final-state quarks or gluons.

The structure of the paper is as follows. 
We begin in section~\ref{sec:factorizationElements} by briefly reviewing the state-of-the-art knowledge of multi-Regge emission vertices. We emphasise there the motivation to determine these vertices in view of the goal of extending the BFKL equation and related cross-section calculations to higher logarithmic accuracy.
In section~\ref{sec:minvar}, building on the work of Ref.~\cite{Byrne:2022wzk}, we define a minimal set of lightcone variables (MSLCVs), derive their relation with spinor products and hence show how they can be used to 
express $n$-point amplitudes, as well as MREVs. 
In section~\ref{sec:PEVCEVdef}, we define MREVs at the colour-dressed and colour-ordered level, and outline the strategy to determine them in terms of our MSLCVs. In section~\ref{resultsof CEVPEV}, we present explicit expressions of the central-emission vertices (CEVs) for the emission of up to three partons. 
These expressions are derived  based on the GGT package~\cite{Dixon:2010ik}, which computes colour-ordered tree amplitudes starting from those in ${\cal N}=4$ sYM theory.
Specifically, in section~\ref{Sec:three_parton_CEV} we present the CEV for the emission of a $q\bar q$ pair and a gluon, which is a new result.
In section~\ref{sec:4partonCEVsPEVs}, we introduce the \texttt{\href{https://github.com/YuyuMo-UoE/Multi-Regge-Emission-Vertices}{Multi-Regge-Emission-Vertices}} ({\tt{MREV}}) Mathematica library~\cite{yuyu_2017_github}, which is again derived from the GGT package. In particular, we explain how to use the library in order to obtain the explicit expressions for QCD MREVs for the emission of four partons, many of which are new results. 
Exploiting the relations between colour-ordered amplitudes, which are naturally defined in the sYM context, in
section~\ref{relationsPEVCEV} we present relations between colour-ordered MREVs.
These relations  are both interesting in their own right, and 
provide robust consistency checks for our results. 
We present our concluding remarks in section~\ref{sec:discuss}.
The paper is furnished with five appendices. In Appendix~\ref{frame} we describe the multi-parton kinematics including our conventions for the spinor helicity variables.
In Appendix~\ref{Appendix:Fact_in_limits} we study the factorization properties of MREVs in various kinematic limits, which provides additional checks for the results in our {\tt{MREV}} library. In Appendix~\ref{sec:BriefKKrelation}
we briefly summarise one of the derivations of the Kleiss-Kuijf relation.
In Appendix~\ref{colourdressingExample} we provide
examples of colour-dressed PEVs with three final-state partons. Finally in Appendix~\ref{sec:mrev} we provide further guidance and examples regarding the use of the {\tt{MREV}} library.

\section{Motivation: elements of factorization in  rapidity}
\label{sec:factorizationElements}

In the Regge limit, $s \gg -t$, in which the squared centre-of-mass energy $s$ is much larger than the momentum transfer $-t$, any $2 \to 2$ scattering process in either QCD or supersymmetric Yang-Mills (sYM) theory is dominated by the exchange of a gluon in the $t$ channel.
The dominant radiative corrections to the amplitude Reggeize~\cite{Lipatov:1976zz}, that is, they exponentiate as~\((s/(-t))^{\alpha_g(t)}\), where $\alpha_g$ is the gluon Regge trajectory~\cite{Lipatov:1976zz,Fadin:1995km,Fadin:1996tb,Fadin:1995xg,Blumlein:1998ib,DelDuca:2001gu,Fadin:2015zea,Fadin:2006bj,Falcioni:2021buo,DelDuca:2021vjq,Caola:2021izf,Falcioni:2021dgr}.
Building upon this phenomenon, the Balitsky-Fadin-Kuraev-Lipatov (BFKL) equation describes strong-interaction processes with two large and disparate scales, $s \gg (-t)$, by resumming the radiative corrections to parton-parton scattering to all orders at leading logarithmic (LL)~\cite{Lipatov:1976zz,Kuraev:1976ge,Kuraev:1977fs,Balitsky:1978ic} and next-to-leading logarithmic (NLL) accuracy~\cite{Fadin:1998py,Ciafaloni:1998gs,Kotikov:2000pm,Kotikov:2002ab} in $\log(s/(-t))$.

At LL, BFKL resums the tower of~${\cal O}([\alpha_s\ln(s/(-t))]^n)$ corrections to the squared amplitude to all orders.
Since the logarithm $\log(s/(-t))$ is related to the rapidity interval in the $2 \to 2$ scattering process,
the BFKL equation~\cite{Lipatov:1976zz,Kuraev:1976ge,Kuraev:1977fs,Balitsky:1978ic} 
effectively describes the rapidity evolution of a gluon ladder exchanged in the $t$ channel, in terms of an integral over transverse momentum of the gluon ladder propagator. The kernel of this integral equation is given by the emission of a gluon through a vertex termed a~\emph{central-emission vertex} (CEV), or Lipatov vertex, which first occurs in the tree-level $2\to 3$ amplitude in multi-Regge kinematics (MRK), where the outgoing partons are strongly ordered in light-cone momentum, or equivalently in rapidity. In the soft limit,
the gluon CEV yields infrared (IR) divergences. Upon integration over phase space,
these singularities cancel against the Regge trajectory, which first occurs in the one-loop $2\to 2$ scattering amplitude at LL accuracy.

\begin{figure}[b]
    \centering
    \subfigure[]{\includegraphics[width=0.24\textwidth]{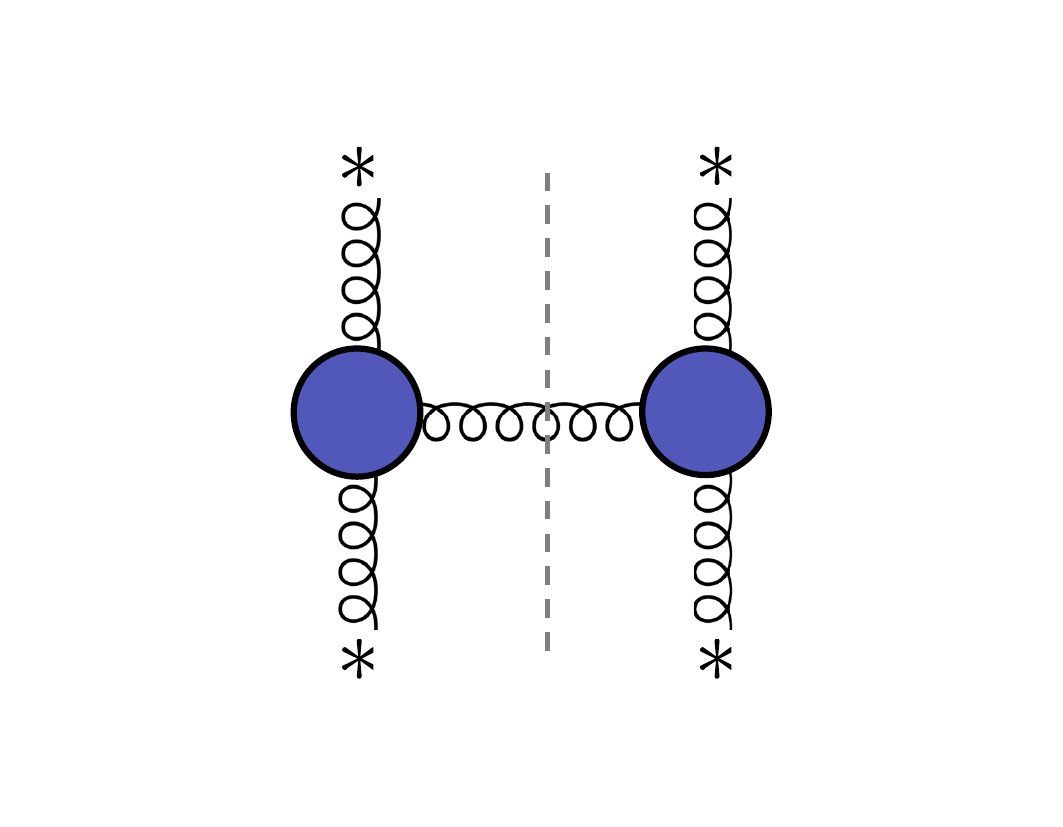}}\label{subfig:C0g}
    \subfigure[]{\includegraphics[width=0.24\textwidth]{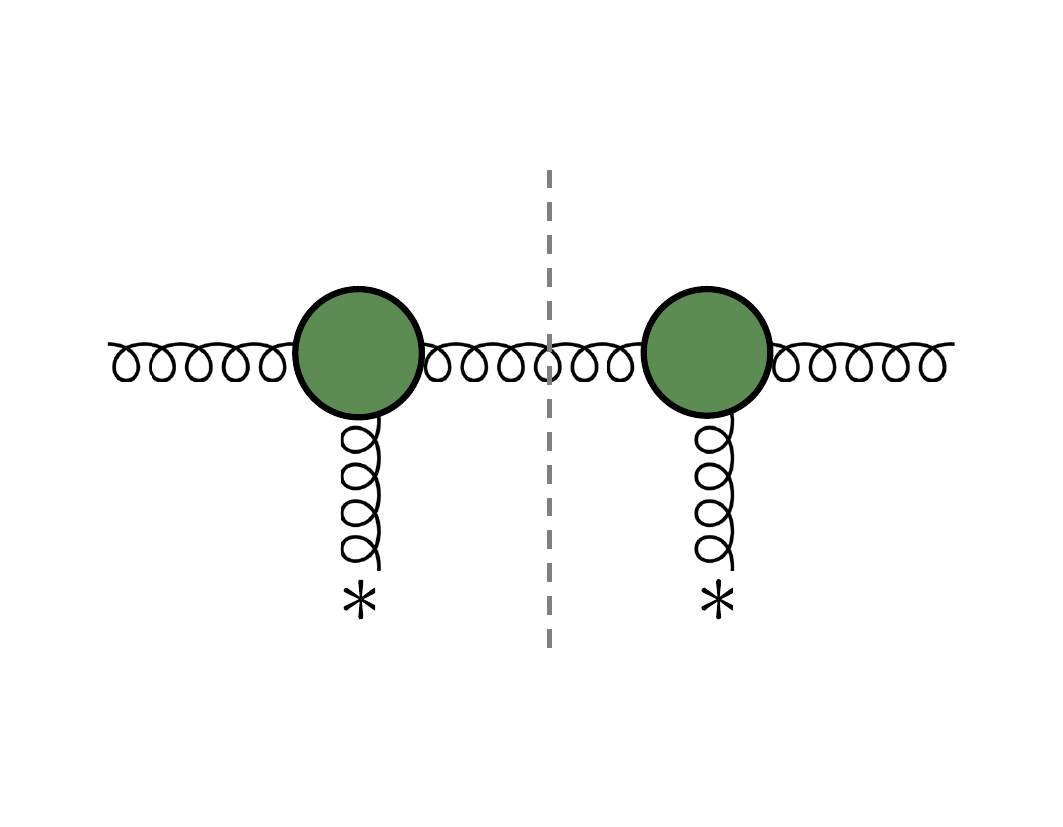}\label{subfig:P0g}} 
    \subfigure[]{\includegraphics[width=0.24\textwidth]{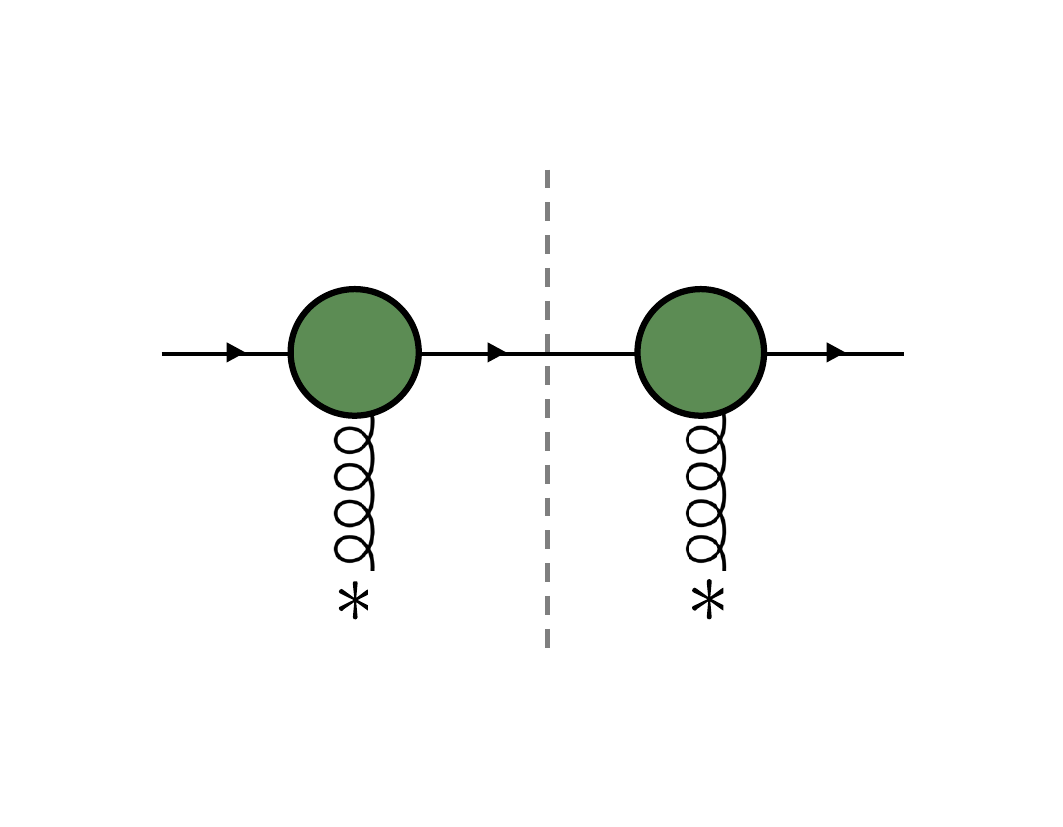}\label{subfig:P0q}}
    \caption{(a) The LO BFKL kernel, consisting of a squared gluon CEV, or Lipatov vertex; (b)-(c) Ingredients of the LO jet impact factors, consisting of squared PEVs. The grey dashed line indicates the final-state cut.}
    \label{fig:BFKL_LO}
\end{figure}

The gluon ladder exchanged in the $t$ channel is \emph{universal}, i.e. independent of the nature of the two particles being scattered.
These particles, which we call  ``target'' and ``projectile'', interact with the gluon ladder at its ends and generate the so-called \emph{peripheral-emission vertices} (PEV), which are therefore process-dependent.
The multi-leg amplitudes which source the BFKL equation are then given by the convolution of two PEVs, one on the target side and the other on the projectile side, with the cut gluon ladder bridging over the large rapidity span between them.
The cross section, or production rate, is proportional to the square of this amplitude. In addition to the BFKL ladder, it involves two \emph{impact factors}, each of which is obtained from a squared PEV, integrated over the phase space of the outgoing partons on either end of the rapidity span. The leading-order (LO) BFKL kernel, and the LO jet impact factors are depicted in Figure~\ref{fig:BFKL_LO}.

\begin{figure}[ht]
    \centering
    \subfigure[]{\includegraphics[width=0.24\textwidth]{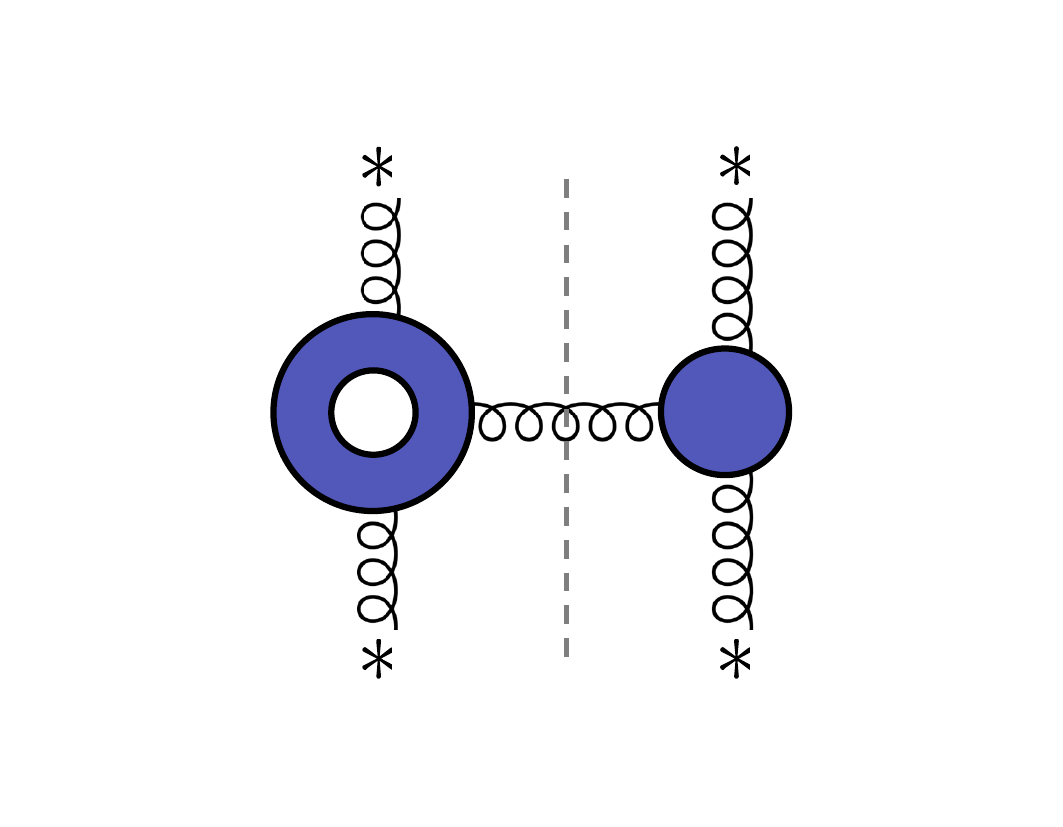}}\label{subfig:C1g}
    \subfigure[]{\includegraphics[width=0.24\textwidth]{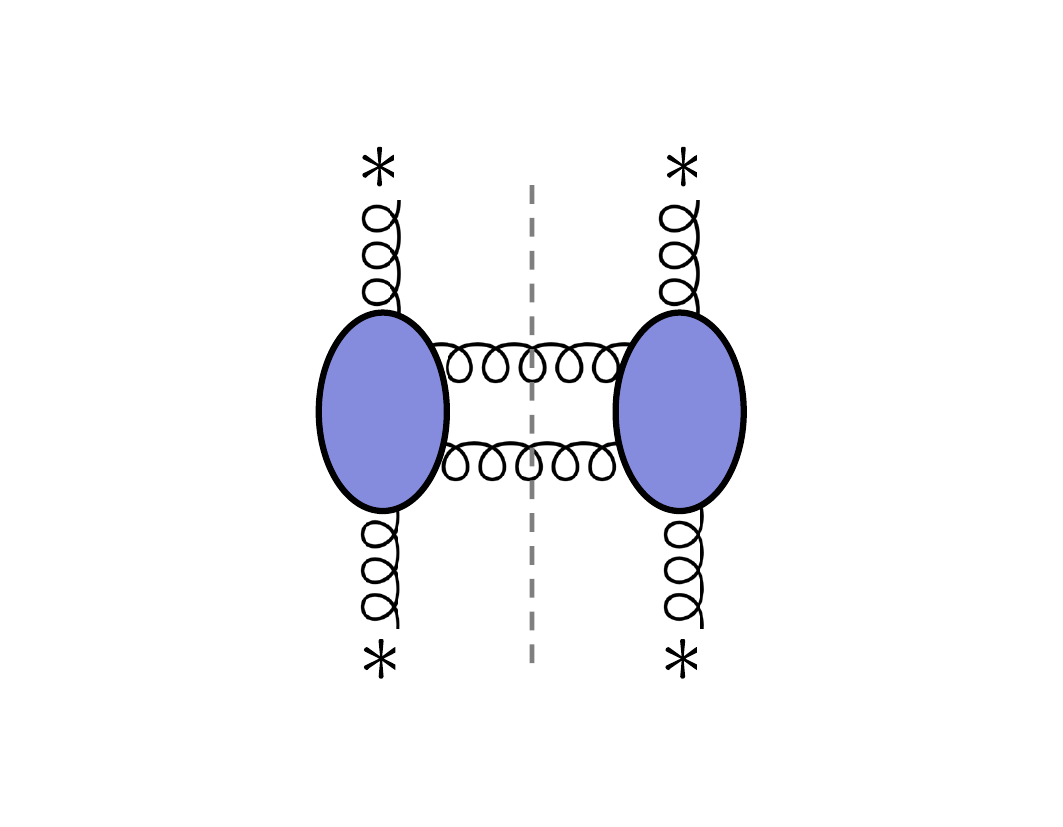}\label{subfig:C0gg}} 
    \subfigure[]{\includegraphics[width=0.24\textwidth]{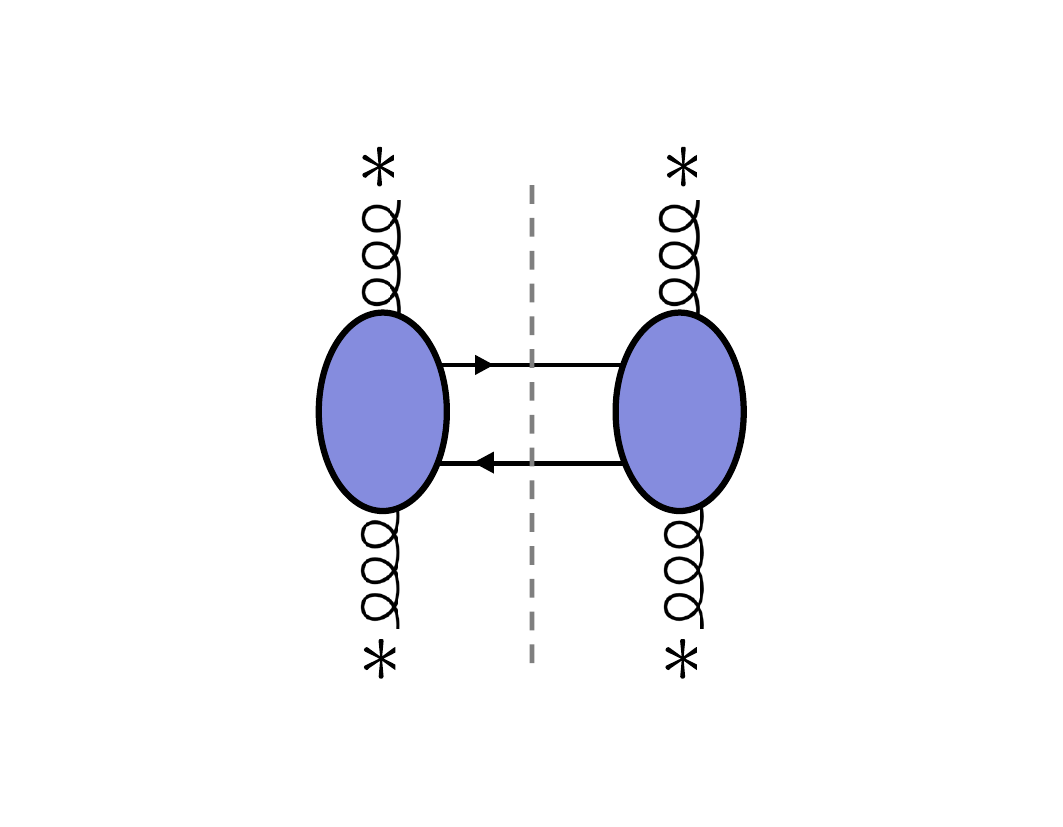}\label{subfig:C0qq}}
    \caption{Ingredients of the NLO BFKL kernel: (a) virtual correction; (b)-(c) real corrections. Punctures in blobs denote loop corrections.}
    \label{fig:Kernel_NLO}
\end{figure}

It is possible to extend the BFKL equation to NLL accuracy~\cite{Fadin:1998py,Ciafaloni:1998gs,Kotikov:2000pm,Kotikov:2002ab}, i.e. to resum the terms of ${\cal O}( \alpha_s [\alpha_s\ln(s/(-t))]^n)$ to all orders, by considering the radiative corrections to the leading-order kernel. These corrections, depicted in Figure~\ref{fig:Kernel_NLO}, are sourced by the CEV for the emission of two gluons, or a $q\bar q$ pair, close in rapidity along the gluon ladder~\cite{Fadin:1989kf,DelDuca:1995ki,Fadin:1996nw,DelDuca:1996nom,DelDuca:1996km}, and the one-loop (virtual) corrections to the single gluon CEV~\cite{Fadin:1993wh,Fadin:1994fj,Fadin:1996yv,DelDuca:1998cx,Bern:1998sc}.
The infrared divergences of the next-to-leading-order (NLO) kernel, which result from integrating the momenta of the partons emitted along the gluon ladder over their phase space, cancel against the singularities of the two-loop gluon Regge trajectory~\cite{Fadin:1995xg,Fadin:1996tb,Fadin:1995km,Blumlein:1998ib,DelDuca:2001gu}.

\begin{figure}[ht]
    \centering
    \subfigure[]{\includegraphics[width=0.24\textwidth]{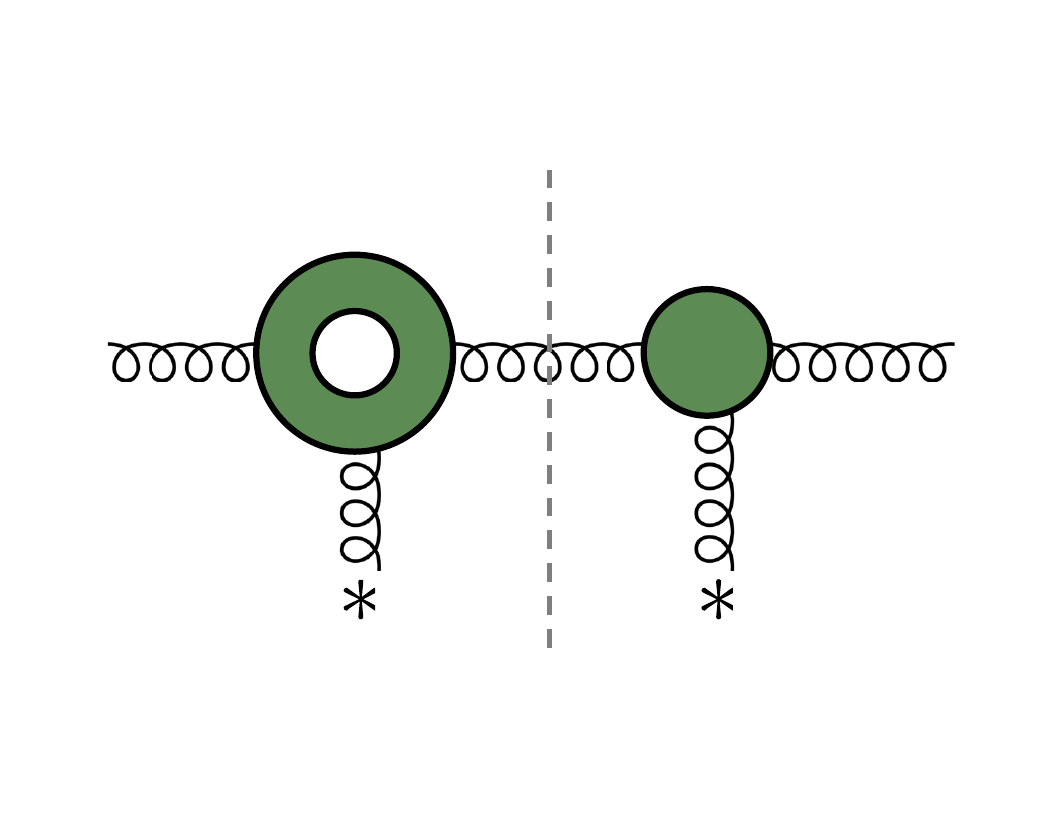}}\label{subfig:P1g}
    \subfigure[]{\includegraphics[width=0.24\textwidth]{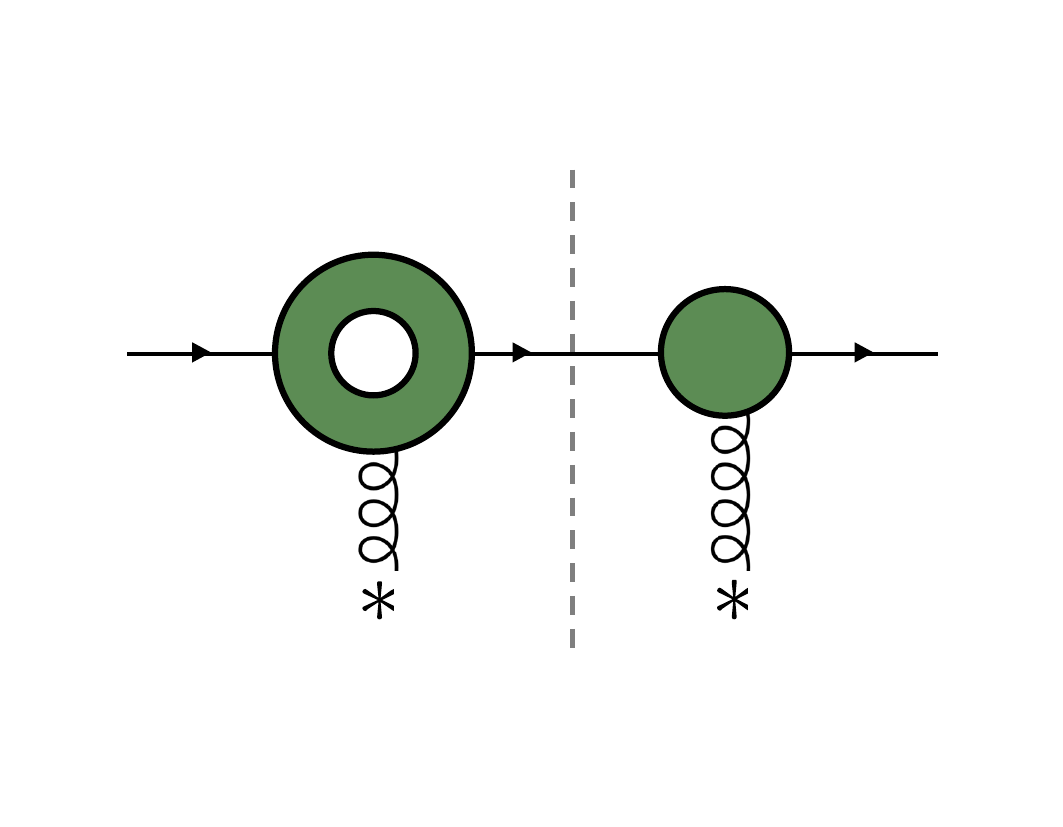}\label{subfig:P1q}} 
    \subfigure[]{\includegraphics[width=0.24\textwidth]{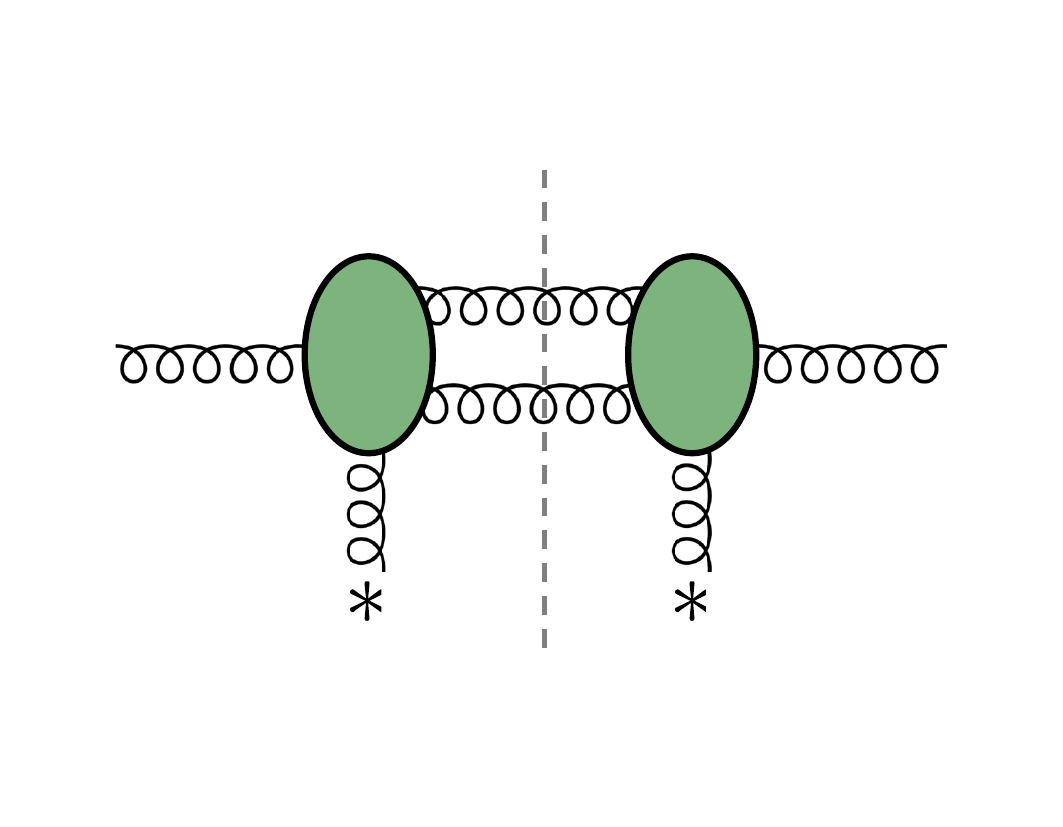}\label{subfig:P0gg}} 
    \subfigure[]{\includegraphics[width=0.24\textwidth]{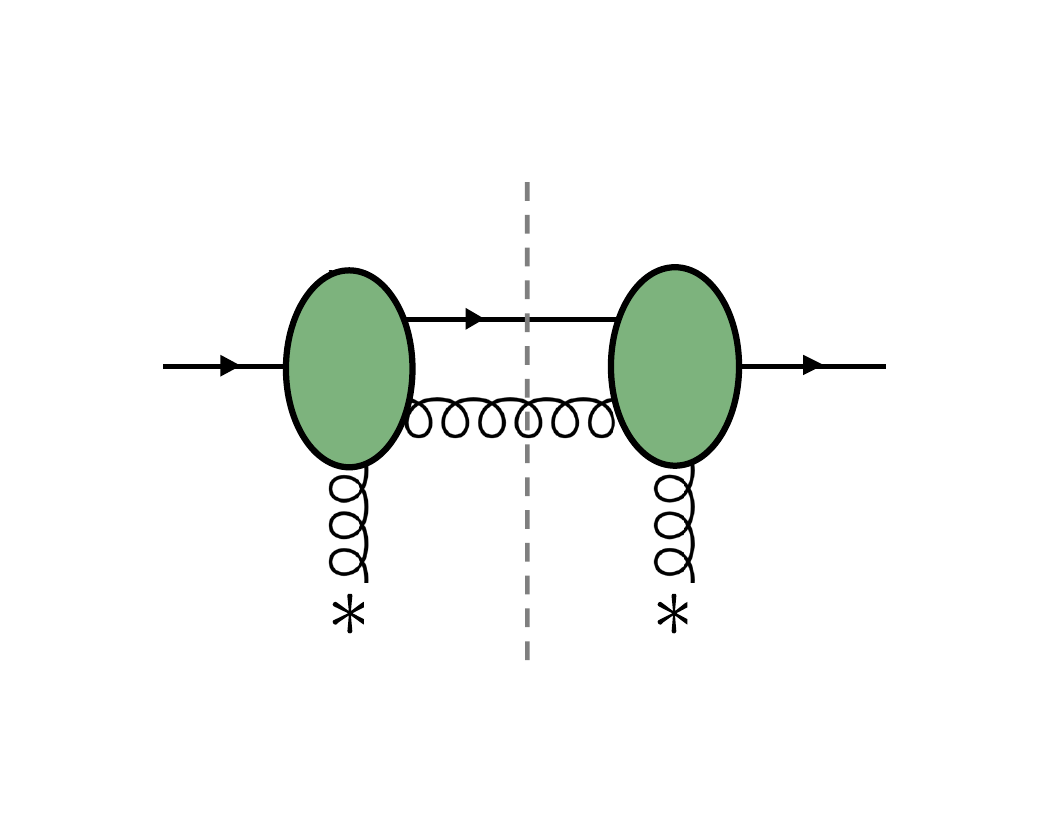}\label{subfig:P0qg}}
    \caption{Ingredients of the NLO jet impact factors: (a)-(b) virtual corrections; (c)-(d) real corrections. Punctures in blobs denote loop corrections.}
    \label{fig:IF_NLO}
\end{figure}

Further, in order to compute jet cross sections at NLL accuracy~\cite{Colferai:2010wu,Ducloue:2013hia} through the BFKL equation, jet impact factors at next-to-leading order (NLO) in $\alpha_s$~\cite{Bartels:2001ge,Bartels:2002yj} are required. They can be computed based on the one-loop single-parton PEV~\cite{Fadin:1993wh,Fadin:1992zt,Fadin:1993qb,DelDuca:1998kx,Bern:1998sc} and the tree-level PEV for the emission of two gluons or of a quark-antiquark pair at one end of the ladder~\cite{Fadin:1989kf,DelDuca:1995ki,Fadin:1996nw,DelDuca:1996nom,DelDuca:1996km,Duhr:2009uxa}. These ingredients are depicted in Figure~\ref{fig:IF_NLO}.
\begin{figure}[t]
    \centering
    \subfigure[]{\includegraphics[width=0.24\textwidth]{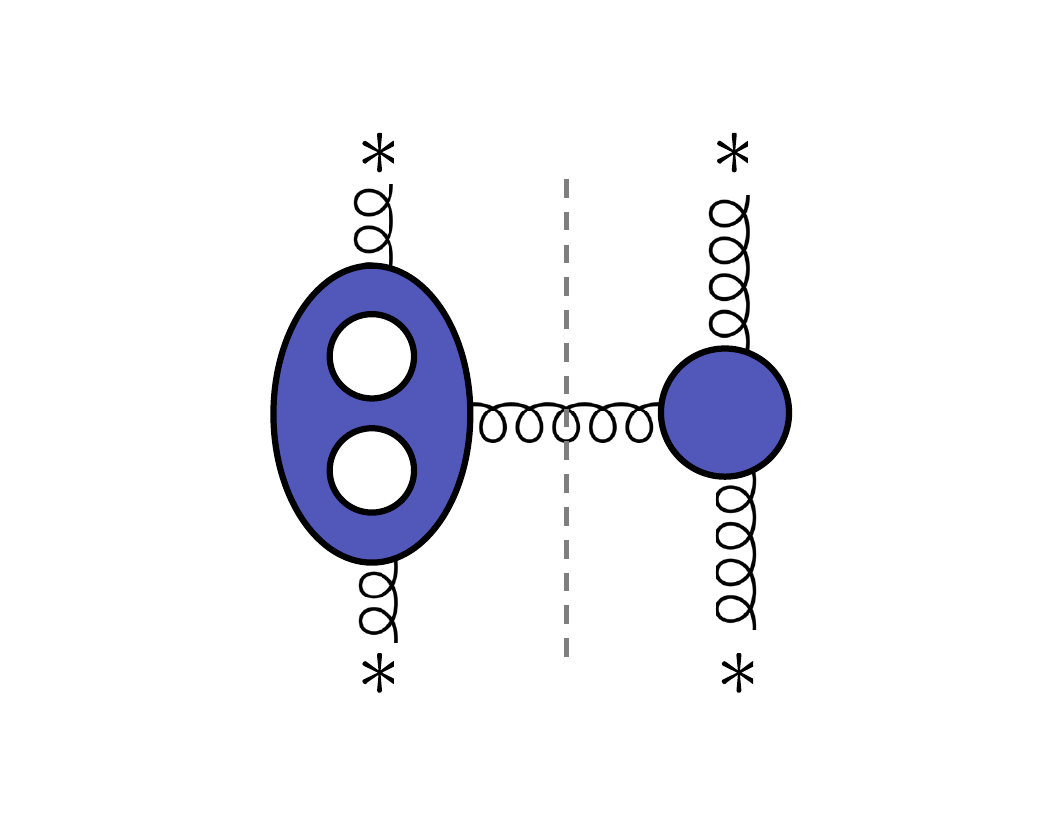}}\label{subfig:C2g}
    \subfigure[]{\includegraphics[width=0.24\textwidth]{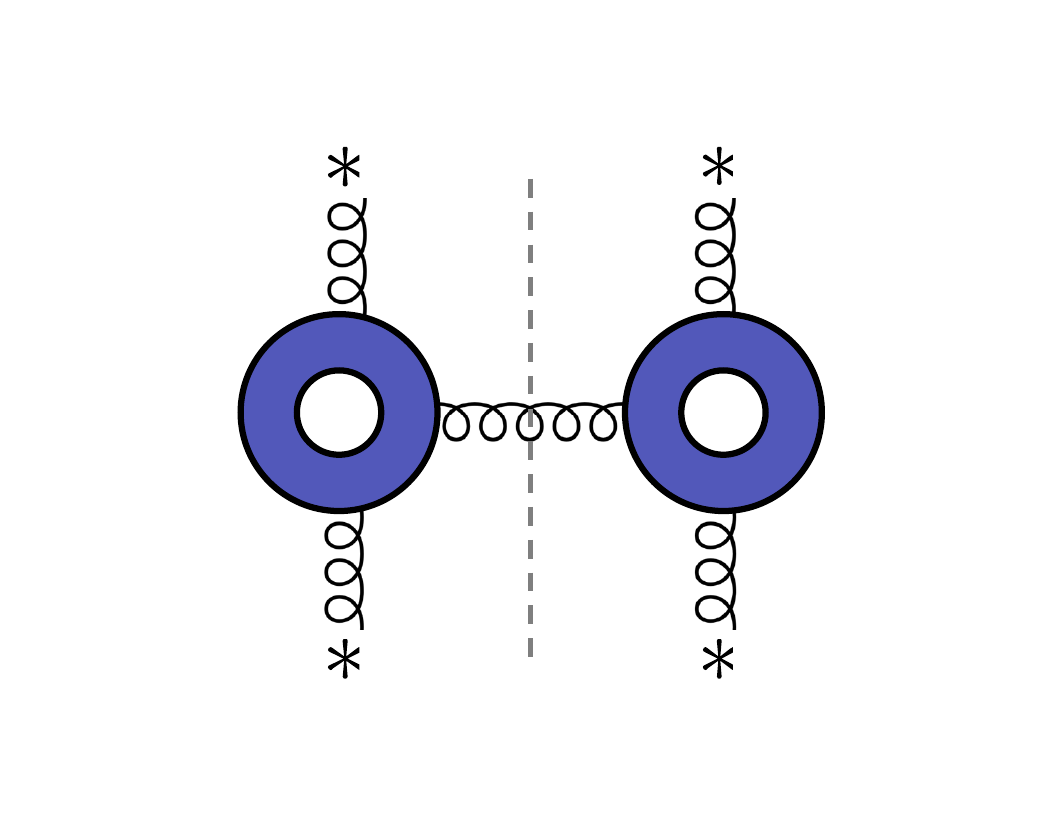}}\label{subfig:C1g_C1g}
    \subfigure[]{\includegraphics[width=0.24\textwidth]{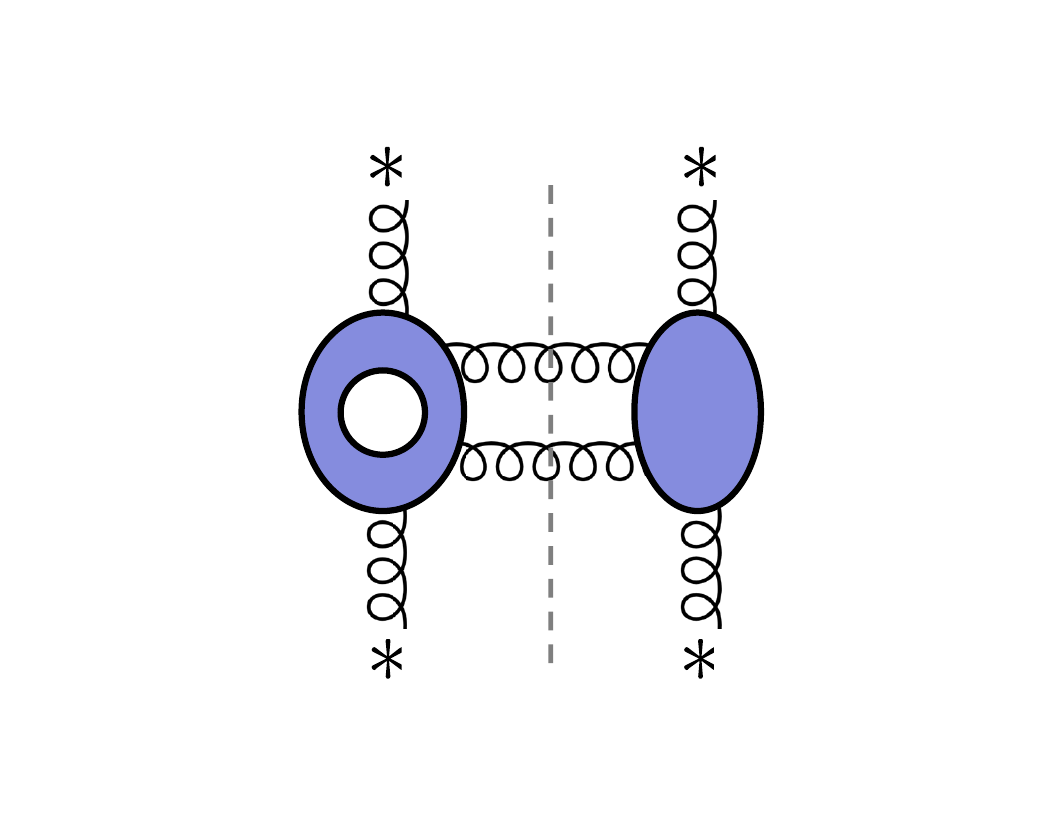}\label{subfig:C1gg}} 
    \subfigure[]{\includegraphics[width=0.24\textwidth]{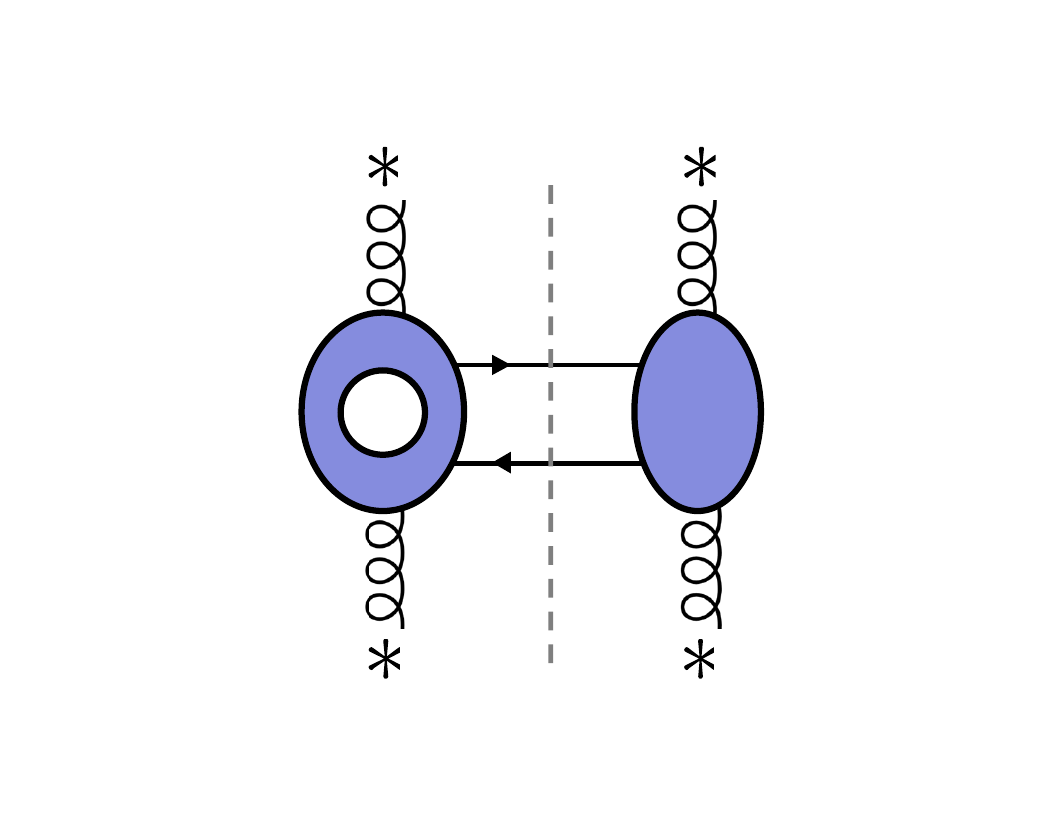}\label{subfig:C1qq}}
    \\
    \subfigure[]{\includegraphics[width=0.24\textwidth]{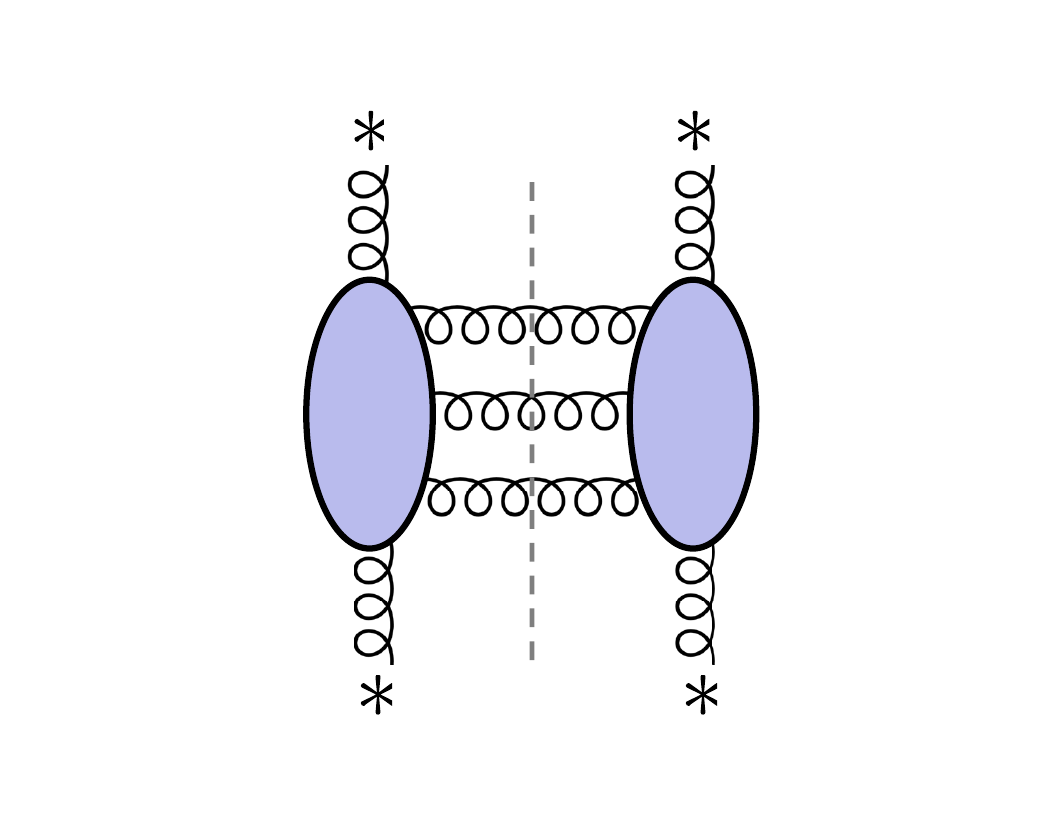}\label{subfig:C0ggg}} 
    \subfigure[]{\includegraphics[width=0.24\textwidth]{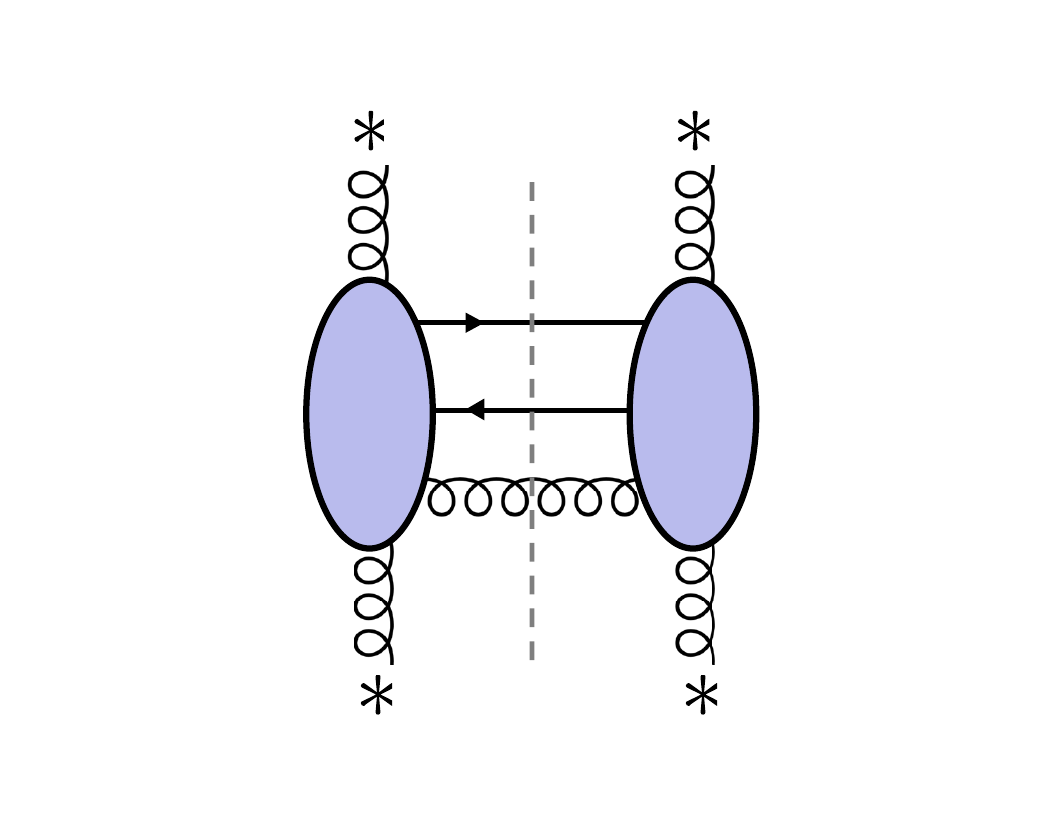}\label{subfig:C0qqg}}
\caption{Ingredients of the NNLO BFKL kernel: (a)-(d) double-virtual corrections; (e)-(g) real-virtual corrections; (h)-(j) double-real corrections.}
\label{fig:Kernel_NNLO}
\end{figure}
\begin{figure}[htb]
    \centering
    \subfigure[]{\includegraphics[width=0.24\textwidth]{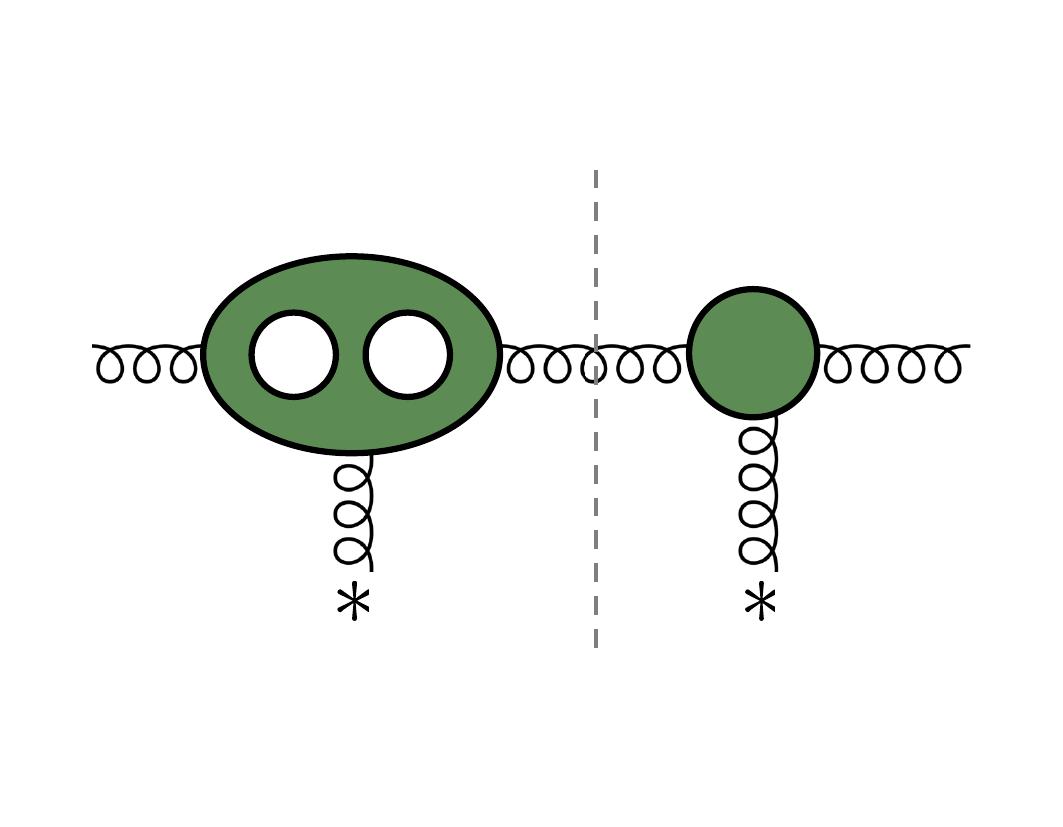}}\label{subfig:P2g}
    \subfigure[]{\includegraphics[width=0.24\textwidth]{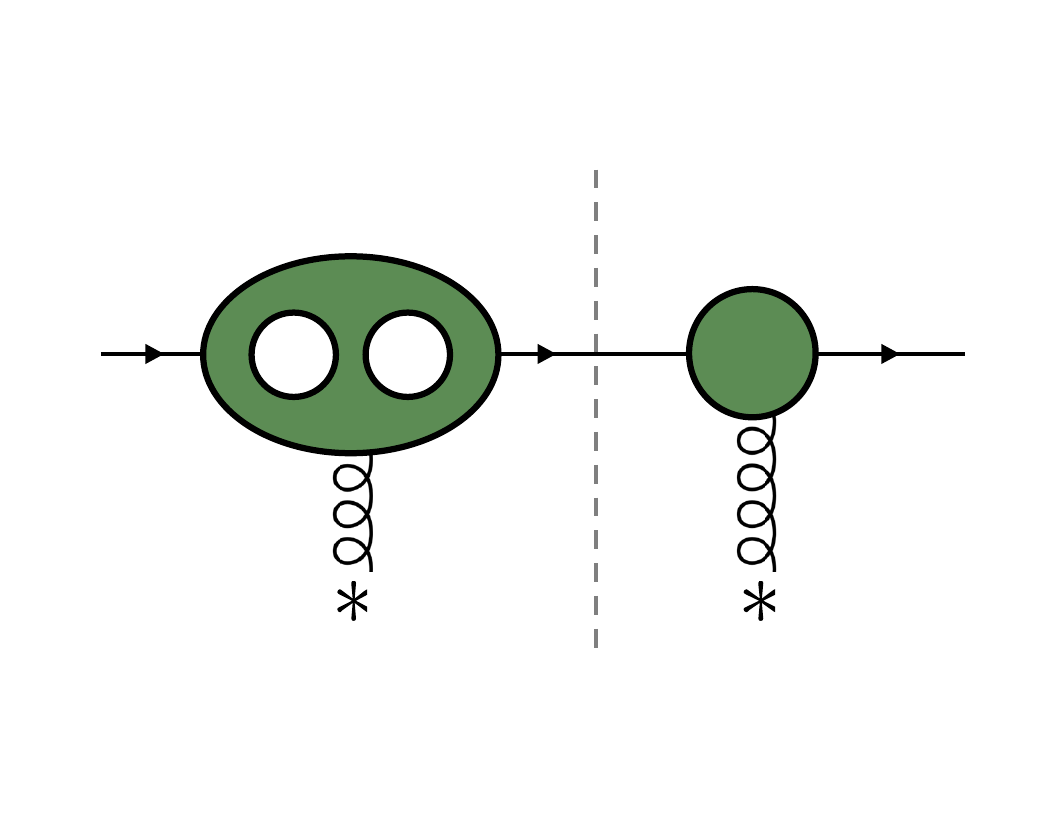}\label{subfig:P2q}} 
    \subfigure[]{\includegraphics[width=0.24\textwidth]{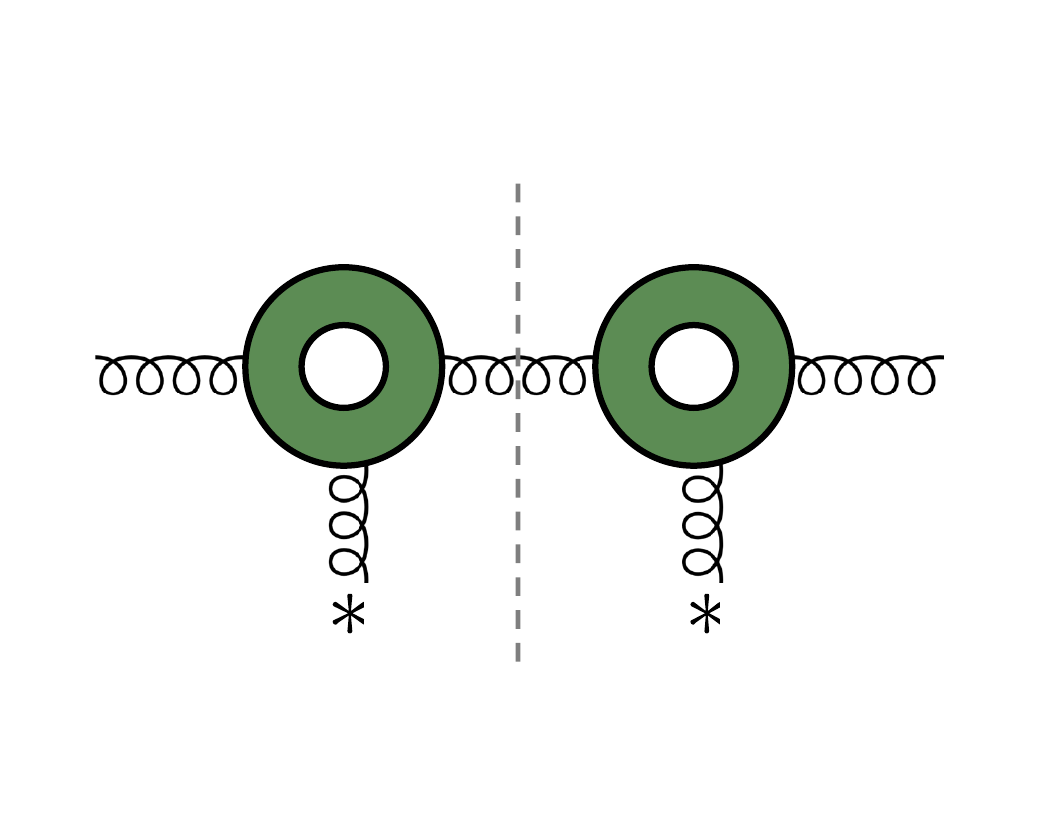}\label{subfig:P1g_P1g}}
    \subfigure[]{\includegraphics[width=0.24\textwidth]{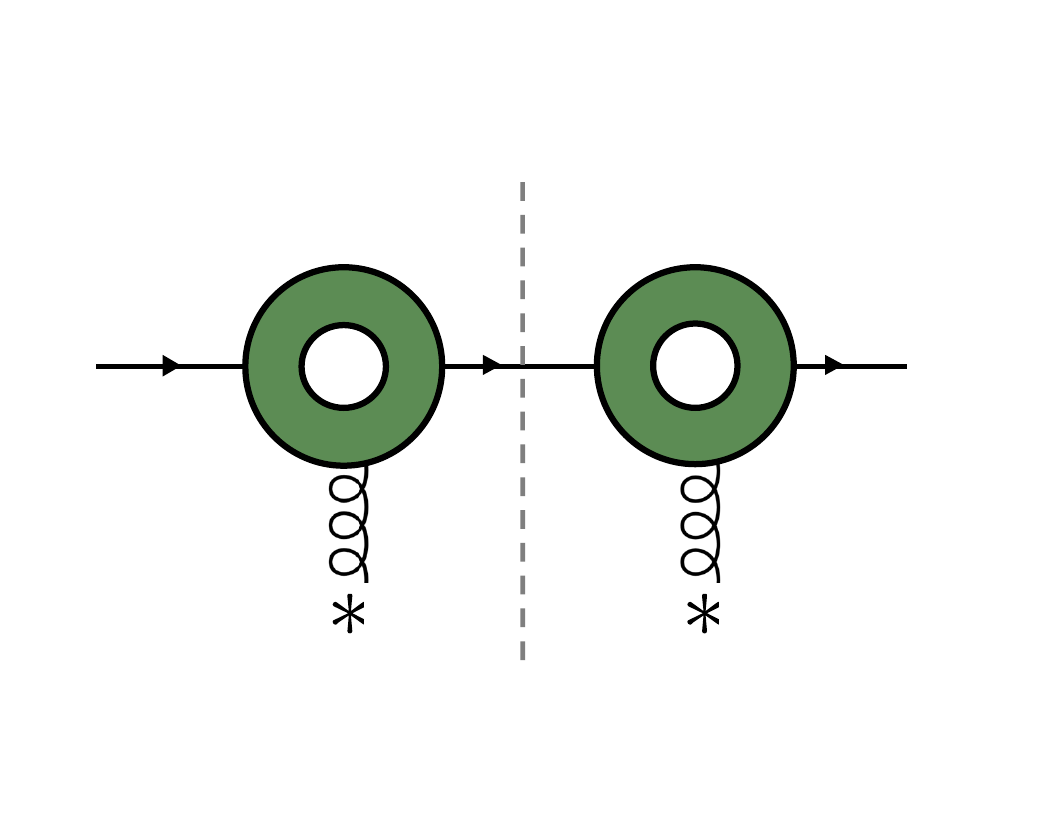}\label{subfig:P1q_P1q}}
    \\
    \subfigure[]{\includegraphics[width=0.24\textwidth]{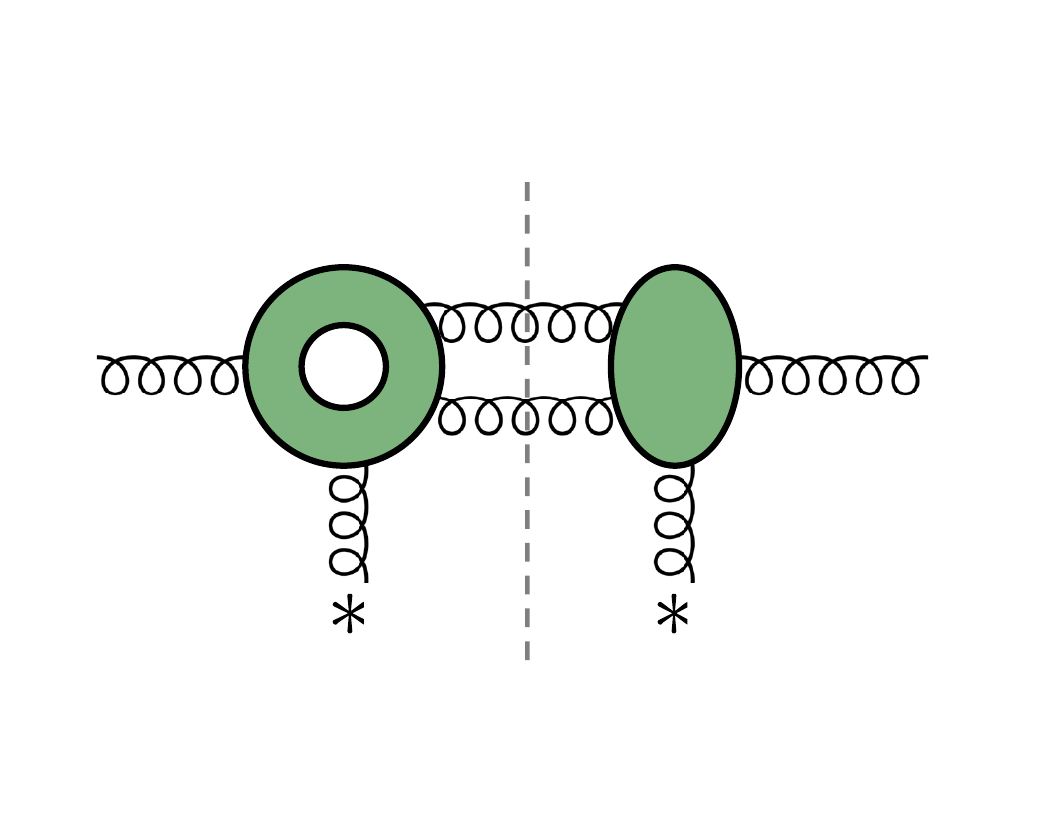}\label{subfig:P1gg}} 
    \subfigure[]{\includegraphics[width=0.24\textwidth]{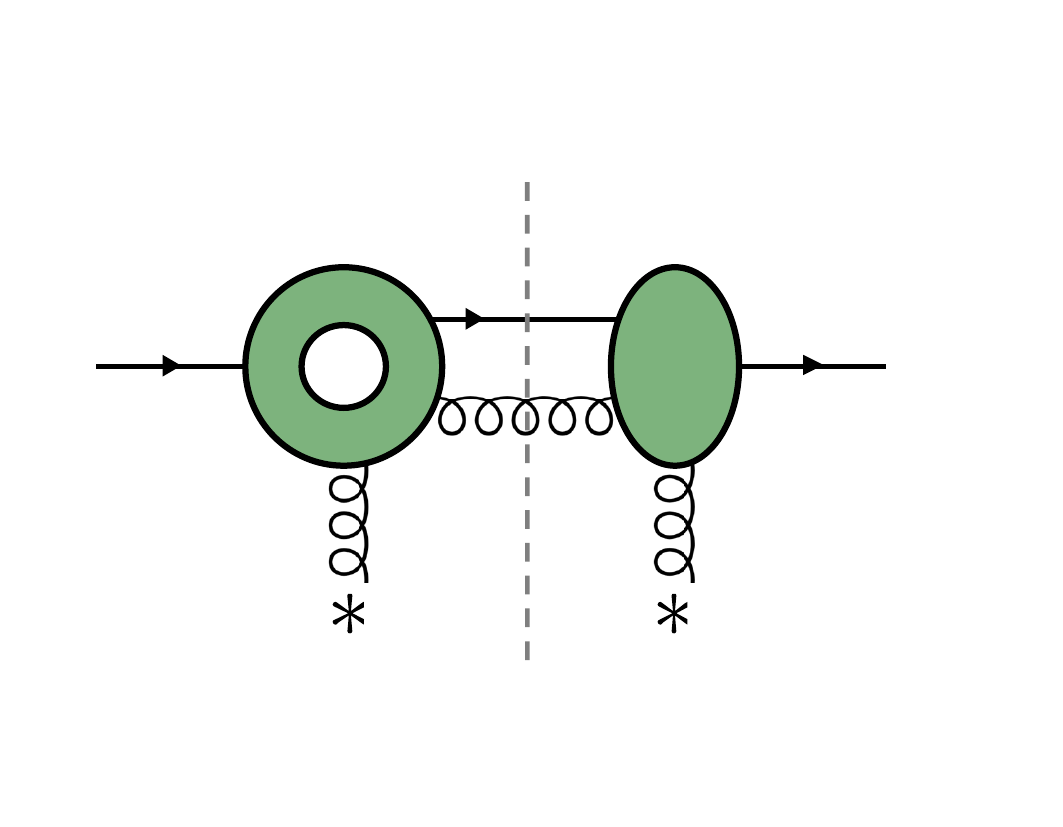}\label{subfig:P1qg}}
    \subfigure[]{\includegraphics[width=0.24\textwidth]{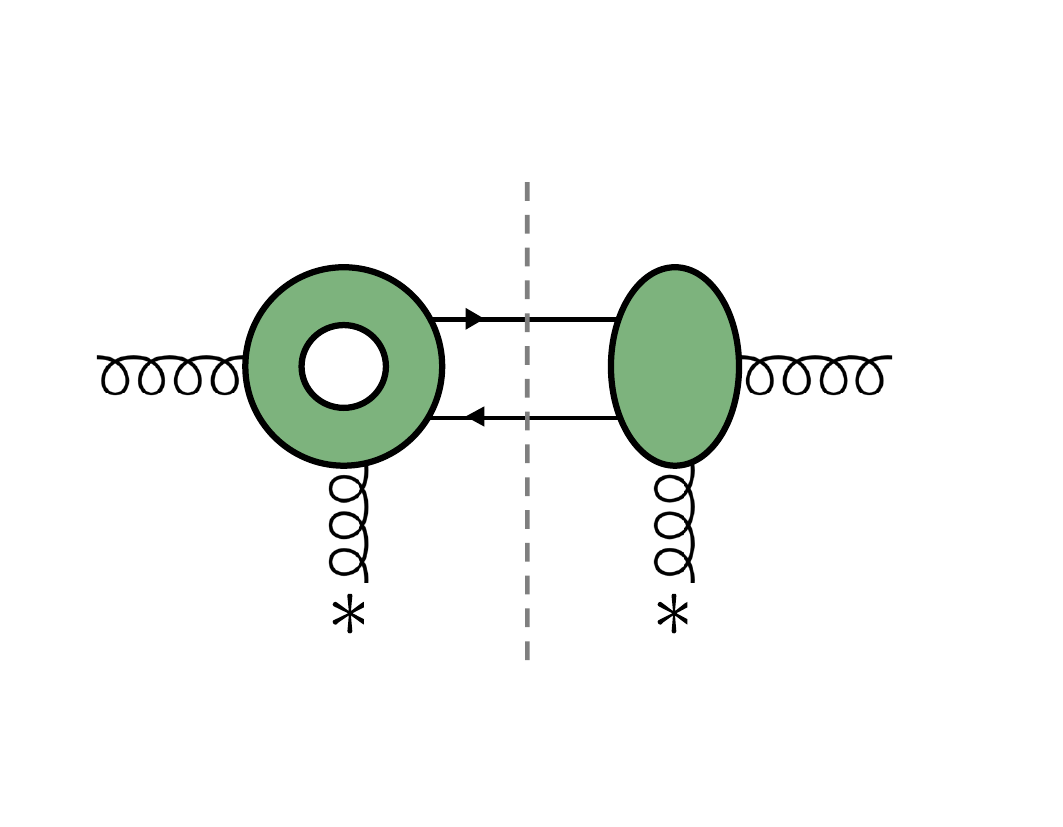}\label{subfig:P1qq}}
    \\
    \subfigure[]{\includegraphics[width=0.24\textwidth]{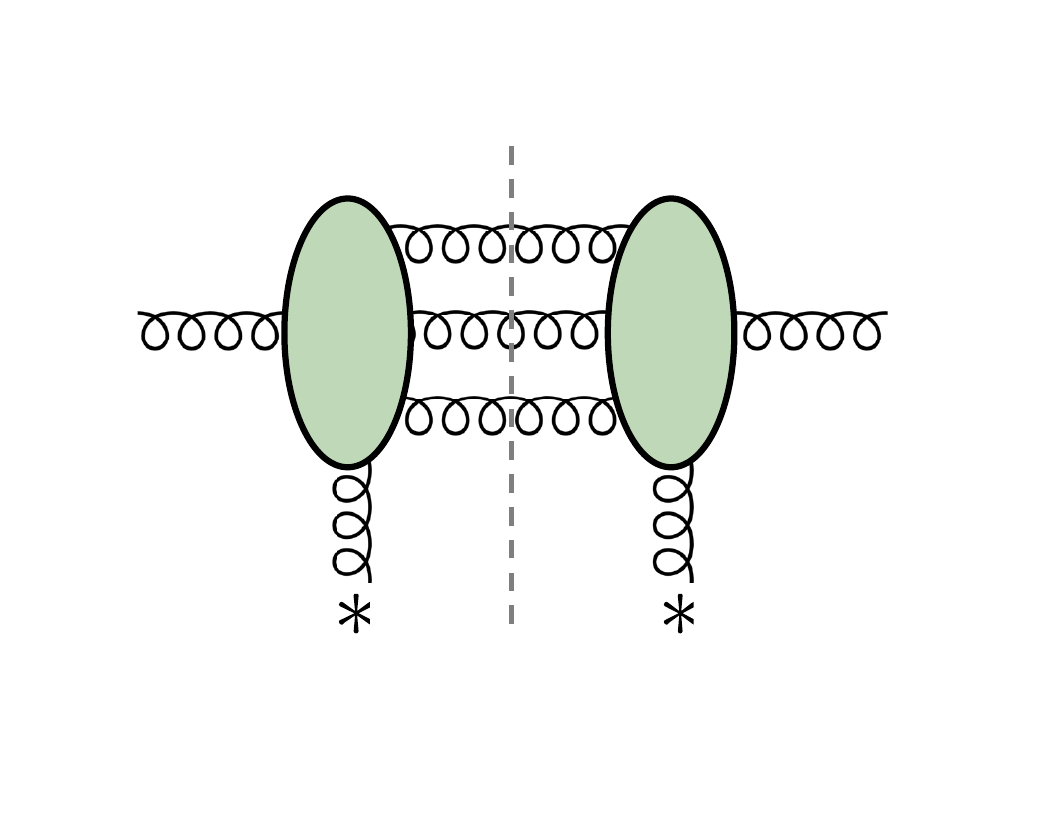}\label{subfig:P0ggg}} 
    \subfigure[]{\includegraphics[width=0.24\textwidth]{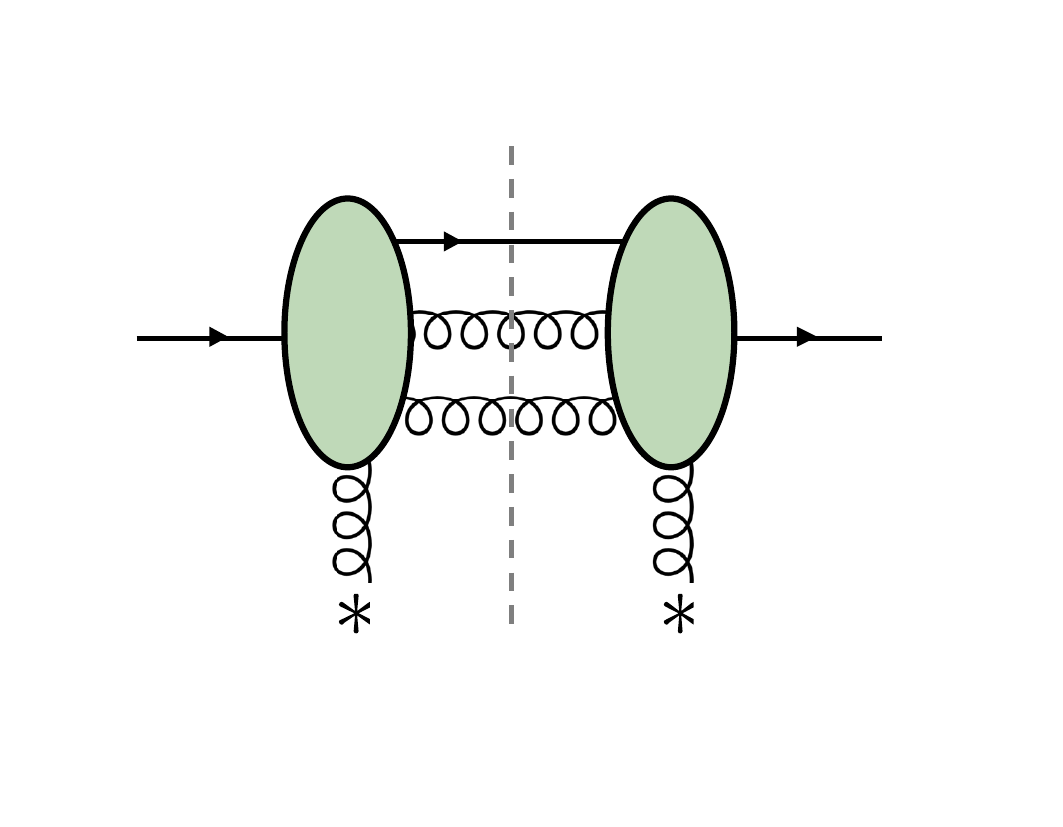}\label{subfig:P0qgg}}
    \subfigure[]{\includegraphics[width=0.24\textwidth]{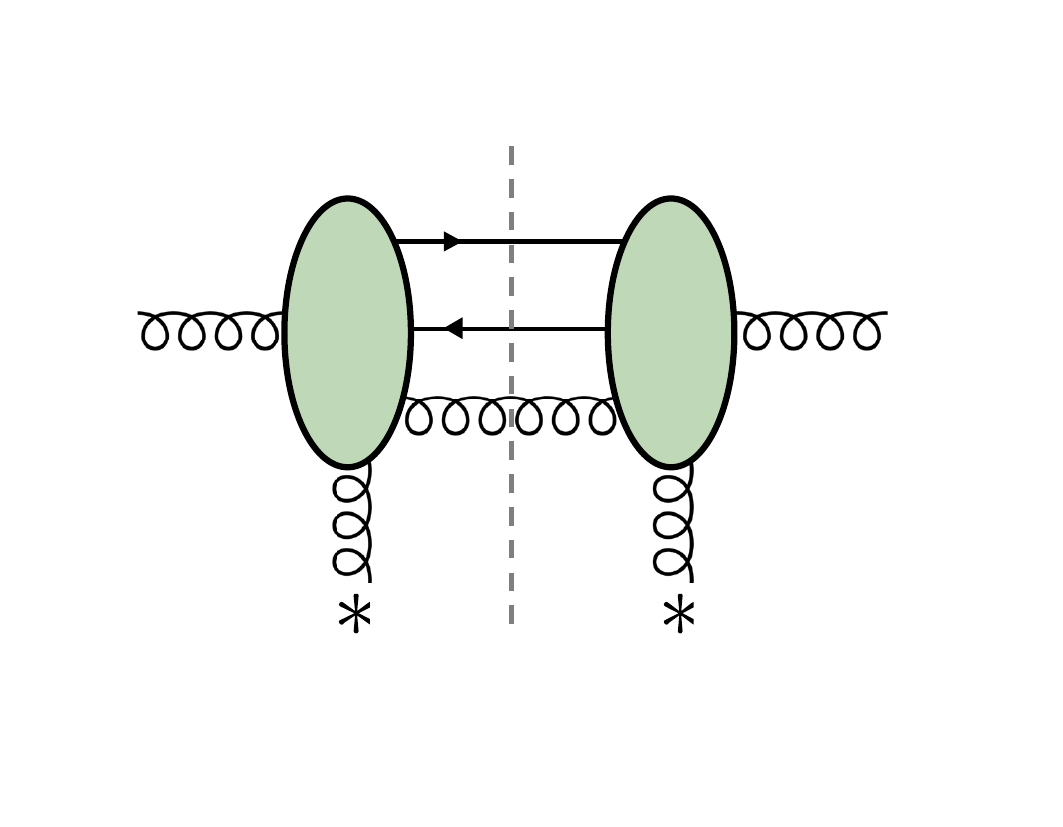}\label{subfig:P0qqg}}
    \caption{Ingredients of the NNLO jet impact factors: (a)-(d) double-virtual corrections; (e)-(g) real-virtual corrections; (h)-(j) double-real corrections. Punctures in blobs denote loop corrections.}
    \label{fig:IF_NNLO}
\end{figure}

Underpinning the BFKL equation at NLL accuracy is the fact that gluon Reggeization holds at that accuracy~\cite{Fadin:2006bj,Fadin:2015zea}.
Gluon Reggeization breaks down beyond NLL accuracy~\cite{DelDuca:2001gu,DelDuca:2011wkl,DelDuca:2011ae,DelDuca:2013ara,DelDuca:2014cya}, because at next-to-next-to-leading logarithmic (NNLL) accuracy also three-Reggeized-gluon exchanges contribute to the amplitude~\cite{Fadin:2016wso,Caron-Huot:2017fxr,Caron-Huot:2017zfo,Fadin:2017nka,Falcioni:2020lvv,Falcioni:2021buo,Falcioni:2021dgr,Buccioni:2024gzo,Abreu:2024xoh}. Provided that Regge factorisation-breaking contributions, 
such as that of three Reggeized gluons, can be evaluated and subtracted, 
there is scope to generalise the BFKL equation to NNLO accuracy and 
resum NNLL radiative corrections of  ${\cal O}( \alpha_s^2 [\alpha_s\ln(s/(-t))]^n)$.
This requires 
evaluating the next-to-next-to-leading-order (NNLO) corrections to the BFKL kernel.
These corrections, illustrated in Figure~\ref{fig:Kernel_NNLO}, are generated by: the two-loop corrections to the single-gluon CEV~\cite{Buccioni:2024gzo,Abreu:2024xoh}; the one-loop corrections to the CEV for the emission of two gluons~\cite{Byrne:2022wzk}, or a $q\bar q$ pair; the tree-level CEV for the emission of three gluons~\cite{DelDuca:1999iql,Antonov:2004hh,Duhr:2009uxa}, or of a $q\bar q$ pair and a gluon. These three-parton tree-level CEVs will all be determined here, the latter being a new result.

Finally, in order to compute jet cross sections at NNLL accuracy through the BFKL 
equation, jet impact factors at next-to-next-to-leading order (NNLO) in $\alpha_s$ will be needed. They will be based on the two-loop gluon or quark PEV~\cite{DelDuca:2014cya,Caron-Huot:2017fxr,DelDuca:2021vjq,Falcioni:2021dgr,Caola:2021izf}; the one-loop PEV for the emission of two gluons~\cite{Canay:2021yvm,Byrne:2023nqx} or of a quark-antiquark pair~\cite{Byrne:2023nqx}; and the tree-level PEV for the emission of three partons~\cite{DelDuca:1999iql}. Further, one must include the square of the one-loop helicity-violating PEV~\cite{DelDuca:1998kx}, as well as the contributions of Regge factorization violating effects. These contributions to the NNLO jet impact factors are depicted in Figure~\ref{fig:IF_NNLO}.

\begin{figure}[ht]
    \centering
    \subfigure[]{\includegraphics[width=0.24\textwidth]{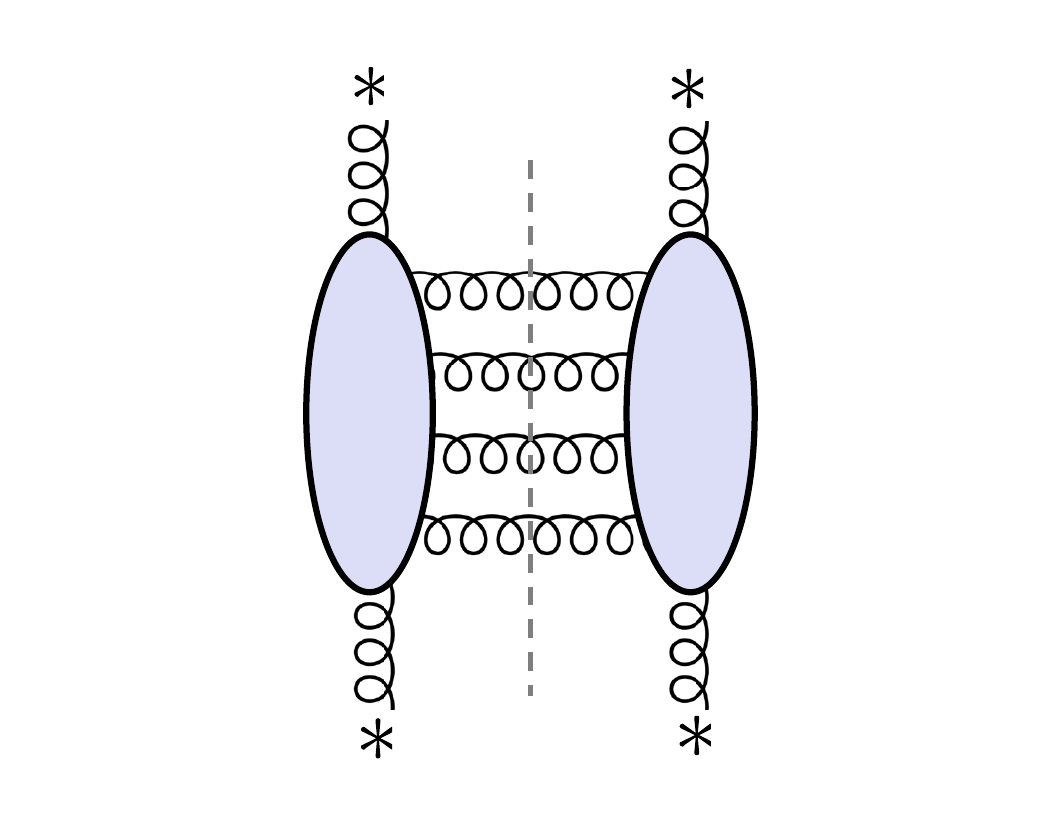}}\label{subfig:C0gggg}
    \subfigure[]{\includegraphics[width=0.24\textwidth]{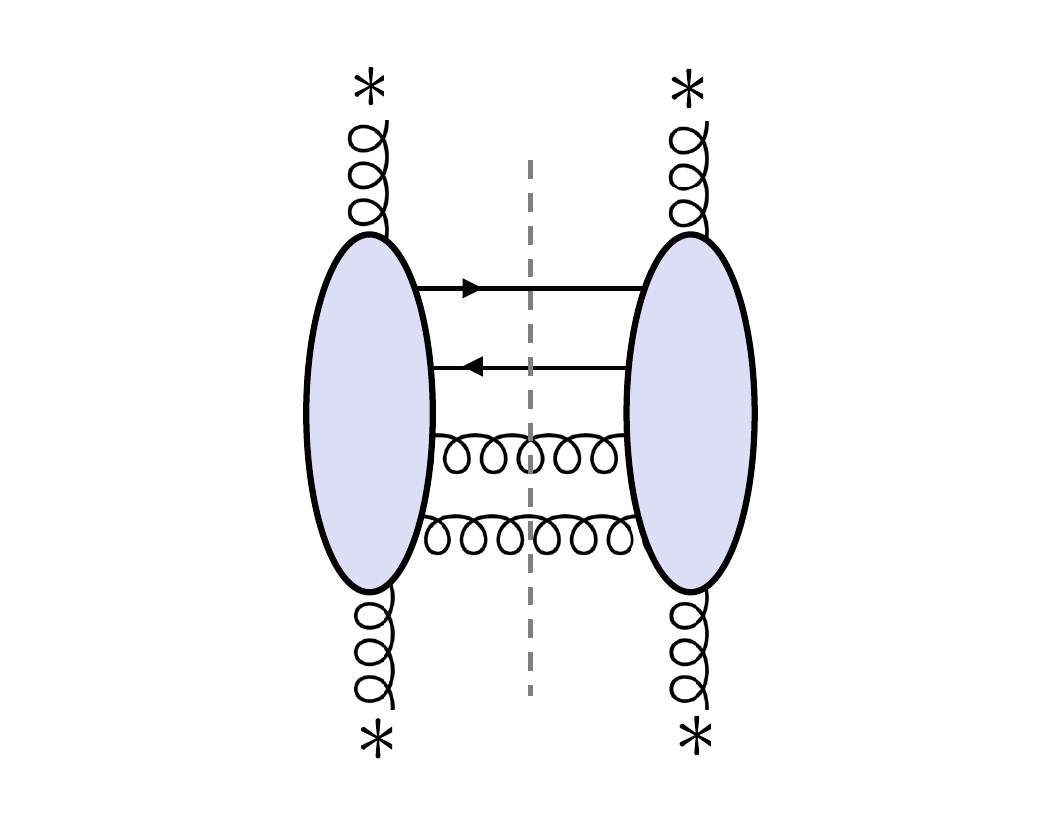}}\label{subfig:C0qqgg}
    \subfigure[]{\includegraphics[width=0.24\textwidth]{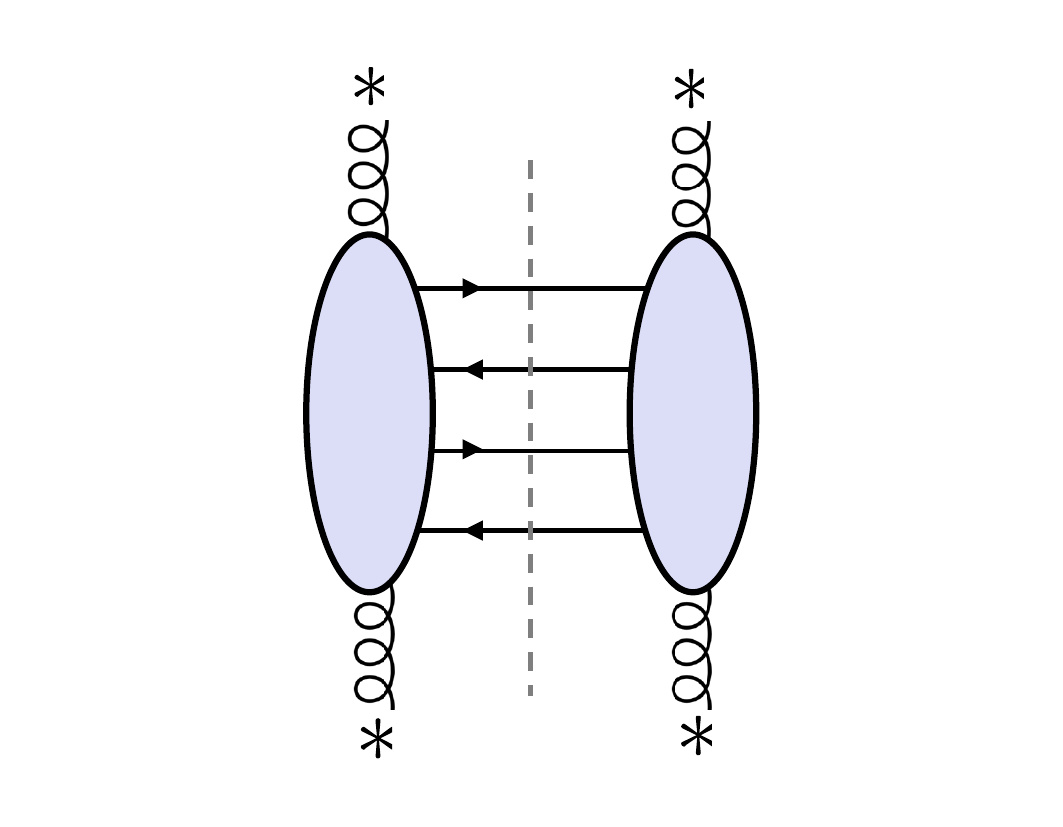}\label{subfig:C0qqqq}} 
    \subfigure[]{\includegraphics[width=0.24\textwidth]{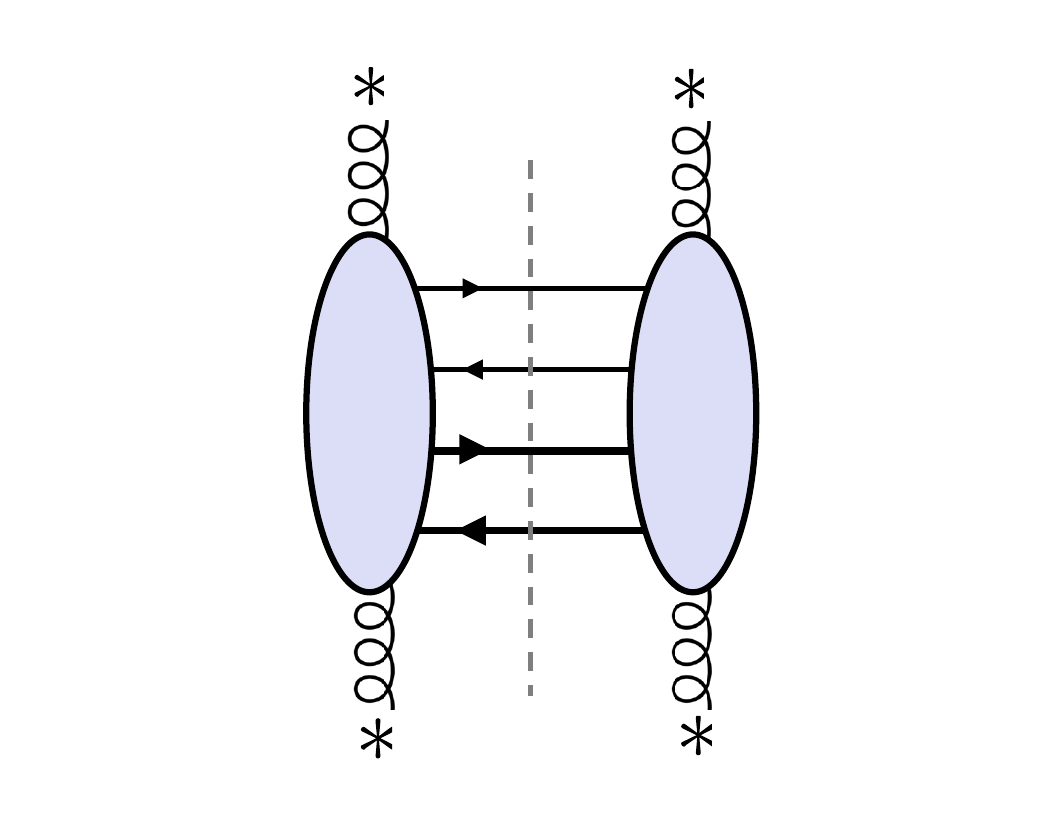}\label{subfig:C0qqxx}}
    \\
    \subfigure[]{\includegraphics[width=0.24\textwidth]{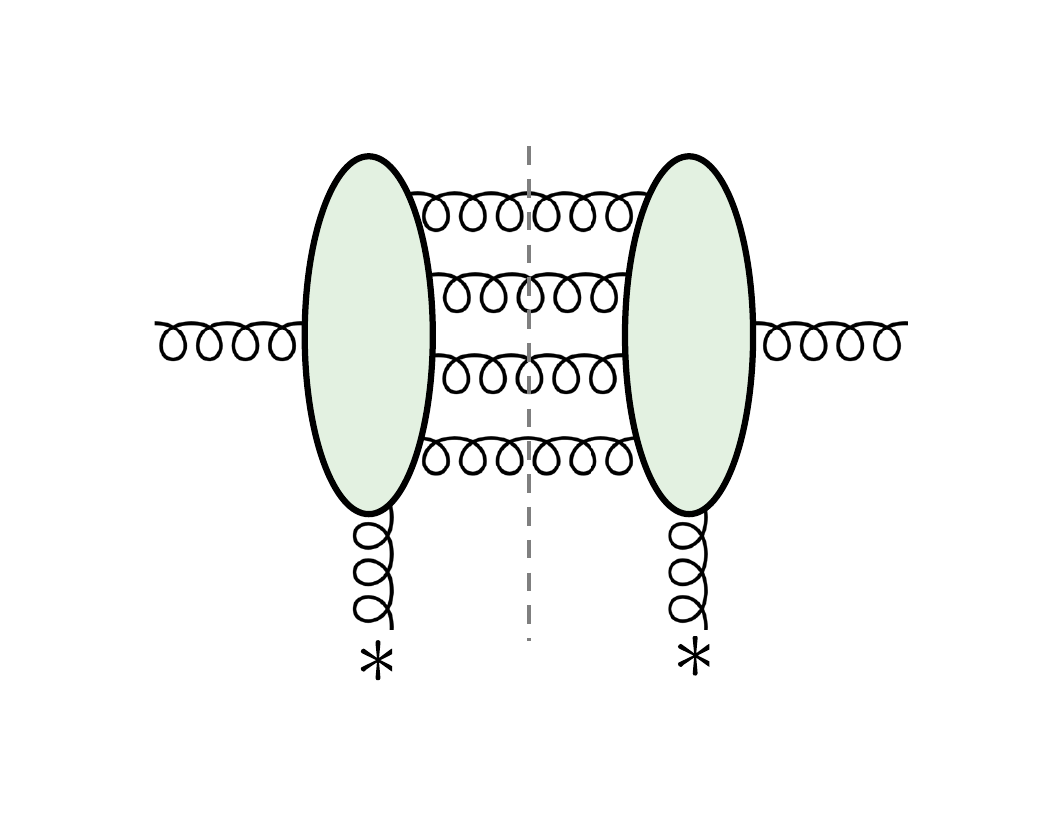}}\label{subfig:P0gggg}
    \subfigure[]{\includegraphics[width=0.24\textwidth]{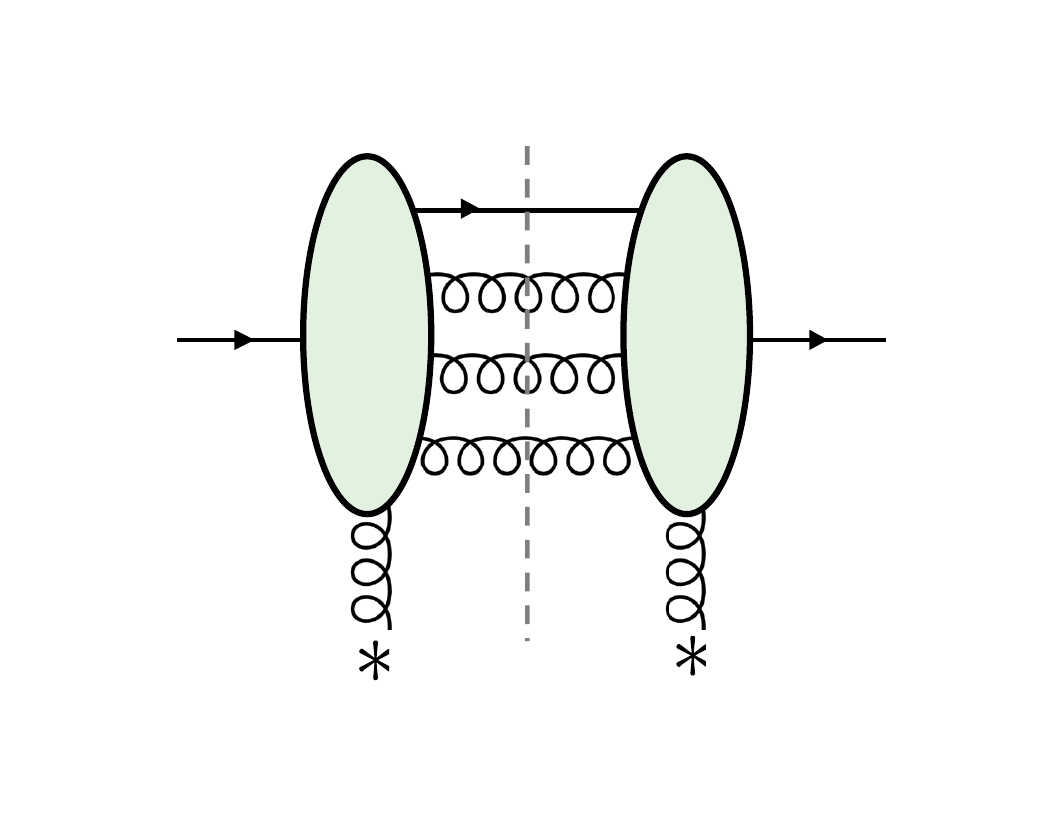}}\label{subfig:P0qggg}
    \subfigure[]{\includegraphics[width=0.24\textwidth]{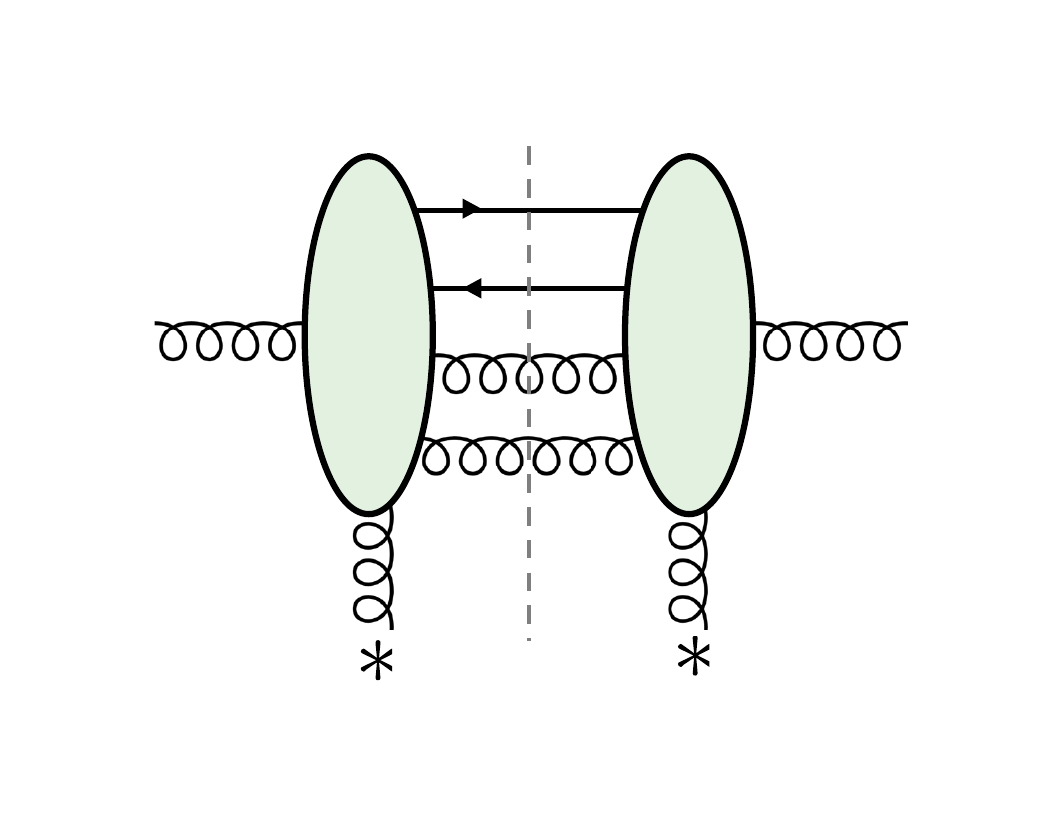}\label{subfig:P0qqgg}} 
    \subfigure[]{\includegraphics[width=0.24\textwidth]{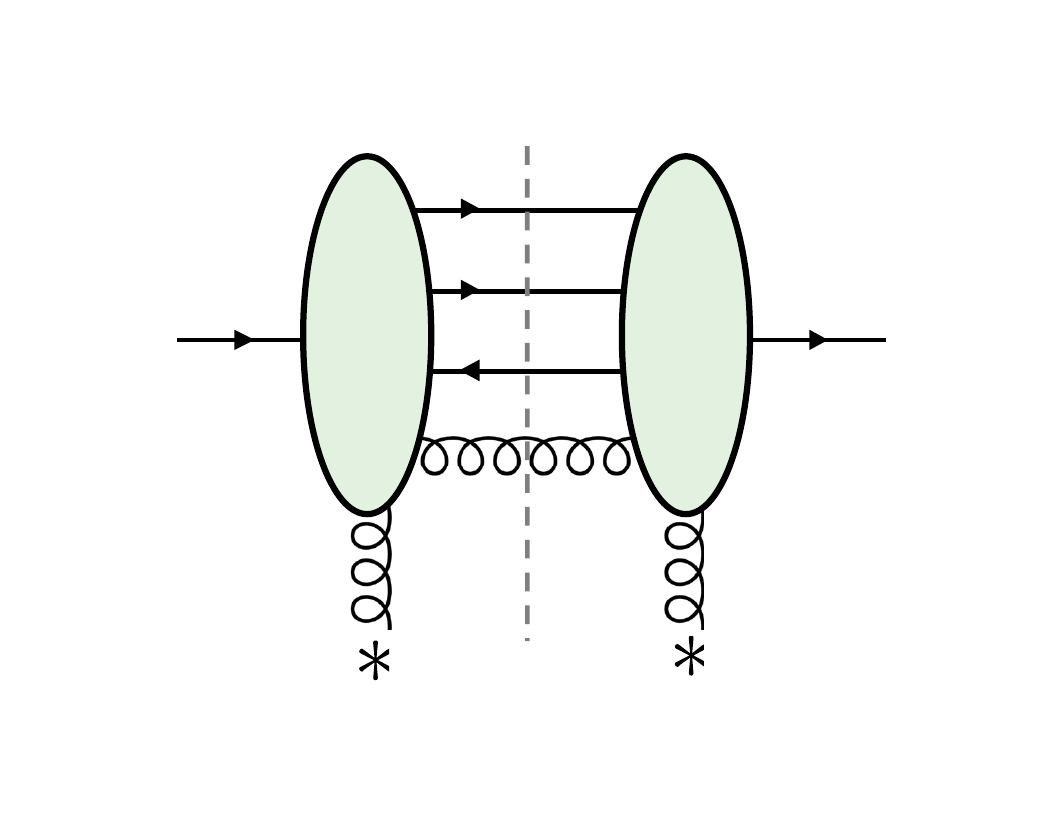}\label{subfig:P0qqqg}}
    \\
    \subfigure[]{\includegraphics[width=0.24\textwidth]{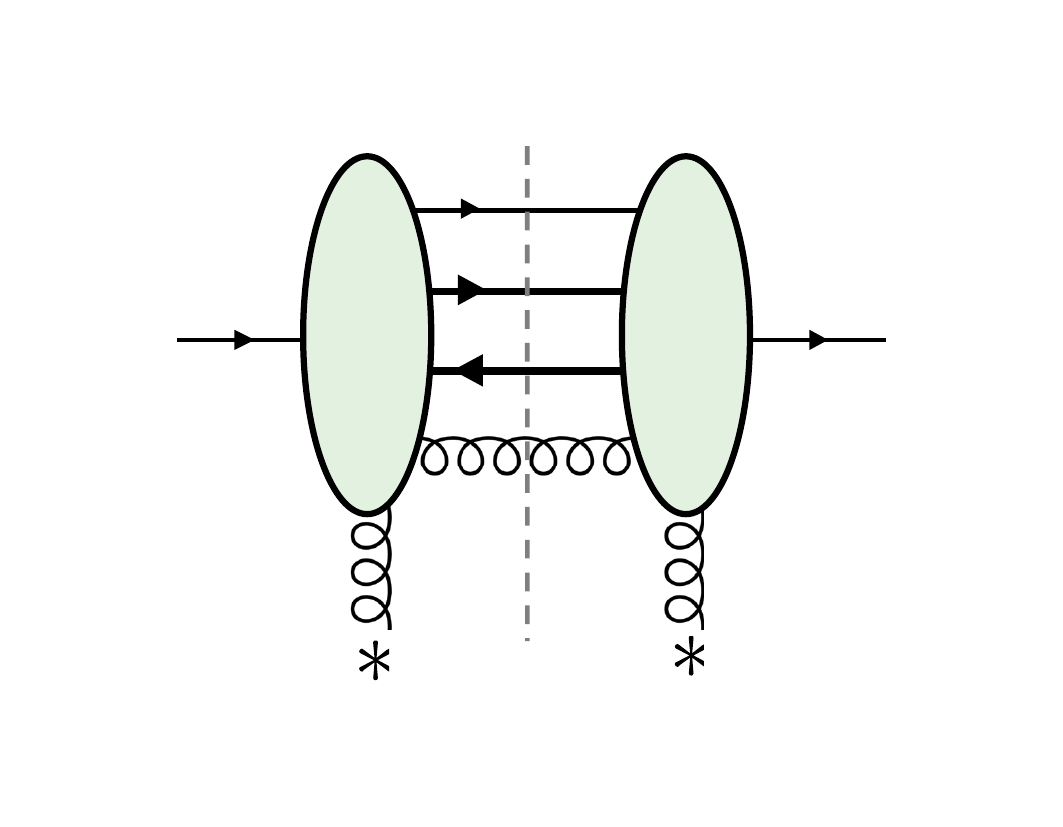}}\label{subfig:P0qxxg}
    \subfigure[]{\includegraphics[width=0.24\textwidth]{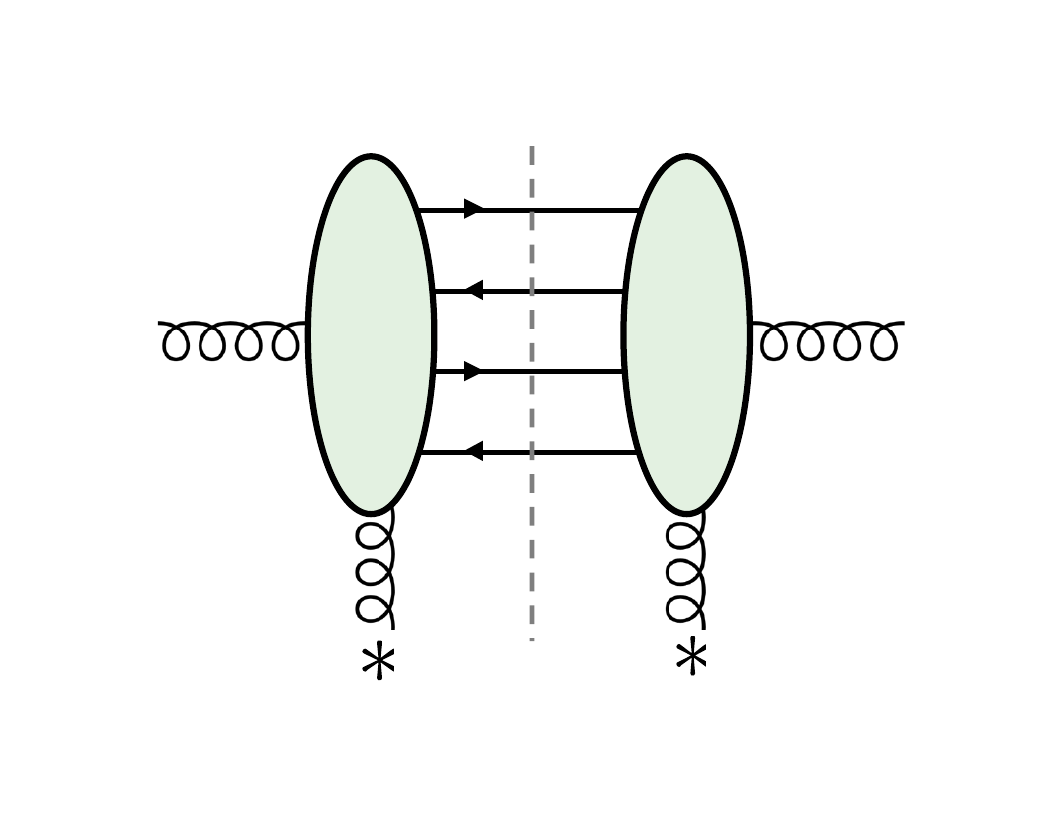}}\label{subfig:P0qqqq}
    \subfigure[]{\includegraphics[width=0.24\textwidth]{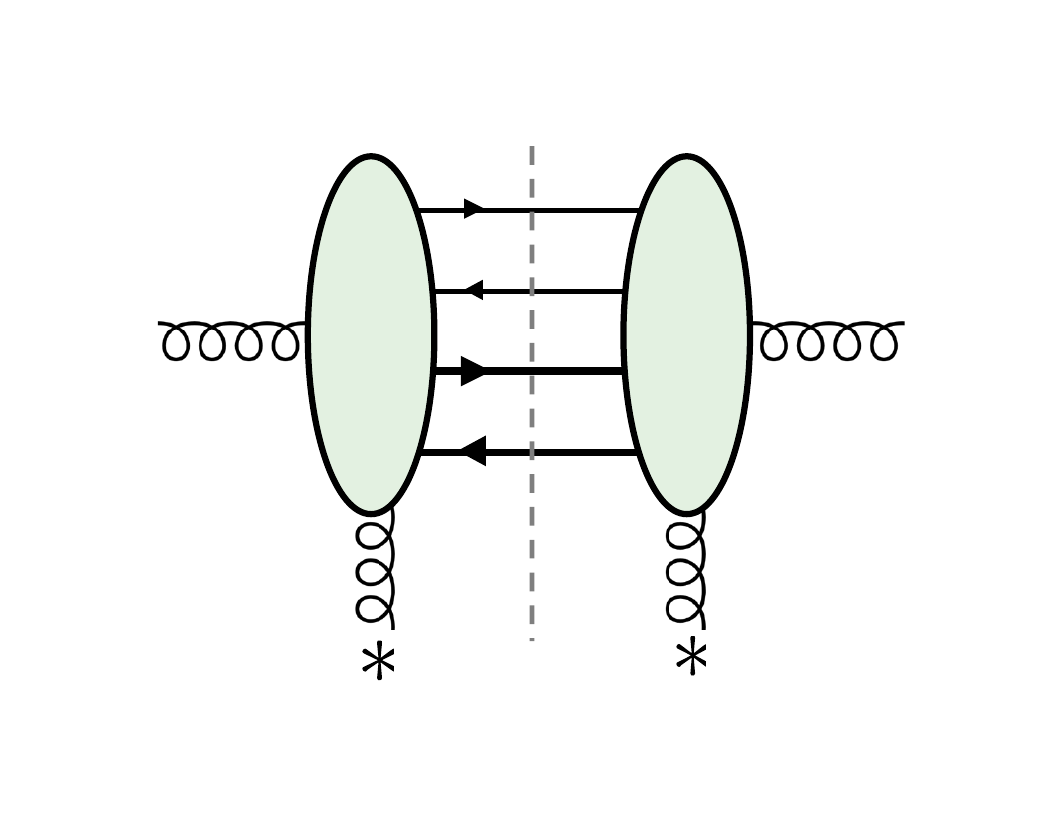}\label{subfig:P0qqxx}}
    \caption{(a)-(d) Tree-level contributions to the N$^3$LO BFKL kernel; (e)-(k) tree-level contributions to the N$^3$LO jet impact factors. The bold fermion lines denote a distinct quark flavour.}
    \label{fig:NNNLO_real}
\end{figure}

At next-to-next-to-next-to-leading logarithmic (${\rm N^3}$LL) accuracy, only the PEV for the emission of four gluons~\cite{DelDuca:1999iql} and the CEV for the emission of four gluons~\cite{Duhr:2009uxa} were previously known.

In this paper we systematically extract all tree-level MREVs, both PEVs and CEVs, up to final-state multiplicity of four partons. As noted above, several of these have been determined in the past, and we were able to compare some of our results with the literature. Many of the high-multiplicity results are new.
In particular, at multiplicity three we compute for the first time the CEV for the emission of a $q\bar q$ pair and a gluon (depicted in Figure~\ref{subfig:C0qqg}), which is required for the NNLO corrections to the BFKL kernel, and thus to solve the BFKL equation at NNLL accuracy. 
In addition, we compute all the tree level vertices which are required at ${\rm N^3}$LL accuracy, and that were previouly unknown, namely the PEV for the emission of four partons and the CEV for the emission of a $q\bar q$ pair and two gluons, and of two $q\bar q$ pairs. The corresponding tree-level contributions to the next-to-next-to-next-to-leading order (N$^3$LO) BFKL kernel and N$^3$LO jet impact factors are depicted in Figure~\ref{fig:NNNLO_real}.

\section{Minimal set of lightcone variables}
\label{sec:minvar}

Let us begin by discussing the kinematic variables needed to express scattering amplitudes with $n$ massless legs. 
It is instructive to first consider momentum degrees of freedom, and only then see how helicity can be represented.
For our purposes it is useful to consider alternative expressions for amplitudes, either 
in terms of Lorentz invariants or in terms of (ratios of) lightcone and transverse momentum components in a special frame.
Recall first that an (on-shell) amplitude with $n$ massless legs in $d=4$ space-time dimensions depends on 
\begin{equation}
\label{ninv}
   n^{(d=4)}_{\rm inv.} = 3n-10
\end{equation}
independent kinematic variables. To see this one may first note that each on-shell particle involves $d-1=3$ independent degrees of freedom, accounting for the on-shell condition.  Translation invariance implies overall momentum conservation, hence fixing four relations among these $3n$ momentum components. Similarly, Lorentz invariance fixes six more relations, ending up with the result of Eq.~(\ref{ninv}). These parameters may be chosen to be any set of $3n-10$ independent Lorentz invariants.

Notwithstanding the availability of a Lorentz-invariant description, it is often convenient to describe a scattering process in specific (classes of) Lorentz frames and refer separately to lightcone and transverse momentum components. When using such kinematic variables, it may be natural to include additional variables, going beyond the number of Lorentz invariants of Eq.~(\ref{ninv}).  

Let us illustrate this in the simplest case of $p_1 p_2 \to p_3 p_4$ scattering. Here $n=4$, so there are $n^{(d=4)}_{\rm inv.}=3n-10 = 2$ Lorentz invariant parameters, which may be chosen as $s = (p_1+p_2)^2$ and $t = (p_2-p_3)^2$. Starting with an arbitrary frame, we can apply boosts and rotations so as to align $p_1$ and $p_2$ along the so-called beam axis, which we denote by $z$. This amounts to fixing four of the six parameters characterizing the Lorentz frame (effectively setting the $x$ and $y$ components of both $p_1$ and $p_2$ to zero). The remaining parameters can be identified as the azimuthal angle $\theta$ which the two outgoing particles $p_3$ and $p_4$ make with respect to an arbitrary direction in the plane perpendicular to the beam axis, and the centre of rapidity, $\bar{y}$, which boosts the centre-of-mass frame along the beam axis. 

Using light-cone coordinates, defined in Appendix~\ref{frame} (such that the initial-state particles $p_1$ and $p_2$ define, respectively, the minus and plus lightcone directions) and considering transverse momentum conservation, $p_{3\perp} = - p_{4\perp} \equiv p_{\perp}$,  the outgoing momenta in this class of Lorentz frames can be expressed in terms of four parameters, 
\begin{equation}
\label{2to2vars}
    \Big\{|p_{\perp}|,\,\, \theta, \,\, X \equiv \frac{p_3^+}{p_4^+},\,\, 
    \frac{p_2^+}{p_1^-}\Big\}\,,
\end{equation} 
but only two of them, $|p_{\perp}|$ and $X$, parametrise the Lorentz invariants in this scattering process. This can be seen using the relation between these two parameters and the Mandelstam invariants, namely
\begin{align}
\label{eq:st1}
\begin{split}
s &\equiv (p_1+p_2)^2 =|p_{\perp}|^2 \left( \sqrt{X } + 1/\sqrt{X } \right)^2 
\,=\, 4 |p_{\perp}|^2 \cosh^2 y^\ast \,, \\
t &\equiv (p_1+p_4)^2 = - |p_{\perp}|^2 \left( 1 + 1/X \right)\,=\,-2 |p_{\perp}|^2 \cosh y^\ast e^{-y^\ast}\,, 
\\
u &\equiv (p_1+p_3)^2
= - |p_{\perp}|^2 \left( 1 + X\right) \,=\,
-2 |p_{\perp}|^2 \cosh y^\ast e^{y^\ast}\,,
\end{split}
\end{align}
where $u = - s -t$ because of momentum conservation. Thus, $X$ and $|p_{\perp}|^2$ acquire a Lorentz invariant meaning,
\begin{align}
X=\frac{u}{t},\qquad 
|p_{\perp}|^2=\frac{ut}{s}=-t\frac{X}{1+X}\,.
\end{align}
Equivalently, in the second expression given in Eq.~(\ref{eq:st1}) for each Mandelstam invariant, we traded $X$ for the rapidity span variable 
\begin{equation}
    y^\ast \equiv  \frac{y_3-y_4}{2}=
    \frac14 \log\left(\frac{p_3^+ p_4^-}{p_3^- p_4^+}\right)
    =\frac12 \log\frac{p_3^+}{p_4^+} =\frac12\log X \,.
\end{equation}
So also $y^\ast$ acquires a Lorentz invariant meaning. This stands in contrast to the remaining two variables in Eq.~(\ref{2to2vars}) namely $\theta$, the azimuthal angle of the final state particles and~$p_2^+/p_1^-$, which both depend on the particular Lorentz-frame used. Note that the latter momentum ratio is related to 
the centre-of-rapidity variable,
\begin{equation}
    \bar{y} \equiv  \frac{y_3+y_4}{2}
    =
    \frac14 \log\left(\frac{p_3^+ p_4^+}{p_3^- p_4^-}\right)
    =
    \frac12 \log \frac{p_2^+}{p_1^-}\,,
\end{equation}
which parametrises how the centre-of-mass frame is boosted along the beam axis, e.g. in a hadron collision, how the parton centre-of-mass frame is boosted with respect to the hadron centre-of-mass frame.

Eq.~(\ref{2to2vars}) generalises straightforwardly to the $n$-point case. Rotating and boosting a $2\to n-2$ scattering process to a Lorentz frame where the two incoming particles $p_1$ and $p_2$ are aligned along the beam axis, we are left with $3n-8$ parameters, consisting of the $n^{(d=4)}_{\rm inv.}=3n-10$ Lorentz invariants, a reference azimuthal angle $\theta$ and the centre of rapidity~$\bar{y}$. 
For example, for five-point scattering, the $3n-10=5$ invariants may be chosen as the sequential ones, $s_{12}$, $s_{23}$, $s_{34}$, $s_{45}$ and $s_{51}$. However, for $n>5$ one must distinguish between the number of independent Lorentz invariants in a general (sufficiently high) number of spacetime dimensions $d$, and the (smaller) number of independent ones in $d=4$ dimensions.  To compute the former, consider 
an $n$-point massless on-shell scattering amplitude in $d$ dimensions, where we may  count the Lorentz invariants by  considering only 
two-particle invariants, $s_{ij}=2p_i\cdot p_j$ with $i<j$. Fixing then one momentum, say $p_n$, through total momentum conservation, and noting that the mass-shell condition on $p_n$ yields a quadratic relation among the remaining momenta, $p_n^2= (p_1+...+p_{n-1})^2 = 0$, the number of independent Lorentz invariants is
\begin{equation}
\label{ninvd}
n^{(d)}_{\mathrm {inv}.} = \frac{(n-1)(n-2)}{2}-1 =\frac{n(n-3)}{2}\,,
\end{equation}
where we assume that $d$ is sufficiently large that $p_i^{\mu}$ are all independent.
Conversely, working in fixed spacetime dimension and increasing the number of external particles $n$, some momenta will eventually be expressible in terms of others. This is reflected in some \emph{vanishing} Gram determinants 
\[
\text{Gram} \left(
\left\{p_i\right\}_{i\in S}\right) = {\text{det}}_{j,k\in S}(p_j\cdot p_k)
\]
where $S$ is a non-empty subset of up to $n-1$ external momenta.
For instance, for six-point scattering, the
$n^{(d)}_{\mathrm {inv}.} =\frac {6\cdot 3}{2}=9$ Lorentz invariants 
$\{ s_{12}, s_{23}, s_{34}, s_{45}, s_{56}, s_{61}, s_{123}, s_{234}, s_{345}\}$ are all independent for any $d\geq 5$, while in four spacetime dimensions $n^{(d=4)}_{\mathrm {inv}.} =8$ (using Eq.~(\ref{ninv})), implying that in $d=4$ there is a single non-trivial relation between these invariants. This relation is precisely the Gram determinant constraint, $G(p_1,p_2,p_3,p_4,p_5) = 0$.
Generally, the number of \emph{independent} Gram determinant constraints is given by the difference  between Eqs.~(\ref{ninv}) and (\ref{ninvd}), namely
\begin{equation}
n_{\text{Gram}}=
n^{(d)}_{\mathrm {inv}.} 
-n^{(d=4)}_{\mathrm {inv}.} 
=
\frac{n(n-3)}{2}-(3n-10)
=\frac{(n-4)(n-5)}{2}\,.
\end{equation}

In this paper, we will be interested in scattering amplitudes of $n$  massless on-shell particles in four spacetime dimensions, and we
will be making use of 
\begin{equation}
n_{\mathrm{par.}}=
n^{(d=4)}_{\mathrm {inv}.} +1=
3n-9,
\end{equation}
parameters
consisting of the $3n-10$ independent Lorentz invariants and one azimuthal reference phase~$\theta$.
The latter will be essential for representing helicity amplitudes using spinor variables.
To represent these $3n-9$ variables we 
adopt the variables proposed in Ref.~\cite{Byrne:2022wzk} (used there in the context of six-point amplitudes) which we generalise to 
$n$-point amplitudes. 
We choose a Minimal set of Light-Cone Variables, $3n-9$ real parameters in total, as follows:
\begin{equation}
\label{minival:nonlorentzinvdd}
\text{MSLCV}\equiv \left\{X_i \equiv \frac{p_{i+2}^{+}}{p_{i+3}^{+}} ,\quad z_j\equiv-\frac{q_{j+1}^{\perp}}{p_{j+3}^{\perp}},\quad  q_{1}^\perp 
\right\}
\,,
\end{equation}
with $i=1\ldots n-3$ and $j=1\ldots n-4$.
We note that  Eq.~(\ref{minival:nonlorentzinvdd}) is a straightforward generalisation of the aforementioned set of three parameters, $q^{\perp}=|q^{\perp}|{\rm e}^{i\theta}$ and \hbox{$X\equiv X_1=\frac{p_3^+}{p_4^+}$},
for $n=4$ in Eq.~(\ref{2to2vars}). 
We emphasise that the MSLCVs so defined are invariant with respect to Lorentz boost along the beam axis ($y_i\to y_i+\Delta y$ in (\ref{kijew})): the $X_i$ are invariant because the boost factor cancels in the ratio of lightcone plus components, and the transverse momenta are themselves invariant.

\begin{figure}
	\centering
	\includegraphics[width=1\textwidth]{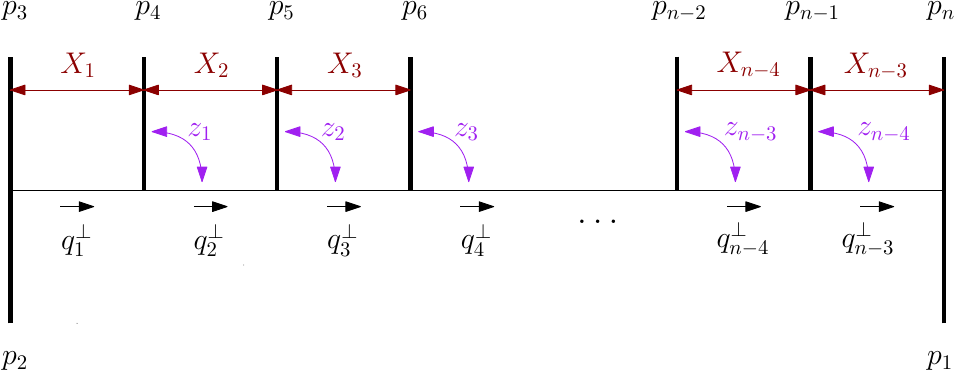}
	\caption{An illustration of the minimal set of Light-Cone Variables (MSLCV) defined in Eq.~(\ref{minival:nonlorentzinvdd}). 
The ratios of plus momentum components, $X_i\equiv {p_{i+2}^+}/{p_{i+3}^+}$, between pairs of sequential emissions are real, positive (dimensionless) numbers, while the ratios of transverse components between each $t$-channel momentum and the preceding emission, $z_i=-{{q}_{(i+1)}^\perp}/{{p}_{(i+3)}^\perp}$, are 
complex-valued  (dimensionless) numbers. Thus, we have $n-3$ real ratios, plus $n-4$ complex ratios, so $3n-11$ real parameters. Adding one complex dimensionful  transverse momentum,  
$q_{1}^\perp = |q_{1}^\perp| {\rm e}^{i\theta}$, fixes the overall scale and the reference azimuthal angle, completing the $3n-9$ set.
}
\label{fig:minimalvargeneral}
	\end{figure}

Clearly, there are multiple ways to pick such sets which span the same kinematic space.  Our specific choice is convenient to describe Regge limits of multi-leg amplitudes.
The chosen ratios of lightcone momentum components in Eq.~(\ref{minival:nonlorentzinvdd}) are illustrated schematically in Figure~\ref{fig:minimalvargeneral}. 
    In Eq.~(\ref{minival:nonlorentzinvdd}), transverse momentum conservation holds at each emission vertex,
\begin{equation}
\label{eq:duequattro}
  q_1^\perp = - p_3^\perp\,, \qquad q_{j+1}^\perp = q_j^\perp - p_{j+3}^\perp\,, \qquad q_{n-3}^\perp = p_n^\perp \,,
\end{equation}
with $j=1\ldots n-4$. 
In particular this means that 
$1-z_j=\frac{q_j^\perp}{p_{j+3}^\perp}
$, which allows us to express 
\begin{align}
\label{p_j_perp_general}
p_4^\perp=q_1^\perp\frac{1}{1-z_1},\quad\qquad 
p_j^\perp=q_1^\perp\frac{(-z_1)(-z_2)\ldots (-z_{j-4})}{(1-z_1)(1-z_2)\ldots(1-z_{j-3})},\quad \forall j\geq 5\,.
\end{align}

Much like the $n=4$ case, the MSLCV of Eq.~(\ref{minival:nonlorentzinvdd}), although defined using lightcone coordinates in a given frame, can be attributed a Lorentz covariant  meaning via
\begin{eqnarray}
\label{minival:lorentzinvdd}
  \text{MSLCV}\equiv  \left\{X_i = \frac{s_{1(i+2)}}{s_{1(i+3)}} \,,\quad z_j = \frac{\left\langle 2\left|\sum_{k=3}^{j+3} k\right| 1\right]}{\left\langle 2\left|(j+3)\right| 1\right]} \,,\quad  q_{1}^\perp  = \frac{\langle 2|3|1]}{[21]}\right\} \,, 
\end{eqnarray} 
with $i=1\ldots n-3$ and $j=1\ldots n-4$, where we used the spinor conventions in Appendix~\ref{frame}.
To show that the definition in Eq.~(\ref{minival:lorentzinvdd}) is equivalent to that in Eq.~(\ref{minival:nonlorentzinvdd}), one needs to use Eq.~(\ref{eq:duequattro}) along with the expressions for the spinors in Eq.~(\ref{spinoconven}). 
For example, starting with the expression in Eq.~(\ref{minival:lorentzinvdd}) for $z_j$ we find\footnote{Here one first  uses~(\ref{spinoridentities_2}), then convert the spinors to lightcone components using~(\ref{spinoconven}), observing that the $+$ and $-$ lightcone momentum factors cancel. Finally one applies momentum conservation (\ref{eq:duequattro}) to recover the original definition in~(\ref{minival:nonlorentzinvdd}).
}
\begin{equation}
\label{eq:z_j}
z_j
= \sum_{k=3}^{j+3} 
\frac{\left\langle 2| k\right\rangle 
\left[ k | 1\right]}{\left\langle 2| j+3\right\rangle 
\left[ j+3 | 1\right]}
=
\frac{p_3^{\perp}+p_4^{\perp}+\ldots+p_{j+3}^{\perp}}{p_{j+3}^{\perp}}= 
-\frac{q_{j+1}^\perp}{p_{j+3}^\perp}\,. 
\end{equation}
We note in passing that this representation of the transverse components of the MSLCVs in terms of ratios of spinor products makes manifest their relation with kinematic variables used in the literature in different contexts -- see for example the treatment of the transverse dynamics in spectator-spectator interactions in Ref.~\cite{Schwartz:2018obd} (see Eq.~(115) there versus (\ref{eq:z_j}) above in the case $j=1$).

Here we have seen that our MSLCVs, originally defined in Eq.~(\ref{p_j_perp_general}) as fractions of momentum components, can also be written in terms of spinor products. Our next observation is that the \emph{complete set of spinor products} $\langle i j\rangle$ with $i,j\in 1\ldots n$, can be expressed in terms of the MSLCVs. To see this explicitly, recall first that every $p_i^\perp$ can be written in terms of the MSLCVs, as in Eq.~(\ref{p_j_perp_general}). Then, since spinor products are invariant with respect to boost along the beam direction, $(p_i^{+},p_i^{-}, p_i^\perp)
\to 
(\lambda p_i^{+},\lambda^{-1} p_i^{-}, p_i^\perp)$ for all $i$,
one may express them using 
\eqref{spinoconven} in terms of the set $\left\{X_i, \,p_i^\perp\right\}$, i.e. in terms of the MSLCVs. 
 To this end $p_1^-$ and $p_2^+$ are expressed in terms of the outgoing particle momenta using momentum conservation, and subsequently the on-shell conditions are used to eliminate all minus momentum components, $p_i^-$, obtaining functions of $\left\{p_i^\perp\right\}$ along with \emph{ratios} of outgoing plus momentum components, $X_i$, which are invariant under said boost. 

With the variables fixed we may write any $n$-point colour-dressed amplitude of massless particles as 
\begin{eqnarray}
\label{eq:dueotto}
\!\!\!\!\mathcal{A}_{2\to n-2}\left(p_1^{h_1},p_2^{h_2},...,p_n^{h_n}\right)= \mathcal{A}_{2\to n-2}(q_{1}^\perp,X_1,X_2,...X_{n-3},z_1,z_2,...,z_{n-4}) \,,
\end{eqnarray}
where the notation on the left indicates that the amplitude depends not only on the particle momenta, but also on their helicities. Since we know that any such amplitude can be expressed in terms of spinor products, and we have shown above that all spinor products can be expressed in terms of the MSLCVs, we have established that the amplitude can be expressed in terms of the MSLCVs, as indicated on the 
right-hand side of~(\ref{eq:dueotto}).

It is important to emphasise that the amplitude, Eq.~(\ref{eq:dueotto}), depends on $3n-9$ real parameters, which amounts to the $3n-10$ Lorentz invariants (which would suffice to express an amplitude of scalars) plus \emph{one} parameter: the azimuthal reference phase $\theta$. 
In our choice of variables, Eq.~(\ref{p_j_perp_general}), this is just the phase of the transverse momentum $q_1^\perp$. 

The physical significance of the additional degree of freedom can be understood through the covariant transformation properties of amplitudes of (massless) particles with spin under Lorentz transformations.  Under the little group transformation for particle 
$i$ in the spinor-helicity formalism, the variables transform as $\lambda_i \rightarrow t \lambda_i$ and $\tilde{\lambda}_i \rightarrow t^{-1} \tilde{\lambda}_i$. This implies that the amplitude transforms as
\begin{align}
\label{calAtransLittleGroup}
\mathcal A_n\bigg( \Big\{ \lvert 1 \rangle,\, \lvert 1 ],\, h_1 \Big\},\, \dots,\, \Big\{ t_i \lvert i \rangle,\, t_i^{-1} \lvert i ],\, h_i \Big\},\, \dots \bigg) = t_i^{-2h_i} \,\mathcal A_n\bigg( \dots \Big\{ \lvert i \rangle,\, \lvert i ],\, h_i \Big\} \dots \bigg).
\end{align}
Thus, any amplitude with an arbitrary helicity configuration can be written as an overall spinor string that encapsulates all the little-group scaling properties, multiplied by a  function depending solely on Lorentz invariants. This implies that the overall phase of the amplitude depends on the little group scaling of all particles which carry spin.
This overall phase constitutes the single additional degree of freedom characterizing the amplitude going beyond the $3n-10$ Lorentz invariant parameters.

Finally, let us provide a couple of examples of amplitudes written in terms of the MSLCVs.
We choose simple, Maximally Helicity Violating (MHV) examples, just for illustration.  
The simplest case is the four-gluon amplitude:
\begin{align}
	{A}_{\mathrm{MHV}}(g_1^\ominus,g_2^\ominus,g_3^\oplus,g_4^\oplus)=\frac{\langle 1 2\rangle^4}{\langle 12\rangle\langle 23\rangle \langle 34\rangle\langle 41\rangle}=\frac{\bar q_1^\perp}{q_1^\perp}\left(1+X_1\right),
	\label{mhv4pt}
 \end{align}
We note that the overall phase, parameterized by $\theta$ as in Eq.~(\ref{2to2vars}), is contained in the ratio $e^{-2i\theta} =\bar q_1^\perp/q_1^\perp$. The next example is the five-gluon amplitude
\begin{align}
\label{mhv5pt}
&	{A}_{\mathrm{MHV}}(g_1^\ominus,g_2^\ominus,g_3^\oplus,g_4^\oplus,g_5^\oplus)\nonumber \\ 
&=	\frac{(\bar{q}_1^\perp)^2 (1-z_1)}{(q_1^\perp)^2 (1-\bar{z}_1)}
	\cdot
	\frac{\left[\left(X_1+1\right)X_2+1\right]\left[|1-z_1|^2+X_1\left(X_2|z_1|^2+1\right)\right]}
	{\bar{q}_1^\perp\left(1-z_1+X_1\right)\left(1+X_2z_1\right)}.
\end{align}

To conclude this section let us spell out the key advantages of using the MSLCV instead of a set of Mandelstam invariants $\{s_{ij}\}$ (or spinors). The first issue stems from our interest in physical momenta in $d=4$ spacetime dimensions, as opposed to a generic (and sufficiently large)~$d$. That is, the fact that to obtain an independent set of $\{s_{ij}\}$ in $d=4$ (i.e. $3n-10$ of them) one needs to impose Gram determinant constraints, taking the form of high-order polynomial relations. Instead,
the MSLCV in Eq.~(\ref{minival:nonlorentzinvdd}) provide, by construction, a set of independent variables parametrizing the kinematic space of interest. 
Indeed, 
one may verify that all the Gram determinants evaluated in terms of Eq.~(\ref{minival:nonlorentzinvdd}) vanish identically.

The second issue stems from the need to describe spin-dependent amplitudes, which amounts to using spinors. While expressions for amplitudes in terms of spinors can be fairly compact, the price is that such expressions use overcomplete sets of spinors, which are related via
Schouten and other identities. As a result, comparing expressions for multi-leg amplitudes between different representations (e.g.~such as those obtained by BCFW and Berends-Giele recursion) or expanding them in certain limits becomes cumbersome (and hence one often resorts to numerical treatment). This is in contrast to the MSLCV, which are indeed a minimal set, and hence amplitude expressions can naturally be dealt with analytically. The third and final advantage of the MSLCV is that they can readily be used to expand around multi-Regge limits,  which are cleanly given by the limits where the $X_i$ variables tend to zero or infinity.

\section{Peripheral and central emission vertices}
\label{sec:PEVCEVdef}

\begin{figure}
    \centering
    \includegraphics[width=1\linewidth]{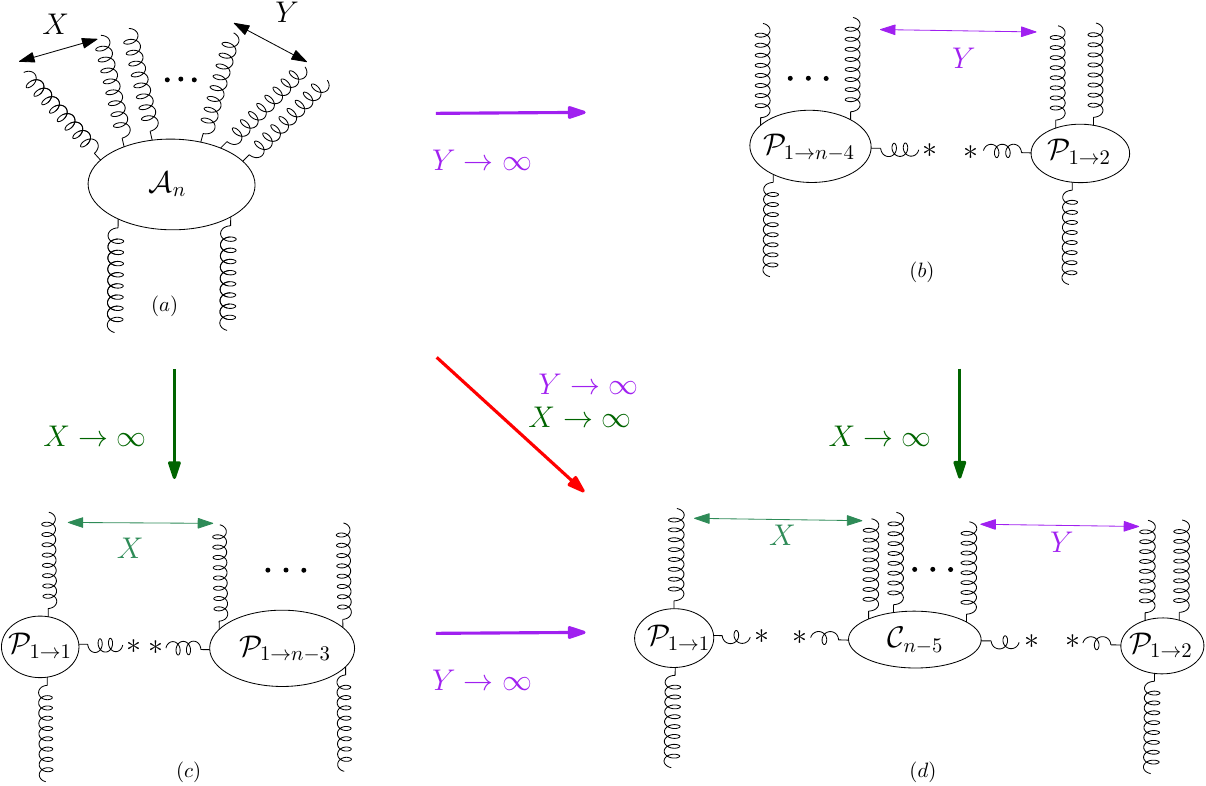}
    \caption{An example of the high-energy factorisation property discussed in the text.  Upon taking one lightcone momentum ratio, i.e. $X$ or $Y$ to be large (as shown by the green vertical and purple horizontal arrows, respectively) we have the amplitude factorizing into a positive-rapidity PEV and a negative-rapidity PEV ((b) and (c)), while upon simultaneously taking $X$ and $Y$ to be large (as shown by the diagonal red arrow) we obtain factorization into a positive-rapidity PEV, a negative-rapidity PEV and a CEV (d). The $*$ denotes an off-shell $t$-channel gluon.
    }
    \label{fig:high energy factorization}
\end{figure}

In this section, we discuss the strategy for extracting tree-level MREVs using the MSLCV from the previous section.
We begin with the definitions of the MREVs, which follow from the relevant Regge limit of an $n$-point amplitude. 
MREVs describe sets of outgoing (on-shell) particles clustered in rapidity, i.e. emitted particles whose rapidities are all comparable and very different from the rapidities of all the other outgoing particles in the amplitude.
In the appropriate multi-Regge limit, the particles forming the MREV 
are connected to remaining particles in the amplitude by off-shell (Reggeized) gluons in the $t$ channel.
In any limit in which the rapidity span between the particles separated by such a gluon is large, 
the amplitude is expected to factorize~\cite{Fadin:2006bj,Fadin:2015zea}. 
The rapidity factorisation properties, which we illustrate in what follows, can be seen to be a consequence of the factorization property of the functional integral \cite{Balitsky:1995ub,Caron-Huot:2013fea}, and hence have an all-order generalization \cite{Kuraev:1976ge}, which we shall not discuss in this paper. The rapidity factorisation properties amount to \emph{universality}, namely the fact that the same MREVs appear as building blocks of suitably chosen Regge limits of amplitudes of different multiplicities and particle content: the MREVs depend only on the degrees of freedom (particle type, colour, helicity and momenta) of the set of particles they contain, as opposed to the rest of the process.
In this factorization 
the PEVs are the clusters of emitted particles at extremal rapidities at either end (the target on one end and the projectile on the other), as depicted in Fig.~\ref{fig:high energy factorization}($b$) 
and~($c$), while the CEVs corresponds to such clusters occurring in between these extremes, separated by large rapidity spans from either of them, as shown in Fig.~\ref{fig:high energy factorization}($d$).

\subsection{Definition}

\label{defPEVandCEV}

Let us now discuss the way the MREVs are defined. The basic idea is to consider the simplest amplitude, which in the appropriate limit, contains a given MREV to define it. 
To this end we will consider factorized amplitudes, where each of the outgoing particles of momentum~$p_i$ and helicity $h_i$ can be either a quark or a gluon, denoted respectively by $q^{h_i}_{p_i}$
and $g^{h_i}_{p_i}$, such that at leading power in the appropriate Regge limit it would be a $t$-channel (virtual) gluon -- rather than a quark -- that is exchanged across the large rapidity span. 
In particular, the simplest, single-emission PEV \cite{DelDuca:1995zy,DelDuca:1996km}, ${\cal P}_{1\to 1}$, is defined through the four-point amplitude in the Regge limit, which factorises into two one-particle PEVs,
 \begin{equation}
 \label{pevlo} 
 \lim_{y_3 \gg y_4}
\cA_{2\to 2}\left(
 1, 2, 3, 4
 \right)= s\, \cP_{1\to 1}\left(p_2^{h_2}, p_3^{h_3},g_{q_1^\perp}^*
 \right)
 \frac1{t}
 \cP_{1\to 1}\left(g_{q_1^\perp}^*,p_4^{h_4}, p_1^{h_1} \right)
 \,,
\end{equation}
with $s = (p_1+p_2)^2$ and $t = (p_2+p_3)^2\,\simeq\,
-|q_1^{\perp}|^2$.
Following the spinor-helicity conventions of Appendix~\ref{frame} (and Refs.~\cite{DelDuca:1999iql,Byrne:2022wzk}), the explicit expressions for the positive rapidity PEVs of gluon and quark are, respectively:
\begin{subequations}
\label{positivePEV}
\begin{alignat}{2}
\label{positivePEV_g}
\mathcal{P}_{1 \to 1}(g_2^{\ominus}, g_3^{\oplus}, g_{q_1^\perp}^*)&=(F^{c_1})_{a_2a_3}&\qquad\mathcal{P}_{1 \to 1}(g_2^{\oplus}, g_3^{\ominus}, g_{q_1^\perp}^*)&=(F^{c_1})_{a_2a_3}
\\
\mathcal{P}_{1 \to 1}(\bar q_2^{\ominus}, q_3^{\oplus}, g_{q_1^\perp}^*) 
  &= i T^{c_1}_{\bar{\imath}_3 i_2}
  &\qquad 
  \mathcal{P}_{1 \to 1}(q_2^{\oplus}, \bar q_3^{\ominus}, g_{q_1^\perp}^*) 
  &= i T^{c_1}_{i_3 \bar{\imath}_2}\,,
\end{alignat}
\end{subequations}
while the negative-rapidity ones are:
\begin{subequations}
\label{negativePEV}
\begin{alignat}{2}
\mathcal{P}_{1 \to 1}(g_{q_1^\perp}^*, g_4^{\ominus}, g_1^{\oplus}) 
  &= \frac{\bar q_1^\perp}{q_1^\perp}( F^{c_1 })_{a_4 a_1}
  &\qquad 
  \mathcal{P}_{1 \to 1}(g_{q_1^\perp}^*, g_4^{\oplus}, g_1^{\ominus}) 
  &= \frac{q_1^\perp}{\bar q_1^\perp}( F^{c_1 })_{a_4 a_1}
  \\
\mathcal{P}_{1 \to 1}(g_{q_1^\perp}^*, \bar q_4^{\ominus}, q_1^{\oplus}) 
  &= -i \sqrt{\frac{q_1^\perp}{\bar q_1^\perp}} T^{c_1}_{i_1 \bar{\imath}_4}
  &\qquad 
  \mathcal{P}_{1 \to 1}(g_{q_1^\perp}^*, q_4^{\oplus}, \bar q_1^{\ominus}) 
  &= -i \sqrt{\frac{\bar q_1^\perp}{q_1^\perp}} T^{c_1}_{\bar{\imath}_1 i_4} \,.
\end{alignat}
\end{subequations}
Here, $T^{c_1}$ denotes the generator in fundamental representation and $F^{c_1}$ the adjoint one. The normalization of these are discussed in subsection~\ref{colourdressPEVCEV} below along with other aspects of colour for MREVs. A further comment is due regarding the terminology used above, of positive versus negative rapidity. In order to distinguish between the two PEVs in Eq.~(\ref{pevlo}), one corresponding to the ``target'' side and the other to the ``projectile'',  we refer to $\mathcal{P}_{1 \to 1}(p_2, p_3, g_{q_1^\perp}^*)$ as the positive-rapidity one-particle PEV and to $\mathcal{P}_{1\to 1}(g_{q_1^\perp}^*, p_4, p_1)$ as the negative-rapidity one. This terminology is based on the rapidity definition in Eq.~\eqref{rapidity}
in the collider setup, where $p_3$ has a large positive rapidity while $p_4$ a large negative one. We shall use this terminology in the remainder of the paper also when referring to higher-point PEVs. 
Comparing between Eqs.~(\ref{negativePEV}) and~(\ref{positivePEV}), we note that they differ just by an overall kinematically-dependent phase. The same is true also for higher-point PEVs, and therefore we will focus on the definition and determination of positive rapidity PEVs only.

The one-gluon CEV $\cC_1$, also called the Lipatov vertex \cite{Lipatov:1976zz,Lipatov:1991nf}, is defined through the five-point amplitude in multi-Regge kinematics (MRK), which factorises into two one-particle PEVs and a one-gluon CEV, thus leveraging the knowledge of the one-particle PEVs acquired at four points in Eq.~(\ref{pevlo}),
\begin{equation}
\label{cevlo}
  \begin{aligned}
\lim_{y_3 \gg y_4 \gg y_5}
\cA_{2\to 3}\left(
 1, 2, 3, 4, 5
 \right) = s\, \cP_{1 \to 1}\left(p_2, p_3,g_{q_1^\perp}^*
 \right)
 \frac1{t_1}
\cC_1\left(g_{q_1^\perp}^*,p_4,g_{q_2^\perp}^*\right)
 \frac1{t_2}
 \cP_{1 \to 1}\left(g_{q_2^\perp}^*,p_5, p_1 \right)
 , 
\end{aligned}
\end{equation}
with $t_1 = (p_2+p_3)^2$ and $t_2 = (p_5+p_1)^2$. The explicit expression for the one-gluon CEV is given in section~\ref{Sec:one_gluon_CEV}.
Note that in Eq.~(\ref{cevlo}) we simplified the notation by suppressing the helicity indices. We will do the same below whenever this information is not essential.

Multi-particle MREVs are obtained in a similar manner in the appropriate limit where the number of emissions is larger than the number of large rapidity spans.
Since MRK is defined as the limit where all emissions are separated by large rapidity spans, it corresponds to the situation in which there is exactly one more emission than large rapidity span. Following this logic ``next-to'' MRK is the limit where the number of emissions is two more than the number of large rapidity spans, etc. Thus, the positive-rapidity $m$-particle PEV, with $m = n-3$, is defined through the N$^{m-1}$MRK limit involving a single large rapidity span of the $n$-particle amplitude (with $n-2$ final-state particles), which factorises into a positive-rapidity $(n-3)$-particle PEV and a negative-rapidity one-particle PEV as follows:
  \begin{equation}
  \begin{aligned}
 \lim_{y_3 \,\simeq\,  \ldots\, \simeq\, y_{n-1} \gg \,y_n}&
 \cA_{2\to n-2}\left(1,
 2,3, \ldots, n-1, n
 \right) 
 \\
 &= s\,
 \cP_{1\to n-3}\left(p_2, p_3, \ldots, 
 p_{n-1},
g_{ q_{n-3}^\perp}^*
 \right)
 \frac1{t_{n-3}}
\cP_{1 \to 1}\left(g_{q_{n-3}^\perp}^*,p_n, p_1 \right), \label{pevbtypedef2} 
 \end{aligned}
\end{equation}
with $t_{n-3} = (p_n+p_1)^2$. A similar formula defines the $m$-emission PEV on the other extreme of negative rapidities. 

The $m$-emission CEV, with $m=n-4$, is defined through another N$^{m-1}$MRK limit of the amplitude, which factorises into an $m$-emission CEV and two single-emission PEVs,
\begin{equation}
\begin{aligned}
&
\lim_{y_3 \gg \, y_4\, \simeq \, \ldots\, \simeq \,y_{n-1} \gg\,  y_n}
\cA_{2\to n-2}\left( 1,2, 3, 4, \ldots n-1,n \right)  \\
& ~~~~~~~~~ =
s\, \cP_{1 \to 1}\left(p_2, p_3,g_{q_{1}^\perp}^* \right)
\frac{1}{t_1}
\cC_{n-4}\left(g_{q_{1}^\perp}^*,p_4, \ldots ,p_{n-1} ,g_{q_{n-3}^\perp}^*\right)  \frac{1}{t_{n-3}}
\cP_{1 \to 1}\left(g_{q_{n-3}^\perp}^*,p_n, p_1 \right)
\,,
\end{aligned}
\label{cevmulti} 
\end{equation}
with $t_1 = (p_2+p_3)^2$ and $t_{n-3} = (p_n+p_1)^2$.

In Eqs.~\eqref{pevbtypedef2} and \eqref{cevmulti}, 
we displayed the general formulae for the extraction of MREVs. Upon expressing the amplitudes using the minimal set of lightcone variables~(MSLCV), according to Fig. \ref{fig:minimalvargeneral}, these factorisation formulae become 
  \begin{eqnarray}
   \lim\limits_{X_{n-3}\to \infty } &&
 \cA_{2\to n-2}\left(q_1^\perp 
 ,X_1,\ldots X_{n-3},
 z_1,\ldots,
 z_{n-4}
 \right)
 \non &=& s   \cP_{1\to n-3}\left(q_1^\perp 
 ,X_1,\ldots X_{n-4},
 z_1,\ldots,
 z_{n-4}
 \right)\frac{1}{t_{n-3}}\cP_{1 \rightarrow 1}\left(q_{n-3}^\perp  \right)
,\non \label{pevbtypedef1} 
\end{eqnarray}
and
 \begin{eqnarray}
   \lim\limits_{X_{1},X_{n-3}\to \infty }
   &&
 \mathcal{A}_{2\to n-2}\left(q_1^{\perp}, X_1, \ldots X_{n-3}, z_1, \ldots, z_{n-4}\right)= 
 \non
 &&
s 
\cP_{1 \rightarrow 1}\left(q_{1}^\perp \right) 
\frac{1}{t_{1}}
\cC_{n-4}\left(q_1^\perp 
 ,X_1\ldots X_{n-4},
{z}_{1},\ldots ,
{z}_{n-4}
 \right)
\frac{1}{t_{n-3}}
\cP_{1 \rightarrow 1}\left(q_{n-3}^\perp \right)\,,\non \label{cevbtypedef2} 
\end{eqnarray}
respectively. The argument of the second PEV here, $q_{n-3}^\perp$, may be expressed in terms of $q_{1}^\perp$ and the set of $\{z_i\}$, but this dependence has been left implicit above for brevity.

\subsection{Colour structure}
    \label{colourdressPEVCEV}

In QCD, and more generally in \(SU(N_c)\) gauge theories, amplitudes can be expressed as linear combinations of colour factors multiplied by colour-ordered amplitudes. The linear independence of colour factors and gauge invariance of the full amplitude then ensure that the colour-ordered amplitudes themselves are gauge invariant~\cite{Bern:1990ux}.

In the fundamental representation of $SU(N_c)$, the algebraic structure is given by the relation
$[t^a,t^b]=i f^{abc} t^c$, where the traceless hermitian $N_c \times N_c$ generator matrices $(t^a )^j_i$ carry quark indices $i, j = 1, .., N_c$ and a gluon index $a = 1, \ldots, N_c^2-1$. The generator matrices are usually normalised by $\operatorname{Tr}\left(t^a t^b\right)= T_F \delta^{a b}$
with $T_F = 1/2$. However, in order to avoid a proliferation of $\sqrt{2}$ factors in the colour-ordered amplitudes, we rescale the generator matrices, $T^a = \sqrt{2}\, t^a$ such that $\operatorname{Tr}\left({T}^{{a}} {T}^{{b}}\right)=\delta^{a b}$. Likewise, we rescale the structure constants, $(F^b)_{ac}=i \sqrt{2} f^{abc} = \operatorname{Tr}\left(\left[T^a, T^b\right] T^c\right)$.

The colour decomposition of the $n$-gluon amplitude in the Del Duca-Dixon-Maltoni (DDM) basis is~\cite{DelDuca:1999iql,DelDuca:1999rs}
\begin{align}
\label{gluonampcolour}
\begin{split}
&\mathcal{A}_{2\to n-2}\left(\left\{p_i, h_i, a_i\right\}\right) = g^{n-2}\sum_{\sigma \in S_{n-2}} \left(F^{a_{\sigma_3}} \ldots F^{a_{\sigma_{n}}}\right)_{a_2a_1} A_{2\to n-2}\left( 1,2, \sigma_3, 
\ldots, \sigma_{n}\right) \,,
\end{split}
\end{align}
where the sum runs over gluon label permutations \(\sigma \in S_{n-2}\).
The structure-constant based amplitude defines $(n-2)!$ independent colour-ordered amplitudes and it is suitable for the high-energy limit because the colour factorization is aligned with high-energy factorisation \cite{DelDuca:1999iql,Byrne:2022wzk,DelDuca:1995hf,Byrne:2023nqx}. 

For amplitudes with a single quark-antiquark pair, the colour decomposition (using the Feynman rules and the colour algebra \( (F^b)_{ac}T^c = [T^a, T^b] \)) yields:
\begin{align}
   & \mathcal{A}_{2\to n-2}\left(\bar{q}_1^{h_1}, q_2^{h_2}, \{p_i, h_i, a_i\}\right)
   \nonumber\\
  =  &
   g^{n-2}\! \sum_{\sigma \in S_{n-2}} \left(T^{a_{\sigma_3}} \cdots T^{a_{\sigma_n}}\right)_{j_2\bar{\imath}_1} A_{2\to n-2}(\bar{q}_1^{h_1}, q_2^{h_2}, \sigma_3, \ldots, \sigma_n),
\label{onequark_pair}
\end{align}
where, without loss of generality, we assign particle 1 as the antiquark \(\bar{q}_1\) and particle 2 as the quark \(q_2\), carrying the fundamental $SU (N_c)$ indices \(\bar{\imath}_1\) and \(j_2\), respectively.

For QCD amplitudes with general colour decomposition involving multiple quark anti-quark pairs, we refer the reader to~\cite{Mangano:1990by,Melia:2013epa,Melia:2013bta,Melia:2015ika,Ochirov:2019mtf,Johansson:2015oia}; in this work we consider up to two pairs. At tree level, from the colour basis perspective, the two quark pairs result in two generator chains -- as opposed to the single generator chain of Eq.~(\ref{onequark_pair}) -- and these two are connected by the adjoint indices carried by gluon interaction blocks between them. Using the commutator and the colour Fierz identity
\begin{align}
    \sum_{a=1}^{N_c^2-1}\left(T^a\right)_{i_1 \bar{\jmath}_1}\left(T^a\right)_{i_2 \bar{\jmath}_2} = \delta_{i_1 \bar{\jmath}_2} \delta_{i_2 \bar{\jmath}_1} - \frac{1}{N_c} \delta_{i_1 \bar{\jmath}_1} \delta_{i_2 \bar{\jmath}_2}\,,
\end{align}
they can be written as follows:
    \begin{equation}
        \begin{aligned}
&  \mathcal{A}_{2\to n-2}\left(\bar{q}_1^{h_1}, q_2^{h_2}, \bar{q}_3^{h_3}, q_4^{h_4} , g_5, \ldots, g_{n}\right)=i g^{n-2} \sum_{k=0}^{n-4} \sum_{\sigma \in S_k} \sum_{\rho \in S_{n-4-k}} \\
&  \biggl[\left(T^{\sigma_1} \ldots T^{\sigma_k}\right)_{\bar{\imath}_1j_4}\left(T^{\rho_1} \ldots T^{\rho_{n-4-k}}\right)_{\bar{\imath}_3j_2} 
\\& 
\qquad \qquad \times A_{2\to  n-2}\left(q_4^{h_4}, g_{\sigma_1}, \ldots, g_{\sigma_k} , \bar{q}_1^{h_1}  ,q_2^{h_2}, g_{\rho_1}, \ldots, g_{\rho_{n-4-k}},\bar{q}_3^{h_3}\right) \\
& -\frac{1}{N_c}\left(T^{\sigma_1} \ldots T^{\sigma_k}\right)_{\bar{\iota}_1j_2}\left(T^{\rho_1} \ldots T^{\rho_{n-4-k}}\right)_{\bar{\imath}_3j_4}  \\
& \qquad \qquad \times  B_{2\to n-2}\left(q_2^{h_2} , g_{\sigma_1}, \ldots, g_{\sigma_k},  \bar{q}_1^{h_1} ,q_4^{h_4}, g_{\rho_1}, \ldots, g_{\rho_{n-4-k}},\bar{q}_3^{h_3}\right)\biggr],\label{doubleqqbarpair}
\end{aligned}
    \end{equation}
where $\sigma$ and $\rho$ label the two subsets forming a partition of the gluon set $\{ g_5, \dots, g_{n} \}$, and 
 $\sum_{\sigma \in S_k} \sum_{\rho \in S_{n-4-k}}$
implies that the sums are taken over all possible bipartitions of the gluon set. For the null partition (e.g. $k=0$), the generator string $\left(T^{\sigma_1} \ldots T^{\sigma_k}\right)_{\bar{\imath}_1j_2}$ reduces to the Kronecker delta $\delta_{\bar{\imath}_1j_2}$, and similarly for the other generator strings.

\begin{figure}
    \centering
    \includegraphics[width=1\linewidth]{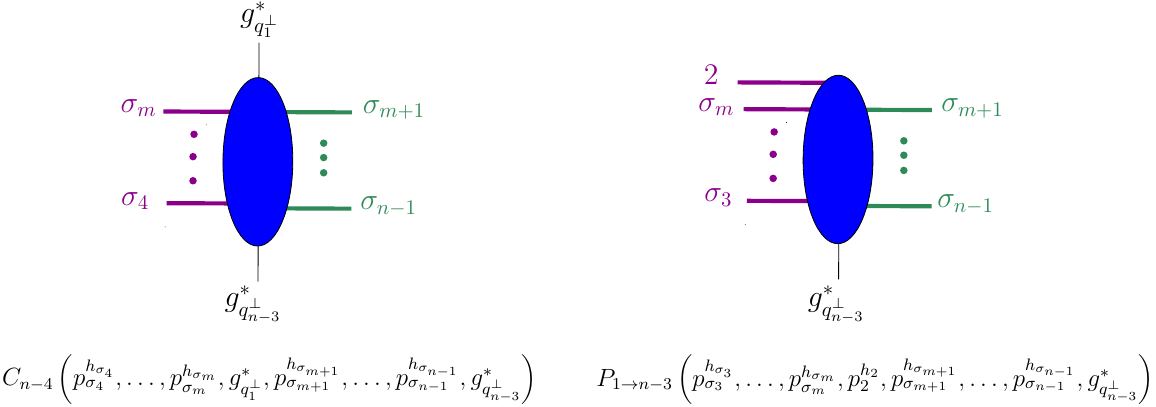}
    \caption{Colour-ordered CEV (left) and PEV (right) for general multiplicities, where $g^*$ denote off-shell $t$-channel gluons, and the $\sigma_i$ represent emissions into the final state; in the case of the PEV, the particle labelled $2$ corresponds to the incoming parton. These diagrams feature (clockwise) colour ordering, hence some final state partons (in purple) appear to the left of the $t$-channel gluon(s) while others (in green) to the right of it.  
}
    \label{fig:ddpevcev}
\end{figure}

The MREVs inherit a similar colour structure to that of the amplitudes from which they are derived.
Following Eqs.~(\ref{gluonampcolour}) and (\ref{cevmulti}), the tree-level pure gluon CEV is expressed as:
\begin{align}
\begin{split}
    &  \cC_{n-4}\big(g_{q_{1}^\perp}^*,g_{4}, \ldots, g_{n-1}, g_{q_{n-3}^\perp}^*\big) \\
    &\qquad \quad \, 
   = g^{n-4}\sum_{\sigma \in S_{n-4}}{C}_{n-4}\big(g_{q_{1}^\perp}^*,g_{\sigma_4}, \ldots, g_{\sigma_{n-1}},g_{q_{n-3}^\perp}^*\big)
   \left(F^{a_{\sigma_4}} \cdots F^{a_{\sigma_{n-1}}}\right)_{c_1 c_{n-3}}\label{CEVDDM},
    \end{split}
\end{align}
where $c_1$ and $c_{n-3}$ represent adjoint indices that contract with blocks of high-energy amplitudes {adjacent on either side of the CEV}, which may themselves be composed of other PEVs or CEVs.

Similarly, the colour-dressed PEV for pure gluon configurations takes the form:
\begin{align}
\begin{split}
&\cP_{1\to n-3}\big(g_2, g_{\sigma_3}, \ldots, g_{\sigma_{n-1}},g_{q_{n-3}^\perp}^*\big)
  \\
    &\qquad \quad \, 
   = g^{n-3} \,
      \sum_{\sigma \in S_{n-3}} P_{1\to n-3}\big(g_2, g_{\sigma_3}, \ldots, g_{\sigma_{n-1}},g_{q_{n-3}^\perp}^*\big)
       \Bigl(F^{a_{\sigma_3}} \cdots F^{a_{\sigma_{n-1}}}\Bigr)_{a_2 c_{n-3}}\,,
       \end{split}
       \label{PEVDDM}
\end{align}
where $a_2$ denotes the adjoint colour index of the incoming gluon “2”, and $c_{n-3}$ corresponds to the colour index of the off-shell (Reggeized) $t$-channel gluon that contracts with a subsequent PEV or CEV.

In the presence of a single quark-antiquark pair, and following Eq.~(\ref{onequark_pair}), we have the following colour decomposition for the PEV:
\begin{align}
\begin{split}
&\cP_{1\to n-3}\big(q_2,\bar{q}_3, g_4,...,g_{n-1}, g^*_{q^\perp_{n-3}}\big) \\
    &\qquad \quad \, 
   =  g^{n-3} \sum_{k=3}^{n-1} \sum_{\substack{\sigma \in S_{n-4} }}
P_{1\to n-3}\big(\underbrace{g_{\sigma_{k+1}},...,g_{\sigma_{n-1}}}_{n-k-1\ \text{gluons}} ,q_2,\bar{q}_3, \underbrace{g_{\sigma_4},...,g_{\sigma_{k}}}_{k-3\ \text{gluons}}, g^*_{q^\perp_{n-3}}\big)  \\
&\quad\qquad\quad\quad\qquad\quad\quad\qquad\quad  \times \Biggl(
\Big(\prod_{m_1=4}^{k} T^{a_{\sigma_{m_1}}}\Big)  
T^{c_{n-3}} 
\Big( \prod_{m_2=k+1}^{n-1} T^{a_{\sigma_{m_2}}}\Big) \Biggr)_{\bar{\jmath}_3i_2}\,,
\end{split}
\label{PEVgeneratpor}
\end{align}  
where $\sigma$ denotes a permutation of the $n-4$ on-shell gluons, and we sum over all possible colour-ordered positions of the off-shell gluon $g^*_{q^\perp_{n-3}}$. The ordering of the generators follows that of the gluons in the colour-ordered PEV.

One can dress the CEV in a similar manner:
\begin{align}
\label{CEVgeneratpor}
\begin{split}
&\hspace*{1pt}\cC_{n-4}\big(g^*_{q^\perp_1}, \bar{q}_{4}, q_{5}, g_6, \ldots, g_{n-1}, g^*_{q^\perp_{n-3}}\big) 
\\
&
=g^{{n-4}} \sum_{k=5}^{{n-1}} \sum_{l=0}^{n-k-1} \sum_{\substack{\sigma \in S_{n-6}}}
\Biggl( 
    \Big(\prod_{m_3=l+k+1}^{n-1} T^{a_{\sigma_{m_3}}}\Big)  T^{c_{n-3}}\Big(\prod_{m_1=6}^{k} T^{a_{\sigma_{m_1}}}\Big)
    T^{c_{1}} 
    \Big(\prod_{m_2=k+1}^{k+l} T^{a_{\sigma_{m_2}}}\Big)
  \Biggr)_{\bar{\imath}_4 j_5} \\
&\quad\qquad \quad\times 
 C_{n-4}\Bigl( 
  \underbrace{g_{\sigma_6}, \ldots, g_{\sigma_{k}}}_{k-5\ \text{gluons}},
  g^*_{q^\perp_1},
  \underbrace{g_{\sigma_{k+1}}, \ldots, g_{\sigma_{l+k}}}_{l\ \text{gluons}},\bar{q}_{4}, q_{5}, \underbrace{g_{\sigma_{l+k+1}}, \ldots, g_{\sigma_{n-1 }}}_{n-k-l-1\ \text{gluons}},
 g^*_{q^\perp_{n-3}}
\Bigr)
\\
&\quad\qquad \quad +(   g^*_{q^\perp_1}\leftrightarrow g^*_{q^\perp_{n-3}},\, c_{1}\leftrightarrow c_{n-3} )
  \end{split}
\end{align}
For MREVs with two quark pairs, the colour decomposition closely follows that of the amplitude counterpart, Eq.~(\ref{doubleqqbarpair}); however, off-shell gluons can enter either $\rho$ or $\sigma$, or they can be split, with one in the $\rho$ and one in the $\sigma$ permutations. For the sake of brevity, we do not display the general dressing rules here, but instead refer to Eq.~(\ref{wqgewqwqwqq}) as a representative example.

Being closely related to amplitudes, the Multi-Regge emission vertices, MREVs, have similar colour structure to amplitudes. Just as colour stripped amplitudes are not independent, but rather satisfy  certain relations such as the photon decoupling identity, the colour stripped MREVs also have such relations. These relations will be briefly introduced in section~\ref{resultsof CEVPEV} in the context of CEVs and then discussed in detail in section~\ref{relationsPEVCEV} for both CEVs and PEVs. The primary goal of  section~\ref{resultsof CEVPEV} is to present the explicit structure of CEVs using the MSLCV, and then in section \ref{sec:4partonCEVsPEVs} we explain how to use the \texttt{\href{https://github.com/YuyuMo-UoE/Multi-Regge-Emission-Vertices}{Multi-Regge-Emission-Vertices}} ({\tt{MREV}}) library~\cite{yuyu_2017_github} which tabulates all MREVs up to multiplicity 4.

\section{Results for central-emission vertices}
\label{resultsof CEVPEV}

In this section, we present explicit expressions for the central-emission vertex (CEV) results for up to three-parton emissions.
The results are derived directly from suitable Regge limits of colour-ordered amplitudes computed using the~\texttt{GGT} package~\cite{Dixon:2010ik}. To ensure their validity, we performed a range of checks of their factorization properties by taking additional kinematic limits, namely soft, collinear, and further high-energy limits (see Appendix~\ref{Appendix:Fact_in_limits}). We also performed comparisons with known results where available~\cite{Duhr:2009uxa,DelDuca:1999iql}. Finally, we verified amplitude-like identities, such as photon decoupling, Kleiss-Kuijf relations, and reversal identities. For brevity, we make direct use of these relations to present the results, deferring a detailed discussion of their derivation to section~\ref{relationsPEVCEV}.

The structure of this section is as follows. We begin in section~\ref{qcd_and_sym_diff} by explaining how we use the~\texttt{GGT} package~\cite{Dixon:2010ik}. The package provides colour-ordered $\mathcal{N}=4$ sYM amplitudes for gluon and gluino scattering; these are sufficient for obtaining all colour-ordered QCD amplitudes. By taking the appropriate high-energy limit following section~\ref{defPEVandCEV} we extract the CEVs (and PEVs) in QCD. In section~\ref{relationCEVPEV} we briefly summarise the relations between colour-ordered CEVs, which we make use of in the remainder of this section to concisely present the colour-dressed CEVs. In sections~\ref{Sec:one_gluon_CEV},~\ref{Sec:two_parton_CEV} and~\ref{Sec:three_parton_CEV} respectively, we present the results for one-, two- and three-parton CEVs in different helicity configurations. Multiplicity-four CEVs as well as PEVs will be discussed in section~\ref{sec:4partonCEVsPEVs} using a dedicated Mathematica library we created, {\tt{MREV}}.

\subsection{Extraction of QCD MREVs from \texorpdfstring{$\mathcal{N}=4$}{N=4} sYM amplitudes}

\label{qcd_and_sym_diff}

We extract all our colour-ordered  QCD MREVs from \texttt{GGT}~\cite{Dixon:2010ik}, which provides the tree-level amplitudes in $\mathcal{N}=4$ sYM for gluon and gluino scattering. As discussed in section~3 of Ref.~\cite{Dixon:2010ik}, translating from $\mathcal{N}=4$ sYM results to QCD needs some care. 
There are two separate issues that need to be addressed. The first concerns the fact that the fermions in ${\cal N}=4$ sYM, the gluinos, are in the adjoint colour representation, while in QCD the quarks are in the fundamental representation. The second, more subtle issue is that amplitudes in ${\cal N}=4$ sYM with two or more gluino flavours receive contributions from scalars, interactions which do not exist in QCD. We consider the impact of these two aspects on colour-ordered amplitudes (and hence on MREVs) in turn.

First, colour dressing of physical QCD amplitudes containing quark-antiquark pairs features only orderings where a quark and an antiquark appear adjacent to each other~\cite{Mangano:1990by}. This can be seen in Eqs.~\eqref{onequark_pair} and \eqref{doubleqqbarpair} for the cases of one and two quark pairs, respectively. The same applies for MREVs, as seen for example in the case of a single quark pair in Eqs.~\eqref{PEVgeneratpor} and \eqref{CEVgeneratpor} for PEVs and CEVs, respectively.
This means that partial amplitudes in which a quark-antiquark pair are interleaved with gluons — for instance, $A(\ldots, g, q, g, \bar{q}, g, \ldots)$ — do not contribute to physical QCD amplitudes. Such configurations instead exist in theories with adjoint fermions, such as sYM theory. Then all particles are in the adjoint representation of the colour group, so the trace basis colour decomposition applies:
\begin{eqnarray}
\hspace*{-8pt}
\mathcal{A}_{2\to n-2}^{\mathcal{N}=4\, \mathrm {sYM}}\left(\left\{p_i, h_i, a_i\right\}\right)
    = g^{n-2} \sum_{\sigma \in S_n / Z_n}
    \operatorname{Tr}\big(T^{a_{\sigma_1}} \cdots T^{a_{\sigma_n}}\big)
A_{2\to n-2}^{\mathcal{N}=4\, \mathrm {sYM}}\big(p_{\sigma_1}^{h_{\sigma_1}}, \ldots, p_{\sigma_n}^{h_{\sigma_n}}\big)\,,
    \label{tracebasis2}
\end{eqnarray}
with $p_i$ being a gluon or a gluino. Note that the cyclicity of the trace implies cyclicity of the colour-ordered amplitudes; we shall discuss this property in section~\ref{relationCEVPEV} below.
The trace basis used in Eq.~(\ref{tracebasis2}) is usually a redundant representation of the amplitude because of the Kleiss-Kuijf relation~\cite{Kleiss:1988ne} (which we discuss in Appendix~\ref{sec:BriefKKrelation}
and in section~\ref{kk}) and can be rewritten in terms of the DDM basis~\cite{DelDuca:1999rs} via
\begin{equation}
	F^{a_1 a_2 x_1} F^{x_1 a_3 x_2} \cdots F^{x_{n-3} a_{n-1} a_n}= \operatorname{tr}\left(T^{a_1}\left[T^{a_2},\left[T^{a_3}, \ldots,\left[T^{a_{n-1}}, T^{a_n}\right] \ldots\right]\right]\right) ,
    \label{fffftoTTTT}
\end{equation}
giving
\begin{align}
\begin{split}
&\mathcal{A}_{2\to n-2}^{\mathcal{N}=4\, \mathrm {sYM}}\left(\left\{p_i, h_i, a_i\right\}\right) 
= g^{n-2} \!\!\!\sum_{\sigma \in S_{n-2}} \!\!\left(F^{a_{\sigma_3}} \ldots F^{a_{\sigma_{n}}}\right)_{a_2 a_1}  A_{2\to n-2}^{\mathcal{N}=4\, \mathrm {sYM}}\left(  p_1^{h_1},p_2^{h_2}, p_{\sigma_3}^{h_{\sigma_3}}, \ldots, p_{\sigma_n}^{h_{\sigma_n}} \right),
\end{split}
\label{ddmcolouradjoint}
\end{align}
similar to the representation of the pure gluon amplitude in~Eq.~(\ref{gluonampcolour}) above.
The colour-ordered amplitudes appearing in Eqs.~\eqref{tracebasis2} and \eqref{ddmcolouradjoint} are the identical.

In conclusion, while in QCD only colour-ordered amplitudes in which the quark and antiquark are adjacent contribute to tree amplitudes, in sYM all permutations enter, including ones where a quark and an antiquark are interleaved by gluons.  
We note that the colour-ordered amplitudes $A_{2\to n-2}$ depend only on the colour-ordered Feynman rules, not on the colour representation, so there is scope to relate them between theories involving fermions of different representations, as Ref.~\cite{Dixon:2010ik} has indeed done.

This brings us to the second aspect that needs to be addressed: in relating QCD to sYM colour-ordered amplitudes one must be careful about the additional scalar interactions present in the latter. It was explicitly shown in Ref.~\cite{Dixon:2010ik} that all QCD tree amplitudes with up to four external, arbitrary-flavour, massless quark-antiquark pairs can be expressed as linear combinations of $\mathcal{N}=4$ sYM tree-level amplitudes. 
These linear combinations are constructed to eliminate scalar interactions, which are absent in QCD. 
If there are only two different flavours, the situation is simpler — no linear combination is required. Nevertheless, some subtleties do arise.
In what follows, we review the relevant case of possible scalar propagation. The discussion starts from pure gluonic amplitudes or ones involving quarks of a single flavour, and then extends to those involving two distinct quark-antiquark pairs.

The first observation is that tree-level colour-ordered QCD amplitudes involving either only gluons, or gluons plus quark-antiquark pairs of a single flavour, coincide exactly with those in $\mathcal{N}=4$ sYM, as the $\mathcal{N}=1$ sYM subsector remains closed. For example,
\begin{align}
\begin{split}
&A^{\mathrm{QCD}}_{2\to n-2}\big(
    g_{1}^{h_{1}},\,g_{2}^{h_{2}},\,g_{3}^{h_{4}},\, {\color{blue} q_{4}^{\oplus}},\, {\color{blue}\bar q_{5}^{\ominus}},\, g_{6}^{h_{6}},\, \ldots,\, g_{m}^{h_{m}},\,
    {\color{blue} q_{m+1}^{\oplus}},\,{\color{blue}\bar{q}_{m+2}^{\ominus}},\,g_{m+3}^{h_{m+3}},\, \ldots,\, g_{n}^{h_{n}}
\big)
\\
&\hspace*{22pt}=A^{\mathcal{N}=4\, \mathrm {sYM}}_{2\to n-2}\big(
    g_{1}^{h_{1}},\, g_{2}^{h_{2}},\,g_{3}^{h_{4}},\,  {\color{blue} \tilde{g}_{A, 4}^{\oplus}},\,   {\color{blue}\tilde{g}_{A, 5}^{\ominus}},\, g_{6}^{h_{6}},\, \ldots,\, g_{m}^{h_{m}},\,
{\color{blue}\tilde{g}_{A, m+1}^{\oplus}},\, {\color{blue}\tilde{g}_{A, m+2}^{\ominus}},\,g_{m+3}^{h_{m+3}},\, \ldots,\, g_{n}^{h_{n}}
\big)   ,
\end{split}
\end{align}
where ${\tilde{g}}_{A,m}$ represents a gluino with flavour $A$ and momentum $p_m$.
 These amplitudes can be directly used in extracting the QCD MREVs according to the definitions in section~\ref{sec:PEVCEVdef}.

However, caution is required when dealing with amplitudes involving quark-antiquark pairs of two distinct flavours. The difference between the two theories arises because $\mathcal{N}=4$ sYM amplitudes in certain configurations include scalar exchanges that are absent in QCD. To clarify this, we first analyse a representative four-point amplitude and then straightforwardly extend the discussion to MREVs.

The colour-ordered Feynman rules in Figure \ref{fig:qcd_vertex} for the vertices $\phi_{[AB]}\tilde{g}_A\tilde{g}_B$ (scalar-gluino interaction) and $\bar{q} \slashed{A} q$ (quark-antiquark gluon coupling) reveal that, at tree level, the $\mathcal{N}=4$ sYM amplitude $A^{\mathcal{N}=4 \; \mathrm {sYM}}_{2\to 2}
\left(
{\color{purple}\tilde{g}_{B,1}^{\oplus}}, {\color{blue}\tilde{g}_{A,2}^{\oplus}}, {\color{purple}\tilde{g}_{B,3}^{\ominus}}, {\color{blue}\tilde{g}_{A,4}^{\ominus}}
\right)
$ is mediated exclusively by an $s$-channel scalar exchange,
	\begin{eqnarray}
    \label{qfojiwqwf}
			A^{\mathcal{N}=4 \; \mathrm {sYM}}_{2\to 2}
\left(
{\color{purple}\tilde{g}_{B,1}^{\oplus}}, {\color{blue}\tilde{g}_{A,2}^{\oplus}}, {\color{purple}\tilde{g}_{B,3}^{\ominus}}, {\color{blue}\tilde{g}_{A,4}^{\ominus}}
\right)
\eqn  
		\begin{tikzpicture}[baseline=(v1.base)]
			\begin{feynman}
				\vertex (v1) at (-1,0);
				\vertex (v2) at (1,0);
				
				\vertex [above left=1.5cm and 0.5cm of v1] (Q1) {$\tilde{g}_{A,2}^{\oplus}$};
				\vertex [below left=1.5cm and 0.5cm of v1] (q2) {$\tilde{g}_{B,1}^{\oplus}$};
				\vertex [above right=1.5cm and 0.5cm of v2] (Q3) {$\tilde{g}_{B,3}^{\ominus}$};
				\vertex [below right=1.5cm and 0.5cm of v2] (q4) {$\tilde{g}_{A,4}^{\ominus}$};
				
				\diagram*{
					(Q1) -- [fermion] (v1),
					(q2) -- [fermion, line width=1.5pt] (v1),
					(v2) -- [fermion, line width=1.5pt] (Q3),
					(v2) -- [fermion] (q4),
					(v1) -- [scalar] (v2)
				};
			\end{feynman}
		\end{tikzpicture}
	\end{eqnarray}
whereas its QCD counterpart vanishes, $A^{\mathrm{QCD}}_{2\to 2}\left(
{\color{purple}Q_1^{\oplus}},\,{\color{blue} q_2^{\oplus}},\, {\color{purple}\bar{Q}_3^{\ominus}},\, {\color{blue}\bar{q}_{4}^{\ominus}}\right) = 0$. In our diagrammatic representations here and below, a thin and bold fermion lines denotes distinct quark flavours in QCD amplitudes, and distinct $SU(4)$ R-charge indices in $\mathcal{N}=4$ sYM amplitudes.

\begin{figure}
	\centering
	\begin{subfigure}
		\centering
		\begin{tikzpicture}[baseline=(a.base)]
			\begin{feynman}
				\vertex (a) at (0,0);
				\vertex [above left=2cm of a] (gA) {$\tilde{g}_A^{h_A}$};
				\vertex [below left=1.5cm and 1.5cm of a] (gB) {$\tilde{g}_B^{h_B}$};
				\vertex [right=2cm of a] (phi) {$\phi$};
				\diagram* {
					(a) -- [fermion, line width=0.5pt] (gA),
					(a) -- [fermion, line width=1.5pt] (gB),
					(a) -- [scalar] (phi)
				};
			\end{feynman}
		\end{tikzpicture}
		\caption*{(a):  Vertex for \(\phi_{[AB]}\tilde{g}_A\tilde{g}_B\) in ${\cal N}=4$ sYM. The antisymmetry of \(\phi_{[AB]}\) in the \(SU(4)\) R-charge indices requires \(A \neq B\). When the gluinos are on-shell, angular momentum conservation further imposes \(h_A = h_B\) for their helicities.}
		\label{subfig:phi_gg}
	\end{subfigure}
	\quad
	\begin{subfigure}
		\centering
		\begin{tikzpicture}[baseline=(a.base)]
			\begin{feynman}
				\vertex (a) at (0,0);
				\vertex [above left=2cm of a] (q) {$q^{h_A}$};
				\vertex [below left=1.5cm and 1.5cm of a] (qbar) {$\bar{q}^{h_B}$};
				\vertex [right=2cm of a] (g) {$g$};
				\diagram* {
					(q) -- [fermion] (a) -- [fermion] (qbar),
					(a) -- [gluon] (g)
				};
			\end{feynman}
		\end{tikzpicture}
		\caption*{(b): Vertex for fermion-gluon interaction \(\bar{q}\,\slashed{A}\,q\). The flavour ($SU(4)$ index) along the quark (gluino) line is the same for both the incoming and outgoing fermions. When the fermions are massless and on-shell, they must have opposite helicities, \(h_A =- h_B\), for the interaction to take place, even if the gluon is off-shell.}
		\label{subfig:qgq}
	\end{subfigure}
	\caption{Relevant Feynman rules in $\mathcal{N}=4$ sYM (a and b) and in QCD (b)}
\label{fig:qcd_vertex}
\end{figure}

This generalises to amplitudes with an arbitrary number of additional gluon legs~\cite{Dixon:2010ik}. Therefore, QCD MREVs derived from the colour-ordered amplitudes of these flavour-flipping configurations vanish identically, i.e. 
\begin{align}
\label{Vanishing_QCD_CEVPEV}
\begin{split}
&C^{\mathrm{QCD}}
\big(\ldots,
{\color{purple} Q_1^{h_1}}, {\color{blue} q_2^{h_1}}, 
{\color{purple}\bar{Q}_3^{(-h_1)}}, 
{\color{blue}\bar{q}_4^{(-h_1)}},\ldots\big)=0,\\ 	&P^{\mathrm{QCD}}\big(\ldots,
{\color{purple} Q_1^{h_1}}, {\color{blue} q_2^{h_1}}, 
{\color{purple}\bar{Q}_3^{(-h_1)}}, 
{\color{blue}\bar{q}_4^{(-h_1)}},\ldots\big)=0,
\end{split}
\end{align} 
whereas their $\mathcal{N}=4$ sYM counterparts remain non-vanishing.
  
After excluding this case, one may expect that upon replacing gluinos by quarks
\hbox{$\{q\leftrightarrow \tilde{g}_A,\,\,  Q\leftrightarrow \tilde{g}_B\}$}, $\mathcal N=4$ sYM MREVs might map directly onto the QCD ones. However, there is another 
notable configuration that should be dealt with carefully, namely 
\begin{align}
\label{Scalar_effect}
\begin{split}
	A_{2\to 2}^{\mathrm{QCD}}
\big(
{\color{blue}\bar{q}_1^{\ominus}}, 
{\color{blue} q_2^{\oplus}}, 
{\color{purple}Q_3^{\oplus}}, {\color{purple}\bar{Q}_4^{\ominus}}
\big) 
	&= A_{2\to 2}^{\mathcal{N}=4 \,\mathrm {sYM}}
\bigl(
{\color{blue}\tilde{g}_{A,1}^{\ominus}},\, {\color{blue}\tilde{g}_{A,2}^{\oplus}},\, {\color{blue}\tilde{g}_{A,3}^{\oplus}},\, {\color{blue}\tilde{g}_{A,4}^{\ominus}}
\bigr) \\
	&\neq A^{\mathcal{N}=4 \, \mathrm {sYM}}_{2\to 2}\bigl(
{\color{blue}\tilde{g}_{A,1}^{\ominus}},\, {\color{blue}\tilde{g}_{A,2}^{\oplus}},\, {\color{purple}\tilde{g}_{B,3}^{\oplus}},\, {\color{purple}\tilde{g}_{B,4}^{\ominus}}\bigr).
    \end{split}
\end{align}  
This disparity is because, despite identical helicity configurations, the latter amplitude $A^{\mathcal{N}=4 \; \mathrm {sYM}}\big({\color{blue}\tilde{g}_{A,1}^{\ominus}},\, {\color{blue}\tilde{g}_{A,2}^{\oplus}},\, {\color{purple}\tilde{g}_{B,3}^{\oplus}},\, {\color{purple}\tilde{g}_{B,4}^{\ominus}}\big)$ involves scalar exchanges through the mismatch of $SU(4)$ R-charge indices (\(A \neq B\) in \(\tilde{g}_{A,2}^{\oplus}\) and \(\tilde{g}_{B,3}^{\oplus}\)), whereas the equality in the first line relies purely on $s$-channel gluon mediation. 
Indeed, the first line in Eq.~(\ref{Scalar_effect}) is described by 
\begin{eqnarray}\label{eq45}
A_{2\to 2}^{\mathrm{QCD}}\big(
{\color{blue}\bar{q}_1^{\ominus}}, 
{\color{blue} q_2^{\oplus}}, 
{\color{purple}Q_3^{\oplus}}, {\color{purple}\bar{Q}_4^{\ominus}}
\big)
\eqn
 \begin{tikzpicture}[baseline=(v1.base)]
  	\begin{feynman}
  		\vertex (v1) at (-1,0);
  		\vertex (v2) at (1,0);
  		
  		\vertex [above left=1.5cm and 0.5cm of v1] (qbar1) {$q_2^{\oplus}$};
  		\vertex [below left=1.5cm and 0.5cm of v1] (q2) {$\bar{q}_1^{\ominus}$};
  		
  		\vertex [above right=1.5cm and 0.5cm of v2] (Q3) {$Q_3^{\oplus}$};
  		\vertex [below right=1.5cm and 0.5cm of v2] (Qbar4) {$\bar{Q}_4^{\ominus}$};
  		
  		\diagram* {
  			(v1) -- [fermion] (q2),
  			(qbar1)    -- [fermion]      (v1),
  			(v1)    -- [gluon]        (v2),
  			(Q3)    -- [fermion, line width=1.5pt]      (v2),
  			(v2)    -- [fermion, line width=1.5pt]   (Qbar4)
  		};
  	\end{feynman}
  \end{tikzpicture}=
  \begin{tikzpicture}[baseline=(v1.base)]
  	\begin{feynman}
  		\vertex (v1) at (-1,0);
  		\vertex (v2) at (1,0);
  		\vertex [above left=1.5cm and 0.5cm of v1] (qbar1) {$\tilde{g}_A^{\oplus}$};
  		\vertex [below left=1.5cm and 0.5cm of v1] (q2) {$\tilde{g}_A^{\ominus}$};
  		\vertex [above right=1.5cm and 0.5cm of v2] (Q3) {$\tilde{g}_A^{\oplus}$};
  		\vertex [below right=1.5cm and 0.5cm of v2] (Qbar4) {$\tilde{g}_A^{\ominus}$};
  		
  		\diagram* {
  			(v1) -- [fermion] (q2),
  			(qbar1)    -- [fermion]      (v1),
  			(v1)    -- [gluon]        (v2),
  			(Q3)    -- [fermion]      (v2),
  			(v2)    -- [fermion]   (Qbar4)
  		};
  	\end{feynman}
  \end{tikzpicture}
\nonumber \\
&
  =& A_{2\to 2}^{\mathcal{N}=4 \; \mathrm {sYM}}\big(
{\color{blue}\tilde{g}_{A,1}^{\ominus}},\, {\color{blue}\tilde{g}_{A,2}^{\oplus}},\, {\color{blue}\tilde{g}_{A,3}^{\oplus}},\, {\color{blue}\tilde{g}_{A,4}^{\ominus}}
  \big),
\end{eqnarray}
while the second line by 
\begin{equation}\label{eq46ee}
    A_{2\to 2}^{\mathcal{N}=4 \mathrm {sYM}}\big(
{\color{blue}\tilde{g}_{A,1}^{\ominus}},\, {\color{blue}\tilde{g}_{A,2}^{\oplus}},\, {\color{purple}\tilde{g}_{B,3}^{\oplus}},\, {\color{purple}\tilde{g}_{B,4}^{\ominus}}   
\big)=
      \begin{tikzpicture}[baseline=(v00.base)]
  	\begin{feynman}
  		\vertex (v1) at (0,1);
  		\vertex (v2) at (0,-1);
  		\vertex (v00) at (0,0);
  		
  		\vertex [above left=0.5cm and 1.5cm of v1] (qbar1) {$\tilde{g}_A^{\oplus}$};
  		\vertex [below left=0.5cm and 1.5cm of v2] (q2) {$\tilde{g}_A^{\ominus}$};
  		
  		\vertex [above right=0.5cm and 1.5cm of v1] (Q3) {$\tilde{g}_B^{\oplus}$};
  		\vertex [below right=0.5cm and 1.5cm of v2] (Qbar4) {$\tilde{g}_B^{\ominus}$};
  		\diagram* {
  			(v2) -- [fermion] (q2),
  			(qbar1)    -- [fermion]      (v1),
  			(v1)    -- [scalar]        (v2),
  			(Q3)    -- [fermion, line width=1.5pt]      (v1),
  			(v2)    -- [fermion, line width=1.5pt]   (Qbar4)
  		};
  	\end{feynman}
  \end{tikzpicture}+
    \begin{tikzpicture}[baseline=(v00.base)]
  	\begin{feynman}
  		\vertex (v1) at (-1,0);
  		\vertex (v2) at (1,0);
  		\vertex (v00) at (0,0);
  		\vertex [above left=1.5cm and 0.5cm of v1] (qbar1) {$\tilde{g}_A^{\oplus}$};
  		\vertex [below left=1.5cm and 0.5cm of v1] (q2) {$\tilde{g}_A^{\ominus}$};
  		
  		\vertex [above right=1.5cm and 0.5cm of v2] (Q3) {$\tilde{g}_B^{\oplus}$};
  		\vertex [below right=1.5cm and 0.5cm of v2] (Qbar4) {$\tilde{g}_B^{\ominus}$};
  		
  		\diagram* {
  			(v1) -- [fermion] (q2),
  			(qbar1)    -- [fermion]      (v1),
  			(v1)    -- [gluon]        (v2),
  			(Q3)    -- [fermion, line width=1.5pt]      (v2),
  			(v2)    -- [fermion, line width=1.5pt]   (Qbar4)
  		};
  	\end{feynman}
  \end{tikzpicture}.
\end{equation}
This generalises to amplitudes with any number of additional gluon emissions. It also applies to the corresponding  MREVs:
\begin{align}
\label{MREVadjacentPlus}
\begin{split}
  	&
C^{\mathrm{QCD}}\big(
\ldots,
{\color{blue}\bar{q}_1^{\ominus}}, 
{\color{blue} q_2^{\oplus}}, 
{\color{purple}Q_3^{\oplus}}, {\color{purple}\bar{Q}_4^{\ominus}},\ldots\big)
=C^{\mathcal{N}=4\,  \mathrm {sYM}}\big(\ldots,{\color{blue}\tilde{g}_{A, 1}^{\ominus}}, {\color{blue}\tilde{g}_{A, 2}^{\oplus}}, 
{\color{blue}\tilde{g}_{A, 3}^{\oplus}}, {\color{blue}\tilde{g}_{A, 4}^{\ominus}},\ldots\big)\\
    &
P^{\mathrm{QCD}}
\big(
\ldots,
{\color{blue}\bar{q}_1^{\ominus}},
{\color{blue} q_2^{\oplus}}, 
{\color{purple}Q_3^{\oplus}}, {\color{purple}\bar{Q}_4^{\ominus}},\ldots\big)
=P^{\mathcal{N}=4\,  \mathrm {sYM}}\big(\ldots,{\color{blue}\tilde{g}_{A, 1}^{\ominus}}, 
{\color{blue}\tilde{g}_{A, 2}^{\oplus}}, 
{\color{blue}\tilde{g}_{A, 3}^{\oplus}}, 
{\color{blue}\tilde{g}_{A, 4}^{\ominus}},\ldots\big)\,,
\end{split}
  \end{align}
where the dots represent any number of gluons.
  
The amplitudes discussed so far, namely Eqs.~(\ref{qfojiwqwf}) and (\ref{eq46ee}), 
represent cases where sYM amplitudes differ from the corresponding QCD ones due to scalar propagation in helicity configurations where the two positive helicity fermions are adjacent (possibly interleaved by gluons).  Another configuration to consider is the alternating helicity case, where there are no such equal helicity fermions which are adjacent in the colour ordering, i.e. configurations of the form $\oplus, \ominus, \oplus, \ominus$ or $\ominus, \oplus, \ominus, \oplus$. These helicity configurations are incompatible with the propagation of a scalar, i.e. the interaction in Figure~\ref{fig:qcd_vertex} (a) does not occur. Therefore, such colour-ordered amplitudes in QCD can readily be identified  with their $\mathcal{N}=4$ sYM counterparts by interchanging $q \leftrightarrow \tilde{g}_A$ and $Q \leftrightarrow \tilde{g}_B$. This carries over to the corresponding MREVs, namely,
  \begin{align}
\label{MREVAlternatingHelicity}
  \begin{split}
  	&
    C^{\mathrm{QCD}}
    \big(\ldots, 
    {\color{blue} q_1^{\oplus}}, {\color{blue}\bar{q}_2^{\ominus}},
    {\color{purple}Q_3^{\oplus}}, {\color{purple}\bar{Q}_4^{\ominus}},\ldots\big)=C^{\mathcal{N}=4\,  \mathrm {sYM}}
    \big(\ldots,
    {\color{blue}\tilde{g}_{A, 1}^{\oplus}},
    {\color{blue}\tilde{g}_{A, 2}^{\ominus}},  {\color{purple}\tilde{g}_{B, 3}^{\oplus}}, {\color{purple}\tilde{g}_{B, 4}^{\ominus}},\ldots\big)
    \\
    &
    P^{\mathrm{QCD}}
    \big(\ldots, 
    {\color{blue} q_1^{\oplus}}, {\color{blue}\bar{q}_2^{\ominus}},
    {\color{purple}Q_3^{\oplus}}, {\color{purple}\bar{Q}_4^{\ominus}},\ldots\big)
    =P^{\mathcal{N}=4\,  \mathrm {sYM}}\big(\ldots,
    {\color{blue}\tilde{g}_{A, 1}^{\oplus}},
    {\color{blue}\tilde{g}_{A, 2}^{\ominus}},  {\color{purple}\tilde{g}_{B, 3}^{\oplus}}, {\color{purple}\tilde{g}_{B, 4}^{\ominus}},\ldots\big),
    \end{split}
  \end{align}
where the dots represent again any number of gluons.

The full set of configurations of flavour and helicity orderings of two quark-antiquark pairs of distinct flavours in QCD, and their correspondence with ${\cal N}=4$ sYM (where each quark-antiquark pair is replaced by a gluino pair with a distinct R-charge index) is summarised in Table~\ref{tab:twoFlavoursQCDsYM} based on the discussion above. In each case the table specifies the interaction, namely gluon exchange, scalar exchange, both (or none, where the result is trivially zero). In every case the QCD amplitude (MREV) can be inferred from the sYM one; where the scalar need to be eliminated it can be done by  identifying the R-charge index of the two pairs of gluinos in the sYM amplitude (MREV).
\begin{table}[h!]
  \centering
  \begin{tabular}{>{}l|c|c}
    \toprule
    & {Flavour alternating} & {Flavour adjacent} \\
    \midrule
    {Equal helicity adjacent flavours} & 
    \begin{tabular}{l}
    ${\color{purple} Q^{h}} {\color{blue} q^{h}} 
{\color{purple}\bar{Q}^{-h}} 
{\color{blue}\bar{q}^{-h}}$ 
\\
sYM: scalar \\
QCD: zero
\end{tabular}
& 
\begin{tabular}{l}
${\color{blue}\bar{q}^{-h}} 
{\color{blue} q^{h}}
{\color{purple}Q^{h}}{\color{purple}\bar{Q}^{-h}}$ 
\\
sYM: scalar, gluon \\
QCD: gluon 
\end{tabular}
\\
\hline\\
\vspace*{-18pt}\\
{Opposite helicity adjacent flavours} & 
\begin{tabular}{l}
    ${\color{purple} Q^{h}} {\color{blue} q^{-h}} 
{\color{purple}\bar{Q}^{-h}} 
{\color{blue}\bar{q}^{h}}$
\\
sYM: scalar \\
QCD: zero 
\end{tabular}
& 
\begin{tabular}{l}
${\color{blue} q^{h}} {\color{blue}\bar{q}^{-h}}
    {\color{purple}Q^{h}} {\color{purple}\bar{Q}^{-h}}$
    \\
sYM: gluon \\
QCD: gluon 
\end{tabular}
\\
    \bottomrule
  \end{tabular}
  \caption{All configurations of QCD amplitudes (or MREVs) involving two distinct quark flavours and their relation with the corresponding ${\cal N}=4$ sYM amplitudes (or MREVs) where the quark flavours are replaced by gluinos with distinct R-charge indices.  \label{tab:twoFlavoursQCDsYM}}
\end{table}

For configurations involving three or more distinct quark flavours, it becomes necessary to construct linear combinations of $\mathcal{N}=4$ sYM colour-ordered amplitudes to cancel scalar exchanges mediated by the Yukawa interaction $\phi_{AB}\tilde{g}_A\tilde{g}_B$; however, this systematic elimination lies beyond the scope of this work. We direct readers to \cite{Dixon:2010ik}  for further details.

Having seen the straightforward relation between QCD and ${\cal N}=4$ sYM colour-ordered amplitudes and MREVs for the cases of interest here -- the only special cases corresponding to Eqs. (\ref{Vanishing_QCD_CEVPEV}), (\ref{MREVadjacentPlus}) and
~(\ref{MREVAlternatingHelicity})
-- in what follows we will drop the superscript identifying the theory wherever it is unambiguous. In doing so we shall also avoid the making a distinction, at the level of the colour-ordered amplitude, between quarks and gluinos: fermion flavours will be denoted by $q$ and $Q$ throughout, even if these represent gluino flavours. This will be applied also in cases where a ``quark'' and an ``antiquark'' are interleaved by gluons in the colour ordering, which are only relevant in the sYM case, or in QCD, beyond tree level. We note that such colour-ordered amplitudes and MREVs do enter relations with QCD ones already at tree level, which will be useful in what follows.

\subsection{Relations among colour-ordered CEVs}
\label{relationCEVPEV}

\subsubsection*{Cyclicity}

Given that the external particles in amplitudes of gluon and gluino scattering are all in the adjoint representation, the full amplitudes can be written in the trace basis as follows:
\begin{eqnarray}
\hspace*{-8pt}
\mathcal{A}_n^{\mathcal{N}=4\,\mathrm{sYM}}\left(\left\{p_i, h_i, a_i\right\}\right)
    = g^{n-2} \sum_{\sigma \in S_n / Z_n}
    \operatorname{Tr}\big(T^{a_{\sigma_1}} \cdots T^{a_{\sigma_n}}\big)
    A_n^{\mathcal{N}=4\,\mathrm{sYM}}\big(p_{\sigma_1}^{h_{\sigma_1}}, \ldots, p_{\sigma_n}^{h_{\sigma_n}}\big)\,.
    \label{tracebasis}
\end{eqnarray}
Due to the cyclic symmetry of the trace basis, the colour-ordered amplitudes $A_n^{\mathcal{N}=4\,\mathrm{sYM}}$ also obey cyclic relations. 
For pure gluon amplitudes, Bose symmetry of the full, colour-dressed amplitudes implies that $A_n^{\mathcal{N}=4\,\mathrm{sYM}}$ are cyclically symmetric. 
For gluino amplitudes (and hence for corresponding quark amplitudes),  the fermionic nature of the fields, and hence the colour-dressed amplitude,
requires that a minus sign must be included whenever a cyclic permutation results in an odd permutation of quarks. Here, an odd permutation refers to the distribution of quarks of a given flavour, in $\{p_{\sigma_1}^{h_{\sigma_1}}, \ldots, p_{\sigma_n}^{h_{\sigma_n}}\}$ relative to the standard ordering $\{p_{1}^{h_{1}}, \ldots, p_{n}^{h_{n}}\}$  (where one ignores the ordering with respect to gluons or quarks of other flavours).
For example,
\begin{equation}
\begin{aligned}
A^{\mathcal{N}=4\,\mathrm{sYM}}\big(\tilde{g}_{A, 1}^{\ominus},\, \tilde{g}_{A, 2}^{\oplus},\, g_{3}^{\oplus},\, g_{4}^{\ominus}\big)
&= -A^{\mathcal{N}=4\, \mathrm{sYM}}\big(\tilde{g}_{A, 2}^{\oplus},\, g_{3}^{\oplus},\, g_{4}^{\ominus},\, \tilde{g}_{A, 1}^{\ominus}\big) \\
&=\,\,\,\,A^{\mathcal{N}=4\, \mathrm{sYM}}\big(g_{3}^{\oplus},\, g_{4}^{\ominus},\, \tilde{g}_{A, 1}^{\ominus},\, \tilde{g}_{A, 2}^{\oplus}\big)\\
&=\,\,\,\,A^{\mathcal{N}=4\, \mathrm{sYM}}\big( g_{4}^{\ominus},\, \tilde{g}_{A, 1}^{\ominus},\, \tilde{g}_{A, 2}^{\oplus},\,g_{3}^{\oplus}\big).
\end{aligned}
\end{equation}
From the discussion in section~\ref{qcd_and_sym_diff}, all non-vanishing QCD MREVs with up to two quark pairs have a one-to-one correspondence with their $\mathcal{N}=4\,\mathrm{sYM}$ counterparts. Thus, they are also cyclically symmetric, up to an additional sign for odd permutations of fermions.

\subsubsection*{Flipping helicities}
\label{fffff}
In the spinor-helicity formalism, flipping all helicities of an amplitude corresponds to exchanging the spinor components \( |p] \) and \( |p\rangle \). For real momenta, these spinors are related by Hermitian conjugation,
\begin{align}
\label{condgugateSpinors}
[p_i| = \mathrm{sign}(p_i^0)
\bigl(|p_i\rangle\bigr)^\dagger \quad \text{and} \quad 
\langle p_i| = 
\mathrm{sign}(p_i^0)
\bigl(|p_i]\bigr)^\dagger.
\end{align}
This relation implies that reversing the helicity of all particles transforms angle brackets \( \langle ij \rangle \) into square brackets, as in Eq.~(\ref{spinoridentities_1}), and vice versa.
By leveraging the spinor-to-momentum transformation, Eq.~(\ref{spinoconven}), and the mapping to MSLCV, Eq.~(\ref{minival:nonlorentzinvdd}), we observe that helicity reversal in the MSLCV corresponds to interchanging \( q_1^{\perp} \leftrightarrow \bar{q}_1^{\perp} \) and \( z_i \leftrightarrow \bar{z}_i \); that is, complex conjugation. For instance:
\begin{align}
C_{2}\Bigl(g_{q_{1}^\perp}^*,\,g_4^{\ominus},\,g_5^{\oplus},\,g_{q_{3}^\perp}^*\Bigr)
=
C_{2}\Bigl(g_{q_{1}^\perp}^*,\,g_4^{\oplus},\,g_5^{\ominus},\,g_{q_{3}^\perp}^*\Bigr)
\Bigg|_{z_i \leftrightarrow \bar{z}_i,\; q_1^{\perp} \leftrightarrow \bar{q}_1^{\perp}}.
\end{align}
We note that the sign of the energy entering (\ref{condgugateSpinors}) has no effect on the transformation of the MSLCVs under helicity flip. This can be readily seen from the definition in Eq.~(\ref{p_j_perp_general}), where the spinors of the incoming partons $1$ and $2$ enters in both the numerator and denominator of the transverse components, so the $\mathrm{sign}(p_i)$ cancels.

\subsubsection*{Charge conjugation}
\label{cc}
At tree level, the emission of a massless quark-antiquark pair within a CEV, just as in a full amplitude, is symmetric under charge conjugation. This is realised by simply reversing the direction of the quark electric charge flow~\cite{Bern:1994fz}. That is,
\begin{align}
\label{charge_conjugation_qqbar_first}
    C_{n-4}(\ldots, q_i^\oplus, \bar{q}_j^\ominus, \ldots) =  C_{n-4}(\ldots, \bar{q}_i^\oplus, q_j^\ominus, \ldots).
\end{align}

\subsubsection*{Photon decoupling identities}
We will also make use of photon decoupling identities for colour-ordered CEVs. That is, by inserting an on-shell gluon ${\color{blue} g_{\sigma_4}^{h_{\sigma_4}}}$ into all possible colour-ordered positions, we obtain
\begin{eqnarray}
0   &=&  C_{n-4}\left( 
{\color{blue}g_{\sigma_4}^{h_{\sigma_4}}}, g_{\sigma_5}^{h_{\sigma_5}},\ldots, g_{\sigma_m}^{h_{\sigma_m}}, 
g_{q_1^\perp}^*,g_{\sigma_{m+1}}^{h_{\sigma_{m+1}}},\ldots,  g_{\sigma_{n-1}}^{h_{\sigma_{n-1}}}, g_{q_{n-3}^\perp}^*
\right) \nonumber \\
&&
+ C_{n-4}\left( 
g_{\sigma_5}^{h_{\sigma_5}}, {\color{blue}g_{\sigma_4}^{h_{\sigma_4}}},\ldots, g_{\sigma_m}^{h_{\sigma_m}},g_{q_1^\perp}^*,
g_{\sigma_{m+1}}^{h_{\sigma_{m+1}}},\ldots,  g_{\sigma_{n-1}}^{h_{\sigma_{n-1}}},g_{q_{n-3}^\perp}^* \right) + \ldots \nonumber \\
&&
+ C_{n-4}\left( g_{\sigma_5}^{h_{\sigma_5}},\ldots,g_{\sigma_m}^{h_{\sigma_m}},g_{q_1^\perp}^*,g_{\sigma_{m+1}}^{h_{\sigma_{m+1}}},\ldots, g_{\sigma_{n-1}}^{h_{\sigma_{n-1}}}, {\color{blue}g_{\sigma_4}^{h_{\sigma_4}}},g_{q_{n-3}^\perp}^*\right).
\label{gluonphotondecoupling}
\end{eqnarray}

This is in direct analogy to the one identity for colour-ordered amplitudes. More complicated cases are discussed in section~\ref{ppdd}.

\subsubsection*{Reversal identity}

Finally, we note that colour-ordered CEVs satisfy reversal identities that originate from the Feynman rules. In particular, the identities we will use are, for pure gluonic CEVs,
\begin{align}
&C_{n-4}\Big(
   g_{\sigma_m}^{h_{\sigma_m}},
   \ldots,
   g_{\sigma_4}^{h_{\sigma_4}},
   g_{q_1^\perp}^*,
   g_{\sigma_{m+1}}^{h_{\sigma_{m+1}}},
   \ldots,
   g_{\sigma_{n-1}}^{h_{\sigma_{n-1}}},
   g_{q_{n-3}^\perp}^*
\Big)
\nonumber \\[1ex]
=&\ 
(-1)^{n} C_{n-4}\Big(
   g_{\sigma_{n-1}}^{h_{\sigma_{n-1}}},
   \ldots,
   g_{\sigma_{m+1}}^{h_{\sigma_{m+1}}},
   g_{q_1^\perp}^*,
   g_{\sigma_4}^{h_{\sigma_4}},
   \ldots,
   g_{\sigma_m}^{h_{\sigma_m}},
   g_{q_{n-3}^\perp}^*
\Big),
\end{align}
and for CEVs with one quark anti-quark pair
\begin{align}
\label{reversal_identity_qqbar_first}
&
C_{n-4}\Bigl(
   g_{\sigma_m}^{h_{\sigma_m}},
   \ldots,
   g_{\sigma_6}^{h_{\sigma_6}},
   \bar{q}_{5}^{h_{5}},
   q_{4}^{h_{4}},
   g_{q_1^\perp}^*,
   g_{\sigma_{m+1}}^{h_{\sigma_{m+1}}},
   \ldots,
   g_{\sigma_{n-1}}^{h_{\sigma_{n-1}}},
   g_{q_{n-3}^\perp}^*
\Bigr)
\nonumber\\[1ex]
=&\
(-1)^{n-1}C_{n-4}\Bigl(
   g_{\sigma_{n-1}}^{h_{\sigma_{n-1}}},
   \ldots,
   g_{\sigma_{m+1}}^{h_{\sigma_{m+1}}},
   g_{q_1^\perp}^*,
   q_{4}^{h_{4}},
   \bar{q}_{5}^{h_{5}},
   g_{\sigma_6}^{h_{\sigma_6}},
   \ldots,
   g_{\sigma_m}^{h_{\sigma_m}},
   g_{q_{n-3}^\perp}^*
\Bigr).
\end{align}
These identities are discussed in greater detail in section~\ref{reversalk}.

\subsection{The \texorpdfstring{$g^*g^* \to g$}{g* g* → g} central-emission vertex}
\label{Sec:one_gluon_CEV}
The $g^*g^*\to g$ CEV, often referred to as the one-gluon CEV or the Lipatov vertex~\cite{Lipatov:1976zz,Lipatov:1991nf}, is a fundamental building block for high-energy scattering processes involving Reggeized gluon exchange. It captures the leading-order gluon-Reggeon interactions in the MRK limit, and plays a central role in the Balitsky–Fadin-Kuraev-Lipatov (BFKL) formalism, which systematically resums logarithmic corrections in high-energy amplitudes~\cite{Kuraev:1977fs,Balitsky:1978ic}.

\begin{figure}
    \centering
    \includegraphics[width=0.3\linewidth]{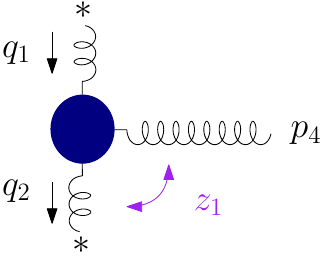}
    \caption{Illustration of the single--gluon CEV, the Lipatov vertex, with the relevant MSLCV.}
    \label{fig:lipatov}
\end{figure}

The colour-dressed form for the one-gluon CEV,  extracted from a five-point amplitude as described in section~\ref{defPEVandCEV}, is
\begin{equation}
\cC_1\left(g_{q_{1}^\perp}^*,g_4^{\oplus},g_{q_{2}^\perp}^*\right)= \big(F^{ d_4 }\big)_{c_1c_2}{C}_{1}\left(g_{q_{1}^\perp}^*,g_4^{\oplus},g_{q_{2}^\perp}^*\right), \label{eq:fulllipatov}
\end{equation}
where we recall that
\( \big(F^{ d_4}\big)_{c_1   c_2} =i\sqrt{2}f^{c_1 d_4 c_2} \) (see section~\ref{colourdressPEVCEV}) and where $c_i$ are the colour indices of the virtual $t$-channel gluons $g^*_{q^\perp_i}$ while $d_4$ is the colour index of the emitted on-shell gluon~$g_4$. The colour-ordered CEV in Eq.~(\ref{eq:fulllipatov}) is given by
\begin{equation}
{C}_{1}\left(g_{q_{1}^\perp}^*,g_4^{\oplus},g_{q_{2}^\perp}^*\right)= z_1 \bar{q}_{1}^\perp= -\frac{q_{2}^\perp}{p_4^\perp} \bar{q}_{1}^\perp\,,\label{lipatov}
\end{equation}
where $z_1$ is defined as in Eq.~(\ref{minival:nonlorentzinvdd}) or Eq.~(\ref{minival:lorentzinvdd}).
The kinematic dependence of the CEV is illustrated in Fig.~\ref{fig:lipatov}.  It only involves transverse momenta and diverges as the emitted gluon becomes soft, $p_4^\perp\to 0$. The emission of a negative-helicity gluon, $C_{1}\left(g_{q_1^\perp}^*, g_4^{\ominus}, g_{q_2}^*\right)$ is obtained by complex conjugation (see~section~\ref{relationCEVPEV}).

\subsection{Two-parton central-emission vertices}
\label{Sec:two_parton_CEV}

\begin{figure}[H]
        \centering
    \includegraphics[width=0.8 \linewidth]{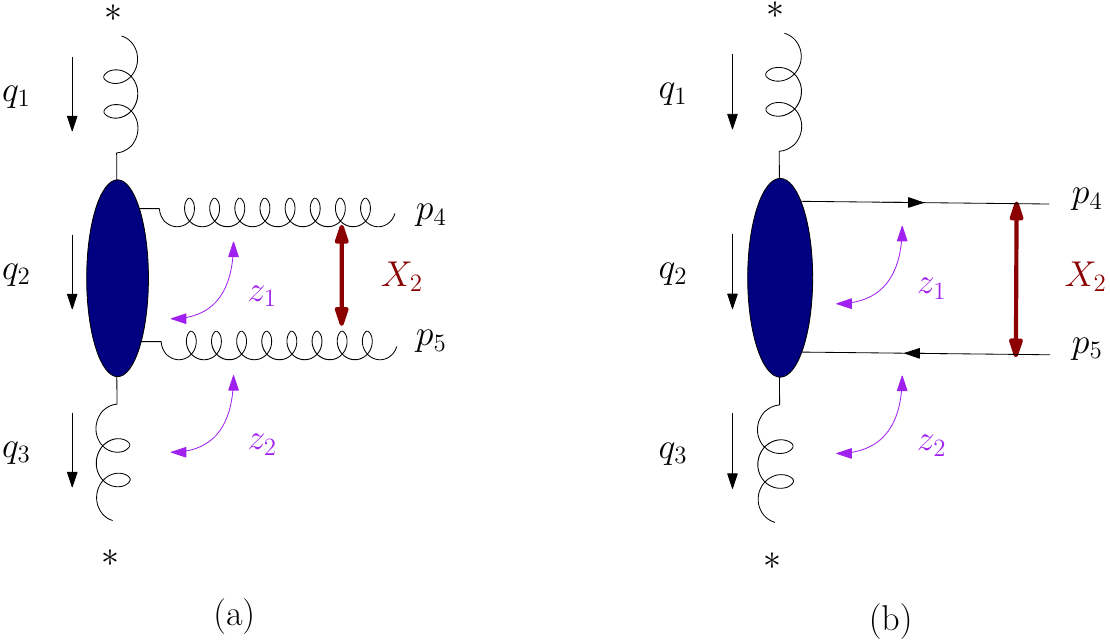}
    \caption{ Illustration of the two two-parton CEVs  with the relevant MSLCV: 
    (a) $g ^*g^* \rightarrow gg$; (b) $g ^*g^* \rightarrow q\bar{q}$.}
    \label{fig:qqbarggCEV} 
\end{figure}
The two-parton CEV is the natural first generalization of the one-gluon CEV in the factorization of amplitudes within the Next-to-Multi-Regge-Kinematics (NMRK) limit, where there is no strong rapidity ordering between the two emissions.
This vertex contributes to the real next-to-leading logarithmic (NLL) corrections to the Balitsky-Fadin-Kuraev-Lipatov (BFKL) kernel~\cite{Fadin:1989kf,DelDuca:1995ki,Fadin:1996nw}.
In section~\ref{ggCEV}, we discuss the two-gluon case, illustrated in Fig.~\ref{fig:qqbarggCEV} (a). In section~\ref{ggCEV}, we discuss the quark-antiquark CEV; see Fig.~\ref{fig:qqbarggCEV} (b).

\subsubsection{The \texorpdfstring{$g^*g^*\to gg$}{g* g* → gg} central-emission vertex}
\label{ggCEV}
 
In contrast to the one-gluon CEV in Eq.~\eqref{eq:fulllipatov}, the slight relaxation of the kinematic conditions in the two-parton CEV allows for additional colour channels. The colour dressing of this configuration is similar to that of the four-gluon amplitude. Using the DDM basis (\ref{gluonampcolour}), we have
\begin{eqnarray}
&&\hspace{-30pt}\cC_2\left(g_{q_{1}^\perp}^*,g_4^{h_4},g_5^{h_5},g_{q_{3}^\perp}^*\right) 
\label{qjk}
\\ \notag
\eqn 
\big(F^{d_4}F^{d_5}\big)_{c_1c_3}
{C}_{2}\left(g_{q_{1}^\perp}^*,g_4^{h_4},g_5^{h_5},g_{q_{3}^\perp}^*\right) 
+
\big(F^{d_5}F^{d_4}\big)_{c_1c_3}
{C}_{2}\left(g_{q_{1}^\perp}^*,g_5^{h_5},g_4^{h_4},g_{q_{3}^\perp}^*\right) \,,
\end{eqnarray}
where the off-shell gluons (Reggeons) carry adjoint colour indices \( c_1 \) and \( c_3 \) at the two ends of the rail of a half (cut) ladder graph (see Fig.~\ref{fig:qqbarggCEV}), while the two emissions, labelled by indices \( 4 \) and \( 5 \), form the rungs of this ladder.

Next, let us discuss the colour-ordered CEVs entering Eq.~(\ref{qjk}). These depend on the helicities of the emitted gluons, $h_4$ and $h_5$. 
There are two helicity configurations that we need to consider: the all-plus case, $h_4 = \oplus$ and $h_5 = \oplus$, and the mixed-helicity case, $h_4 = \oplus$ and $h_5 = \ominus$. The other helicity configurations can be obtained from these two by complex conjugation.

For the all-plus case, $h_4=h_5=\oplus$, the partial CEV can be extracted from MHV amplitudes giving
\begin{subequations}
\begin{align}
{C}_{2}\left(g_{q_{1}^\perp}^*,g_4^{\oplus},g_5^{\oplus},g_{q_{3}^\perp}^*\right)
&=
-\frac{\sqrt{X_2} z_1 z_2 \bar{q}_{1}^\perp}{(1-z_2) \langle 45 \rangle}, 
\label{high_energy_limit_1} \\ 
{C}_{2}\left(g_{q_{1}^\perp}^*,g_5^{\oplus},g_4^{\oplus},g_{q_{3}^\perp}^*\right) 
&=
-\frac{z_2 \bar{q}_{1}^\perp}{\sqrt{X_2} \langle 45 \rangle}\,,
\label{high_energy_limit_2}
\end{align}
\end{subequations}
with 
\begin{align}
\label{Angle45}
\langle 45\rangle
  =
  \frac{\left(1-z_2+X_2 z_1\right) q_{1}^\perp}{\sqrt{X_2} (1-z_1) (1-z_2)}
\,.
\end{align}
It is interesting to note that a momentum permutation operation can be defined directly in terms of the minimal set of light-cone variables (MSLCV). From the Lorentz-invariant definition of the MSLCV in~Eq.(\ref{minival:lorentzinvdd}), swapping the momenta $p_4$ and $p_5$ corresponds to
\begin{equation}
    \mathbb{P}_{54} = \left\{ X_2 \to \frac{1}{X_2},\quad z_1 \to  z_2
    +\frac{1- z_2}{z_1},\quad z_2 \to -\frac{z_1 z_2}{1-z_2} \right\}.
    \label{p4545}
\end{equation}
One may verify that~\eqref{high_energy_limit_1} and \eqref{high_energy_limit_2} are related by this transformation: \begin{equation}{C}_{2}\left(g_{q_{1}^\perp}^*,g_5^{\oplus},g_4^{\oplus},g_{q_{3}^\perp}^*\right)
    =
{C}_{2}\left(g_{q_{1}^\perp}^*,g_4^{\oplus},g_5^{\oplus},g_{q_{3}^\perp}^*\right)
    \Big|_{\mathbb{P}_{54}}.
\end{equation}
One may also observe that the first term in Eq.~(\ref{qjk}), namely $C_2\bigl(g_{q_1^\perp}^*, g_4^{\oplus}, g_5^{\oplus}, g_{q_3^\perp}^*\bigr)$ of Eq.~(\ref{high_energy_limit_1}), is leading upon taking the further high-energy limit $y_4 \gg y_5$, where the rapidity span between the two emitted gluons is getting large such that $X_2 \to \infty$. Conversely, the second term in Eq.~(\ref{qjk}), namely $C_{2}\bigl(g_{q_{1}^\perp}^*, g_5^{\oplus}, g_4^{\oplus}, g_{q_{3}^\perp}^*\bigr)$ of Eq.~(\ref{high_energy_limit_2}), is leading in the opposite limit, $y_5 \gg y_4$, where $X_2 \to 0$. In both these cases the dominant terms in Eqs.~\eqref{high_energy_limit_1} and~\eqref{high_energy_limit_2}, do not feature a collinear pole $\ab{4}{5}$, and yield a product of two one-gluon CEVs, Eq.~(\ref{lipatov}). This provides an example of  factorization properties which characterize CEVs under further high-energy limits.

The anti-MHV case, $h_4 = h_5 = \ominus$, is given by complex conjugation of in Eqs.~(\ref{high_energy_limit_1}) and~(\ref{high_energy_limit_2}).

For the mixed helicity configuration, where $h_4=\oplus$ and $h_5=\ominus$,
the partial CEV can be extracted from Next-to-MHV (NMHV) amplitudes, giving: 
\begin{eqnarray}
\label{C2mixed_hel}
\hspace*{-30pt}{C}_{2}\left(g_{q_{1}^\perp}^*,g_4^{\oplus},g_5^{\ominus},g_{q_{3}^\perp}^*\right) \label{eqgq}
  &=& \frac{X_2^{3/2} z_1^2 \bar{z}_1^2 \bar{z}_2^2 q_{1}^\perp (1-\bar{z}_2)}{\Omega_{\left\langle 5|q_3^\perp|+\right]} \Omega _{\langle+|P_{45}|-]} r_{+45} \langle45\rangle} \\ \notag
 &&
 + \, 
 \frac{z_2 q_{1}^\perp \bar{q}_{1}^\perp (1-\bar{z}_2)}{X_2^2 \Omega_{\left\langle 5|q_3^\perp|+\right]} \Omega_{\left\langle -|q_1|4\right]} s_{q_14}}
 +
 \frac{X_2^{3/2} z_2 \bar{q}_{1}^\perp}{\Omega_{\left\langle -|q_1^\perp|4\right]} \Omega _{\langle+|P_{45}|-]} r_{45-} [45]}\,,
\end{eqnarray}
where, in terms of our MSLCV, we have physical poles at $\langle45\rangle\to 0$ (see Eq.~(\ref{Angle45})) and also where any of the following kinematic variables vanish
\begin{subequations}
    \begin{align}
s_{q_1 4}
&= -\frac{\left(1+X_2 |z_1|^2 \right) |q_{1}^{\perp}|^2 }{X_2 |1-z_1|^2 }
\\[1ex]
r_{+45}
&\equiv
\lim_{X_1\to \infty}\frac{s_{345}}{s_{34}}
= 1+ \frac{X_2 |z_1|^2}{|1-z_2|^2}
\\[1ex]
r_{45-}
&\equiv
\lim_{X_3\to \infty}\frac{s_{456}}{s_{56}}
= 1+X_2\,.
\end{align} \label{eq:invratios2}
\end{subequations}
The $s_{q_1 4}$ is the invariant mass-squared of the internal propagator within the CEV. The pole \(r_{+45}\) represents a factorization channel involving particles 4 and 5 and any particle with significantly larger positive lightcone component than these two, which we label generically as `+'.  
The large rapidity gap separating between particles 4 and 5 and those with larger `+' components projects out all momentum dependence for this \(+\) particle, meaning \(r_{+45}\) depends only on the kinematics \textit{within} the CEV. 
The other factorization channel pole, $r_{45-}$, is defined in a similar manner. 

Within the physical kinematic region (\( p_i^0 > |p_i^z| > 0 \) for $i>2$), we have $p_i^+ \equiv p_i^0+p_i^z > 0$ and therefore \( X_2 > 0 \), which means that the poles associated with \( s_{q_{14}} \), \( r_{45-} \), and \( r_{+45} \) remain kinematically inaccessible.
However, when specific emissions ($i=4$ or $5$) are analytically continued to the incoming region, the energy component becomes negative, and since it dominates over spatial $z$ momentum component (\( p_i^0 < 0 \) with \( |p_i^0| > p_i^z \)) one can reach points with \( X_2=\frac{p_4^+}{p_5^+} < 0 \). Under this analytic continuation, the previously unreachable poles become accessible. 

In contrast to the above physical poles, the three \(\Omega\) factors appearing in the denominators of Eq.~(\ref{C2mixed_hel}), which we define explicitly below, correspond to \emph{spurious poles} which we have chosen to normalise in such a way that they asymptotically scale as \( \mathcal{O}(1) \) in the MRK limit where $X_2\to \infty$. Spurious poles in tree amplitudes can be eliminated by bringing them into a common denominator form which, however, usually leads to significantly more complicated expressions. 
Unlike physical singularities, these poles are artifacts of the BCFW shift used to derive the amplitude, and they originate from spinor strings in the shifted lower-point amplitude~\cite{Britto:2004ap,Britto:2005fq}. They are:
\begin{eqnarray}
\label{eq:Omegas1}
\Omega_{\left\langle 5|q_3^\perp|+\right]} 
\eqn
\bar{z}_1 \bar{z}_2\left(1-\bar{z}_1\right)\left|1-z_2\right|^2 
\underset{\text{\tiny
  \begin{tabular}{l}
    $X_1 \to \infty$ \\
    $X_3 \to \infty$
  \end{tabular}
}}{\text{lim}}
\frac{\jc{5}{6}{1}{2}}{\ab{5}{3}\sbb{3}{2}}
\non
\eqn 
-\frac{\bar{z}_1 \bar{z}_2 |1-z_2|^2  }{X_2 z_1}\left( \frac{X_2 s_{q_14} |1-z_1|^2) }{|q_{1}^\perp |^2}+z_2\right),
\end{eqnarray}
\begin{eqnarray}
\label{eq:Omegas2}
\Omega_{\left\langle -|q_1^\perp|4\right]}
\eqn 
\frac{\left(1-\bar{z}_1\right)}{\left|1-z_2\right|^2}
\underset{\text{\tiny
  \begin{tabular}{l}
    $X_1 \to \infty$ \\
    $X_3 \to \infty$
  \end{tabular}
}}{\text{lim}}\frac{\jc{1}{2}{3}{4}}{\ab{1}{6}\sbb{6}{4}}
\non \eqn 
\frac{\left(1+X_2 \bar{z}_1\right) (1-\bar{z}_1)}{X_2 \bar{z}_1 \bar{z}_2 (1-z_2)},
\end{eqnarray}
and
\begin{eqnarray}
\Omega _{\langle+|P_{45}|-]}
\eqn 
\frac{\left|1-z_2\right|^2}{\left(1-\bar{z}_1\right) }
\underset{\text{\tiny
  \begin{tabular}{l}
    $X_1 \to \infty$ \\
    $X_3 \to \infty$
  \end{tabular}
}}{\text{lim}}
\frac{\jc{3}{4}{5}{6}}{\ab{3}{5}\sbb{5}{6}}
\non
\eqn 
-\frac{|1-z_2|^2 \left(1-z_1-z_2\right)}{z_1 (1-\bar{z}_1)}. \label{eq:Omegas3}
\end{eqnarray}
Spurious poles typically cancel out pairwise, as they do in Eq.~(\ref{C2mixed_hel}), 
leaving only the physically meaningful singularities as previously discussed.

In the full MRK limit, $X_2\to \infty$, we find $\langle 45\rangle\sim \sqrt{X}_2$, $r_{+45},r_{45-}\sim X_2$, while~$s_{q_1 4}\sim 1$ and all $\Omega \sim 1$, and hence the first and the third terms of Eq.~(\ref{eqgq}) are leading. Adding them together indeed yields the product of two one-gluon CEVs,  Eq.~\eqref{lipatov}, with emission $g_4^\oplus$ and $g_5^\ominus$ respectively.

To construct the colour-dressed two-gluon CEV, Eq.~(\ref{qjk}), we also need 
the colour-ordered CEV 
$C_{2}\big(g_{q_{1}^\perp}^*, g_5^{\ominus}, g_4^{\oplus},g_{q_3^\perp}^*\big)$, which is related to $C_{2}\big(g_{q_{1}^\perp}^*, g_4^{\oplus}, g_5^{\ominus},g_{q_3^\perp}^*\big)$ by complex conjugation followed by momentum permutation ${\mathbb{P}_{54}}$. That is
\begin{align}
C_{2}\bigl(g_{q_{1}^\perp}^*, g_4^{\ominus}, g_5^{\oplus}, g_{q_3^\perp}^*\bigr)
 &=  C_{2}\bigl(g_{q_{1}^\perp}^*, g_4^{\oplus}, g_5^{\ominus}, g_{q_3^\perp}^*\bigr)
      \Big|_{\substack{
          q_1^{\perp}\leftrightarrow \bar{q}_1^{\perp} \\
          z_i \leftrightarrow \bar{z}_i
        }}
\nonumber\\[1ex]
C_{2}\bigl(g_{q_{1}^\perp}^*,g_5^{\ominus},g_4^{\oplus},g_{q_{3}^\perp}^*\bigr)
  &= C_{2}\bigl(g_{q_{1}^\perp}^*,g_4^{\ominus},g_5^{\oplus},g_{q_{3}^\perp}^*\bigr)
      \Big|_{\mathbb{P}_{54}}.
\end{align}
The remaining helicity configurations of colour-dressed two-gluon CEVs, $$\left\{\mathcal{C}_2\big(g_{q_1^\perp}^*, g_4^{\ominus}, g_5^{\ominus}, g_{q_3^\perp}^*\big), \mathcal{C}_2\big(g_{q_1^\perp}^*, g_5^{\ominus}, g_4^{\ominus}, g_{q_3^\perp}^*\big), \mathcal{C}_2\big(g_{q_1^\perp}^*, g_5^{\oplus}, g_4^{\ominus}, g_{q_3^\perp}^*\big)\right\},$$ may be obtained, respectively, by complex conjugation of~$$\left\{\mathcal{C}_2\big(g_{q_1^\perp}^*, g_4^{\oplus}, g_5^{\oplus}, g_{q_3^\perp}^*\big), \mathcal{C}_2\big(g_{q_1^\perp}^*, g_5^{\oplus}, g_4^{\oplus}, g_{q_3^\perp}^*\big), \mathcal{C}_2\big(g_{q_1^\perp}^*, g_5^{\ominus}, g_4^{\oplus}, g_{q_3^\perp}^*\big)\right\}.$$ 

\subsubsection{The \texorpdfstring{$g^*g^*\to q\bar{q}$}{g* g* → q\textbackslash{}overline\{q\}} central-emission vertex}
We now analyze the central emission of a quark-antiquark pair, another real-emission building block in the NLL QCD BFKL kernel~\cite{Fadin:1996nw,DelDuca:1996nom,Fadin:1989kf}. The interaction dynamics here differ from pure gluonic emissions due to the quark’s fundamental representation of $SU(N_c)$.

Specifically, the colour dressing of the CEV is analogous to that of the $ggq\bar{q}$ amplitude (i.e. the $n=4$ case of Eq.~(\ref{onequark_pair})), with the quarks taken on-shell but the gluons (Reggeons) off-shell:
\begin{eqnarray}
&&
\hspace*{-30pt}\mathcal{C}_2^{ \bar{q} q}\big( g_{q_{1}^\perp}^*,{q}_4^{h_4}, \bar{q}_5^{h_5},g_{q_{3}^\perp}^*\big)
  \\
  \eqn
  \left(T^{c_1} T^{c_3} \right)_{i_5 \bar{\imath}_4} C_{2}\big(g_{q_{1}^\perp}^*,{q}_4^{h_4}, \bar{q}_5^{h_5} ,g_{q_{3}^\perp}^*\big)+\big(T^{c_3} T^{c_1} \big)_{i_5 \bar{\imath}_4} C_{2}\big({q}_4^{h_4},\bar{q}_5^{h_5}, g_{q_{1}^\perp}^*,g_{q_{3}^\perp}^* \big),\notag
\end{eqnarray}
where $T^c$ denotes the $SU(N_c)$ generators in the fundamental representation, normalised as explained in section~\ref{colourdressPEVCEV}. The quark and anti-quark must have opposite helicities, so these CEVs are extracted from NMHV amplitudes.

For the helicity assignment $h_4 = \oplus$ and $h_5 = \ominus$, we find
\begin{align}
\label{qqb}
\begin{split}
C_{2}\big(g_{q_{1}^\perp}^*,q_4^{\oplus},\bar q_5^{\ominus},g_{q_{3}^\perp}^*\big) 
 =
 &- \frac{z_1^2 z_2^2 \bar{q}_{1}^\perp (1-z_2) (1-\bar{z}_2)^3}{\bar{\Omega }_{\left\langle 5|q_3|+\right]} \bar{\Omega }_{\langle+|P_{45}|-]} \bar{z}_1 r_{+45} [45]}
\\ & - 
 \frac{z_1^3 \bar{z}_2 q_{1}^\perp \bar{q}_{1}^\perp (1-z_2)}{\sqrt{X_2} \bar{\Omega }_{\left\langle 5|q_3|+\right]} \bar{\Omega }_{\left\langle -|q_1|4\right]} s_{q_14}}
 - 
 \frac{\bar{z}_2 q_{1}^\perp}{\bar{\Omega }_{\left\langle -|q_1|4\right]} \bar{\Omega }_{\langle+|P_{45}|-]} r_{45-} \langle45\rangle},
 \end{split}
\end{align}
where the $\bar{\Omega}$ factors are obtained from Eqs.~(\ref{eq:Omegas1}) to~(\ref{eq:Omegas3}) 
by complex conjugation. To complete the colour dressing we need
\begin{eqnarray}
&&
 C_{2}\big(g_{q_{1}^\perp}^*,{q}_4^{\ominus},\bar{q}_5^{\oplus}, g_{q_{3}^\perp}^* \big)= C_{2}\big(g_{q_{1}^\perp}^*,q_4^{\oplus},\bar q_5^{\ominus},g_{q_{3}^\perp}^*\big) \big|_{z_i\leftrightarrow \bar{z}_i,\;q_1^\perp \leftrightarrow \bar{q}_1^\perp}\;,
\non
&& C_{2}\big({q}_4^{\oplus},\bar{q}_5^{\ominus}, g_{q_{1}^\perp}^*,g_{q_{3}^\perp}^* \big)
  =
  C_{2}\big(g_{q_{1}^\perp}^*,\bar{q}_5^{\ominus},{q}_4^{\oplus}, g_{q_{3}^\perp}^* \big)
=
  C_{2}\big(g_{q_{1}^\perp}^*,{q}_4^{\ominus},\bar{q}_5^{\oplus}, g_{q_{3}^\perp}^* \big)\big|_{\mathbb P_{54}},
  \label{eq46}
\end{eqnarray}
Notably, all terms in Eq.~(\ref{qqb}) exhibit power suppression in the MRK limit $X_2\to\infty$, in contrast to the  previous gluonic CEVs. This is to be expected from Regge scaling~\cite{Regge:1959mz}, where the $t$-channel exchange of a spin-$\frac{1}{2}$ quark is power suppressed compared to the $t$-channel exchange of a spin-$1$ gluon. 

The remaining helicity configurations: $\mathcal{C}_2^{\bar{q} q}\big(g_{q_1^\perp}^*, q_4^{\ominus}, \bar{q}_5^{\oplus}, g_{q_3^\perp}^*\big)$, $\mathcal{C}_2^{\bar{q} q}\big(g_{q_1^\perp}^*, \bar q_4^{\ominus}, {q}_5^{\oplus}, g_{q_3^\perp}^*\big)$ and 
$\mathcal{C}_2^{\bar{q} q}\big(g_{q_1^\perp}^*, \bar q_4^{\oplus}, {q}_5^{\ominus}, g_{q_3^\perp}^*\big)$ can be obtained through complex conjugation or charge conjugation.

\subsection{Three-parton central-emission vertices}
\label{Sec:three_parton_CEV}

\begin{figure}[H]
    \centering
\includegraphics[width=0.8\linewidth]{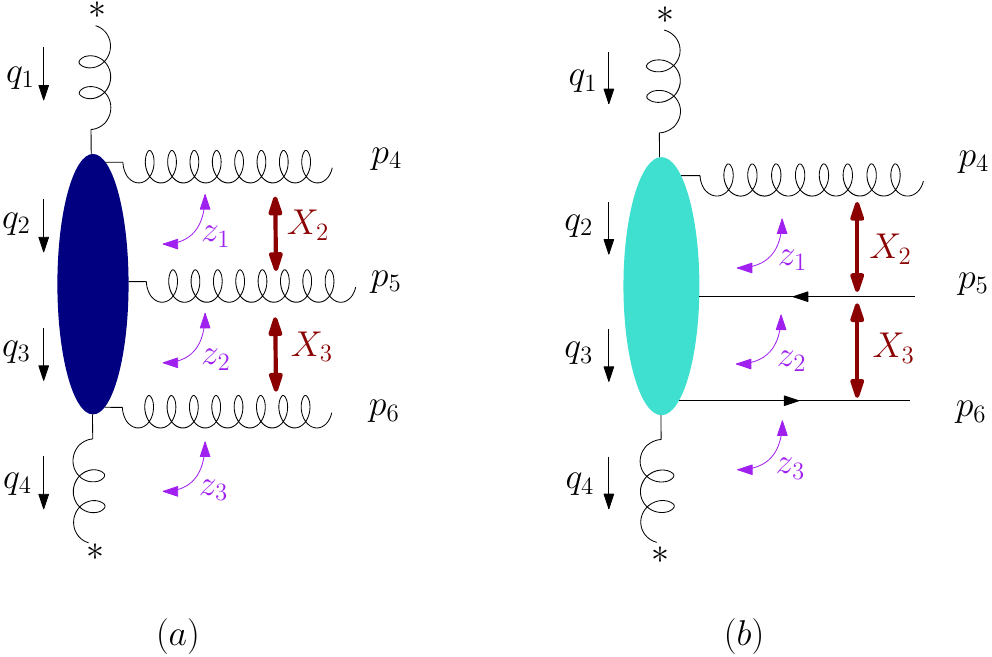}
    \caption{Illustration of the two three-parton CEVs with the relevant MSLCV: (a) $g ^*g^* \rightarrow g gg$;  (b) $g ^*g^* \rightarrow g \bar{q} q$. }
    \label{puregluonCEVggg}
\end{figure}

Having catalogued all the QCD CEVs contributing to the NLL BFKL real corrections in the previous subsection, we now present the three-parton CEVs, which are building blocks for purely real corrections to the NNLL BFKL equation. These vertices are derived from the Next-to-Next-to-Multi-Regge-Kinematic (NNMRK) limit of QCD amplitudes, where the requirement of strong ordering is dropped among a group of three emissions.

There are two types of colour-dressed three-parton CEVs: the pure gluon case, $\mathcal{C}_3^{ggg}$, as shown in Figure~\ref{puregluonCEVggg}$(a)$, and the case with a quark-antiquark pair and a single gluon, $\mathcal{C}_3^{g \bar{q} q}$, as shown in Figure~\ref{puregluonCEVggg}$(b)$. The colour structure is similar to that for five-parton amplitudes,
 \begin{equation}
     \begin{aligned}
\mathcal{C}_3^{ggg}
\big(
g_{q_{1}^\perp}^*,
g_4^{h_4}, 
g_5^{h_5},
g_6^{h_6},
g_{q_{4}^{\perp}}^*
\big)= \,
 g^3\bigg[&
\big(F^{d_4} F^{d_5} F^{d_6}\big)_{c_1c_4} C_{3}\big( g_{q_{1}^\perp}^*,g_4^{h_4}, g_5^{h_5}, g_6^{h_6},g_{q_{4}^{\perp}}^*\big)\\
&
+
\big(F^{d_4} F^{d_6} F^{d_5}\big)_{c_1c_4} C_{3}\big( g_{q_{1}^\perp}^*,g_4^{h_4}, g_6^{h_6}, g_5^{h_5},g_{q_{4}^{\perp}}^*\big)\\
&
+
\big(F^{d_5} F^{d_4} F^{d_6}\big)_{c_1c_4} C_{3}\big( g_{q_{1}^\perp}^*,g_5^{h_5},g_4^{h_4}, g_6^{h_6}, g_{q_{4}^{\perp}}^*\big)\\
&
+
\big(F^{d_5} F^{d_6} F^{d_4}\big)_{c_1c_4} C_{3}\big( g_{q_{1}^\perp}^*,g_5^{h_5},g_6^{h_6},g_4^{h_4},  g_{q_{4}^{\perp}}^*\big)\\
&
+
\big(F^{d_6} F^{d_4} F^{d_5}\big)_{c_1c_4} C_{3}\big( g_{q_{1}^\perp}^*, g_6^{h_6},g_4^{h_4},g_5^{h_5}, g_{q_{4}^{\perp}}^*\big)\\
&
+
\big(F^{d_6} F^{d_5} F^{d_4}\big)_{c_1c_4} C_{3}\big( g_{q_{1}^\perp}^*,g_6^{h_6},g_5^{h_5},g_4^{h_4},  g_{q_{4}^{\perp}}^*\big)
\bigg]
\end{aligned}
\label{quiq2}
 \end{equation} 
and  \begin{equation} \begin{aligned}
\mathcal{C}_3^{g \bar{q} q}\big(g_{q_{1}^\perp}^*,g_4^{\nu_4}, \bar{q}_5^{\nu_5}, q_6^{\nu_6},g_{q_{4}^{\perp}}^*\big)= 
 g^3\bigg[&\big(T^{c_4} T^{c_1} T^{d_4}\big)_{i_6 \bar{\imath}_5} C_{3}\big( g_{q_{1}^\perp}^*,g_4^{\nu_4}, \bar{q}_5^{\nu_5}, q_6^{\nu_6},g_{q_{4}^{\perp}}^*\big) \\
& 
+\big(T^{c_1} T^{c_4} T^{d_4}\big)_{i_6 \bar{\imath}_5}C_{3}\big(g_4^{\nu_4}, \bar{q}_5^{\nu_5}, q_6^{\nu_6},g_{q_{1}^\perp}^*,g_{q_{4}^{\perp}}^*\big) \\
& +\big(T^{c_4} T^{d_4} T^{c_1}\big)_{i_6 \bar{\imath}_5} C_3\big(g_4^{\nu_4}, g_{q_1^{\perp}}^*, \bar{q}_5^{\nu_5}, q_6^{\nu_6}, g_{q_4^{\perp}}^*\big) \\
& +\big(T^{c_1} T^{d_4} T^{c_4}\big)_{i_6 \bar{\imath}_5} C_3\big(\bar{q}_5^{\nu_5}, q_6^{\nu_6}, g_{q_1^{\perp}}^*, g_4^{\nu_4}, g_{q_4^{\perp}}^*\big)\\
&+\big(T^{d_4} T^{c_4} T^{c_1}\big)_{i_6 \bar{\imath}_5} C_{3}\big(g_{q_{1}^\perp}^*,\bar{q}_5^{\nu_5}, q_6^{\nu_6},g_4^{\nu_4},g_{q_{4}^{\perp}}^*\big) \\
& +\big(T^{d_4} T^{c_1} T^{c_4}\big)_{i_6 \bar{\imath}_5} C_{3}\big(\bar{q}_5^{\nu_5}, q_6^{\nu_6}, g_4^{\nu_4},g_{q_{1}^\perp}^*,g_{q_{4}^{\perp}}^*\big) \bigg],
\end{aligned}
\label{ciuwe}
 \end{equation}
which should be compared with the $n$=5 cases of Eq.~(\ref{gluonampcolour}) and Eq.~(\ref{onequark_pair}) respectively.
We proceed to formulate these colour-ordered CEVs in the MSLCV following a brief overview of three-parton momentum permutations in section~\ref{pp}.
 \subsubsection{Permutation of three final-state momenta in the MSLCV}
\label{pp}

To construct the colour-dressed three-parton CEVs, one must permute both the helicities and momenta of the three external legs of the colour-ordered CEVs. There are no relations between different helicity configurations other than complete reversal of the spins, which corresponds to complex conjugation; hence, all independent configurations must be determined. In contrast, momentum permutations adhere to the dihedral group \( D_3 \) (the symmetry group of an equilateral triangle encompassing 2 rotations and 3 reflections). 

Let \(\mathbb{P}_{ijk}\) denote the permutation operator that maps the reference indices \(456\) to the configuration \(ijk\), namely
$4\rightarrow i,\quad 5\rightarrow j,\quad 6\rightarrow k$.
By swapping momentum labels in the Lorentz-invariant definition of MSLCV, Eq.~(\ref{minival:lorentzinvdd}), we derive the following momentum permutation rules implemented in MSLCV, which correspond to the generators of the dihedral group~$D_3$:
\begin{itemize}
	\item 
Cyclic permutation $4 \rightarrow 5, \quad 5 \rightarrow 6, \quad 6 \rightarrow 4,$
	\begin{equation}
	    \mathbb P_{564}=\left\{\begin{array}{l}
	X_2\to X_3,X_3\to \frac{1}{X_2 X_3},\\ z_1\to \frac{z_1 z_2-z_2+1}{z_1},z_2\to \frac{z_1 z_3 z_2-z_3 z_2+z_2+z_3-1}{z_1 z_2},z_3\to \frac{z_1 z_2 z_3}{\left(z_2-1\right)
		\left(z_3-1\right)}
	\end{array}\right\}.
    \label{per31}
	\end{equation}
	\item 
Cyclic permutation $
4 \rightarrow 6, \quad 5 \rightarrow 4, \quad 6 \rightarrow 5,
$
\begin{align}
\mathbb P_{645}=\left\{\begin{array}{l}
	X_2\to \frac{1}{X_2 X_3},X_3\to X_2,\\z_1\to \frac{z_2 z_3 z_1-z_3 z_1+z_1+z_2-z_2 z_3+z_3-1}{z_1 z_2}, z_2\to \frac{z_1 \left(z_2 z_3-z_3+1\right)}{\left(z_2-1\right)
		\left(z_3-1\right)},z_3\to \frac{z_2 z_3}{z_3-1}
	\end{array}\right\}.
    \label{per32}
\end{align}
\item Reflection $ 4 \leftrightarrow 5,$ 
\begin{equation}
    \mathbb P_{546}=
{ {\left\{\begin{array}{l}
		X_2\to \frac{1}{X_2},X_3\to X_2 X_3,\\ z_1\to \frac{z_1
			z_2-z_2+1}{z_1},z_2\to \frac{z_1 z_2}{z_2-1},z_3\to z_3
		\end{array}\right\} }}.
    \label{per33}
\end{equation}
      This is closely related to the two-parton momentum permutation \(\mathbb{P}_{54}\), Eq.~(\ref{p4545}), defined when deriving the two-parton CEVs. 
	\item 
Reflection $5 \leftrightarrow 6,$  
\begin{equation}
    \mathbb P_{465}=\left\{\begin{array}{l}
	X_2\to X_2 X_3,X_3\to \frac{1}{X_3},\\z_1\to z_1,z_2\to \frac{z_2 z_3-z_3+1}{z_2},z_3\to \frac{z_2 z_3}{z_3-1}
	\end{array}\right\}.
    \label{per34}
\end{equation}
	\item Reflection $ 4 \leftrightarrow 6,$
\begin{equation}
\mathbb P_{654}=\left\{\begin{array}{l}
	X_2\to \frac{1}{X_3},X_3\to \frac{1}{X_2},\\z_1\to \frac{(z_2-1) z_3 z_1+z_1+z_2-z_2 z_3+z_3-1}{z_1 z_2},
	z_2\to \frac{(z_1-1) z_3 z_2+z_2+z_3-1}{z_1
		\left(z_3-1\right)},z_3\to \frac{z_1 z_2 z_3}{\left(z_2-1\right) \left(z_3-1\right)}
	\end{array}\right\}.
    \label{exanscwp}
\end{equation}
	
\end{itemize}

Since these permutations only involve outgoing particles, the MSLCV implementations of such momentum exchanges do not induce mixing between transverse and longitudinal components. This is evident in Eqs.~(\ref{per31}) through~\eqref{exanscwp}, which never mix the $X$-type and $z$-type variables. 
This may be contrasted with a momentum swap involving an incoming particle, which generally results in mixing between longitudinal and transverse MSLCVs, for instance, \( z \to f(z, X) \)~\cite{Byrne:2023nqx}.

With these transformations of momenta  established, we now proceed to present the complete set of colour-ordered three-parton CEVs. We begin with pure-gluon configurations, followed by CEVs containing a quark-antiquark pair. 

\subsubsection{Colour-ordered \texorpdfstring{$g^*g^*\to ggg$}{g* g* → ggg} central-emission vertices}

\noindent 
\textbf{Equal-helicity three-gluon CEVs}

We begin with the {all-plus configuration},
$h_4 = h_5 = h_6 = \oplus$,
which can be derived from an MHV Yang-Mills amplitude.
The colour-ordered CEVs required for the dressed CEV in Eq.~\eqref{quiq2} in this case are given by
\begin{subequations}
\begin{align}
C_{3}(g_{q_{1}^\perp}^*,g^\oplus_4,g^\oplus_5 ,g^\oplus_6,g_{q_{4}^{\perp}}^*)
&=
\frac{\sqrt{X_2 X_3} z_1 z_2 z_3 \bar{q}_{1}^\perp}{ \langle45\rangle  \langle56\rangle (1-z_2) (1-z_3)}
\label{qwkqwk1} \\ 
C_{3}(g_{q_{1}^\perp}^*,g^\oplus_4,g^\oplus_6,g^\oplus_5 ,g_{q_{4}^{\perp}}^*)
&= 
-\frac{\sqrt{X_2} z_1 z_2 z_3 \bar{q}_{1}^\perp}{ \langle46\rangle  \langle56\rangle (1-z_2) (1-z_3)}
\label{qwkqwk2} \\ 
C_{3}(g_{q_{1}^\perp}^*,g^\oplus_5,g^\oplus_4,g^\oplus_6 ,g_{q_{4}^{\perp}}^*)
&= 
\frac{\sqrt{X_3} z_2 z_3 \bar{q}_{1}^\perp}{ \langle45\rangle  \langle46\rangle (1-z_3)}
\label{qwkqwk3} \\ 
C_{3}(g_{q_{1}^\perp}^*,g^\oplus_5,g^\oplus_6,g^\oplus_4 ,g_{q_{4}^{\perp}}^*)
&= 
\frac{z_2 z_3 \bar{q}_{1}^\perp}{  \langle46\rangle  \langle56\rangle (1-z_3) \sqrt{X_2}}
\label{qwkqwk4} \\ 
C_{3}(g_{q_{1}^\perp}^*,g^\oplus_6,g^\oplus_4,g^\oplus_5 ,g_{q_{4}^{\perp}}^*)
&= 
-\frac{z_3 \bar{q}_{1}^\perp}{ \langle45\rangle  \langle46\rangle \sqrt{X_3} }
\label{qwkqwk5} \\ 
C_{3}(g_{q_{1}^\perp}^*,g^\oplus_6,g^\oplus_5,g^\oplus_4 ,g_{q_{4}^{\perp}}^*)
&= 
\frac{z_3 \bar{q}_{1}^\perp}{  \langle45\rangle  \langle56\rangle \sqrt{X_2 X_3}}.
\label{qwkqwk}
\end{align}
\end{subequations}
We note that these can be obtained from each other by applying the momentum permutations in Eqs.~(\ref{per31})--(\ref{exanscwp}).
 The collinear poles in the MSLCV representation are given by:
 \begin{subequations}
     \begin{align}
    \label{cccppp}
 \langle46\rangle= & \frac{q_{1}^\perp  \left(1-z_2-z_3+z_2 z_3-X_2 X_3 z_1 z_2\right)}{\sqrt{ X_2 X_3} (1-z_3) (1-z_2) (1-z_1)} \\
 \langle45\rangle =& \frac{q_{1}^\perp \left(1-z_2+X_2 z_1\right)}{\sqrt{X_2} (1-z_2) (1-z_1)} \\
 \langle56\rangle = &-\frac{q_{1}^\perp  \left(1-z_3+ X_3 z_2\right) z_1}{\sqrt{X_3} (1-z_3) (1-z_2) (1-z_1)} .
\end{align}
 \end{subequations}
In the next-to-MRK (NMRK) limit with an additional large rapidity gap between emissions 4 and 5, these collinear poles scale as $\langle 45\rangle\sim \sqrt{X}_2,\ \langle 56\rangle\sim 1, \ \langle 46\rangle\sim \sqrt{X}_2$.  Similarly, in the alternative NMRK limit with an additional large rapidity gap between 5 and 6, the scaling is $\langle 45\rangle\sim 1,\ \langle 56\rangle\sim \sqrt{X_3}, \ \langle 46\rangle\sim \sqrt{X_3}$.  These are  consistent with the full MRK limit, where the rapidity gaps between emissions 4, 5, and 6 are all large, and the poles scale as $\langle 45\rangle\sim \sqrt{X_2},\ \langle 56\rangle\sim \sqrt{X_3}, \ \langle 46\rangle\sim \sqrt{X_2 X_3}$.

Finally, the all-minus helicity configuration is obtained from the above by complex conjugation.

\noindent
\textbf{Mixed-helicity three-gluon CEVs} 

In order to compactly present the kinematic dependence of mixed-helicity CEVs we introduce shorthand notation for both physical and spurious poles, in analogy with those used in the context of the two-parton CEV in section~\ref{ggCEV}. 

Firstly, the physical poles are given by the following multi-parton invariant masses:
\begin{eqnarray}
 s_{q_14}=-\frac{|q_1^{\perp}|^2}{|1-z_1|^2}\bigg[\frac{  1}{X_2 X_3 }+\frac{ 1}{X_2  }+|z_1|^2\bigg]\,,
 \label{eq440}
\end{eqnarray}
\begin{eqnarray}
 s_{6q_4}=- \frac{|q_1^{\perp}|^2}{|1-z_1|^2}\bigg[\frac{1}{X_2 X_3 }+
 \frac{|z_1|^2}{ X_3  |1-z_2|^2}+
 \frac{|z_1|^2 |z_2|^2}{ |1-z_2|^2}\bigg]
\end{eqnarray}
and
\begin{equation}
    \begin{aligned}
      s_{456} 
      =  \,\,    \frac{|q_1^{\perp}|^2}{|1-z_1|^2}\bigg[ &
        \frac{1}{X_2}\left(1+\frac{1}{X_3}\right)
      +\frac{|z_1|^2 }{  |1-z_2|^2}
 \left(X_2+\frac{1}{X_3}\right)
      +
         \frac{(X_2+1) X_3 |z_1|^2|z_2|^2 }{ |1-z_2|^2 |1-z_3|^2}
        \\
        -& \frac{\left(\bar{z}_1 \left(\bar{z}_2+\bar{z}_3-1\right) (1-z_2) (1-z_3)-z_1 \bar{z}_1 \bar{z}_2 (1-z_3)+c.c.\right) }{ |1-z_2|^2 |1-z_3|^2}\bigg]\,,
    \end{aligned}
\end{equation}
where $c.c.$ means $z_i \leftrightarrow \bar{z}_i$. The relevant invariant-mass ratios (analogous to Eq.~(\ref{eq:invratios2})) are given by:
\begin{subequations}
  \begin{align}
  r_{56-} & = 1 + X_3 
  & \qquad
  r_{456-} & = 1 + X_3\,(1+X_2) \\
  r_{+45} & = 1+ \frac{X_2 |z_1|^2 }{ |1-z_2|^2} 
  & \qquad
  r_{+456} & = 1+\frac{X_2 |z_1|^2}{|1-z_2|^2}\left(1+ \frac{X_3|z_2|^2}{ |1-z_3|^2} \right).
\end{align}  
\end{subequations}
Note that all these physical poles are real and positive in the physical region considered. 
The spurious poles, instead, are complex. They are given by:      
\begin{align}
\label{eq444}
    \begin{split}
 \Omega_{\langle4|q_{1}^\perp|-]}&= -\frac{\left(1-z_2\right) \left(1-z_3\right) \bar{z}_2 \left(X_2 X_3 z_1+X_3+1\right)}{X_2 X_3 z_1 z_3}
\\
    \Omega_{\langle+|{q_{4}^\perp}|6]}
    &
    =
  \frac{\bar{z}_2 (1-z_1) (1-z_2)|1-z_2|^2 |1-z_3|^2  }{X_2 X_3 z_2 (1-\bar{z}_1)}  
    \\
    &
    \hspace*{30pt}\times{\Big[\left(1-\bar{z}_3\right) |1-z_2|^2+X_2 \left(1-\bar{z}_3\right) |z_1|^2+X_2 X_3 |z_1|^2 |z_2|^2\Big]}
\\
  \Omega _{\langle5|P_{q_{1}^\perp4}\text{}|-]}&= -\frac{z_1 \left(1-z_3\right) \left(1+X_3 z_2\right) (1-\bar{z}_1){}^2}{X_3 z_2 \bar{z}_2}
\\
    \Omega _{\langle-|P_{56}|4]}&=  \frac{(1-z_3)}{X_2 X_3 \bar{z}_1}{\Big[X_2 X_3 \bar{z}_1 \left(\bar{z}_2+\bar{z}_3-1\right)-\left(X_3+1\right) \left(1-\bar{z}_2\right) \left(1-\bar{z}_3\right)\Big]
}
\\
         \Omega _{\langle+|P_{45}|6]}&=
     -\frac{\bar{z}_1}{X_2 X_3}\Big[X_2 \bar{z}_1 \left(X_3 \left(z_2-1\right) \bar{z}_2+z_1 \left(X_3 
\bar{z}_2-\bar{z}_3+1\right)\right)
+|1-z_2|^2\left(1-\bar{z}_3\right)\Big]
\\
    \Omega _{\langle-|P_{456}|+]}&= \bar{z}_1 \bar{z}_2 
    \left(\bar{z}_2 \bar{z}_1+\bar{z}_3 \bar{z}_1+\bar{z}_2 \bar{z}_3-\bar{z}_1-\bar{z}_2-\bar{z}_3+1\right)
\\
     \Omega _{\langle+|P_{56{q_{4}^\perp}}P_{56}|-\rangle}&=\frac{\bar{z}_1 (1-z_1)}{X_2 X_3}{\Big[X_2 X_3 |z_1|^2 \left(\bar{z}_2+\bar{z}_3-1\right)-\left(X_3+1\right) \left(1-\bar{z}_2\right) \left(1-\bar{z}_3\right)\Big],
}
\end{split}
 \end{align}
where the notation is as shown in Eqs.~(\ref{eq:Omegas1}) to~(\ref{eq:Omegas3}).

In the process of determining the full set of colour-ordered CEVs entering Eq.~\eqref{quiq2} for NMHV helicity configurations we need to permute both helicities and momenta of the three emitted gluons. 
The required helicity permutations are
\[
\{\,\ominus\,\oplus\,\oplus  ,\quad \oplus\,\ominus\,\oplus,\quad \oplus\,\oplus\,\ominus\,\}.
\]
We shall refer to these three helicity assignments for $g_4$, $g_5$ and $g_6$, respectively, as the primitive CEVs. We shall present them explicitly below.
Starting with these three, we shall apply the momentum permutations introduced in section \ref{pp} to generate the complete set of colour-ordered CEVs. The first primitive CEV is:
\begin{align}
\begin{split}
C_{3}(g_{q_1^\perp}^*,g^\ominus_4,g^\oplus_5 ,g^\oplus_6,g_{q_4^\perp}^*)    = 
&\frac{[56]^3 z_3 z_2 z_1 (1-\bar{z}_3){}^2 (1-\bar{z}_2){}^2 (1-\bar{z}_1){}^2}{[45] s_{456} \Omega _{\langle-|P_{56}|4]} \Omega _{\langle+|P_{45}|6]} \bar{q}_{1}^\perp \sqrt{X_3} \sqrt{X_2}} \\
&+ \frac{X_3 \sqrt{X_2} \bar{z}_2^4 \bar{z}_1^4 z_3 (1-z_2){}^4 (1-z_1){}^2}{\langle45\rangle \Omega _{\langle+|P_{45}|6]} \Omega _{\langle+|{q_{4}^\perp}|6]} \Omega _{\langle5|P_{+4}P_{+456}|-\rangle} r_{+456} r_{+45}} 
\\
&- \frac{ \sqrt{X_3} \left(\bar{z}_3-(1-\bar{z}_2)\right){}^4 \bar{z}_1^4 z_3 z_2 (1-\bar{z}_1) (1-z_1){}^2}{\langle56\rangle \Omega _{\langle-|P_{56}|4]} \Omega _{\langle-|P_{456}|+]} \Omega _{\langle5|P_{+4}P_{+456}|-\rangle} \Omega _{\langle+|P_{56{q_{4}^\perp}}P_{56}|-\rangle}} \\
&- \frac{q_{1}^\perp  X_2^{3/2} X_3^{3/2} \bar{z}_3 \bar{z}_2^3 \bar{z}_1^2 z_2}{\langle45\rangle \langle56\rangle \Omega _{\langle4|q_{1}^\perp|-]} \Omega _{\langle-|P_{456}|+]} r_{456-}} \\
&- \frac{q_{1}^\perp \bar{q}_{1}^\perp r_{56-}^3  \bar{z}_3 \bar{z}_2 \bar{z}_1^2 (1-\bar{z}_1){}^2 (1-z_3) (1-z_2)}{\langle56\rangle s_{q_14} \Omega _{\langle4|q_{1}^\perp|-]} \Omega _{\langle5|P_{q_{1}^\perp4}\text{}|-]} \Omega _{\langle+|P_{56{q_{4}^\perp}}P_{56}|-\rangle} X_2^2 X_3^{5/2}} \\
&- \frac{q_{1}^\perp \bar{q}_{1}^\perp  \bar{z}_3 \bar{z}_2 \bar{z}_1 (1-\bar{z}_3) (1-\bar{z}_2) (1-\bar{z}_1) (1-z_3){}^2 (1-z_2){}^4}{\langle45\rangle s_{6q_4} \Omega _{\langle+|{q_{4}^\perp}|6]} \Omega _{\langle5|P_{q_{1}^\perp4}\text{}|-]} X_2^{3/2} X_3^2 z_2^2} . 
\end{split}
\end{align}
The remaining two primitives are as follows.
\begin{align}
\begin{split}
C_{3}(g_{q_1^\perp}^*,g^\oplus_4,g^\ominus_5 ,g^\oplus_6,g_{q_4^\perp}^*)  = &
\frac{[46]^4 z_3 z_2 z_1 (1-\bar{z}_3){}^2 (1-\bar{z}_2){}^2 (1-\bar{z}_1){}^2}{[45] [56] s_{456} \Omega _{\langle-|P_{56}|4]} \Omega _{\langle+|P_{45}|6]} \bar{q}_{1}^\perp \sqrt{X_3} \sqrt{X_2}} \\
&+ \frac{X_2^{5/2} X_3 \bar{z}_2^4 \bar{z}_1^4 z_3 z_1^4 (1-z_1){}^2}{\langle45\rangle \Omega _{\langle+|P_{45}|6]} \Omega _{\langle+|{q_{4}^\perp}|6]} \Omega _{\langle5|P_{+4}P_{+456}|-\rangle} r_{+456} r_{+45}} \\
&- \frac{q_{1}^\perp  X_3^{3/2} \bar{z}_3 \bar{z}_2^3 \bar{z}_1^2 z_2}{\langle45\rangle \langle56\rangle \Omega _{\langle4|q_{1}^\perp|-]} \Omega _{\langle-|P_{456}|+]} r_{456-} \sqrt{X_2}} \\
&- \frac{ \sqrt{X_3} z_3 z_2 (1-\bar{z}_3){}^4 (1-\bar{z}_2){}^4 (1-\bar{z}_1) (1-z_1){}^2}{\langle56\rangle \Omega _{\langle-|P_{56}|4]} \Omega _{\langle-|P_{456}|+]} \Omega _{\langle5|P_{+4}P_{+456}|-\rangle} \Omega _{\langle+|P_{56{q_{4}^\perp}}P_{56}|-\rangle} X_2^2} \\
&- \frac{q_{1}^\perp \bar{q}_{1}^\perp  X_3^{3/2} \bar{z}_3 \bar{z}_2 \bar{z}_1^2 z_1^4 (1-\bar{z}_1){}^2 (1-z_3) (1-z_2)}{\langle56\rangle s_{q_14} \Omega _{\langle4|q_{1}^\perp|-]} \Omega _{\langle5|P_{q_{1}^\perp4}\text{}|-]} \Omega _{\langle+|P_{56{q_{4}^\perp}}P_{56}|-\rangle} r_{56-}} \\
&- \frac{q_{1}^\perp \bar{q}_{1}^\perp  \sqrt{X_2} \bar{z}_3 \bar{z}_2 \bar{z}_1 z_1^4 (1-\bar{z}_3) (1-\bar{z}_2) (1-\bar{z}_1) (1-z_3){}^2}{\langle45\rangle s_{6q_4} \Omega _{\langle+|{q_{4}^\perp}|6]} \Omega _{\langle5|P_{q_{1}^\perp4}\text{}|-]} X_3^2 z_2^2}  \end{split}
\end{align}
and
\begin{align}
\label{CEVgggPPM}
\begin{split}    C_{3}(g_{q_1^\perp}^*,g^\oplus_4,g^\oplus_5,g^\ominus_6,g_{q_4^\perp}^* )    = &
\frac{r_{+45}^3 z_3 (1-\bar{z}_3){}^4 (1-\bar{z}_2){}^4 (1-z_2){}^4 (1-z_1){}^2}{\langle45\rangle \Omega _{\langle+|P_{45}|6]} \Omega _{\langle+|{q_{4}^\perp}|6]} \Omega _{\langle5|P_{+4}P_{+456}|-\rangle} r_{+456} X_2^{3/2} X_3} \\
&- \frac{q_{1}^\perp \bar{q}_{1}^\perp  \bar{z}_3 \bar{z}_2 \bar{z}_1^2 z_1^4 (1-\bar{z}_1){}^2 (1-z_3) (1-z_2)}{\langle56\rangle s_{q_14} \Omega _{\langle4|q_{1}^\perp|-]} \Omega _{\langle5|P_{q_{1}^\perp4}\text{}|-]} \Omega _{\langle+|P_{56{q_{4}^\perp}}P_{56}|-\rangle} r_{56-} \sqrt{X_3}}\\
&- \frac{q_{1}^\perp \bar{q}_{1}^\perp  \sqrt{X_2} \bar{z}_3 \bar{z}_2 \bar{z}_1 z_2^2 z_1^4 (1-\bar{z}_3) (1-\bar{z}_2) (1-\bar{z}_1) (1-z_3){}^2}{\langle45\rangle s_{6q_4} \Omega _{\langle+|{q_{4}^\perp}|6]} \Omega _{\langle5|P_{q_{1}^\perp4}\text{}|-]}} \\
&+ \frac{[45]^3 z_3 z_2 z_1 (1-\bar{z}_3){}^2 (1-\bar{z}_2){}^2 (1-\bar{z}_1){}^2}{[56] s_{456} \Omega _{\langle-|P_{56}|4]} \Omega _{\langle+|P_{45}|6]} \bar{q}_{1}^\perp \sqrt{X_3} \sqrt{X_2}} \\
&- \frac{q_{1}^\perp  \bar{z}_3 \bar{z}_2^3 \bar{z}_1^2 z_2}{\langle45\rangle \langle56\rangle \Omega _{\langle4|q_{1}^\perp|-]} \Omega _{\langle-|P_{456}|+]} r_{456-} \sqrt{X_3} \sqrt{X_2}} 
\\
&- \frac{ z_3 z_2 (1-\bar{z}_3){}^4 (1-\bar{z}_2){}^4 (1-\bar{z}_1) (1-z_1){}^2}{\langle56\rangle \Omega _{\langle-|P_{56}|4]} \Omega _{\langle-|P_{456}|+]} \Omega _{\langle5|P_{+4}P_{+456}|-\rangle} \Omega _{\langle+|P_{56{q_{4}^\perp}}P_{56}|-\rangle} X_2^2 X_3^{3/2}}.
\end{split}
\end{align}

Considering now the colour-dressed CEV with the helicity configuration $h_4=\ominus  $, $h_5=\oplus$, and $h_6=\oplus$, $\mathcal C^{ggg}\big(g_{q_1^\perp}^*, g_4^{\ominus}, g_5^{\oplus}, g_6^{\oplus}, g_{q_4^\perp}^*\big)$, we see that according to Eq.~(\ref{quiq2}), in addition to the primitive $C_{3}\big(g_{q_1^\perp}^*, g_4^{\ominus}, g_5^{\oplus}, g_6^{\oplus}, g_{q_4^\perp}^*\big)$, 
five more colour-ordered CEVs are required. They are related to the above three primitive CEVs through momentum permutations, which are defined in section~\ref{pp} in terms of MSLCV:
\begin{eqnarray}
\begin{array}{ll}
    C_{3}(g^\ominus_4,g^\oplus_6,g^\oplus_5 )
=
    C_{3}(g^\ominus_4,g^\oplus_5 ,g^\oplus_6)\bigg| _{\mathbb P_{465}} ,
    \quad
   & C_{3}(g^\oplus_5,g^\ominus_4,g^\oplus_6 )
  =
    C_{3}(g^\oplus_4,g^\ominus_5,g^\oplus_6 )\bigg|_{\mathbb P_{546} },\\[.6cm]
C_{3}\left(g_5^{\oplus}, g_6^{\oplus}, g_4^{\ominus}\right)=C_{3}\left(g_4^{\oplus}, g_5^{\oplus}, g_6^{\ominus}\right)\bigg|_{\mathbb P_{564}},
    \quad
& C_{3}\left(g_6^{\oplus}, g_4^{\ominus}, g_5^{\oplus}\right)=C_{3}\left(g_4^{\oplus}, g_5^{\ominus}, g_6^{\oplus}\right)\bigg|_{\mathbb P_{564}},\hspace{20pt} \\[.6cm]
 C_{3}\left(g_6^{\oplus}, g_5^{\oplus}, g_4^{\ominus}\right)
= C_{3}\left(g_4^{\oplus}, g_5^{\oplus}, g_6^{\ominus}\right)\bigg|_{\mathbb P_{654}},
&
\end{array}
\end{eqnarray}
where, for a compact notation, we omitted the off-shell gluons \(g_{q_1^{\perp}}^*\) and \(g_{q_4^\perp}^*\) which do not participate in the permutation. 
Finally, we note that the colour-dressed CEV with helicity configuration with one positive and two negative helicities~$\mathcal{C}_3^{g g g}\big(g_{q_1^\perp}^*, g_4^{\oplus}, g_5^{\ominus}, g_6^{\ominus}, g_{q_4^\perp}^*\big)$ is obtained from results here by complex conjugation.

\subsubsection{Colour-ordered \texorpdfstring{$g^*g^*\to g\bar{q} q$}{g* g* → g qbar q} central-emission vertices}

Compared to pure gluonic colour-ordered CEVs, those with a quark-antiquark pair exhibit an additional layer of complexity: even under identical helicity permutations, distinct colour-ordered arrangements of quarks and gluons yield new configurations. 
In particular, each of the three helicity configurations --- \(\oplus\oplus\ominus\), \(\oplus\ominus\oplus\), and \(\ominus\oplus\oplus\) --- can be associated with two distinct quark-antiquark-gluon configurations, resulting in six primitives.
All colour-ordered CEVs with one $q \bar q$ pair are obtained from these primitives by momentum permutation and the use of reversal and photon decoupling identities. 
We note that in colour-ordered QCD amplitudes, the quark and antiquark are always colour-adjacent, so some of these primitives, in which the pair is interleaved by a gluon would not directly appear. Nonetheless, all six primitives will be useful to express relevant CEVs using said relations.
We first present the primitives below. Subsequently we show how to use identities to derive the complete set of colour ordered CEVs and build the colour-dressed ones.

\noindent
\textbf{Six primitive $g^*g^*\to g\bar{q} q$ CEVs}

The explicit expressions for the six primitive colour-ordered CEVs containing a quark-antiquark pair and a gluon are listed below, always adhering to colour ordering of the final-state particles as 4, 5, 6. This is motivated by the fact that we have at our disposal the complete set of momentum permutations starting from this ordering in section~\ref{pp}. Having fixed the order of momentum labels, we just need to consider the three helicity configurations \(\oplus\oplus\ominus\), \(\oplus\ominus\oplus\), and \(\ominus\oplus\oplus\), and for each of these consider two independent flavour assignments. We choose the primitive CEVs such that 
the antiquark is always the (only) negative helicity particle, so the two primitives are obtained by choosing the two positive helicity particles to be gluon and quark in the two possible orders. To summarise, our set of primitive CEVs are those where the final-state particles have the following momentum, helicity and flavour assignments:
\begin{align*}
\nonumber
\Big\{
&
\big(
g_4^{\oplus}, {q}_5^{\oplus},\bar q_6^{\ominus}
\big),
\quad
\big(
q_4^{\oplus}, g_5^{\oplus},\bar{q}_6^{\ominus}
\big),
\quad
\big(
g_4^{\oplus}, \bar{q}_5^{\ominus},q_6^{\oplus}
\big),
\\
&
\big(
q_4^{\oplus}, \bar{q}_5^{\ominus},g_6^{\oplus}
\big),
\quad
\big(
\bar{q}_4^{\ominus}, g_5^{\oplus},q_6^{\oplus}
\big),
\quad
\big(
\bar{q}_4^{\ominus}, q_5^{\oplus},g_6^{\oplus}
\big)
\Big\}\,.
\end{align*}

We begin with the two primitive CEVs 
corresponding to the $\oplus\oplus\ominus$ helicity configuration:
 \begin{align}
 \label{gqqb} 
 \begin{split}
 C_{3}\big(g_{q_{1}^\perp}^*,g_4^{\oplus}, {q}_5^{\oplus},\bar q_6^{\ominus},g_{q_{4}^{\perp}}^*\big)   = &
 -\frac{q_{1}^\perp  \bar{z}_3 \bar{z}_2^3 \bar{z}_1^2 z_2}{\langle45\rangle \langle56\rangle \Omega _{\langle4|q_{1}^\perp|-]} \Omega _{\langle-|P_{456}|+]} r_{456-} \sqrt{X_2}} \\
 &- \frac{[45]^2 [46] z_3 z_2 z_1 (1-\bar{z}_3){}^2 (1-\bar{z}_2){}^2 (1-\bar{z}_1){}^2}{[56] s_{456} \Omega _{\langle-|P_{56}|4]} \Omega _{\langle+|P_{45}|6]} \bar{q}_{1}^\perp \sqrt{X_3} \sqrt{X_2}} \\
 &+ \frac{r_{+45}^2 \bar{z}_2 \bar{z}_1 z_3 z_1 (1-\bar{z}_3){}^3 (1-\bar{z}_2){}^3 (1-z_2){}^3 (1-z_1){}^2}{\langle45\rangle \Omega _{\langle+|P_{45}|6]} \Omega _{\langle+|{q_{4}^\perp}|6]} \Omega _{\langle5|P_{+4}P_{+456}|-\rangle} r_{+456} \sqrt{X_3} \sqrt{X_2}} \\
 &- \frac{q_{1}^\perp \bar{q}_{1}^\perp  \bar{z}_3 \bar{z}_2 \bar{z}_1^2 z_1^4 (1-\bar{z}_1){}^2 (1-z_3) (1-z_2)}{\langle56\rangle s_{q_14} \Omega _{\langle4|q_{1}^\perp|-]} \Omega _{\langle5|P_{q_{1}^\perp4}\text{}|-]} \Omega _{\langle+|P_{56{q_{4}^\perp}}P_{56}|-\rangle} r_{56-}}
 \\
 &+ \frac{q_{1}^\perp \bar{q}_{1}^\perp \sqrt{X_2} \bar{z}_3 \bar{z}_2 \bar{z}_1 z_2 z_1^4 (1-\bar{z}_3) (1-\bar{z}_2) (1-\bar{z}_1) (1-z_3){}^2}{\langle45\rangle s_{6q_4} \Omega _{\langle+|{q_{4}^\perp}|6]} \Omega _{\langle5|P_{q_{1}^\perp4}\text{}|-]} \sqrt{X_3}} \\
 &-
 \frac{z_3 z_2 (1-\bar{z}_3){}^4 (1-\bar{z}_2){}^4 (1-\bar{z}_1) (1-z_1){}^2}{\langle56\rangle \Omega _{\langle-|P_{56}|4]} \Omega _{\langle-|P_{456}|+]} \Omega _{\langle5|P_{+4}P_{+456}|-\rangle} \Omega _{\langle+|P_{56{q_{4}^\perp}}P_{56}|-\rangle} X_2^2 X_3}   
 \end{split}
 \end{align}
and
\begin{align} 
\label{eq4.31}    
 \begin{split}
  C_{3}^{\text{}}\big(g_{q_{1}^\perp}^*, {q}_4^{\oplus},g_5^{\oplus}, \bar q_6^{\ominus},g_{q_{4}^\perp}^*\big)    = &
   -\frac{q_{1}^\perp  \bar{z}_3 \bar{z}_2^3 \bar{z}_1^2 z_2}{\langle45\rangle \langle56\rangle \Omega _{\langle4|q_{1}^\perp|-]} \Omega _{\langle-|P_{456}|+]} r_{456-}}\\
   &+ \frac{[45]^2 z_3 z_2 z_1 (1-\bar{z}_3){}^2 (1-\bar{z}_2){}^2 (1-\bar{z}_1){}^2}{s_{456} \Omega _{\langle-|P_{56}|4]} \Omega _{\langle+|P_{45}|6]} \bar{q}_{1}^\perp \sqrt{X_3} \sqrt{X_2}}\\
   &+ \frac{\left(\bar{z}_3-(1-\bar{z}_2)\right) \bar{z}_1 z_3 z_2 (1-\bar{z}_3){}^3 (1-\bar{z}_2){}^3 (1-\bar{z}_1) (1-z_1){}^2}{\langle56\rangle \Omega _{\langle-|P_{56}|4]} \Omega _{\langle-|P_{456}|+]} \Omega _{\langle5|P_{+4}P_{+456}|-\rangle} \Omega _{\langle+|P_{56{q_{4}^\perp}}P_{56}|-\rangle} X_2^{3/2} X_3} \\
   &- \frac{r_{+45}^2  \bar{z}_2 \bar{z}_1 z_3 (1-\bar{z}_3){}^3 (1-\bar{z}_2){}^3 (1-z_2){}^4 (1-z_1){}^2}{\langle45\rangle \Omega _{\langle+|P_{45}|6]} \Omega _{\langle+|{q_{4}^\perp}|6]} \Omega _{\langle5|P_{+4}P_{+456}|-\rangle} r_{+456} X_2 \sqrt{X_3}} \\
   &+ \frac{q_{1}^\perp \bar{q}_{1}^\perp \bar{z}_3 \bar{z}_2 \bar{z}_1^2 z_1^3 (1-\bar{z}_1){}^2 (1-z_3) (1-z_2)}{\langle56\rangle s_{q_14} \Omega _{\langle4|q_{1}^\perp|-]} \Omega _{\langle5|P_{q_{1}^\perp4}\text{}|-]} \Omega _{\langle+|P_{56{q_{4}^\perp}}P_{56}|-\rangle} X_3 \sqrt{X_2}} \\
   &- \frac{q_{1}^\perp \bar{q}_{1}^\perp  \bar{z}_3 \bar{z}_2 \bar{z}_1 z_2 z_1^3 (1-\bar{z}_3) (1-\bar{z}_2) (1-\bar{z}_1) (1-z_3){}^2 (1-z_2)}{\langle45\rangle s_{6q_4} \Omega _{\langle+|{q_{4}^\perp}|6]} \Omega _{\langle5|P_{q_{1}^\perp4}\text{}|-]} \sqrt{X_3}}\,.   
   \end{split}
 \end{align}
Next we present the two primitive CEVs with  the $\oplus\ominus\oplus$ helicity configuration:   
\begin{align} 
\label{gPqbMqP}
 \begin{split}
C_{3}\big(g_{q_{1}^\perp}^*,g_4^{\oplus}, \bar {q}_5^{\ominus},q_6^{\oplus},g_{q_{4}^{\perp}}^*\big) 
 = &
 \frac{z_3 z_2 (1-\bar{z}_3){}^4 (1-\bar{z}_2){}^4 (1-\bar{z}_1) (1-z_1){}^2}{\langle56\rangle \Omega _{\langle-|P_{56}|4]} \Omega _{\langle-|P_{456}|+]} \Omega _{\langle5|P_{+4}P_{+456}|-\rangle} \Omega _{\langle+|P_{56{q_{4}^\perp}}P_{56}|-\rangle} X_2^2} 
 \\
 &+ \frac{q_{1}^\perp X_3 \bar{z}_3 \bar{z}_2^3 \bar{z}_1^2 z_2}{\langle45\rangle \langle56\rangle \Omega _{\langle4|q_{1}^\perp|-]} \Omega _{\langle-|P_{456}|+]} r_{456-} \sqrt{X_2}} \\
 &+ \frac{q_{1}^\perp \bar{q}_{1}^\perp X_3 \bar{z}_3 \bar{z}_2 \bar{z}_1^2 z_1^4 (1-\bar{z}_1){}^2 (1-z_3) (1-z_2)}{\langle56\rangle s_{q_14} \Omega _{\langle4|q_{1}^\perp|-]} \Omega _{\langle5|P_{q_{1}^\perp4}\text{}|-]} \Omega _{\langle+|P_{56{q_{4}^\perp}}P_{56}|-\rangle} r_{56-}}
 \\
 &- \frac{ X_2^{3/2} \sqrt{X_3} \bar{z}_2^3 \bar{z}_1^3 z_3 z_1^3 (1-\bar{z}_3) (1-\bar{z}_2) (1-z_2) (1-z_1){}^2}{\langle45\rangle \Omega _{\langle+|P_{45}|6]} \Omega _{\langle+|{q_{4}^\perp}|6]} \Omega _{\langle5|P_{+4}P_{+456}|-\rangle} r_{+456}} \\
 &- \frac{q_{1}^\perp \bar{q}_{1}^\perp  \sqrt{X_2} \bar{z}_3 \bar{z}_2 \bar{z}_1 z_1^4 (1-\bar{z}_3) (1-\bar{z}_2) (1-\bar{z}_1) (1-z_3){}^2}{\langle45\rangle s_{6q_4} \Omega _{\langle+|{q_{4}^\perp}|6]} \Omega _{\langle5|P_{q_{1}^\perp4}\text{}|-]} X_3^{3/2} z_2} 
 \\
 &+ \frac{[46]^3 z_3 z_2 z_1 (1-\bar{z}_3){}^2 (1-\bar{z}_2){}^2 (1-\bar{z}_1){}^2}{[56] s_{456} \Omega _{\langle-|P_{56}|4]} \Omega _{\langle+|P_{45}|6]} \bar{q}_{1}^\perp \sqrt{X_3} \sqrt{X_2}} 
 \end{split}
 \end{align}
and
 \begin{align} 
 \label{qPqbMgP}
 \begin{split}
 C_{3}\big(g_{q_{1}^\perp}^*,{q}_4^{\oplus},\bar q_5^{\ominus},g_6^{\oplus}, g_{q_{4}^{\perp}}^*\big)    = &
 - \frac{q_{1}^\perp  X_3^{3/2} \bar{z}_3 \bar{z}_2^3 \bar{z}_1^2 z_2}{\langle45\rangle \langle56\rangle \Omega _{\langle4|q_{1}^\perp|-]} \Omega _{\langle-|P_{456}|+]} r_{456-}} \\
 &- \frac{[46]^3  z_3 z_2 z_1 (1-\bar{z}_3){}^2 (1-\bar{z}_2){}^2 (1-\bar{z}_1){}^2}{[45] s_{456} \Omega _{\langle-|P_{56}|4]} \Omega _{\langle+|P_{45}|6]} \bar{q}_{1}^\perp \sqrt{X_3} \sqrt{X_2}} \\
 &- \frac{ X_2^2 X_3 \bar{z}_2^4 \bar{z}_1^4 z_3 z_1^3 (1-z_2) (1-z_1){}^2}{\langle45\rangle \Omega _{\langle+|P_{45}|6]} \Omega _{\langle+|{q_{4}^\perp}|6]} \Omega _{\langle5|P_{+4}P_{+456}|-\rangle} r_{+456} r_{+45}}\\
 &+ \frac{\sqrt{X_3} \left(\bar{z}_3-(1-\bar{z}_2)\right) \bar{z}_1 z_3 z_2 (1-\bar{z}_3){}^3 (1-\bar{z}_2){}^3 (1-\bar{z}_1) (1-z_1){}^2}{\langle56\rangle \Omega _{\langle-|P_{56}|4]} \Omega _{\langle-|P_{456}|+]} \Omega _{\langle5|P_{+4}P_{+456}|-\rangle} \Omega _{\langle+|P_{56{q_{4}^\perp}}P_{56}|-\rangle} X_2^{3/2}} \\
 &+ \frac{q_{1}^\perp \bar{q}_{1}^\perp \sqrt{X_3} \bar{z}_3 \bar{z}_2 \bar{z}_1^2 z_1^3 (1-\bar{z}_1){}^2 (1-z_3) (1-z_2)}{\langle56\rangle s_{q_14} \Omega _{\langle4|q_{1}^\perp|-]} \Omega _{\langle5|P_{q_{1}^\perp4}\text{}|-]} \Omega _{\langle+|P_{56{q_{4}^\perp}}P_{56}|-\rangle} \sqrt{X_2}} \\
 &+ \frac{q_{1}^\perp \bar{q}_{1}^\perp \bar{z}_3 \bar{z}_2 \bar{z}_1 z_1^3 (1-\bar{z}_3) (1-\bar{z}_2) (1-\bar{z}_1) (1-z_3){}^2 (1-z_2)}{\langle45\rangle s_{6q_4} \Omega _{\langle+|{q_{4}^\perp}|6]} \Omega _{\langle5|P_{q_{1}^\perp4}\text{}|-]} X_3^2 z_2^2} \,.
 \end{split}
 \end{align}

Finally, we present the two primitive CEVs with the $\ominus\oplus\oplus$ helicity configuration: 
 \begin{align}
  \begin{split}
  &C_{3}^{\text{}} \big(g_{q_{1}^\perp}^*, \bar {q}_4^{\ominus},g_5^{\oplus}, q_6^{\oplus},g_{q_{4}^\perp}^*\big)    \\= &\frac{q_{1}^\perp X_2 X_3 \bar{z}_3 \bar{z}_2^3 \bar{z}_1^2 z_2}{\langle45\rangle \langle56\rangle \Omega _{\langle4|q_{1}^\perp|-]} \Omega _{\langle-|P_{456}|+]} r_{456-}} \\
  &- \frac{[56]^2  z_3 z_2 z_1 (1-\bar{z}_3){}^2 (1-\bar{z}_2){}^2 (1-\bar{z}_1){}^2}{s_{456} \Omega _{\langle-|P_{56}|4]} \Omega _{\langle+|P_{45}|6]} \bar{q}_{1}^\perp \sqrt{X_3} \sqrt{X_2}} 
  \\
  &+ \frac{\sqrt{X_3} \bar{z}_2^3 \bar{z}_1^3 z_3 (1-\bar{z}_3) (1-\bar{z}_2) (1-z_2){}^4 (1-z_1){}^2}{\langle45\rangle \Omega _{\langle+|P_{45}|6]} \Omega _{\langle+|{q_{4}^\perp}|6]} \Omega _{\langle5|P_{+4}P_{+456}|-\rangle} r_{+456}} 
  \\
  &- \frac{  \big(\bar{z}_3-(1-\bar{z}_2)\big){}^3 \bar{z}_1^3 z_3 z_2 (1-\bar{z}_3) (1-\bar{z}_2) (1-\bar{z}_1) (1-z_1){}^2}{\langle56\rangle \Omega _{\langle-|P_{56}|4]} \Omega _{\langle-|P_{456}|+]} \Omega_{\langle5|P_{+4}P_{+456}|-\rangle} \Omega _{\langle+|P_{56{q_{4}^\perp}}P_{56}|-\rangle} \sqrt{X_2}} 
 \\
  &- \frac{q_{1}^\perp \bar{q}_{1}^\perp r_{56-}^2 \bar{z}_3 \bar{z}_2 \bar{z}_1^2 z_1 (1-\bar{z}_1){}^2 (1-z_3) (1-z_2)}{\langle56\rangle s_{q_14} \Omega _{\langle4|q_{1}^\perp|-]} \Omega _{\langle5|P_{q_{1}^\perp4}\text{}|-]} \Omega _{\langle+|P_{56{q_{4}^\perp}}P_{56}|-\rangle} X_2^{3/2} X_3^2} 
  \\
  &+ \frac{q_{1}^\perp \bar{q}_{1}^\perp \bar{z}_3 \bar{z}_2 \bar{z}_1 z_1 (1-\bar{z}_3) (1-\bar{z}_2) (1-\bar{z}_1) (1-z_3){}^2 (1-z_2){}^3}{\langle45\rangle s_{6q_4} \Omega _{\langle+|{q_{4}^\perp}|6]} \Omega _{\langle5|P_{q_{1}^\perp4}\text{}|-]} X_2 X_3^{3/2} z_2}  ,    \end{split}
 \end{align}
and
\begin{align} 
\label{qbMqPgP}
 \begin{split}
   C_{3} \big(g_{q_{1}^\perp}^*, \bar {q}_4^{\ominus}, q_5^{\oplus},g_6^{\oplus},g_{q_{4}^\perp}^*\big)    
   = &
   \frac{q_{1}^\perp X_2 X_3^{3/2} \bar{z}_3 \bar{z}_2^3 \bar{z}_1^2 z_2}{\langle45\rangle \langle56\rangle \Omega _{\langle4|q_{1}^\perp|-]} \Omega _{\langle-|P_{456}|+]} r_{456-}} \\
   &+ \frac{[46] [56]^2 z_3 z_2 z_1 (1-\bar{z}_3){}^2 (1-\bar{z}_2){}^2 (1-\bar{z}_1){}^2}{[45] s_{456} \Omega _{\langle-|P_{56}|4]} \Omega _{\langle+|P_{45}|6]} \bar{q}_{1}^\perp \sqrt{X_3} \sqrt{X_2}} 
   \\
   &+ \frac{X_2 X_3 \bar{z}_2^4 \bar{z}_1^4 z_3 z_1 (1-z_2){}^3 (1-z_1){}^2}{\langle45\rangle \Omega _{\langle+|P_{45}|6]} \Omega _{\langle+|{q_{4}^\perp}|6]} \Omega _{\langle5|P_{+4}P_{+456}|-\rangle} r_{+456} r_{+45}} \\
   &- \frac{ \sqrt{X_3}  \big(\bar{z}_3-(1-\bar{z}_2)\big){}^3 \bar{z}_1^3 z_3 z_2 (1-\bar{z}_3) (1-\bar{z}_2) (1-\bar{z}_1) (1-z_1){}^2}{\langle56\rangle \Omega _{\langle-|P_{56}|4]} \Omega _{\langle-|P_{456}|+]} \Omega _{\langle5|P_{+4}P_{+456}|-\rangle} \Omega _{\langle+|P_{56{q_{4}^\perp}}P_{56}|-\rangle} \sqrt{X_2}} \\
   &- \frac{q_{1}^\perp \bar{q}_{1}^\perp r_{56-}^2  \bar{z}_3 \bar{z}_2 \bar{z}_1^2 z_1 (1-\bar{z}_1){}^2 (1-z_3) (1-z_2)}{\langle56\rangle s_{q_14} \Omega _{\langle4|q_{1}^\perp|-]} \Omega _{\langle5|P_{q_{1}^\perp4}\text{}|-]} \Omega _{\langle+|P_{56{q_{4}^\perp}}P_{56}|-\rangle} X_2^{3/2} X_3^{3/2}} \\
   &- \frac{q_{1}^\perp \bar{q}_{1}^\perp  \bar{z}_3 \bar{z}_2 \bar{z}_1 z_1 (1-\bar{z}_3) (1-\bar{z}_2) (1-\bar{z}_1) (1-z_3){}^2 (1-z_2){}^3}{\langle45\rangle s_{6q_4} \Omega _{\langle+|{q_{4}^\perp}|6]} \Omega _{\langle5|P_{q_{1}^\perp4}\text{}|-]} X_2 X_3^2 z_2^2} \,. 
   \end{split}
 \end{align}
Utilising these primitive CEVs, we now systematically demonstrate how colour-dressed three-parton-emission CEVs involving a quark-antiquark pair can be constructed from colour-ordered CEVs using the relations in sections~\ref{pp} and~\ref{relationCEVPEV}.

 \subsubsection*{Helicity configuration $\nu_4=\oplus,\nu_5=\ominus ,\nu_6= \oplus$}
 
Here we present the colour-dressed CEV $\mathcal{C}_3^{gq \bar q} \big(  g_{q_{1}^\perp}^*, g_4^{\oplus}, \bar{q}_5^{\ominus}, q_6^{\oplus}, g_{q_{4}^\perp}^* \big) $. Beyond the primitive colour-ordered CEV $C_{3} \big(  g_{q_{1}^\perp}^*, g_4^{\oplus}, \bar{q}_5^{\ominus}, q_6^{\oplus}, g_{q_{4}^\perp}^* \big) $ required for Eq.~\eqref{ciuwe}, five additional  CEVs must be incorporated. We now explicitly demonstrate how these additional CEVs relate to the primitive CEVs listed above through momentum permutation,
reversal and photon decoupling identities.
The second colour-ordered CEV appearing in Eq.~(\ref{ciuwe}), namely $C_{3} \big(  g_{q_{4}^\perp}^*,g_4^{\oplus}, \bar{q}_5^{\ominus}, q_6^{\oplus} ,g_{q_{1}^\perp}^* \big)$, 
is related to the primitive $C_{3} \big(  g_{q_{1}^\perp}^*,q_4^{\oplus}, \bar{q}_5^{\ominus}, g_6^{\oplus} ,g_{q_{4}^\perp}^* \big)  $ by the reversal identity followed by the momentum permutation~${\mathbb P_{654}}$,
 \begin{align}
&C_{3} \big(  g_{q_{4}^\perp}^*,g_4^{\oplus}, \bar{q}_5^{\ominus}, q_6^{\oplus} ,g_{q_{1}^\perp}^* \big)  
=
C_{3} \big(  g_{q_{1}^\perp}^*,q_6^{\oplus}, \bar{q}_5^{\ominus}, g_4^{\oplus} ,g_{q_{4}^\perp}^* \big)  
= 
 C_{3} \big(  g_{q_{1}^\perp}^*,q_4^{\oplus}, \bar{q}_5^{\ominus}, g_6^{\oplus} ,g_{q_{4}^\perp}^* \big)  \bigg | _{\mathbb P_{654}.
}
 \end{align}
To represent the third term in Eq.~(\ref{ciuwe}), we first apply the photon decoupling identity, and then determine the relevant colour-ordered partials in terms of the primitive ones using the permutations in section~\ref{pp}. 
Applying the photon decoupling identity we have:
 \begin{align}
 & \hspace*{-15pt}   C_{3} \big(  g_4^{\oplus} ,g_{q_{1}^\perp}^*, \bar{q}_5^{\ominus}, q_6^{\oplus},g_{q_{4}^\perp}^* \big) 
 \nonumber 
 \\
 =&-C_{3} \big(  g_{q_{1}^\perp}^*, g_4^{\oplus}, \bar{q}_5^{\ominus}, q_6^{\oplus},g_{q_{4}^\perp}^* \big) 
 -C_{3}^{\text{}}  \big(  g_{q_{1}^\perp}^*,  \bar{q}_5^{\ominus}, g_4^{\oplus},q_6^{\oplus},g_{q_{4}^\perp}^* \big) -C_{3} \big(  g_{q_{1}^\perp}^*,  \bar{q}_5^{\ominus},q_6^{\oplus},g_4^{\oplus},g_{q_{4}^\perp}^* \big) .
 \end{align}
Here we recognise the first term as one of our primitive CEVs, Eq.~(\ref{gPqbMqP}), and we determine the second and third ones from the primitives by applying momentum permutations:
 \begin{eqnarray}
    C_{3}^{\text{}}  \big(  g_{q_{1}^\perp}^*,  \bar{q}_5^{\ominus}, g_4^{\oplus},q_6^{\oplus},g_{q_{4}^\perp}^* \big) 
\eqn 
C_{3} ^{\text{}} \big(  g_{q_{1}^\perp}^*,  \bar{q}_4^{\ominus}, g_5^{\oplus},q_6^{\oplus},g_{q_{4}^\perp}^* \big) \bigg|_{\mathbb P_{546}},\non 
C_{3} \big(  g_{q_{1}^\perp}^*,  \bar{q}_5^{\ominus},q_6^{\oplus},g_4^{\oplus},g_{q_{4}^\perp}^* \big) 
\eqn 
C_{3} \big(  g_{q_{1}^\perp}^*,  \bar{q}_4^{\ominus},q_5^{\oplus},g_6^{\oplus},g_{q_{4}^\perp}^* \big) 
\bigg|_{\mathbb P_{564}.
}
 \end{eqnarray}
To express the fourth term in Eq.~(\ref{ciuwe}), 
\(
C_{3}\bigl(\bar{q}_5^{\ominus},\, q_6^{\oplus},\, g_{q_{1}^\perp}^*,\, g_4^{\oplus},\, g_{q_{4}^\perp}^*\bigr),
\)
one must first apply the reversal identity, and subsequently apply the same process as above, i.e., use the photon decoupling identity and subsequently relate the CEVs to the primitive ones using momentum permutations. 
Thus, in the first two steps we have:
\begin{align}
\label{gqqPhotonDecoupling}
 \begin{split}
 &\hspace*{-2pt}
 C_{3} \big(  \bar{q}_5^{\ominus}, q_6^{\oplus} ,g_{q_{1}^\perp}^*, g_4^{\oplus},g_{q_{4}^\perp}^* \big) 
 =
  C_{3} \big(   g_4^{\oplus},g_{q_{1}^\perp}^*, q_6^{\oplus} ,\bar{q}_5^{\ominus},g_{q_{4}^\perp}^* \big) \\
 & =
 -C_{3} \big(  g_{q_{1}^\perp}^*, g_4^{\oplus}, q_6^{\oplus}, \bar{q}_5^{\ominus}, g_{q_{4}^\perp}^* \big)  
-C_{3}^{\text{}}  \big(  g_{q_{1}^\perp}^*, q_6^{\oplus}, g_4^{\oplus}, \bar{q}_5^{\ominus}, g_{q_{4}^\perp}^* \big) 
-C_{3} \big(  g_{q_{1}^\perp}^*, q_6^{\oplus}, \bar{q}_5^{\ominus}, g_4^{\oplus}, g_{q_{4}^\perp}^* \big) \,.
\end{split}
 \end{align}
The three CEVs we obtained here are related to the primitives by the following momentum permutations:
 \begin{eqnarray}
     \begin{aligned}
& C_{3} \big(  g_{q_{1}^\perp}^*, g_4^{\oplus}, q_6^{\oplus}, \bar{q}_5^{\ominus}, g_{q_{4}^\perp}^* \big)  
= C_{3} \big(  g_{q_{1}^\perp}^*, g_4^{\oplus}, q_5^{\oplus}, \bar{q}_6^{\ominus}, g_{q_{4}^\perp}^* \big)  \Big|_{\mathbb 
 P_{465}} \\
&C_{3}^{\text{}}  \big(  g_{q_{1}^\perp}^*, q_6^{\oplus}, g_4^{\oplus}, \bar{q}_5^{\ominus}, g_{q_{4}^\perp}^* \big)  = C_{3}^{\text{}}  \big(  g_{q_{1}^\perp}^*, q_4^{\oplus}, g_5^{\oplus}, \bar{q}_6^{\ominus}, g_{q_{4}^\perp}^* \big) \Big|_{\mathbb 
 P_{645}} \\
& C_{3} \big(  g_{q_{1}^\perp}^*, q_6^{\oplus}, \bar{q}_5^{\ominus}, g_4^{\oplus}, g_{q_{4}^\perp}^* \big) = C_{3} \big(  g_{q_{1}^\perp}^*, q_4^{\oplus}, \bar{q}_5^{\ominus}, g_6^{\oplus}, g_{q_{4}^\perp}^* \big) \Big|_{\mathbb 
 P_{654}}.
\end{aligned}
 \end{eqnarray}
The fifth term in Eq.~(\ref{ciuwe}) is obtained simply by applying a momentum permutation to the primitive CEV in Eq.~(\ref{qbMqPgP}):
\begin{eqnarray}
C_{3} \big(  g_{q_{1}^\perp}^*,\bar{q}_5^{\ominus}, q_6^{\oplus}, g_4^{\oplus},g_{q_{4}^\perp}^* \big) =
C_{3} \big(  g_{q_{1}^\perp}^*,\bar{q}_4^{\ominus}, q_5^{\oplus}, g_6^{\oplus},g_{q_{4}^\perp}^* \big) \bigg|_{\mathbb P_{564}}.
\end{eqnarray}
The final term in Eq.~(\ref{ciuwe}) is obtained by applying the reversal identity and then performing a momentum permutation
\begin{align}
C_{3} \big( \bar{q}_5^{\ominus}, q_6^{\oplus}, g_4^{\oplus},g_{q_{1}^\perp}^* , g_{q_{4}^\perp}^*\big) =
C_{3} \big(  g_{q_{1}^\perp}^*, g_4^{\oplus}, q_6^{\oplus},\bar{q}_5^{\ominus},g_{q_{4}^\perp}^* \big) 
=
C_{3} \big(  g_{q_{1}^\perp}^*, g_4^{\oplus}, q_5^{\oplus},\bar{q}_6^{\ominus},g_{q_{4}^\perp}^* \big)  \bigg| _{\mathbb P_{465}}.
\end{align}
Thus, we have successfully expressed all the colour-ordered CEVs in MSLCV which are required for the dressed CEV,
\(\mathcal{C}^{gq\bar{q}} \big(  g_{q_{1}^\perp}^*, g_4^{\oplus}, \bar{q}_5^{\ominus}, q_6^{\oplus}, g_{q_{4}^\perp}^* \big) \),
according to Eq.~(\ref{ciuwe}).

 \subsubsection*{Helicity configuration $\nu_4=\oplus,\nu_5=\oplus ,\nu_6= \ominus$}

We now wish to obtain the colour-dressed CEV for the helicity configuration $\nu_4=\oplus,\nu_5=\oplus ,\nu_6= \ominus$. In this case Eq.~(\ref{ciuwe}) yields
\begin{align}
\label{qwjk}
\begin{split}
&\hspace{-10pt} \mathcal{C}_3^{g \bar{q} q} \big(  g_{q_{1}^\perp}^*,g_4^{\oplus}, \bar{q}_5^{\oplus}, q_6^{\ominus},g_{q_{4}^\perp}^* \big) = \\
& g^3\left[ \big(  T^{c_4} T^{c_1} T^{d_4} \big) _{i_6 \bar{\imath}_5} C_{3} \big(   g_{q_{1}^\perp}^*,g_4^{\oplus}, \bar{q}_5^{\oplus}, q_6^{\ominus},g_{q_{4}^\perp}^* \big) + \big(  T^{c_1} T^{c_4} T^{d_4} \big) _{i_6 \bar{\imath}_5}C_{3} \big(  g_4^{\oplus}, \bar{q}_5^{\oplus}, q_6^{\ominus},g_{q_{1}^\perp}^* ,g_{q_{4}^\perp}^*\big) \right. \\
& \hspace{10pt}+ \big(  T^{c_4} T^{d_4} T^{c_1} \big) _{i_6 \bar{\imath}_5} C_{3} \big(  g_4^{\oplus},g_{q_{1}^\perp}^*,\bar{q}_5^{\oplus}, q_6^{\ominus},g_{q_{4}^\perp}^* \big) + \big(  T^{c_1} T^{d_4} T^{c_4} \big) _{i_6 \bar{\imath}_5} C_{3} \big(  \bar{q}_5^{\oplus}, q_6^{\ominus},g_{q_{1}^\perp}^*, g_4^{\oplus},g_{q_{4}^\perp}^* \big) \\
& \hspace{10pt}\left.+ \big(  T^{d_4} T^{c_4} T^{c_1} \big) _{i_6 \bar{\imath}_5} C_{3} \big(  g_{q_{1}^\perp}^*,\bar{q}_5^{\oplus}, q_6^{\ominus},g_4^{\oplus},g_{q_{4}^\perp}^* \big) + \big(  T^{d_4} T^{c_1} T^{c_4} \big) _{i_6 \bar{\imath}_5} C_{3} \big(\bar{q}_5^{\oplus}, q_6^{\ominus}, g_4^{\oplus},g_{q_{1}^\perp}^*,  g_{q_{4}^\perp}^* \big)  \right]\,.
\end{split}
\end{align}
The configuration containing negative-helicity quarks paired with positive-helicity anti-quarks can be mapped to positive-helicity quarks with negative-helicity anti-quarks through 
charge-conjugation symmetry:
\begin{eqnarray}
    \mathcal{C}_3^{g \bar{q} q} \big(  g_{q_{1}^\perp}^*, g_4^{\oplus}, \bar{q}_5^{\oplus}, q_6^{\ominus},g_{q_{4}^\perp}^* \big) 
    = 
    \mathcal{C}_3^{g \bar{q} q} \big(  g_{q_{1}^\perp}^*, g_4^{\oplus}, {q}_5^{\oplus}, \bar q_6^{\ominus},g_{q_{4}^\perp}^* \big) .
\end{eqnarray}
This yields
\begin{align}
\label{qwjk_after_charge_conj}
\begin{split}
&\hspace{-10pt} \mathcal{C}_3^{g \bar{q} q} \big(  g_{q_{1}^\perp}^*,g_4^{\oplus}, {q}_5^{\oplus}, \bar q_6^{\ominus},g_{q_{4}^\perp}^* \big) = \\
& g^3\left[ \big(  T^{c_4} T^{c_1} T^{d_4} \big) _{ \bar{\imath}_6 i_5} C_{3} \big(   g_{q_{1}^\perp}^*,g_4^{\oplus}, q_5^{\oplus}, \bar{q}_6^{\ominus},g_{q_{4}^\perp}^* \big) + \big(  T^{c_1} T^{c_4} T^{d_4} \big) _{ \bar{\imath}_6 i_5}C_{3} \big(g_4^{\oplus}, q_5^{\oplus}, \bar{q}_6^{\ominus},g_{q_{1}^\perp}^* ,  g_{q_{4}^\perp}^*\big) \right. \\
& \hspace{10pt}+ \big(  T^{c_4} T^{d_4} T^{c_1} \big) _{ \bar{\imath}_6 i_5} C_{3} \big(  g_4^{\oplus},g_{q_{1}^\perp}^*,q_5^{\oplus}, \bar{q}_6^{\ominus},g_{q_{4}^\perp}^* \big) + \big(  T^{c_1} T^{d_4} T^{c_4} \big) _{ \bar{\imath}_6 i_5} C_{3} \big(  q_5^{\oplus}, \bar{q}_6^{\ominus},g_{q_{1}^\perp}^*, g_4^{\oplus},g_{q_{4}^\perp}^* \big) \\
& \hspace{10pt}\left.+ \big(  T^{d_4} T^{c_4} T^{c_1} \big) _{ \bar{\imath}_6 i_5} C_{3} \big(  g_{q_{1}^\perp}^*,q_5^{\oplus}, \bar{q}_6^{\ominus},g_4^{\oplus},g_{q_{4}^\perp}^* \big) + \big(  T^{d_4} T^{c_1} T^{c_4} \big) _{ \bar{\imath}_6 i_5} C_{3} \big(  \bar{q}_5^{\oplus}, q_6^{\ominus}, g_4^{\oplus},g_{q_{1}^\perp}^*,g_{q_{4}^\perp}^* \big)  \right]\,,
\end{split}
\end{align}
where we used Eq.~(\ref{charge_conjugation_qqbar_first}) for each of the colour-ordered CEVs.

Next we wish to express each of the colour-ordered CEVs appearing in Eq.~(\ref{qwjk_after_charge_conj}) in terms of the primitive ones in Eqs.~(\ref{gqqb}) through~(\ref{qbMqPgP}). The first CEV in Eq.~(\ref{qwjk_after_charge_conj}), $C \big(  g_{q_{1}^\perp}^*, g_4^{\oplus}, q_5^{\oplus}, \bar{q}_6^{\ominus},g_{q_{4}^{\perp}}^* \big), $ is just the primitive of Eq.~(\ref{gqqb}).  The second CEV in Eq.~(\ref{qwjk_after_charge_conj}) may be obtained by first applying the reversal identity, and then using a momentum permutation of Eq.~(\ref{qbMqPgP}):
\begin{align}
&C_{3} \big( g_4^{\oplus}, {q}_5^{\oplus}, \bar q_6^{\ominus} ,g_{q_{1}^\perp}^* , g_{q_{4}^\perp}^*\big) 
=
C_{3} \big(  g_{q_{1}^\perp}^*,\bar q_6^{\ominus}, {q}_5^{\oplus}, g_4^{\oplus} ,g_{q_{4}^\perp}^* \big)  
=
 C_{3} \big( g_{q_{1}^\perp}^*,\bar q_4^{\ominus }, {q}_5^{\oplus}, g_6^{\oplus} ,g_{q_{4}^\perp}^* \big)  \bigg | _{\mathbb P_{654}.
}
\end{align}
The third CEV in Eq.~(\ref{qwjk_after_charge_conj}) is obtained by first applying the photon decoupling identity, and then identifying the resulting colour-ordered CEVs as primitive ones after appropriate momentum permutations. The photon decoupling identity yields
 \begin{align}
 \begin{split}
 C_{3} \big( g_4^{\oplus} ,g_{q_{1}^\perp}^*, q_5^{\oplus}, \bar{q}_6^{\ominus},g_{q_{4}^\perp}^* \big) 
=&
-C_{3} \big( g_{q_{1}^\perp}^*, g_4^{\oplus}, q_5^{\oplus}, \bar{q}_6^{\ominus},g_{q_{4}^\perp}^* \big)  \\
&
-C_{3}^{\text{}}  \big( g_{q_{1}^\perp}^*,  q_5^{\oplus}, g_4^{\oplus}, \bar{q}_6^{\ominus},g_{q_{4}^\perp}^* \big)  
-C_{3} \big( g_{q_{1}^\perp}^*,  q_5^{\oplus}, \bar{q}_6^{\ominus}, g_4^{\oplus},g_{q_{4}^\perp}^* \big)\,, 
\end{split}
\end{align}
where the last two terms are obtained from the primitives by applying momentum permutations
 \begin{eqnarray}
    C_{3} ^{\text{}} \big( g_{q_{1}^\perp}^*, q_5^{\oplus}, g_4^{\oplus}, \bar{q}_6^{\ominus}, g_{q_{4}^\perp}^* \big) \eqn
C_{3} ^{\text{}} \big( g_{q_{1}^\perp}^*,  {q}_4^{\oplus}, g_5^{\oplus},\bar q_6^{\ominus},g_{q_{4}^\perp}^* \big) \bigg|_{\mathbb P_{546}}\,,
\non
C_{3} \big( g_{q_{1}^\perp}^*, q_5^{\oplus}, \bar{q}_6^{\ominus}, g_4^{\oplus}, g_{q_{4}^\perp}^* \big) \eqn
C_{3} \big( g_{q_{1}^\perp}^*,  {q}_4^{\oplus}, \bar q_5^{\ominus},g_6^{\oplus},g_{q_{4}^\perp}^* \big) 
\bigg|_{\mathbb P_{564}
}.
 \end{eqnarray}
In the fourth CEV in Eq.~(\ref{qwjk_after_charge_conj}) 
 we first apply the 
 reversal identity and then proceed in a similar fashion to the previous case.
We obtain
 \begin{align}
 \begin{split}
 & \hspace*{-20pt} C_{3} \big( q_5^{\oplus}, \bar{q}_6^{\ominus}, g_{q_{1}^\perp}^*, g_4^{\oplus}, g_{q_{4}^\perp}^* \big) 
= 
C_{3} \big( g_4^{\oplus}, g_{q_{1}^\perp}^*, \bar{q}_6^{\ominus}, q_5^{\oplus}, g_{q_{4}^\perp}^* \big) 
\\  = &
 -C_{3} \big( g_{q_{1}^\perp}^*, g_4^{\oplus}, \bar{q}_6^{\ominus}, q_5^{\oplus}, g_{q_{4}^\perp}^* \big)  
-C_{3}^{\text{}}  \big( g_{q_{1}^\perp}^*, \bar{q}_6^{\ominus}, g_4^{\oplus}, q_5^{\oplus}, g_{q_{4}^\perp}^* \big) 
-
C_{3} \big( g_{q_{1}^\perp}^*, \bar{q}_6^{\ominus}, q_5^{\oplus}, g_4^{\oplus}, g_{q_{4}^\perp}^* \big)\,, 
 \end{split}
 \end{align}
 where 
\begin{eqnarray}
     \begin{aligned}
& C_{3} \big( g_{q_{1}^\perp}^*, g_4^{\oplus}, \bar q_6^{\ominus}, {q}_5^{\oplus}, g_{q_{4}^\perp}^* \big)  
= C_{3} \big( g_{q_{1}^\perp}^*, g_4^{\oplus}, \bar q_5^{\ominus}, {q}_6^{\oplus}, g_{q_{4}^\perp}^* \big)  \Big|_{\mathbb P_{465}} \\
&C_{3} ^{\text{}} \big( g_{q_{1}^\perp}^*, \bar q_6^{\ominus}, g_4^{\oplus}, {q}_5^{\oplus}, g_{q_{4}^\perp}^* \big)  = C_{3} ^{\text{}} \big( g_{q_{1}^\perp}^*, \bar q_4^{\ominus}, g_5^{\oplus}, {q}_6^{\oplus}, g_{q_{4}^\perp}^* \big) \Big|_{\mathbb P_{645}} \\
& C_{3} \big( g_{q_{1}^\perp}^*, \bar q_6^{\ominus}, {q}_5^{\oplus}, g_4^{\oplus}, g_{q_{4}^\perp}^* \big) = C_{3} \big( g_{q_{1}^\perp}^*, \bar q_4^{\ominus}, {q}_5^{\oplus}, g_6^{\oplus}, g_{q_{4}^\perp}^* \big) \Big|_{\mathbb P_{654}}.
\end{aligned}
 \end{eqnarray}
To obtain the fifth CEV in Eq.~(\ref{qwjk_after_charge_conj}),
we directly apply momentum permutation to the primitive in Eq.~(\ref{qPqbMgP}) 
\begin{eqnarray}
C_{3} \big( g_{q_{1}^\perp}^*, q_5^{\oplus}, \bar{q}_6^{\ominus}, g_4^{\oplus}, g_{q_{4}^\perp}^* \big) =
C_{3} \big( g_{q_{1}^\perp}^*,{q}_4^{\oplus}, \bar q_5^{\ominus}, g_6^{\oplus},g_{q_{4}^\perp}^* \big) \bigg|_{\mathbb P_{564}}.
\end{eqnarray}
Finally, to determine the sixth CEV in Eq.~(\ref{qwjk_after_charge_conj}) we first apply the reversal identity and then identify the resulting colour-ordered CEV as a momentum permutation of Eq.~(\ref{gPqbMqP}):
\begin{align}
&
C_{3} \big(q_5^{\oplus}, \bar{q}_6^{\ominus}, g_4^{\oplus}, g_{q_{1}^\perp}^* , g_{q_{4}^\perp}^*\big)  =
C_{3} \big( g_{q_{1}^\perp}^*, g_4^{\oplus}, \bar{q}_6^{\ominus}, q_5^{\oplus}, g_{q_{4}^\perp}^* \big) 
=
C_{3} \big( g_{q_{1}^\perp}^*, g_4^{\oplus}, q_5^{\ominus},\bar{q}_6^{\oplus},g_{q_{4}^\perp}^* \big)  \bigg| _{\mathbb P_{465}}.
\end{align}

The remaining helicity configurations, containing two negative and one positive helicities may be obtained from the results above by complex conjugation. 
Having presented the complete set of colour-dressed CEVs for triple emissions 
containing a single \( q\bar{q} \) pair, we 
now proceed to outline the higher-multiplicity MREVs.

\section{Multi-Regge Emission Vertices (MREV): a Mathematica library }
\label{sec:4partonCEVsPEVs}
\begin{figure}
    \centering
    \includegraphics[width=1\linewidth]{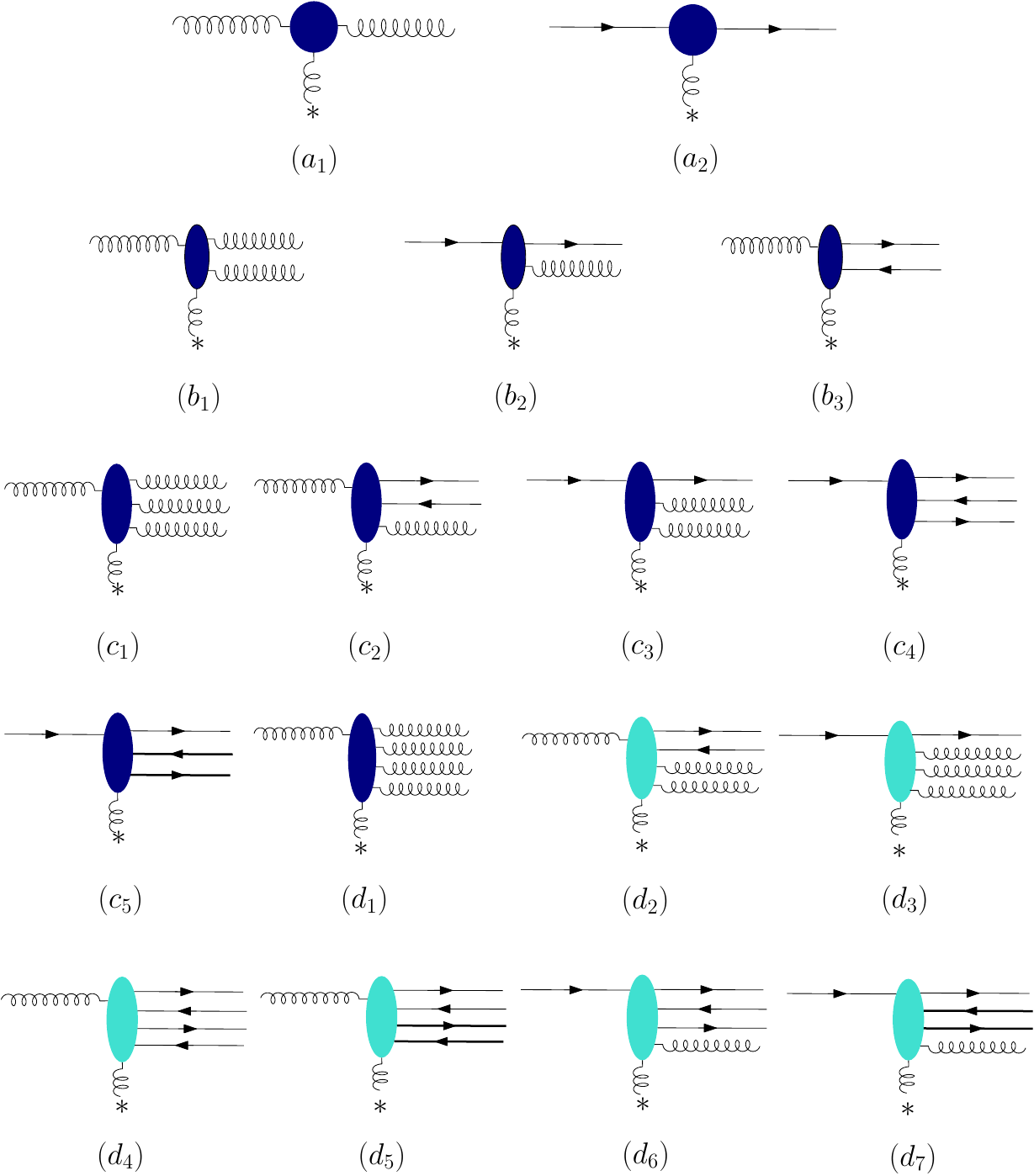}
    \caption{Peripheral Emission Vertices (PEVs) corresponding to final state multiplicities of up to four partons collected in our \texttt{\href{https://github.com/YuyuMo-UoE/Multi-Regge-Emission-Vertices}{Multi-Regge-Emission-Vertices}}  (MREV) library~\cite{yuyu_2017_github}. The diagrams labelled $(a)$, $(b)$, $(c)$, and $(d)$ respectively represent final states with 1, 2, 3, or 4 partons, including cases with pure gluons as well as configurations containing up to two quark--antiquark pairs.
}
    \label{fig:list of results of PEV}
\end{figure}

\begin{figure}
    \centering
    \includegraphics[width=0.8\linewidth]{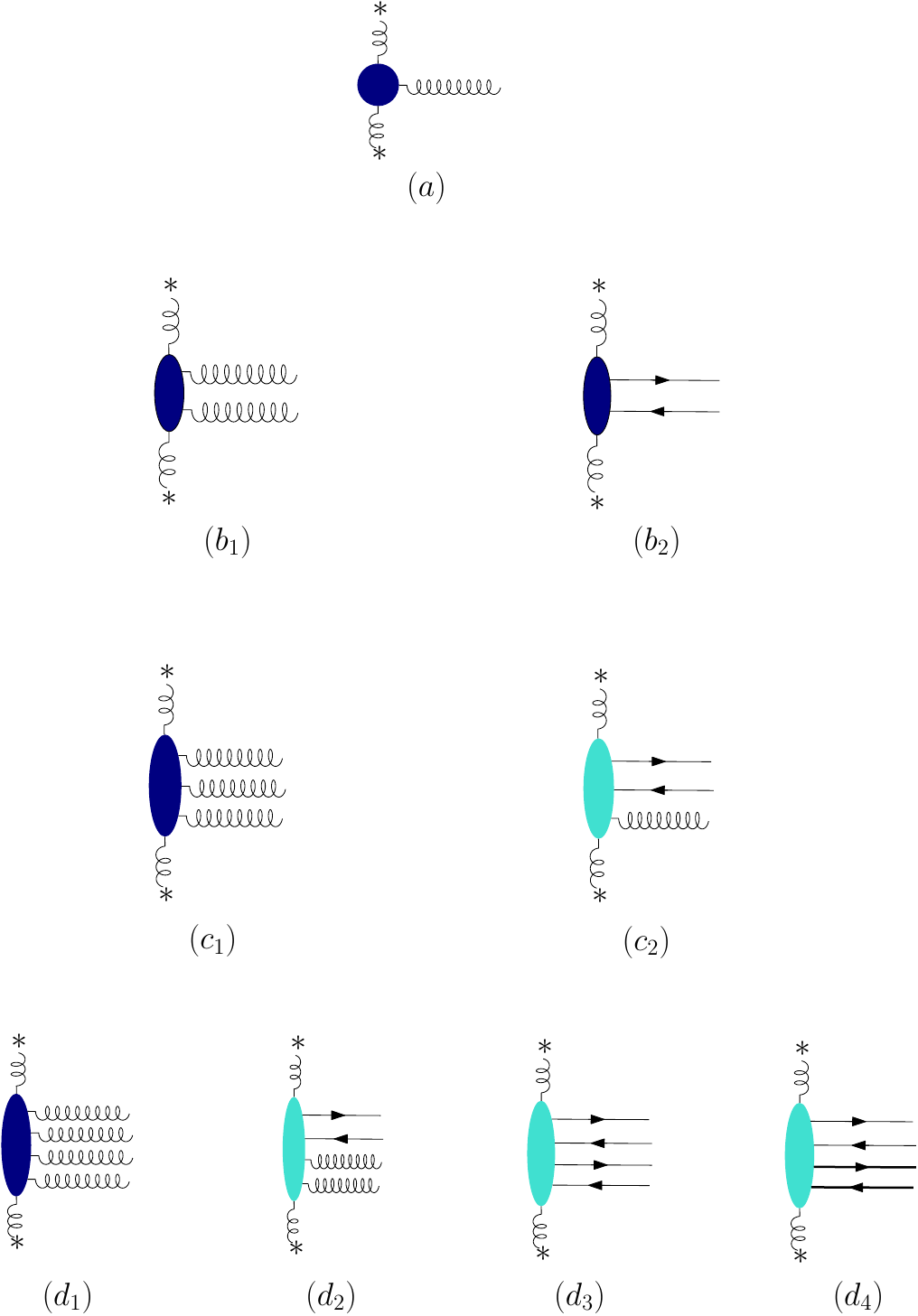}
    \caption{Central-emission vertices (CEVs) corresponding to final state multiplicities of up to four partons collected in our \texttt{\href{https://github.com/YuyuMo-UoE/Multi-Regge-Emission-Vertices}{Multi-Regge-Emission-Vertices}} (MREV) library~\cite{yuyu_2017_github}. The diagrams labelled $(a)$, $(b)$, $(c)$, and $(d)$ respectively represent final states with 1, 2, 3, or 4 partons, including all flavour configurations, ranging from  pure gluons to two quark--antiquark pairs.}
    \label{fig:list of results of CEV}
\end{figure}

In the previous section, we presented explicit expressions for CEVs with multiplicity up to three. The results are rather complex rational functions even in terms of the shorthand notation we adopted here. 
For higher multiplicities, it is necessary to use computer algebra to obtain and present the results. To this end, we have prepared a Mathematica library, \texttt{\href{https://github.com/YuyuMo-UoE/Multi-Regge-Emission-Vertices}{Multi-Regge-Emission-Vertices}} (\texttt{MREV})~\cite{yuyu_2017_github},
which contains results for MREVs with any helicity configuration and up to multiplicity~four.
Figures~\ref{fig:list of results of PEV} and~\ref{fig:list of results of CEV} categorize our PEV and CEV results according to multiplicity; those shown in light blue are new results. The library files containing the corresponding data are listed in Table~\ref{tab:PEVandCEVfiles}. 

The implementation consists of two layers. The lower layer is a library of ${\cal N}=4$ sYM colour-ordered MREVs of gluons and gluinos, which we derived based on amplitudes computed using GGT package~\cite{Dixon:2010ik} by taking the appropriate high-energy limits, and tabulated in several data files. The upper layer is a map between these sYM objects and QCD ones implementing the relations discussed in section~\ref{qcd_and_sym_diff}. This map is only non-trivial starting with MREVs with two distinct quark-antiquark pairs.

We note that the \texttt{MREV} library is fully self-contained: all the MREVs to multiplicity four have already been explicitly extracted and stored. Nevertheless, we present here also the extraction step, so as to make it possible to obtain results for MREVs at higher multiplicities or different (massless) theories from the corresponding amplitudes.

We proceed as follows. First, in subsection~\ref{extractCEVPEV}, we  present an explicit example demonstrating the extraction of both MREVs from a given  amplitude with massless partons, thereby illustrating how expressions in our library have been derived. 
Subsequently, in subsections~\ref{ggggCEVPEV}--\ref{qqQQCEVPEV}, we explain the use of the library by considering MREVs at multiplicity~four. Additional examples involving PEVs with lower multiplicities are provided in Appendix~\ref{colourdressingExample}.

\subsection{Examples of PEV and CEV extraction}
\label{extractCEVPEV}

In this section we provide a quick tutorial on using the \texttt{\href{https://github.com/YuyuMo-UoE/Multi-Regge-Emission-Vertices}{Multi-Regge-Emission-Vertices}} (\texttt{MREV})~\cite{yuyu_2017_github} package and library. We demonstrate  both how we used the \texttt{GGT}~\cite{Dixon:2010ik} package to derive the MREVs and how to retrieve the resulting expressions from the library.
We use two examples. One is the QCD colour-ordered PEV,
$P_{1\to 4}\big(\bar Q_2^\ominus,\, Q_3^\oplus,\, q_4^\oplus,\, \bar{q}_5^\ominus,\, g_6^\oplus,\, g_{q_4^\perp}^*\big)$,
and the other is the QCD colour-ordered CEV
$C_{3}\big(g_{q_1^\perp}^*,\, q_4^\oplus,\, \bar{q}_5^\ominus,\, g_6^\oplus,\, g_{q_4^\perp}^*\big)$.
To determine these MREVs we start, respectively, with the following  
colour-ordered seven-point amplitudes: 
\begin{align}
\label{A2to5:Q}
A_{2\to 5}^{\mathrm{QCD}}\big(g_1^\oplus,\,  \bar Q_2^\ominus,\, Q_3^\oplus,\, q_4^\oplus,\, \bar{q}_5^\ominus,\, g_6^\oplus,\, g_7^\ominus\big)&=A_{2\to 5}^{\mathcal{N}=4\, \text{sYM}}\big(g_1^\oplus,\,  \tilde{g}_{A,2}^\ominus,\, \tilde{g}_{A,3}^\oplus,\, \tilde{g}_{A,4}^\oplus,\, \tilde{g}_{A,5}^\ominus,\, g_6^\oplus,\, g_7^\ominus\big),
\\
\label{A2to5:g}
A_{2\to 5}^{\mathrm{QCD}}\big(g_1^\oplus,\, g_2^\oplus,\, g_3^\ominus,\, q_4^\oplus,\, \bar{q}_5^\ominus,\, g_6^\oplus,\, g_7^\ominus\big)&=A_{2\to 5}^{\mathcal{N}=4\, \text{sYM}}\big(g_1^\oplus,\, g_2^\oplus,\, g_3^\ominus,\, \tilde{g}_{A,4}^\oplus,\, \tilde{g}_{A,5}^\ominus,\, g_6^\oplus,\, g_7^\ominus\big).
\end{align}
Note that we have chosen these
amplitudes to be  helicity conserving in the relevant high-energy limits which define the $P_{1\to4}$ and $C_3$ MREVs under consideration. That is, gluons 1 and 7 have opposite helicities, as do gluons 2 and 3.
Also note that in Eq.~\eqref{A2to5:Q} we use gluino pairs of the same R-charge to replace both quark-antiquark pairs, rather than the na\"ive replacement,
$A_{2\to 5}^{\mathcal{N}=4\, \text{sYM}}\big(g_1^\oplus,\,  \tilde{g}_{B,2}^\ominus,\, \tilde{g}_{B,3}^\oplus,\, \tilde{g}_{A,4}^\oplus,\, \tilde{g}_{A,5}^\ominus,\, g_6^\oplus,\, g_7^\ominus\big)$, which features
gluinos of two distinct R-charge indices.
In this way we eliminate the scalar interaction, and thus directly  obtain the required QCD amplitude, in line with  section~\ref{qcd_and_sym_diff} (see Eq.~\eqref{MREVadjacentPlus}).
In contrast, for 
(\ref{A2to5:g}) there is a single gluino pair so there are no scalar interactions. Thus, here no subtleties arise and the colour-ordered amplitudes with a straightforward replacement of quarks by gluinos are identical in the two theories.

To begin we need to load the \texttt{GGT}~\cite{Dixon:2010ik} and \texttt{S\&M}~\cite{Maitre:2007jq} packages: 
\begin{lstlisting}
SetDirectory["path/to/GGT"]; 
<< GGT.m
\end{lstlisting}
\begin{lstlisting}
$SpinorsPath ="path/to/Spinors-1.0"; 
If[Not[MemberQ[$Path, $SpinorsPath]], $Path = Flatten[{$Path, $SpinorsPath }]]; 
<< Spinors`.
\end{lstlisting}
Then, the following commands initialise our \texttt{MREV} package:
\begin{lstlisting}
SetDirectory["path/to/MREVinit.m"]; 
Import["MREVinit.m"].
\end{lstlisting}
Next, we extract the PEV and CEV by rewriting the amplitudes in terms of the minimal set of light-cone variables (MSLCV),
\begin{equation}
\begin{split}
A_{2\to 5}^{\mathcal{N}=4\, \text{sYM}}\big(g_1^\oplus,\,  \tilde{g}_{A,2}^\ominus,\, \tilde{g}_{A,3}^\oplus,\, \tilde{g}_{A,4}^\oplus,\, \tilde{g}_{A,5}^\ominus,\, g_6^\oplus,\, g_7^\ominus\big)
\to
A_{2\to 5}^\prime\big(q_1^\perp 
 ,X_1,X_{2},X_{3},X_{4},
 z_1,z_2,z_3
 \big)
 \\
 A_{2\to 5}^{\mathcal{N}=4\, \text{sYM}}\big(g_1^\oplus,\, g_2^\oplus,\, g_3^\ominus,\, \tilde{g}_{A,4}^\oplus,\, \tilde{g}_{A,5}^\ominus,\, g_6^\oplus,\, g_7^\ominus\big)
 \to
A_{2\to 5}^{\prime\prime}\big(q_1^\perp 
 ,X_1,X_{2},X_{3},X_{4},
 z_1,z_2,z_3
 \big)
\end{split}
\end{equation}
 by leveraging the spinor-to-momentum transformation, Eq.~(\ref{spinoconven}), and the mapping to MSLCV, Eq.~(\ref{minival:nonlorentzinvdd}). This is implemented in \texttt{MREV} via:
\begin{lstlisting}
Amp1 = GGTnmhv4fermS[7, {3, 4}, {2, 5}] // Expand // Map[GGTtoMSLCV[#, 7] &]; 
Amp2 = GGTnmhv2ferm[7, 3, 4, 5] // Expand // Map[GGTtoMSLCV[#, 7] &]; 
\end{lstlisting}
where, as explained in Appendix~C of \texttt{GGT}~\cite{Dixon:2010ik}, the function  \texttt{GGTnmhv4fermS[7, \{3, 4\}, \{2, 5\}]} generates the $\mathcal{N}=4\, \mathrm{sYM}$ helicity  amplitude for a 7-parton process in which particles 3 and 4 are positive-helicity gluinos, particles 2 and 5 are negative-helicity gluinos, and the remaining particles are gluons, with gluon~7 having negative helicity, while the rest (gluons 1 and 6) have positive helicity. For the second case,
the function~\texttt{GGTnmhv2ferm[7, 3, 4, 5]} generates the one in which particle~4 is a positive-helicity gluino, particle~5 a negative-helicity gluino, and particles~3 and~7 are negative-helicity gluons, while the rest, gluons 1 and 6 have positive helicities.
The function \texttt{GGTtoMSLCV} directly converts helicity amplitudes from the \texttt{GGT} package into the MSLCV representation.

\begin{table}
\centering
\begin{tabular}{|c|l|}
\hline
\textbf{Figure  Label} & \textbf{  File within the MREV library} \\
\hline
\multicolumn{2}{|c|}{Figure~\ref{fig:list of results of PEV}} \\
\hline
$(a_1)$, $(a_2)$, $(b_1)$, $(b_2)$, $(b_3)$ & \texttt{PEV\_1to1\_and\_1to2.m} \\
$(c_1)$, $(c_2)$, $(c_3)$, $(c_4)$, $(c_5)$ & \texttt{PEV\_1to3.m} \\
$(d_1)$ & \texttt{PEV\_1to4\_ggggg.m} \\
$(d_2)$, $(d_3)$ & \texttt{PEV\_1to4\_qqbggg.m} \\
 $(d_4)$, $(d_6)$ & \texttt{PEV\_1to4\_qqbqqbg.m} \\
$(d_5)$, $(d_7)$ & \texttt{PEV\_1to4\_qqbQQbg\_.m} \\
\hline
\multicolumn{2}{|c|}{Figure~\ref{fig:list of results of CEV}} \\
\hline
$(a)$, $(b_1)$, $(b_2)$ & \texttt{CEV\_1\_and\_2.m} \\
$(c_1)$, $(c_2)$ & \texttt{CEV\_3.m} \\
$(d_1)$ & \texttt{CEV\_4\_gggg.m} \\
$(d_2)$ & \texttt{CEV\_4\_qqbgg.m} \\
$(d_3)$ & \texttt{CEV\_4\_qqbqqb.m} \\
$(d_4)$ & \texttt{CEV\_4\_qqbQQb_.m} \\
\hline
\end{tabular}
\caption{Correspondence between the PEVs in figure~\ref{fig:list of results of PEV}, the CEVs in figure~\ref{fig:list of results of CEV}, and the data files within the \texttt{\href{https://github.com/YuyuMo-UoE/Multi-Regge-Emission-Vertices}{Multi-Regge-Emission-Vertices}} (MREV) library.
}
\label{tab:PEVandCEVfiles}
\end{table}

The PEV, $P_{1\to 4}\big(\bar Q_2^\ominus,\, Q_3^\oplus,\, q_4^\oplus,\, \bar{q}_5^\ominus,\, g_6^\oplus,\, g_{q_4^\perp}^*\big)$, is extracted from the amplitude (\ref{A2to5:Q}) according to the following equation
\begin{equation}
    \begin{split}
      &\hspace*{-20pt}\lim_{y_3 \,\simeq\, y_4 \,\simeq\, y_5 \,\simeq\, y_6 \gg y_7}
     A_{2\to 5}^{\mathcal{N}=4\, \text{sYM}}\big(g_1^\oplus,\,  \tilde{g}_{A,2}^\ominus,\, \tilde{g}_{A,3}^\oplus,\, \tilde{g}_{A,4}^\oplus,\, \tilde{g}_{A,5}^\ominus,\, g_6^\oplus,\, g_7^\ominus\big) \\
      &\qquad = s\,\,P_{1\to 4}\big( \tilde{g}_{A,2}^\ominus,\, \tilde{g}_{A,3}^\oplus,\, \tilde{g}_{A,4}^\oplus,\, \tilde{g}_{A,5}^\ominus,\, g_6^\oplus,\, g_{q_4^\perp}^*\big)
      \frac{1}{t_4}
      {P}_{1 \rightarrow 1}\big(g_{q_4^\perp}^*,\, g_7^\ominus,\, g_1^\oplus\big)\\
      &\qquad = s\,\,P_{1\to 4}\big(\bar Q_2^\ominus,\, Q_3^\oplus,\, q_4^\oplus,\, \bar{q}_5^\ominus,\, g_6^\oplus,\, g_{q_4^\perp}^*\big)
      \frac{1}{t_4}
      {P}_{1 \rightarrow 1}\big(g_{q_4^\perp}^*,\, g_7^\ominus,\, g_1^\oplus\big),
    \end{split}
\label{A_P_P}
\end{equation}
where we have used eq.~\eqref{MREVadjacentPlus} in the final line.
In MSLCV it is
  \begin{eqnarray}
   \lim\limits_{X_{4}\to \infty } &&
A^\prime_{2\to 5}\big(q_1^\perp 
 ,X_1,X_{2},X_{3},X_{4},
 z_1,z_2,z_3
 \big)
 \non &=& s   P_{1\to 4}\big(q_1^\perp 
 ,X_1,X_{2},X_{3},
 z_1,z_2,z_3
 \big)\frac{1}{t_{4}}P_{1 \rightarrow 1}\big(q_{n-3}^\perp  \big).
\end{eqnarray}
This extraction is automated using the \texttt{getPEV} function as follows:
\begin{lstlisting}
PEV$ = Amp1 // Map[getPEV[#, 7, {p, m}] &];.
\end{lstlisting}
Here, \texttt{\{p, m\}} specifies the helicities of the gluons $g_1$ and $g_7$, respectively, which are needed to obtain the correct $P_{1 \to 1}$ in Eq.~\eqref{A_P_P}. 

For amplitudes with a different multiplicity, the argument \texttt{7}, which denotes the seven-point amplitude $A_{2 \rightarrow 5}^{\mathcal{N}=4 \,\mathrm{sYM}}\big(g_1^{\oplus}, \tilde{g}_{A, 2}^{\ominus}, \tilde{g}_{A, 3}^{\oplus}, \tilde{g}_{A, 4}^{\oplus}, \tilde{g}_{A, 5}^{\ominus}, g_6^{\oplus}, g_7^{\ominus}\big)$, should, of course, be replaced with the appropriate number of external legs.

The CEV $C_{3}\big(g_{q_1^\perp}^*,\, q_4^\oplus,\, \bar{q}_5^\ominus,\, g_6^\oplus,\, g_{q_4^\perp}^*\big)$ is extracted directly from the amplitude, Eq.~(\ref{A2to5:g}), as follows:
\begin{equation}
    \begin{split}
         &\lim_{y_3 \gg\, y_4 \,\simeq\,\, y_5 \simeq\, y_6 \gg\, y_7}
         A_{2\to 5}^{\mathcal{N}=4\, \text{sYM}}\big(g_1^\oplus,\, g_2^\oplus,\, g_3^\ominus,\, \tilde{g}_{A,4}^\oplus,\, \tilde{g}_{A,5}^\ominus,\, g_6^\oplus,\, g_7^\ominus\big)\\
        &\hspace*{40pt}=\lim_{y_3 \gg\, y_4 \,\simeq\,\, y_5 \simeq\, y_6 \gg\, y_7}
        A_{2\to 5}^{\text{QCD}}\big(g_1^\oplus,\, g_2^\oplus,\, g_3^\ominus,\, q_4^\oplus,\, \bar{q}_5^\ominus,\, g_6^\oplus,\, g_7^\ominus\big)\\
        &\hspace*{40pt}= s\,  {P}_{1 \rightarrow 1}\big( g_2^\oplus,\, g_3^\ominus,\, g_{q_1^\perp}^*\big)\,   \frac{1}{t_1}\, 
        C_{3}\big(g_{q_1^\perp}^*,\, q_4^\oplus,\, \bar{q}_5^\ominus,\, g_6^\oplus,\, g_{q_4^\perp}^*\big)\, 
        \frac{1}{t_4}\, 
        {P}_{1 \rightarrow 1}\big( g_{q_4^\perp}^*,\,g_7^\ominus,\, g_1^\oplus\big)\,.
    \end{split}
\label{cevppcc}
\end{equation}
In MSLCV, it reads
 \begin{eqnarray}
   \lim\limits_{X_{1},X_{4}\to \infty }
   &&
{A}_{2\to 5}\big(q_1^{\perp}, X_1, X_2,X_3,X_4, z_1,z_2,z_3\big)= 
 \non
 &&
s 
P_{1 \rightarrow 1}\big(q_{1}^\perp \big) 
\frac{1}{t_{1}}
C_{3}\big(q_1^\perp 
 ,X_2,X_3,
{z}_{1},{z}_{2},{z}_{3}
 \big)
\frac{1}{t_{n-3}}
P_{1 \rightarrow 1}\big(q_{n-3}^\perp \big)\,.
\end{eqnarray}
This extraction is performed automatically by the \texttt{getCEV} function as:
\begin{lstlisting}
CEV$ = Amp2 // Map[getCEV[#, 7, {p, p, m, m}] &];.
\end{lstlisting}
Here, \texttt{\{p, p, m, m\}} specifies the helicities of the gluons $g_1$, $g_2$, $g_3$ and $g_7$, respectively, which are needed to obtain the correct $P_{1 \rightarrow 1}$ in Eq.~\eqref{cevppcc}.
This extraction 
reproduces the result given in Eq.~\eqref{qPqbMgP}. Furthermore, one can verify that the extracted expressions \texttt{cev} and \texttt{pev}, for a given helicity configuration, are independent of the amplitude from which they are obtained.

As a first consistency check, one may verify that the extracted PEV and CEV coincide with the predetermined ones present in the library. 
To this end one first loads the data:
\begin{lstlisting}
Import["CEV_3.m"];
Import["PEV_1to4_qqbqqbg.m"];.
\end{lstlisting}
Next one may issue the following command and find that \texttt{cev} and \texttt{pev} agree with the previously computed results:
\begin{lstlisting}
cevLib=cev[P, qP, qM][(*@$  \texttt{X}_2, \texttt{X}_3,   \texttt{q}[1,``\perp"], \texttt{q}[1,``\perp*"],   \texttt{z}_1, \texttt{zb}_1, \texttt{z}_2, \texttt{zb}_2,   \texttt{z}_3, \texttt{zb}_3$@*)]  //  Factor;
In[1]:= CEV$ - cevLib
Out[1]= 0

pevLib=pev[{qM, Q}, {qP, Q}, {qP, q}, {qM, q}, {P, g}][(*@$  \texttt{X}_1, \texttt{X}_2, \texttt{X}_3,   \texttt{q}[1,``\perp"], \texttt{q}[1,``\perp*"],   \texttt{z}_1, \texttt{zb}_1, \texttt{z}_2, \texttt{zb}_2,   \texttt{z}_3, \texttt{zb}_3$@*)] //  Factor; 
In[2]:= PEV$ - pevLib
Out[2]= 0.
\end{lstlisting}
As illustrated in this example, the input for the MSLCV representation of positive rapidity PEVs takes the form
\texttt{pev[helicity labels][MSLCV labels]},
and that of CEVs takes the form
\texttt{cev[helicity labels][MSLCV labels]}.
The dependence on the MSLCV is specified according to Eq.~\eqref{pevbtypedef1} and Eq.~\eqref{cevbtypedef2}, respectively.

As explained above, because of the additional scalar interaction
\begin{equation}
    P_{1 \rightarrow 4}\big(\bar{Q}_{2}^{\ominus}, Q_{\sigma_3}^{\oplus}, q_{\sigma_4}^{\oplus}, \bar{q}_{\sigma_5}^{\ominus}, g_{\sigma_6}^{\oplus}, g_{q_4^{\perp}}^*\big)
    \ne
    P_{1 \rightarrow 4}^{{\mathcal{N}=4\, \mathrm{sYM}}}\big(\tilde{g}_{Q,2}^{\ominus}, \tilde{g}_{Q,3}^{\oplus}, \tilde{g}_{q,4}^{\oplus}, \tilde{g}_{q,5}^{\ominus}, g_{6}^{\oplus}, g_{q_4^{\perp}}^*\big)\,.
\end{equation}
The command \texttt{pev} (\texttt{cev})  directly provides the QCD PEV (CEV)
after applying the map from ${\cal N}=4$ to QCD, which eliminates the scalar interactions. 
If one is interested instead in the ${\cal N}=4$ sYM prior to applying this map, one may use instead  \texttt{PEV} (\texttt{CEV}) defined in capital letters.
The following table provides the complete set of commands used to retrieve MREVs from the library and manipulate the obtained expressions. 
\begin{table}[H]
\centering
\renewcommand{\arraystretch}{1.2}
\begin{tabular}{lll}
\toprule
Theory & Representation & Notation \\
\midrule
QCD & MSLCV & \texttt{pev[helicity][MSLCV labels]} \\
    &           & \texttt{cev[helicity][MSLCV labels]} \\
    & momentum  & \texttt{pevp[helicity][momentum labels]} \\
    &              & \texttt{cevp[helicity][momentum labels]} \\
    & transition code & \texttt{pevtopevp} \\
    &                   & \texttt{cevtocevp} \\
\midrule
$\mathcal{N}=4$ sYM & MSLCV  & \texttt{PEV[helicity][MSLCV]} \\
    &               & \texttt{CEV[helicity][MSLCV labels]} \\
    & momentum   & \texttt{PEVp[helicity][momentum labels]} \\
    &               & \texttt{CEVp[helicity][momentum labels]} \\
    & transition code & \texttt{PEVtoPEVp} \\
    &                   & \texttt{CEVtoCEVp} \\
\bottomrule
\end{tabular}
\caption{Notation of PEVs and CEVs in the \texttt{MREV}~\cite{yuyu_2017_github} library for QCD and $\mathcal{N}=4$ sYM.}
\end{table}

Next we note that $P_{1\to 4}\big(\bar Q_2^\ominus,\, Q_3^\oplus,\, q_4^\oplus,\, \bar{q}_5^\ominus,\, g_6^\oplus,\, g_{q_4^\perp}^*\big)$ and $C_{3}\big(g_{q_1^\perp}^*,\, q_4^\oplus,\, \bar{q}_5^\ominus,\, g_6^\oplus,\, g_{q_4^\perp}^*\big)$ are related by the factorization property in the limit where the rapidity $y_3$ is taken much larger than the others:
\begin{equation}
    \begin{split}
         &\hspace*{-40pt}\lim_{y_3 \, \gg\,y_4 \simeq\,y_5   \simeq\, y_{6}}
 P_{1\to 4}\big(\bar Q_2^\ominus,\, Q_3^\oplus,\, q_4^\oplus,\, \bar{q}_5^\ominus,\, g_6^\oplus,\, g_{q_4^\perp}^*\big)\\
    &=s\,  {P}_{1 \rightarrow 1}\big( \bar Q_2^\ominus,\,  Q_3^\oplus,\, g_{q_1^\perp}^*\big)\,   \frac{1}{t_1}\,C_{3}\big(g_{q_1^\perp}^*,\, q_4^\oplus,\, \bar{q}_5^\ominus,\, g_6^\oplus,\, g_{q_4^\perp}^*\big),
    \end{split}
    \label{check2}
\end{equation}
which provides a second consistency check of the results.
This is implemented as follows:
\begin{lstlisting}
Import["PEV_1to1_and_1to2.m"];
\end{lstlisting}
\begin{lstlisting}
In[2]:= PEV$/(pev[qM, qP][(*@$ \texttt{q}[1,``\perp"], \texttt{q}[1,``\perp*"]$@*)]* 1/((*@$ \texttt{q}[1,``\perp"]* \texttt{q}[1,``\perp*"]$@*))*CEV$)/. Subscript[X, 1] -> 2^32 /. nptxX1;
Out[3]= 1. -2.00577*10^-10 I.
\end{lstlisting}
where the numerical check has been done with a random numerical point \texttt{nptxX1} given in eq. \eqref{npt2}.

As explained in section~\ref{resultsof CEVPEV},
momentum permutations are useful for obtaining the complete set of colour-ordered MREVs required for colour-dressing. 
To implement such permutations it is convenient to use a momentum representation of the MREV expressions. 
Starting from the MSLCV representation, it is straightforward to convert to the momentum representation by inverting Eq.~(\ref{minival:nonlorentzinvdd}). This is implemented by the functions \texttt{pevtopevp} and \texttt{cevtocevp}, which convert the MSLCV representation of \texttt{pev} and \texttt{cev} to momentum representation \texttt{pevp} and \texttt{cevp}, respectively, for up to four final-state particles:
\begin{lstlisting}
pev[{qM, Q}, {qP, Q}, {qP, q}, {qM, q}, {P, g}] // pevtopevp
cev[qP, qM, P] // cevtocevp
\end{lstlisting}
For the PEV under consideration, $P_{1 \rightarrow 4}\big(\bar{Q}_2^{\ominus}, Q_3^{\oplus}, q_4^{\oplus}, \bar{q}_5^{\ominus}, g_6^{\oplus}, g_{q_4^{\perp}}^*\big)$, all permutations of the emitted particle momenta with fixed helicity and particle content can then be called by 
\begin{align}
   & P_{1 \rightarrow 4}\big(\bar{Q}_{2}^{\ominus}, Q_{\sigma_3}^{\oplus}, q_{\sigma_4}^{\oplus}, \bar{q}_{\sigma_5}^{\ominus}, g_{\sigma_6}^{\oplus}, g_{q_4^{\perp}}^*\big)
    \nonumber
    \\
    &\hspace{1cm }=
\mathtt{pevp[\{qM, Q\}, \{qP, Q\}, \{qP, q\}, \{qM, q\}, \{P, g\}][\text{$\sigma$}_3,\text{$\sigma$}_4,\text{$\sigma$}_5,\text{$\sigma$}_6]}.
\label{pevcodeexample1}
\end{align}
Here, we omit the label ``2'' (corresponding to the incoming parton) in $[\sigma_3,\, \sigma_4,\, \sigma_5,\, \sigma_6]$. The momentum dependence of $g_2^\oplus$ and $g_{q_4^\perp}^*$ does not need to be specified in the code, since it is fixed by momentum conservation in the plus and transverse directions, respectively.

Similarly, for the CEV under consideration, $C_{3}\big(g_{q_1^\perp}^*,\, q_{\sigma_4}^\oplus,\, \bar{q}_{\sigma_5}^\ominus,\, g_{\sigma_6}^\oplus,\, g_{q_4^\perp}^*\big)$,  all permutations of the emitted particle momenta 
with fixed helicity and particle content, can be called by 
\begin{align}
   & C_{3}\big(g_{q_1^\perp}^*,\, q_{\sigma_4}^\oplus,\, \bar{q}_{\sigma_5}^\ominus,\, g_{\sigma_6}^\oplus,\, g_{q_4^\perp}^*\big)=\mathtt{cevp[qP,\, qM,\, P][1,\text{$\sigma$}_4,\text{$\sigma$}_5,\text{$\sigma$}_6]}.
   \label{cevcodeexample1}
\end{align}
Here, the label ``1'' denotes the off-shell gluon $g^*_{q^\perp_1}$ in the colour-ordered CEV (\emph{cf.}\ Figure~\ref{fig:ddpevcev}). The momentum dependence of the other off-shell gluon $g_{q_4^\perp}^*$ in Eq.~(\ref{cevcodeexample1}) does not need to be specified in the code. Further details on how to use the library will be provided in the next subsections in the context of the more complex multiplicity-four MREVs.

\subsection{The \texorpdfstring{$g^*g\to gggg$}{g* g → gggg} PEV and the \texorpdfstring{$g^*g^*\to gggg$}{g* g* → gggg} CEV}
\label{ggggCEVPEV}
According to Eq.~\eqref{PEVDDM}, the four-gluon PEV $g^* g \rightarrow g g g g$ presented first in \cite{DelDuca:1999iql} takes the following form
\begin{align}
& \mathcal{P}_{1\to 4} \big( g_{2}^{h_2},\ g_{3}^{h_3},\ g_{4}^{h_4},\ g_{5}^{h_5},\ g_{6}^{h_6},\ g_{q_4^\perp}^* \big)  \nonumber\\
& =
g^4 \sum_{\sigma \in S_4}
 \big( F^{a_{\sigma_2}} F^{a_{\sigma_3}} F^{a_{\sigma_4}} F^{a_{\sigma_5}} F^{a_{\sigma_6}} \big) _{c_1 c_4}
P_{1\to 4} \big( g_{{\sigma_2}}^{h_{\sigma_2}},\ g_{{\sigma_3}}^{h_{\sigma_3}},\ g_{{\sigma_4}}^{h_{\sigma_4}},\ g_{{\sigma_5}}^{h_{\sigma_5}},\ g_{{\sigma_6}}^{h_{\sigma_6}},\ g_{q_4^\perp}^* \big) \,.
\label{eq610}
\end{align}
From this formula it is clear that one needs to compute different permutations of the colour ordered PEV. This may be done by constructing the corresponding momentum permutation replacement list in the MSLCVs, as described in section~\ref{pp}. However, within the code, it is more convenient to first convert the expression into the momentum representation and then directly permute the momentum  labels before returning to the MSLCVs. In the following we explain how this is done using in the library.

To get the data from the library, one first issues the following commands to initialise the \texttt{MREV} package:
\begin{lstlisting}
Quit[];
SetDirectory["path/to/MREVinit.m"];
Import["MREVinit.m"];.
\end{lstlisting}
The next step is to load the four-gluon PEVs $g^* g \rightarrow g g g  g$ expressed in terms of MSLCVs, which are tabulated in \texttt{PEV\_4\_emissions\_purely\_gluon.m}:
\begin{lstlisting}
Import["PEV_1to4_ggggg.m"];.
\end{lstlisting}
Next, one should initialise the momentum representation of the colour-ordered PEVs:
\begin{align*}
&\mathtt{pevtopevp[\#] \& /@} \big\{
\mathtt{pev[P,P,P,M,M],} \, \mathtt{pev[P,P,M,M,P],} \, \mathtt{pev[P,M,P,M,P],}\, ...\big\},
\end{align*}
where \texttt{P} and \texttt{M} represent~$g^\oplus$ and~$g^\ominus$, respectively, and the set in the curly brackets consists of all helicity configurations that are required in order to write the colour dressed PEV of interest. 
Then a given colour-ordered pure gluon PEV in the momentum representation can be called by
\begin{align}
   & {P}_{1\to 4} \big( 
      g_{{2}}^{h_{2}},
      \, g_{{\sigma_3}}^{h_{\sigma_3}},
      \, g_{{\sigma_4}}^{h_{\sigma_4}},
      \, g_{{\sigma_5}}^{h_{\sigma_5}},
      \, g_{{\sigma_6}}^{h_{\sigma_6}},
      \, g_{q_4^\perp}^*
     \big) 
    =
    \mathtt{pevp[h_{2},h_{\text{$\sigma$}_3},h_{\text{$\sigma$}_4},h_{\text{$\sigma$}_5},h_{\text{$\sigma$}_6}][\text{$\sigma$}_3,\text{$\sigma$}_4,\text{$\sigma$}_5,\text{$\sigma$}_6]}
\end{align}
with $\mathtt{h_{\text{$\sigma$}_i}=P \text{ or } M}$. For example, one instance is given by
\begin{align}
    {P}_{1\to 4}\big( g_{{2}}^{\oplus},\, g_{5}^{\ominus},\, g_{6}^{\oplus},\, g_{4}^{\ominus},\, g_{{3}}^{\oplus},\, g_{q_4^\perp}^*\big)
 = \mathtt{pevp[P,M,P,M,P][5,6,4,3]}.
\end{align}

In a similar manner we now consider the colour-dressed four-gluon CEV. According to Eq.~\eqref{CEVDDM} it has the following form~\cite{DelDuca:1999iql}
\begin{align}
&\mathcal{C}_4 \big( g_{q_1^\perp}^*,\, g_{4}^{h_4},\, g_{5}^{h_5},\, g_{6}^{h_6},\, g_{7}^{h_7},\, g_{q_5^\perp}^* \big) 
\nonumber
\\
& =
 g^{4} \sum_{\sigma \in S_{4}} \big( F^{a_{\sigma_4}} F^{a_{\sigma_5}}F^{a_{\sigma_6}}F^{a_{\sigma_7}} \big) _{c_1 c_2} 
C_4 \big( g_{q_1^\perp}^*,\, g_{{\sigma_4}}^{h_{\sigma_4}},\, g_{{\sigma_5}}^{h_{\sigma_5}},\, g_{{\sigma_6}}^{h_{\sigma_6}},\, g_{{\sigma_7}}^{h_{\sigma_7}},\, g_{q_5^\perp}^* \big) .
\end{align}
As with the PEV case, to begin one loads the
four-gluon CEVs $g^* g^* \rightarrow g g g  g$ in terms of MSLCVs, which are 
tabulated in \texttt{CEV\_4\_gggg.m}:
\begin{lstlisting}
Import["CEV_4_gggg.m"].
\end{lstlisting}
Next, one should initialise the momentum representation of the colour-ordered CEVs
\begin{align*}
&\mathtt{cevtocevp[\#] \& /@} \big\{\mathtt{cev[P,P,M,M],}\, \mathtt{cev[P,M,M,P],}\, 
\mathtt{cev[M,P,M,P],}\, \mathtt{cev[P,M,P,M],}\, ...\big\}
\end{align*}
where \texttt{P, M} in the arguments represent $g^\oplus$ and $g^\ominus$ respectively.
Then a given colour-ordered pure-gluon CEV in the momentum representation can be called by
\begin{align}
   & {C}_4 \big( g_{q_1^\perp}^*,\, g_{{\sigma_4}}^{h_{\sigma_4}},\, g_{{\sigma_5}}^{h_{\sigma_5}},\, g_{{\sigma_6}}^{h_{\sigma_6}},\, g_{{\sigma_7}}^{h_{\sigma_7}},\, g_{q_5^\perp}^* \big) =\mathtt{cevp[h_{\text{$\sigma$}_4},h_{\text{$\sigma$}_5},h_{\text{$\sigma$}_6},h_{\text{$\sigma$}_7}][1,\text{$\sigma$}_4,\text{$\sigma$}_5,\text{$\sigma$}_6,\text{$\sigma$}_7]}
\end{align}
with $\mathtt{h_{\text{$\sigma$}_i}=P \text{ or } M}$ and where ``1'' denotes the off-shell gluon $g^*_{q^\perp_1}$ in the colour-ordered CEV (cf.\ Figure~\ref{fig:ddpevcev}). The momentum dependence of the other off-shell gluon $g_{q_5^\perp}^*$ does not need to be specified in the code because of momentum conservation. For example, one permutation is given by 
\begin{align}
    {C}_4\bigl(g_{q_1^\perp}^*,\, g_{6}^{\ominus},\, g_{7}^{\oplus},\, g_{4}^{\ominus},\, g_{5}^{\oplus},\, g_{q_5^\perp}^*\bigr)
 = \mathtt{cevp[M,P,M,P][1,6,7,4,5]}.
\end{align}
In the following subsections we present multiplicity four MREVs with quarks.

\subsection{The \texorpdfstring{$g^*\bar{q}\to q g gg$}{g* qbar → q g gg} PEV and the \texorpdfstring{$g^*g^*\to \bar{q} q gg$}{g* g* → qbar q gg} CEV}

In analogy with the pure-gluon case discussed so far, the \texttt{MREV} library contains the full set of MREVs with quarks. These MREVs have not been computed before, and are presented here for the first time. 

The complete set of MREVs with gluons and quarks is presented in Figures~\ref{fig:list of results of PEV} and~\ref{fig:list of results of CEV}. Table~\ref{tab:PEVandCEVfiles} summarises these, pointing to the relevant files in the library. In the present subsection and in the following one we will discuss the structure of the colour-dressed MREVs containing respectively one and two quark-antiquark pairs, and show how to use the library to determine the relevant ingredients.

For PEVs of the type $g^*\bar{q}\to q g gg$, according to Eq.~\eqref{PEVgeneratpor} we consider the colour-dressing example of ${\cal P}_4 \big( \bar{q}_{2}^{\ominus}, q_{3}^{\oplus}, g_{4}^{\ominus}, g_{5}^{\oplus}, g_{6}^{\oplus}, g_{q_4^\perp}^* \big) $:
\begin{align}
     {\cal P}_{1\to 4} \big( \bar{q}_{2}^{\ominus}, q_{3}^{\oplus}, g_{4}^{\ominus}, g_{5}^{\oplus}, g_{6}^{\oplus}, g_{q_4^\perp}^* \big) 
    =\,\, & \big( T^{a_4}T^{a_5}T^{a_6} T^{c_4}  \big) _{i_3 {\bar{\imath}}_2} {P}_{1\to 4} \big( \bar{q}_{2}^{\ominus}, q_{3}^{\oplus}, g_{4}^{\ominus}, g_{5}^{\oplus}, g_{6}^{\oplus}, g_{q_4^\perp}^* \big) 
        \nonumber 
    \\
    &+\, \big( T^{a_4}T^{a_5} T^{c_4}T^{a_6}  \big) _{i_3 {\bar{\imath}}_2} {P}_{1\to 4} \big(  g_{6}^{\oplus} ,\bar{q}_{2}^{\ominus}, q_{3}^{\oplus}, g_{4}^{\ominus}, g_{5}^{\oplus}, g_{q_4^\perp}^*\big) 
        \nonumber 
    \\
    &+\, \big( T^{a_4} T^{c_4}T^{a_5}T^{a_6}  \big) _{i_3 {\bar{\imath}}_2} {P}_{1\to 4} \big(  g_{5}^{\oplus}, g_{6}^{\oplus},\bar{q}_{2}^{\ominus}, q_{3}^{\oplus}, g_{4}^{\ominus}, g_{q_4^\perp}^* \big) 
        \nonumber 
    \\
    &+\,  \big( T^{c_4}T^{a_4}T^{a_5} T^{a_6}  \big) _{i_3 {\bar{\imath}}_2} {P}_{1\to 4} \big( g_{4}^{\ominus}, g_{5}^{\oplus}, g_{6}^{\oplus} , \bar{q}_{2}^{\ominus}, q_{3}^{\oplus}, g_{q_4^\perp}^*\big) 
       \nonumber 
    \\
    &+\, \cdots,
    \label{qwndqqw22222}
\end{align}
where the ellipsis stand for additional colour orderings, where  $g_4^{\ominus},\, g_5^{\oplus}$ and $g_6^{\oplus}$ are permuted in the argument of ${P}_{1\to 4}$ in conjunction with  the corresponding colour indices $a_4,\,a_5$ and $a_6$.
Next we show how the individual colour-ordered PEVs are obtained. One first loads the data in terms of MSLCVs, 
\begin{lstlisting}
Import["PEV_1to4_qqbggg.m"],
\end{lstlisting}
and then initialise the momentum representation as 
\begin{align*}
&\mathtt{pevtopevp[\#] \& /@} \nonumber\\[1mm]
&\begin{array}{ll}
\big\{\mathtt{pev[qM,qP,M,P,P]}, \quad&\mathtt{pev[qM,qP,M,P,Pc]}, \\[1mm]
\mathtt{pev[qM,qP,M,Pc,Pc]}, \quad& \mathtt{pev[qM,qP,Pc,Pc,Mc]}\big\},
\end{array}\,.
\end{align*}
Here, \texttt{qP} and \texttt{qM} represent $q^\oplus$ and $\bar{q}^\ominus$, respectively, while \texttt{Pc} and \texttt{Mc} denote $g^\oplus$ and $g^\ominus$, respectively; the additional letter \texttt{c} indicates that their colour position is after the off-shell gluon $g_{q_4^{\perp}}^*$, following the clockwise ordering and starting from the incoming parton~2 (\emph{cf.}\ Figure~\ref{fig:ddpevcev}).
The colour-ordered PEVs in Eq.~(\ref{qwndqqw22222}) can be obtained as follows:
\begin{align}
     {P}_{1\to 4} \big( \bar{q}_{2}^{\ominus}, q_{3}^{\oplus}, g_{4}^{\ominus}, g_{5}^{\oplus}, g_{6}^{\oplus}, g_{q_4^\perp}^* \big) =\,\, &\mathtt{pevp[qM,qP,M,P,P][3,4,5,6]}
      \nonumber 
    \\
       {P}_{1\to 4} \big(  g_{6}^{\oplus},\bar{q}_{2}^{\ominus}, q_{3}^{\oplus}, g_{4}^{\ominus}, g_{5}^{\oplus}, g_{q_4^\perp}^* \big) =\,\,&\mathtt{pevp[qM,qP,M,P,Pc][3,4,5,6]}
             \nonumber 
    \\
       {P}_{1\to 4} \big( g_{5}^{\oplus}, g_{6}^{\oplus} , \bar{q}_{2}^{\ominus}, q_{3}^{\oplus}, g_{4}^{\ominus}, g_{q_4^\perp}^*\big) =\,\,&\mathtt{pevp[qM,qP,M,Pc,Pc][3,4,6,5]}
       \nonumber 
    \\
       {P}_{1\to 4} \big(g_{4}^{\ominus}, g_{5}^{\oplus}, g_{6}^{\oplus}, \bar{q}_{2}^{\ominus}, q_{3}^{\oplus}, g_{q_4^\perp}^* \big) =\,\,&\mathtt{pevp[qM,qP,Pc,Pc,Mc][3,6,5,4]}.
\end{align}

We now consider a CEV example. According to Eq.~\eqref{CEVgeneratpor},
the colour-dressed CEV, $\mathcal{C}_4 \big( g_{q_1^\perp}^*, \bar{q}_{4}^{\ominus}, q_{5}^{\oplus}, g_{6}^{\ominus}, g_{7}^{\oplus}, g_{q_5^\perp}^* \big)$, may be expressed as:
\begin{align}
    \mathcal{C}_4 \big( g_{q_1^\perp}^*,\bar{q}_{{4}}^{\ominus} , q_{5}^{\oplus},g_{6}^{\ominus}, g_{7}^{\oplus},g_{q_5^\perp}^* \big) 
    =\,\,& \big( T^{a_6}T^{a_7}T^{c_5} T^{c_1}  \big) _{i_5 \bar{\imath}_4} {C}_4 \big( g_{q_1^\perp}^*,\bar{q}_{{4}}^{\ominus} , q_{5}^{\oplus},g_{6}^{\ominus}, g_{7}^{\oplus},g_{q_5^\perp}^* \big) 
        \nonumber 
    \\
    &+\, \big( T^{a_6}T^{a_7}T^{c_1} T^{c_5}  \big) _{i_5 \bar{\imath}_4} {C}_4 \big( \bar{q}_{{4}}^{\ominus} , q_{5}^{\oplus},g_{6}^{\ominus}, g_{7}^{\oplus},g_{q_1^\perp}^*,g_{q_5^\perp}^* \big) 
      \nonumber 
    \\
    &+\, \big( T^{a_6}T^{c_5}T^{a_7} T^{c_1}  \big) _{i_5 \bar{\imath}_4} {C}_4 \big( g_{7}^{\oplus},g_{q_1^\perp}^*,\bar{q}_{{4}}^{\ominus} , q_{5}^{\oplus},g_{6}^{\ominus},g_{q_5^\perp}^* \big) 
      \nonumber 
    \\
     &+\, \big( T^{a_6}T^{c_1}T^{a_7} T^{c_5}  \big) _{i_5 \bar{\imath}_4} {C}_4 \big( \bar{q}_{{4}}^{\ominus} , q_{5}^{\oplus},g_{6}^{\ominus},g_{q_1^\perp}^*,g_{7}^{\oplus},g_{q_5^\perp}^* \big) 
      \nonumber 
    \\
     &+\, \big( T^{c_1}T^{a_6}T^{a_7} T^{c_5}  \big) _{i_5 \bar{\imath}_4} 
    C_4 \big( \bar{q}_{\bar{q}_4}^{\ominus}, q_{5}^{\oplus} , g_{q_1^\perp}^*,g_{6}^{\ominus}, g_{7}^{\oplus}, g_{q_5^\perp}^*\big) 
      \nonumber 
    \\
     &+\, \big( T^{c_5}T^{a_6}T^{a_7} T^{c_1}  \big) _{i_5 \bar{\imath}_4}
    C_4 \big(  g_{6}^{\ominus}, g_{7}^{\oplus},g_{q_1^\perp}^*, \bar{q}_{4}^{\ominus}, q_{5}^{\oplus}, g_{q_5^\perp}^* \big) 
      \nonumber 
    \\
   &+\, \big( T^{c_5}T^{a_6} T^{c_1}T^{a_7}  \big) _{i_5 \bar{\imath}_4} 
    C_4 \big( g_{6}^{\ominus}, g_{q_1^\perp}^*, g_{7}^{\oplus}, \bar{q}_{4}^{\ominus}, q_{5}^{\oplus}, g_{q_5^\perp}^* \big) 
      \nonumber 
    \\
    &+\, \big( T^{c_1}T^{a_6} T^{c_5}T^{a_7}  \big) _{i_5 \bar{\imath}_4}
    C_4 \big(g_{7}^{\oplus}, \bar{q}_{4}^{\ominus}, q_{5}^{\oplus} ,  g_{q_1^\perp}^*,g_{6}^{\ominus}, g_{q_5^\perp}^*\big) 
        \nonumber 
    \\
    &+\, \big( T^{a_6}T^{c_5} T^{c_1}T^{a_7}  \big) _{i_5 \bar{\imath}_4} C_4 \big( g_{q_1^\perp}^*, g_{7}^{\oplus}, \bar{q}_{4}^{\ominus}, q_{5}^{\oplus}, g_{6}^{\ominus},g_{q_5^\perp}^* \big) 
      \nonumber 
    \\
    &+\, \big( T^{a_6}T^{c_1} T^{c_5}T^{a_7}  \big) _{i_5 \bar{\imath}_4}
    C_4 \big( g_{7}^{\oplus}, \bar{q}_{4}^{\ominus}, q_{5}^{\oplus}, g_{6}^{\ominus},g_{q_1^\perp}^*, g_{q_5^\perp}^* \big) 
           \nonumber 
    \\
     &+\, \big( T^{c_5} T^{c_1}T^{a_6}T^{a_7}  \big) _{i_5 \bar{\imath}_4}
    C_4 \big( g_{q_1^\perp}^*, g_{6}^{\ominus}, g_{7}^{\oplus}, \bar{q}_{4}^{\ominus}, q_{5}^{\oplus}, g_{q_5^\perp}^* \big) 
      \nonumber 
    \\
     &+\, \big( T^{c_1} T^{c_5}T^{a_6}T^{a_7}  \big) _{i_5 \bar{\imath}_4}
    C_4 \big( g_{6}^{\ominus}, g_{7}^{\oplus}, \bar{q}_{4}^{\ominus}, q_{5}^{\oplus} ,g_{q_1^\perp}^*,g_{q_5^\perp}^*\big) 
      \nonumber 
    \\
&+\,\left[g_6^{\ominus}\leftrightarrow g_7^{\oplus},\, a_6\leftrightarrow a_7\right].
    \label{qwndqqw}
\end{align}
Next one loads the CEVs by
\begin{lstlisting}
Import["CEV_4_qqbgg.m"],
\end{lstlisting}
and subsequently initialises the momentum representation using 
\begin{align*}
&\mathtt{cevtocevp[\#] \& /@} \nonumber\\[1mm]
&\begin{array}{ll}
\big\{\mathtt{cev[qM,qP,M,P]}, \quad& \mathtt{cev[Pc,Mc,qPc,qMc]}, \\[1mm]
\mathtt{cev[qM,qP,M,Pc]}, \quad& \mathtt{cev[P,Mc,qPc,qMc]}, \\[1mm]
\mathtt{cev[M,P,qPc,qMc]}, \quad& \mathtt{cev[qM,qP,Pc,Mc]}, \\[1mm]
\mathtt{cev[P,qM,qP,Mc]}, \quad& \mathtt{cev[M,qPc,qMc,Pc]}, \\[1mm]
\mathtt{cev[P,qM,qP,M]}, \quad& \mathtt{cev[Mc,qPc,qMc,Pc]}, \\[1mm]
\mathtt{cev[M,P,qM,qP]}, \quad& \mathtt{cev[qPc,qMc,Pc,Mc]}\big\}\,.
\end{array}
\end{align*}
Finally, one can get the colour-ordered CEVs in Eq.~(\ref{qwndqqw}) using 
\begin{align}
C_4 \big( g_{q_1^\perp}^*, \bar{q}_{4}^{\ominus}, q_{5}^{\oplus}, g_{6}^{\ominus}, g_{7}^{\oplus}, g_{q_5^\perp}^* \big)  
&=\mathtt{cevp[qM,qP,M,P][1,4,5,6,7]}, \nonumber \\
C_4 \big( \bar{q}_{4}^{\ominus}, q_{5}^{\oplus}, g_{6}^{\ominus}, g_{7}^{\oplus} ,g_{q_1^\perp}^*, g_{q_5^\perp}^*\big)  
&= \mathtt{cevp[Pc,Mc,qPc,qMc][1,7,6,5,4]}, \nonumber \\
C_4 \big(  g_{7}^{\oplus},g_{q_1^\perp}^*, \bar{q}_{4}^{\ominus}, q_{5}^{\oplus}, g_{6}^{\ominus}, g_{q_5^\perp}^* \big)  
&= \mathtt{cevp[qM,qP,M,Pc][1,4,5,6,7]}, \nonumber \\
C_4 \big( \bar{q}_{4}^{\ominus}, q_{5}^{\oplus}, g_{6}^{\ominus},g_{q_1^\perp}^*, g_{7}^{\oplus}, g_{q_5^\perp}^* \big)  
&=  \mathtt{cevp[P,Mc,qPc,qMc][1,7,6,5,4]}, \nonumber \\
C_4 \big( \bar{q}_{4}^{\ominus}, q_{5}^{\oplus}, g_{q_1^\perp}^*, g_{6}^{\ominus}, g_{7}^{\oplus}, g_{q_5^\perp}^* \big)  
&= \mathtt{cevp[M, P, qPc, qMc][1, 6, 7, 5, 4]}, \nonumber \\
C_4 \big(  g_{6}^{\ominus}, g_{7}^{\oplus},g_{q_1^\perp}^*, \bar{q}_{4}^{\ominus}, q_{5}^{\oplus}, g_{q_5^\perp}^* \big)  
&= \mathtt{cevp[qM,\, qP,\, Pc,\, Mc][1,\,4,\,5,\,7,\,6]}, \nonumber \\
C_4 \big( g_{6}^{\ominus} ,g_{q_1^\perp}^*, g_{7}^{\oplus}, \bar{q}_{4}^{\ominus}, q_{5}^{\oplus}, g_{q_5^\perp}^*\big)  
&= \mathtt{cevp[P,\, qM,\, qP,\, Mc][1,\,7,\,4,\,5,\,6]}, \nonumber \\
C_4 \big(  g_{7}^{\oplus}, \bar{q}_{4}^{\ominus}, q_{5}^{\oplus} ,g_{q_1^\perp}^*, g_{6}^{\ominus}, g_{q_5^\perp}^*\big)  
&= \mathtt{cevp[M,\, qPc,\, qMc,\, Pc][1,\,6,\,5,\,4,\,7]}, \nonumber \\
C_4 \big( g_{q_1^\perp}^*, g_{7}^{\oplus}, \bar{q}_{4}^{\ominus}, q_{5}^{\oplus}, g_{6}^{\ominus}, g_{q_5^\perp}^* \big)  
&=  \mathtt{cevp[P,\, qM,\, qP,\, M][1,\,7,\,4,\,5,\,6]}, \nonumber \\
C_4 \big(  g_{7}^{\oplus}, \bar{q}_{4}^{\ominus}, q_{5}^{\oplus}, g_{6}^{\ominus},g_{q_1^\perp}^*, g_{q_5^\perp}^* \big)  
&= \mathtt{cevp[Mc,qPc,qMc,Pc][1,6,5,4,7]}, \nonumber \\
C_4 \big( g_{q_1^\perp}^*, g_{6}^{\ominus}, g_{7}^{\oplus}, \bar{q}_{4}^{\ominus}, q_{5}^{\oplus}, g_{q_5^\perp}^* \big)  
&= \mathtt{cevp[M,P,qM,qP][1,6,7,4,5]}, \nonumber \\
C_4 \big( g_{6}^{\ominus}, g_{7}^{\oplus}, \bar{q}_{4}^{\ominus}, q_{5}^{\oplus} ,g_{q_1^\perp}^*, g_{q_5^\perp}^*\big)  
&=  \mathtt{cevp[qPc,qMc,Pc,Mc][1,5,4,7,6]}.
\end{align}

\subsection{The \texorpdfstring{$g^* \bar{q}\to q\bar{Q}Qg$}{g* qbar → q Qbar Q g} PEV  and the \texorpdfstring{$g^*g^*\to \bar{q} q \bar{Q}Q$}{g* g* → qbar q Qbar Q} CEV}
\label{qqQQCEVPEV}

Here we consider MREVs with two quark-antiquark pairs. We consider the general case where these two pairs have different flavours which we denote respectively by $q$ ($\bar{q}$) and $Q$ ($\bar{Q}$).
Using $\mathcal{P}_{1\to 4} \big( 
    \bar{q}_{2}^{\ominus},\,
    q_{3}^{\oplus},\,
    \bar{Q}_{4}^{\ominus},\,
    Q_{5}^{\oplus},\,  g_{6}^{\ominus},\,
    g_{q_4^\perp}^*
 \big) $ as an example, the colour dressing of the PEV $g^*\bar{q}\to q \bar{Q}Qg$ is as follows,
\begin{align}
&\hspace*{-30pt} \mathcal{P}_{1\to 4} \big( 
    \bar{q}_{2}^{\ominus},\,
    q_{3}^{\oplus},\,
    \bar{Q}_{4}^{\ominus},\,
    Q_{5}^{\oplus},\,  g_{6}^{\ominus},\,
    g_{q_4^\perp}^*
 \big) 
\nonumber\\
&= \biggl\{
    (T^{c_5}T^{a_6})_{i_3\,\bar{\jmath}_4}
    \delta_{j_5\,\bar{\imath}_2}
    P_{1\to 4}\big( 
    g_{6}^{\ominus},\, \bar{Q}_{4}^{\ominus},\,
      Q_{5}^{\oplus},\,  \bar{q}_{2}^{\ominus},\,
      q_{3}^{\oplus},\, g_{q_4^\perp}^*
     \big) 
\nonumber\\
&\qquad
  + (T^{c_5}T^{a_6})_{j_5\,\bar{\imath}_2}
  \delta_{i_3\,\bar{\jmath}_4}
  P_{1\to 4}\big( g_{6}^{\ominus},\, 
 \bar{q}_{2}^{\ominus},\, q_{3}^{\oplus},\, \bar{Q}_{4}^{\ominus},\, Q_{5}^{\oplus},\, g_{q_4^\perp}^*
   \big) 
\nonumber\\
&\qquad
  + T^{c_5}_{j_5\,\bar{\imath}_2} T^{a_6}_{i_3\,\bar{\jmath}_4}
  P _{1\to 4}\big( 
 \bar{q}_{2}^{\ominus},\, q_{3}^{\oplus},\,    g_{6}^{\ominus},\, \bar{Q}_{4}^{\ominus},\, Q_{5}^{\oplus},\, g_{q_4^\perp}^*
   \big) 
 -\frac{1}{N_c}
    \left[
      \bar{Q}_{4}^{\ominus} \leftrightarrow \bar{q}_{2}^{\ominus},\;
      \bar{\imath}_2 \leftrightarrow \bar{\jmath}_4
    \right]
  \biggr\}
\nonumber\\
&\,\,
+\biggl\{
    (T^{a_6}T^{c_5})_{i_3\,\bar{\jmath}_4}
    \delta_{j_5\,\bar{\imath}_2}
    P_{1\to 4} \big(  \bar{Q}_{4}^{\ominus},\, Q_{5}^{\oplus},
   \bar{q}_{2}^{\ominus},\, q_{3}^{\oplus},\,    g_{6}^{\ominus},\, g_{q_4^\perp}^*
     \big) 
\nonumber\\
&\qquad
  + (T^{a_6}T^{c_5})_{j_5\,\bar{\imath}_2}
    \delta_{i_3\,\bar{\jmath}_4}
    P_{1\to 4} \big( 
      \bar{q}_{2}^{\ominus},\, q_{3}^{\oplus},\, \bar{Q}_{4}^{\ominus},\, Q_{5}^{\oplus},\, g_{6}^{\ominus},\, g_{q_4^\perp}^*
     \big) 
\nonumber\\
&\qquad
  + T^{a_6}_{j_5\,\bar{\imath}_2} T^{c_5}_{i_3\,\bar{\jmath}_4}
  P_{1\to 4}\big( 
  \bar{Q}_{4}^{\ominus},\, Q_{5}^{\oplus},\, g_{6}^{\ominus},\,   \bar{q}_{2}^{\ominus},\, q_{3}^{\oplus},\, g_{q_4^\perp}^*
   \big) 
  -\frac{1}{N_c}
    \left[
      \bar{Q}_{4}^{\ominus} \leftrightarrow \bar{q}_{2}^{\ominus},\;
      \bar{\imath}_2 \leftrightarrow \bar{\jmath}_4
    \right]
  \biggr\}
\nonumber\\
&\quad
-\delta_{qQ}
  \left[
    Q_{5}^{\oplus} \leftrightarrow q_{3}^{\oplus},\;
    j_5 \leftrightarrow j_3
  \right],
\label{pevqqQQg}
\end{align}
where $\delta_{qQ}$ equals $1$ if the flavours are identical, or $0$ otherwise. We first load the PEVs, written in MSLCVs, using:
\begin{lstlisting}
Import["PEV_1to4_qqbQQbg_.m"],
\end{lstlisting}
and then initialise the momentum representation of the PEV in Eq.~\eqref{pevqqQQg}
by
\begin{align*}
&\mathtt{pevtopevp[\#]\ \& /@}\\[1mm]
\big\{&\mathtt{pevp[\{qM,\,q\},\ \{qP,\,q\},\ \{qPc,\,Q\},\ \{qMc,\,Q\},\ \{Mc,\,gc\}]}\\[1mm]
&\mathtt{pev[\{qM,\,q\},\ \{qP,\,q\},\ \{qM,\,Q\},\ \{qP,\,Q\},\ \{Mc,\,gc\}]} \\[1mm]
&\mathtt{pev[\{qM,\,q\},\ \{qP,\,q\},\ \{M,\,g\},\ \{qM,\,Q\},\ \{qP,\,Q\}]} \\[1mm]
&\mathtt{pev[\{qM,\,q\},\ \{qP,\,q\},\ \{M,\,g\},\ \{qPc,\,Q\},\ \{qMc,\,Q\}]} \\[1mm]
&\mathtt{pev[\{qM,\,q\},\ \{qP,\,q\},\ \{qM,\,Q\},\ \{qP,\,Q\},\ \{M,\,g\}]} \\[1mm]
&\mathtt{pev[\{qM,\,q\},\ \{qP,\,q\},\ \{Mc,\,gc\},\ \{qPc,\,Q\},\ \{qMc,\,Q\}]}\big\}.
\end{align*}
Here, we include additional arguments, $\texttt{q}$, $\texttt{Q}$, $\texttt{g}$, (or $\texttt{gc}$), to label the particle content as a quark of flavour~1, a quark of flavour~2, and a gluon, respectively.
Then, the colour-ordered PEVs appearing in Eq.~(\ref{pevqqQQg}) in the momentum representation are given by 
\begin{align*}
&\hspace*{-40pt}
    P_{1\to 4} \big( g_{6}^{\ominus},\, \bar{Q}_{4}^{\ominus},\, Q_{5}^{\oplus},\, 
    \bar{q}_{2}^{\ominus},\, q_{3}^{\oplus},\, g_{q_4^\perp}^*
   \big), 
   \nonumber\\
\hspace{1.2em} =&\,\,
 \mathtt{pevp[\{qM,\,q\},\ \{qP,\,q\},\ \{qPc,\,Q\},\ \{qMc,\,Q\},\ \{Mc,\,gc\}][3,\,5,\,4,\,6]}
  \\
& \hspace*{-40pt}
  P _{1\to 4}\big( 
    g_{6}^{\ominus},\, \bar{q}_{2}^{\ominus},\, q_{3}^{\oplus},\, \bar{Q}_{4}^{\ominus},\, Q_{5}^{\oplus},\, g_{q_4^\perp}^*
   \big), 
   \nonumber\\
\hspace{1.2em} =&\,\,
 \mathtt{pevp[\{qM,\,q\},\ \{qP,\,q\},\ \{qM,\,Q\},\ \{qP,\,Q\},\ \{Mc,\,gc\}][3,\,5,\,4,\,6]}
 \\
& \hspace*{-40pt}
  P_{1\to 4} \big( 
    \bar{q}_{2}^{\ominus},\, q_{3}^{\oplus},\, g_{6}^{\ominus},\, \bar{Q}_{4}^{\ominus},\, Q_{5}^{\oplus},\, g_{q_4^\perp}^*
   \big), 
   \nonumber\\
\hspace{1.2em} =&\,\,
 \mathtt{pevp[\{qM,\,q\},\ \{qP,\,q\},\ \{M,\,g\},\ \{qM,\,Q\},\ \{qP,\,Q\}][3,\,6,\,4,\,5]}
  \\
& \hspace*{-40pt}
  P_{1\to 4} \big( 
  \bar{Q}_{4}^{\ominus},\, Q_{5}^{\oplus},\,   \bar{q}_{2}^{\ominus},\, q_{3}^{\oplus},\, g_{6}^{\ominus},\, g_{q_4^\perp}^*
   \big), 
 \nonumber\\
\hspace{1.2em} =&\,\,
 \mathtt{pevp[\{qM,\,q\},\ \{qP,\,q\},\ \{M,\,g\},\ \{qPc,\,Q\},\ \{qMc,\,Q\}][3,\,6,\,5,\,4]}
  \\
& \hspace*{-40pt}
  P _{1\to 4}\big( 
    \bar{q}_{2}^{\ominus},\, q_{3}^{\oplus},\, \bar{Q}_{4}^{\ominus},\, Q_{5}^{\oplus},\, g_{6}^{\ominus},\, g_{q_4^\perp}^*
   \big), 
  \nonumber\\
\hspace{1.2em} =&\,\,
 \mathtt{pevp[\{qM,\,q\},\ \{qP,\,q\},\ \{qM,\,Q\},\ \{qP,\,Q\},\ \{M,\,g\}][3,\,4,\,5,\,6]}
  \\
& \hspace*{-40pt}
  P_{1\to 4} \big( \bar{Q}_{4}^{\ominus},\, Q_{5}^{\oplus},\, g_{6}^{\ominus},\, 
    \bar{q}_{2}^{\ominus},\, q_{3}^{\oplus},\, g_{q_4^\perp}^*
   \big), 
    \nonumber\\
\hspace{1.2em} =&\,\,
 \mathtt{pevp[\{qM,\,q\},\ \{qP,\,q\},\ \{Mc,\,gc\},\ \{qPc,\,Q\},\ \{qMc,\,Q\}][3,\,6,\,5,\,4]}.
\end{align*}

The colour-dressed CEV $g^*g^*\to \bar{q} q \bar{Q}Q$ is as follows,
\begin{align}
&   \hspace*{-40pt} 
\mathcal{C}_4 \big( g_{q_1^\perp}^*,\bar{q}_{{4}}^{\ominus} , q_{5}^{\oplus}, \bar{Q}_{6}^{\ominus}, Q_{7}^{\oplus},g_{q_5^\perp}^* \big) \non &= \biggl\{(T^{c_5}T^{c_1})_{{i_5}{\bar{\jmath}_6}} \delta_{{j_7}{\bar{\imath}_4}}C_4 \big( g_{q_1^\perp}^*,\bar{Q}_{6}^{\ominus}, Q_{7}^{\oplus},\bar{q}_{{4}}^{\ominus} , q_{5}^{\oplus} ,g_{q_5^\perp}^* \big)  \non &\quad \,
+ 
(T^{c_5}T^{c_1})_{{j_7}{\bar{\imath}_4}} \delta_{{i_5}{\bar{\jmath}_6}}C _4\big( g_{q_1^\perp}^*,\bar{q}_{{4}}^{\ominus} , q_{5}^{\oplus} ,\bar{Q}_{6}^{\ominus}, Q_{7}^{\oplus},g_{q_5^\perp}^* \big) \non  &\quad  \,
+T^{c_5}_{{j_7}{\bar{\imath}_4}} T^{c_1}_{{i_5}{\bar{\jmath}_6}}C_4 \big( \bar{q}_{{4}}^{\ominus} , q_{5}^{\oplus} ,g_{q_1^\perp}^*,\bar{Q}_{6}^{\ominus}, Q_{7}^{\oplus} ,g_{q_5^\perp}^*\big) -\frac{1}{N_c}\left[\bar{Q}_{6}^{\ominus}\leftrightarrow \bar{q}_{{4}}^{\ominus}, \; \bar{\imath}_4\leftrightarrow\bar{\jmath}_6\right]\biggr\}\non& \,
+\biggl\{ (T^{c_1}T^{c_5})_{{i_5}{\bar{\jmath}_6}} \delta_{{j_7}{\bar{\imath}_4}}C _4\big( \bar{Q}_{6}^{\ominus}, Q_{7}^{\oplus},\bar{q}_{{4}}^{\ominus} , q_{5}^{\oplus} ,g_{q_1^\perp}^*,g_{q_5^\perp}^*\big)  \non & 
\quad 
+(T^{c_1}T^{c_5})_{{j_7}{\bar{\imath}_4}} \delta_{{i_5}{\bar{\jmath}_6}}C _4\big( \bar{q}_{{4}}^{\ominus} , q_{5}^{\oplus} ,\bar{Q}_{6}^{\ominus}, Q_{7}^{\oplus},g_{q_1^\perp}^*,g_{q_5^\perp}^* \big) \non  &\quad 
+T^{c_1}_{{j_7}{\bar{\imath}_4}} T^{c_5}_{{i_5}{\bar{\jmath}_6}}C _4\big(  \bar{Q}_{6}^{\ominus}, Q_{7}^{\oplus} ,g_{q_1^\perp}^*,\bar{q}_{{4}}^{\ominus} , q_{5}^{\oplus},g_{q_5^\perp}^*\big) -\frac{1}{N_c}\left[\bar{Q}_{6}^{\ominus}\leftrightarrow \bar{q}_{{4}}^{\ominus}, \; \bar{\imath}_4\leftrightarrow\bar{\jmath}_6\right]\biggr\}\non &
\quad -\delta_{qQ} \left[
    Q_{7}^{\oplus} \leftrightarrow q_{5}^{\oplus},\;
    j_7 \leftrightarrow j_5
  \right],\non 
\label{wqgewqwqwqq}
\end{align}
where the $\delta$-function in the final line adds the additional terms which arise with identical quark flavours. To use the library, we begin as usual by loading the CEV data in terms of MSLCV,
\begin{lstlisting}
Import["CEV_4_qqbQQb_.m"],
\end{lstlisting}
and initialising the momentum representation by
\begin{align*}
&\mathtt{cevtocevp[\#] \& /@} \nonumber\\[1mm]
&\begin{array}{ll}
\big\{\mathtt{cev[\{qM,Q\},\{qP,Q\},\{qM,q\},\{qP,q\}],}\quad& \mathtt{cev[\{qM,q\},\{qP,q\},\{qM,Q\},\{qP,Q\}],}\\[1mm]
\mathtt{cev[\{qPc,q\},\{qMc,q\},\{qM,Q\},\{qP,Q\}],}\quad &\mathtt{cev[\{qPc,q\},\{qMc,q\},\{qPc,Q\},\{qMc,Q\}],} \\[1mm]
\mathtt{cev[\{qPc,Q\},\{qMc,Q\},\{qPc,q\},\{qMc,q\}],}\quad &\mathtt{cev[\{qM,q\},\{qP,q\},\{qPc,Q\},\{qMc,Q\}]}\big\}.
\end{array}\,
\end{align*}
The colour-ordered CEVs appearing in \eqref{wqgewqwqwqq} in the momentum representation are given by 
\begin{align}
C_4\big(g_{q_1^\perp}^*,\,\bar{Q}_{6}^{\ominus},\, Q_{7}^{\oplus},\,\bar{q}_{{4}}^{\ominus},\, q_{5}^{\oplus},\,g_{q_5^\perp}^*\big)
&= \mathtt{cevp[\{qM,Q\},\{qP,Q\},\{qM,q\},\{qP,q\}][1,6,7,4,5]}, \nonumber\\[1mm]
C_4\big(g_{q_1^\perp}^*,\,\bar{q}_{{4}}^{\ominus},\, q_{5}^{\oplus},\,\bar{Q}_{6}^{\ominus},\, Q_{7}^{\oplus},\,g_{q_5^\perp}^*\big)
&= \mathtt{cevp[\{qM,q\},\{qP,q\},\{qM,Q\},\{qP,Q\}][1,4,5,6,7]}, \nonumber\\[1mm]
C_4\big(\bar{q}_{{4}}^{\ominus},\, q_{5}^{\oplus},\, \,g_{q_1^\perp}^*,\,\bar{Q}_{6}^{\ominus},\, Q_{7}^{\oplus},\,g_{q_5^\perp}^*\big)
&= \mathtt{cevp[\{qPc,q\},\{qMc,q\},\{qM,Q\},\{qP,Q\}][1,5,4,6,7]}, \nonumber\\[1mm]
C_4\big(\bar{Q}_{6}^{\ominus}, \,Q_{7}^{\oplus},\,\bar{q}_{{4}}^{\ominus} ,\, q_{5}^{\oplus},\, g_{q_1^\perp}^*,\,g_{q_5^\perp}^* \big)
&= \mathtt{cevp[\{qPc,q\},\{qMc,q\},\{qPc,Q\},\{qMc,Q\}][1,5,4,7,6]}, \nonumber\\[1mm]
C_4\big(\bar{q}_{{4}}^{\ominus},\, q_{5}^{\oplus},\,\bar{Q}_{6}^{\ominus},\, Q_{7}^{\oplus},\, g_{q_1^\perp}^*,\, g_{q_5^\perp}^*\big)
&= \mathtt{cevp[\{qPc,Q\},\{qMc,Q\},\{qPc,q\},\{qMc,q\}][1,7,6,5,4]}, \nonumber\\[1mm]
C_4\big(\bar{Q}_{6}^{\ominus},\, Q_{7}^{\oplus},\,g_{q_1^\perp}^*,\,\bar{q}_{{4}}^{\ominus},\, q_{5}^{\oplus},\,g_{q_5^\perp}^*\big)
&= \mathtt{cevp[\{qM,q\},\{qP,q\},\{qPc,Q\},\{qMc,Q\}][1,4,5,7,6]}.
\label{wqiujhgq2223}
\end{align}

 \section{Relations among MREVs}\label{relationsPEVCEV}

Many relations between colour-ordered amplitudes are known, including the photon decoupling (dual Ward) identity~\cite{Bern:1990ux,Mangano:1990by,Mangano:1987xk}, the Kleiss-Kuijf relation~\cite{Kleiss:1988ne}, the reversal (reflection) identity~\cite{Bern:1994fz} and the \(\mathcal{N}=1\) SUSY Ward identity~\cite{Grisaru:1977px}. In this section, we study whether  and how these 
translate into 
relations among colour-ordered MREVs. 
Although some of these identities have been investigated for pure-gluon MREVs in \cite{DelDuca:1999iql}, here we extend them to general parton flavours. In particular, we build upon ref.~\cite{DelDuca:1999iql} by including the on-shell and off-shell photon decoupling identities for MREVs, as well as supersymmetric Ward identities at higher multiplicities. Together, these relations provide a robust consistency check of the results. These checks are complementary to the factorization properties of MREVs in the MRK, soft, and collinear limits, which we study in Appendix~\ref{Appendix:Fact_in_limits}.

Usually these relations apply independently of the specific theory considered.
Notably, there is one exception regarding the photon decoupling identity, that when colour-ordered amplitudes or MREVs involve two distinct quark flavours, the identity does not hold within QCD, but it is restored upon including additional contributions from \(\mathcal{N}=4\) super Yang-Mills (sYM) theory. This will be demonstrated in section \ref{ppdd}.

\subsection{Photon Decoupling Identity for MREVs}\label{ppdd}

Colour-ordered amplitudes satisfy photon decoupling identities~\cite{Bern:1990ux,Mangano:1990by,Mangano:1987xk}. Let us first recall how these arise and then see how they apply in the context of MREVs.

Consider first a \(U(N_c)\) gauge theory. A tree-level \(n\)-point amplitude of gauge bosons can be expressed in terms of the gauge-group generators in a trace basis as
\begin{equation}
\begin{aligned}
    &\hspace*{-20pt}\mathcal{A}_{2\to n-2}(\{p_i,\varepsilon_i,a_i\}) 
   \\
     =&\,g^{n-2} \sum_{\sigma \in \mathrm{S}_n / \mathrm{Z}_n} \operatorname{Tr}\big(T^{a_{\sigma(1)}} \ldots T^{a_{\sigma(n)}}\big) \times A_{2\to n-2}\big(p_{\sigma(1)}, \varepsilon_{\sigma(1)}\, ;\, \ldots\, ;\, p_{\sigma(n)}, \varepsilon_{\sigma(n)}\big)\,,
\end{aligned}
\label{ejkqwf}
\end{equation}
where the triplet $(p_i,\varepsilon_i,a_i)$ represents the momentum, polarization vector and colour index of parton $i$, respectively.
This $U(N_c)$ theory can be interpreted as a theory of $N_c^2-1$  ``gluons'' of a \(SU(N_c)\) non-Abelian gauge group and a single ``photon'' of an Abelian \(U(1)\) gauge group. Since the  \(U(1)\) generator has vanishing commutators with the  \(SU(N_c)\) ones, there is no interaction between the \(U(1)\) photon and the \(SU (N_c)\) gluons. Then, any amplitudes involving one photon must vanish, i.e., upon singling out parton $1$ as the photon,
\begin{equation}
\label{photonDecouplingCondition}
\mathcal{A}_{2\to n-2}\left(\left\{p_i, \varepsilon_i, a_i\right\}_{i=2}^{n} ; p_1, \varepsilon_1, a_{1}\right)=0\,.
\end{equation}
Since the \(U(1)\) generator is proportional to the identity and the traces
$
\operatorname{Tr}\bigl(T^{a_{\sigma(2)}} \cdots T^{a_{\sigma(n)}}\bigr)
$
form an independent basis, one may look at just one given trace. In particular, by extracting the coefficient of
$
\operatorname{Tr}\bigl(T^{a_2}\cdots T^{a_{n}}\bigr)
$---where the \(T^{a_{U(1)}}\) is inserted in each possible slot in \(\operatorname{Tr}\bigl(T^{a_2}\cdots T^{a_{n}}\bigr)\)---Eq.~(\ref{photonDecouplingCondition}) implies a decoupling identity for the colour-ordered amplitude:
\begin{align}
& A_{2\to n-2}\bigl(\textcolor{blue}{p_1},\textcolor{blue}{\varepsilon_1};\, p_2,\varepsilon_2;\, p_3,\varepsilon_3;\, \ldots;\, p_n,\varepsilon_n\bigr)
+\, A_{2\to n-2}\bigl(p_2,\varepsilon_2;\, \textcolor{blue}{p_1},\textcolor{blue}{\varepsilon_1};\, p_3,\varepsilon_3;\, \ldots;\, p_n,\varepsilon_n\bigr)
\nonumber\\[1ex]
&\quad +\, \cdots +\, A_{2\to n-2}\bigl(p_2,\varepsilon_2;\, p_3,\varepsilon_3;\, \ldots;\, \textcolor{blue}{p_1},\textcolor{blue}{\varepsilon_1};\, p_n,\varepsilon_n\bigr)
=\, 0\,,
\label{wqdqddqw}
\end{align}
where we highlight the photon in blue. Eq.~\eqref{wqdqddqw} can be interprested as cyclically permuting the photon around the colour-ordered amplitude. The colour-ordered amplitudes in Eq.~(\ref{wqdqddqw}) are identical to amplitudes of $n$ gluons in \(SU(N_c)\) gauge theory, so this establishes a relation among pure gluon amplitudes.
Using the cyclicity property reviewed in section~\ref{relationCEVPEV}, one can also rewrite it as
\begin{align}
    \sum_{\sigma \in Z_{n-1}} A_{2\to n-2}\big(g_1^{h_1}, g_{\sigma_2}^{h_{\sigma_2}}, \ldots, g_{\sigma_n}^{h_{\sigma_n}}\big) = 0,
    \label{gluon_u_1}
\end{align}
where $Z_{n-1}$ represents cyclic permutations.

This conclusion naturally extends to $\mathcal{N}=4$ sYM amplitudes with gluons and gluinos, as both particles are in the adjoint representation of the gauge group, so their colour factors can be represented using the trace basis as in Eq.~(\ref{ejkqwf}). However, due to the fermionic nature of gluinos,
anticommuting them amounts to a minus sign, 
and hence the equivalent of Eq.~(\ref{gluon_u_1}) for amplitudes involving gluinos has a minus sign for terms corresponding to odd permutations of gluinos compared to those having an even permutation (see e.g.~\cite{Luo:2005my}).
Note that this odd or even nature is determined regardless of how many gluons are interleaved, or how the gluons are ordered.
For example, considering six-point colour-ordered amplitudes of two pairs of gluinos and two gluons, one may formulate the following relation:
\begin{equation}
\begin{aligned}
0&=\, A_{2\to 4}(\tilde{g}_{A,1}^{\ominus},\, \tilde{g}_{A,2}^{\oplus},\, \tilde{g}_{B,3}^{\ominus},\, {\color{blue}{\color{blue}\tilde{g}_{B,4}^{\oplus}}},\, g_5^{h_5},\, g_6^{h_6}) - A_{2\to 4}(\tilde{g}_{A,1}^{\ominus},\, \tilde{g}_{A,2}^{\oplus},\, {\color{blue}\tilde{g}_{B,4}^{\oplus}},\, \tilde{g}_{B,3}^{\ominus},\, g_5^{h_5},\, g_6^{h_6}) \\
&\,+ A_{2\to 4}(\tilde{g}_{A,1}^{\ominus},\, {\color{blue}\tilde{g}_{B,4}^{\oplus}},\, \tilde{g}_{A,2}^{\oplus},\, \tilde{g}_{B,3}^{\ominus},\, g_5^{h_5},\, g_6^{h_6}) 
+ A_{2\to 4}(\tilde{g}_{A,1}^{\ominus},\, \tilde{g}_{A,2}^{\oplus},\, \tilde{g}_{B,3}^{\ominus},\, g_5^{h_5},\, g_6^{h_6},\, {\color{blue}\tilde{g}_{B,4}^{\oplus}})
 \\
&\,+ A_{2\to 4}(\tilde{g}_{A,1}^{\ominus},\, \tilde{g}_{A,2}^{\oplus},\, \tilde{g}_{B,3}^{\ominus},\, g_5^{h_5},\, {\color{blue}\tilde{g}_{B,4}^{\oplus}},\, g_6^{h_6})\,.
\end{aligned}
\end{equation} 
Based on the discussion in  Ref.~\cite{Dixon:2010ik} and in section~\ref{qcd_and_sym_diff} above, QCD colour-ordered amplitudes involving gluons and quarks of a single flavour share identical photon decoupling identities with the corresponding $\mathcal{{N}}=4$ sYM amplitudes, where the quarks are traded for gluinos. However, for amplitudes with two or more distinct quark flavours, some photon decoupling identities do not hold in QCD; instead, they intertwine $\mathcal{N}=4$ sYM amplitudes with QCD ones. Next, we will discuss how these decoupling identities extend to PEVs and~CEVs.

\subsubsection{Permuting on-shell partons}

\begin{figure}
    \centering
    \includegraphics[width=1\linewidth]{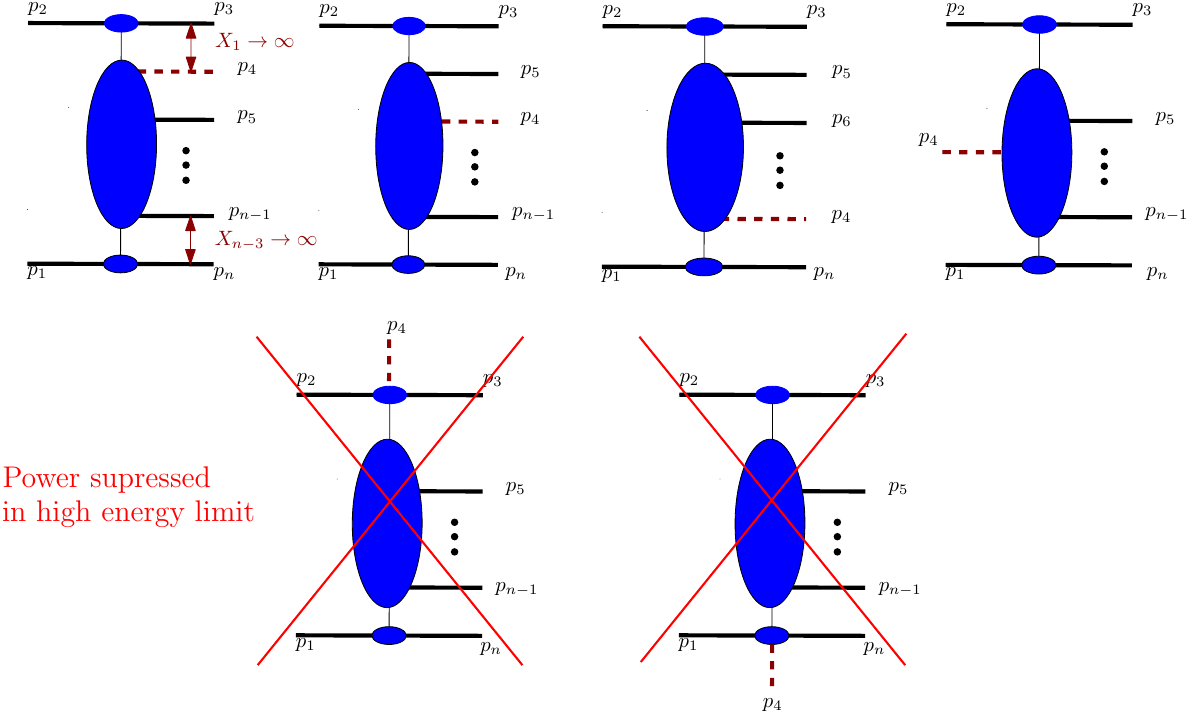}
    \caption{Starting from the photon decoupling identity for particular ``mother'' amplitudes, one can derive the photon decoupling identities for colour-ordered MREVs by considering the high-energy limit where \(X_1, X_{n-3} \to \infty\). 
    The set of diagrams in the upper row contribute to the MREV photon-decoupling identity, while those in the lower row exhibit power suppression, as the corresponding colour-ordered amplitudes have physical poles scaling as \(\sim {1}/{X_1}\) or \(\sim {1}/{X_{n-3}}\) when \(X_1, X_{n-3} \to \infty\).
}
    \label{fig:u(1)}
\end{figure}

As shown in Figure~\ref{fig:u(1)}, by taking the high-energy limit (see Eq.~(\ref{cevmulti})) of the photon decoupling identities of colour-ordered amplitudes as in Eqs.~\eqref{gluon_u_1}, we can extract the photon decoupling identities for MREVs. 
The figure describes the case of \( (n-4) \) emission CEVs, derived from \( n \)-point mother amplitudes in the limit \(X_1, X_{n-3} \to \infty\). In this 
configuration, the particles labelled 2 and 3, which we take to be helicity- and flavour- conserving, represent the \textit{target}, while the particles 
labelled 1 and \( n \) denote the \textit{projectile} (these also conserve helicity and flavour). By permuting gluon 4 through all admissible positions, distinct colour-ordered CEVs are generated. However, no CEV is generated by placing 
gluon~4 
between particles 2 and 3 or between 1 and \( n\), as these configurations correspond to subleading contributions in the high-energy limit taken here. 

We have already seen the photon decoupling identity for pure gluon CEVs in (\ref{gluonphotondecoupling}). We now explore a few examples of these identities for MREVs containing one or two quark pairs. First, consider the colour-ordered CEV $g^* g^* \rightarrow g q \bar{q}$ where we obtain the following photon decoupling identity: 
\begin{eqnarray}
0\eqn C_3({\color{blue}g^\oplus_4},g_{q_1^\perp}^*,q^\oplus_5,\bar{q}^\ominus_6,g_{q_4^\perp}^*)+ C_3(g_{q_1^\perp}^*,{\color{blue}g^\oplus_4},q^\oplus_5,\bar{q}^\ominus_6,g_{q_4^\perp}^*)+ C_3^{\text{}}(g_{q_1^\perp}^*,q^\oplus_5,{\color{blue}g^\oplus_4},\bar{q}^\ominus_6,g_{q_4^\perp}^*)\non&&+ C_3(g_{q_1^\perp}^*,q^\oplus_5,\bar{q}^\ominus_6,{\color{blue}g^\oplus_4},g_{q_4^\perp}^*),
\label{qjfoqfwq}
\end{eqnarray}
where we cyclically inserted ${\color{blue}g^\oplus_4}$ in all relevant positions. 
Note that the third CEV in Eq.~(\ref{qjfoqfwq}), where the quark and antiquark are interleaved by a gluon, does not feature directly in a colour-dressed CEV (\ref{ciuwe}) in QCD. 
These relations are nonetheless useful in expressing such colour-ordered objects in terms of others, as we have done for example in eq.~(\ref{gqqPhotonDecoupling}) above.
In a similar way, considering the colour-ordered PEV $ q g^*\to  \bar{q} q \bar{q} g$ we get:
\begin{eqnarray} 0&=&P_{1\to 4}({\color{blue}g^\oplus_4},q^\oplus_2,\bar{q}^\ominus_3,q^\oplus_5,\bar{q}^\ominus_6,g_{q_4^\perp}^*)+P_{1\to 4}^{\text{}}(q^\oplus_2,{\color{blue}g^\oplus_4},\bar{q}^\ominus_3,q^\oplus_5,\bar{q}^\ominus_6,g_{q_4^\perp}^*)\non &&\hspace{0.8cm}+P_{1\to 4}(q^\oplus_2,\bar{q}^\ominus_3,{\color{blue}g^\oplus_4},q^\oplus_5,\bar{q}^\ominus_6,g_{q_4^\perp}^*)+P_{1\to 4}^{\text{}}(q^\oplus_2,\bar{q}^\ominus_3,q^\oplus_5,{\color{blue}g^\oplus_4},\bar{q}^\ominus_6,g_{q_4^\perp}^*)\non &&\hspace{0.8cm}+P_{1\to 4}(q^\oplus_2,\bar{q}^\ominus_3,q^\oplus_5,\bar{q}^\ominus_6,{\color{blue}g^\oplus_4},g_{q_4^\perp}^*).
\label{qwjoingqg}
\end{eqnarray}
Also here the second and fourth term do not feature in the colour-dressed PEV in QCD. 
The codes for checking Eqs.~\eqref{qjfoqfwq} and \eqref{qwjoingqg} are given in Appendix \ref{u1u1checkcehc}.

Next, upon translating an on-shell quark around the colour-ordered amplitude, we obtain the following identity amongst the colour-ordered CEVs for $g^* g^* \rightarrow  q \bar{q}g$:
\begin{align}
0& = C_3\bigl(g_{q_1^\perp}^*,\,{\color{blue}q^\oplus_4},\,\bar{q}^\ominus_5,\,g_6^\oplus,\,g_{q_4^\perp}^*\bigr)  
-\,
{C_3\bigl(g_{q_1^\perp}^*,\,\bar{q}^\ominus_5,\,{\color{blue}q^\oplus_4},\,g_6^\oplus,\,g_{q_4^\perp}^*\bigr)} \, %
\nonumber \\
&\hspace{1cm}-{C_3^{\text{}}\bigl(g_{q_1^\perp}^*,\,\bar{q}^\ominus_5,\,g_6^\oplus,\,{\color{blue}q^\oplus_4},\,g_{q_4^\perp}^*\bigr)} +  \, C_3^{\text{}}\bigl({\color{blue}q^\oplus_4},\,g_{q_1^\perp}^*,\,\bar{q}^\ominus_5,\,g_6^\oplus,\,g_{q_4^\perp}^*\bigr)
,\label{quifggqwccccc}
\end{align}
where the signs are determined based on the relative ordering of the quarks alone.

Consider now colour-ordered CEVs involving two quark-antiquark pairs of the same flavour, $g^* g^* \rightarrow  q \bar{q}q \bar{q}$. By moving the $q_4^\oplus$ around one obtains the following identity:
\begin{align}
\label{quifggqw:sameflavour}
0 =&\ C_4\bigl(g_{q_1^\perp}^*,\,{\color{blue}q^\oplus_4},\,\bar{q}^\ominus_5,\,q^\oplus_6,\,\bar{q}^\ominus_7,\,g_{q_5^\perp}^*\bigr)  
-\,
{C_4\bigl(g_{q_1^\perp}^*,\,\bar{q}^\ominus_5,\,{\color{blue}q^\oplus_4},\,q^\oplus_6,\,\bar{q}^\ominus_7,\,g_{q_5^\perp}^*\bigr)} \nonumber \\
      &  +\, 
{C_4^{}\bigl(g_{q_1^\perp}^*,\,\bar{q}^\ominus_5,\,q^\oplus_6,\,{\color{blue}q^\oplus_4},\,\bar{q}^\ominus_7,\,g_{q_5^\perp}^*\bigr)} 
-\, C_4\bigl(g_{q_1^\perp}^*,\,\bar{q}^\ominus_5,\,q^\oplus_6,\,\bar{q}^\ominus_7,\,{\color{blue}q^\oplus_4},\,g_{q_5^\perp}^*\bigr) \nonumber \\
& + \, C_4^{\text{}}\bigl({\color{blue}q^\oplus_4},g_{q_1^\perp}^*,\,\bar{q}^\ominus_5,\,q^\oplus_6,\,\bar{q}^\ominus_7,\,g_{q_5^\perp}^*\bigr)\,.
\end{align}
Moving on to similar CEVs for two \emph{distinct} flavours,
$g^* g^* \rightarrow  q \bar{q}Q \bar{Q}$, we have
\begin{align}
\label{quifggqw:differentFlavours}
0 =&\ C_4\bigl(g_{q_1^\perp}^*,\,{\color{blue}q^\oplus_4},\,\bar{q}^\ominus_5,\,Q^\oplus_6,\,\bar{Q}^\ominus_7,\,g_{q_5^\perp}^*\bigr)  
-\,
{C_4\bigl(g_{q_1^\perp}^*,\,\bar{q}^\ominus_5,\,{\color{blue}q^\oplus_4},\,Q^\oplus_6,\,\bar{Q}^\ominus_7,\,g_{q_5^\perp}^*\bigr)} \nonumber \\
      &  +\, 
{C_4^{\mathcal{N}=4\, \mathrm {sYM}}\bigl(g_{q_1^\perp}^*,\,\bar{q}^\ominus_5,\,Q^\oplus_6,\,{\color{blue}q^\oplus_4},\,\bar{Q}^\ominus_7,\,g_{q_5^\perp}^*\bigr)} 
-\, C_4^{\text{}}\bigl(g_{q_1^\perp}^*,\,\bar{q}^\ominus_5,\,Q^\oplus_6,\,\bar{Q}^\ominus_7,\,{\color{blue}q^\oplus_4},\,g_{q_5^\perp}^*\bigr) \nonumber \\
      & + \, C_4^{\text{}}\bigl({\color{blue}q^\oplus_4},g_{q_1^\perp}^*,\,\bar{q}^\ominus_5,\,Q^\oplus_6,\,\bar{Q}^\ominus_7,\,g_{q_5^\perp}^*\bigr),
\end{align}
where we note that the signs remain the same as in Eq.~(\ref{quifggqw:sameflavour}). In contrast to Eq.~(\ref{quifggqw:sameflavour}) however, Eq.~(\ref{quifggqw:differentFlavours}) is not an identity within QCD: recall that in Section~\ref{qcd_and_sym_diff} (see Eq.~(\ref{Vanishing_QCD_CEVPEV}) there), we saw that the CEV $C_4^{\mathrm{QCD}}\big(g_{q_1^\perp}^*,\, \bar{q}_5^{\ominus},\, Q_6^{\oplus},\, q_4^{\oplus},\, \bar{Q}_7^{\ominus},\, g_{q_5^\perp}^*\big)$ vanishes. In turn, in~\hbox{$\mathcal{N}=4$}\, sYM, where the corresponding CEV is non-vanishing thanks to the scalar interaction, 
there is such an identity. Since all but one of the terms in this identity  match QCD CEVs, Eq.~(\ref{quifggqw:differentFlavours}), can be read as one that mixes CEVs of the two theories.
Similarly, for the PEV $\bar q g^* \rightarrow  \bar q q \bar{q} g$, by moving the \(\bar q_3^\ominus\) around we have
\begin{eqnarray}
0 &=& P_{1\to 4}\bigl(q^\oplus_2,{\color{blue}\bar{q}^\ominus_3},q^\oplus_4,\bar{q}^\ominus_5,g^\oplus_6,g_{q_4^\perp}^*\bigr)-P_{1\to 4}^{}\bigl(q^\oplus_2,q^\oplus_4,{\color{blue}\bar{q}^\ominus_3},\bar{q}^\ominus_5,g^\oplus_6,g_{q_4^\perp}^*\bigr) \nonumber \\
  &&+ P_{1\to 4}^{\text{}}\bigl(q^\oplus_2,q^\oplus_4,\bar{q}^\ominus_5,{\color{blue}\bar{q}^\ominus_3},g^\oplus_6,g_{q_4^\perp}^*\bigr)+P_{1\to 4}^{\text{}}\bigl(q^\oplus_2,q^\oplus_4,\bar{q}^\ominus_5,g^\oplus_6,{\color{blue}\bar{q}^\ominus_3},g_{q_4^\perp}^*\bigr)\nonumber \\
  &&- P_{1\to 4}\bigl({\color{blue}\bar{q}^\ominus_3},q^\oplus_2,q^\oplus_4,\bar{q}^\ominus_5,g^\oplus_6,g_{q_4^\perp}^*\bigr),
\end{eqnarray}
while with distinct flavours, $\bar q g^* \rightarrow  \bar q Q \bar{Q} g$, we obtain
\begin{eqnarray}
\label{PEVqbargstar}
0 &=& P_{1\to 4}\bigl(q^\oplus_2,{\color{blue}\bar{q}^\ominus_3},Q^\oplus_4,\bar{Q}^\ominus_5,g^\oplus_6,g_{q_4^\perp}^*\bigr)-P_{1\to 4}^{\mathcal{N}=4\, \mathrm {sYM}}\bigl(q^\oplus_2,Q^\oplus_4,{\color{blue}\bar{q}^\ominus_3},\bar{Q}^\ominus_5,g^\oplus_6,g_{q_4^\perp}^*\bigr) \nonumber \\
  &&+ P_{1\to 4}^{\text{}}\bigl(q^\oplus_2,Q^\oplus_4,\bar{Q}^\ominus_5,{\color{blue}\bar{q}^\ominus_3},g^\oplus_6,g_{q_4^\perp}^*\bigr)+P_{1\to 4}^{\text{}}\bigl(q^\oplus_2,Q^\oplus_4,\bar{Q}^\ominus_5,g^\oplus_6,{\color{blue}\bar{q}^\ominus_3},g_{q_4^\perp}^*\bigr)\nonumber \\
  &&- P_{1\to 4}\bigl({\color{blue}\bar{q}^\ominus_3},q^\oplus_2,Q^\oplus_4,\bar{Q}^\ominus_5,g^\oplus_6,g_{q_4^\perp}^*\bigr)\,,
\end{eqnarray}
where here again there is no corresponding identity strictly within QCD.
These examples, along with several others, are provided in the accompanying Mathematica notebook~\texttt{Examples_v0.nb}.
We have explicitly verified the photon decoupling identities for all MREVs of up to four final-state partons. Next we move on to discuss the extension of the photon decoupling identities to off-shell gluons.

\subsubsection{Permuting off-shell gluons}\label{OffshellUone}

Following the argument presented at the beginning of the section, the decoupling identity can be seen as a redundancy in expressing the amplitudes in the trace basis, regardless of whether the particles are on-shell or off-shell. Therefore, the off-shell gluons $g^*_{{q}_1^{\perp}}$ or $g^*_{q_{n+1}^\perp}$ can also play the role of a ``decoupling photon''.

For CEVs, permuting $g_{{q}_1^{\perp}}^*$ gives rise to:
\begin{eqnarray}
0 & =&C_{n-4}\big(  {\color{blue}g^*_{q_{1}^\perp}},\,p_{\sigma_{4}}^{h_{\sigma_{4}}},\,p_{\sigma_{5}}^{h_{\sigma_{5}}},\,\ldots,\,p_{\sigma_{n-1}}^{h_{\sigma_{n-1}}},\,g^*_{q_{n-3}^\perp}\big) + C_{n-4}\big( p_{\sigma_{4}}^{h_{\sigma_{4}}},\, {\color{blue}g^*_{q_{1}^\perp}},\,p_{\sigma_{5}}^{h_{\sigma_{5}}},\,\ldots,\,p_{\sigma_{n-1}}^{h_{\sigma_{n-1}}},\,g^*_{q_{n-3}^\perp}\big) 
\nonumber\\
&&+\ldots + \,C_{n-4}\big( p_{\sigma_{4}}^{h_{\sigma_{4}}},\,p_{\sigma_{5}}^{h_{\sigma_{5}}},\,\ldots,\,p_{\sigma_{n-1}}^{h_{\sigma_{n-1}}},\,{\color{blue}g^*_{q_{1}^\perp}},\,g^*_{q_{n-3}^\perp} \big)
.
\label{permuting_q1}
\end{eqnarray}
Thus, the permutation of the off-shell gluon $g^*_{q_{1}^\perp}$ in the CEV is handled in the same way as an on-shell gluon, with no additional sign introduced.
For example, for the colour-ordered CEV process $g^* g^* \to g\,\bar{q}\,q$, as in Eq.~\eqref{permuting_q1}, permuting $g^*_{q_{1}^\perp}$ we get
\begin{align}
0=&\ C_{3}\big({\color{blue}g^*_{q_{1}^\perp}},\, g_4^\oplus,\, \bar{q}_5^{\ominus},\, q_6^{\oplus},\, g^*_{q_{4}^\perp}\big)
+ C_{3}\big(g_4^\oplus,\, {\color{blue}g^*_{q_{1}^\perp}},\, \bar{q}_5^{\ominus},\, q_6^{\oplus},\, g^*_{q_{4}^\perp}\big) 
\nonumber \\
& + C_{3}^{\text{}}\big(g_4^\oplus,\, \bar{q}_5^{\ominus},\, {\color{blue}g^*_{q_{1}^\perp}},\, q_6^{\oplus},\, g^*_{q_{4}^\perp}\big)
+ C_{3}\big(g_4^\oplus,\, \bar{q}_5^{\ominus},\, q_6^{\oplus},\, {\color{blue} g^*_{q_{1}^\perp}},\, g^*_{q_{4}^\perp}\big).
\end{align}
We may straightforwardly generalise these relations to higher-multiplicity CEVs. For example, let us consider the colour-ordered CEV for $g^* g^* \to x\,q\,\bar{q}\,y$, with $x$ and $y$ being either gluons ($x = g_4$, $y = g_7$), quarks of the same flavour ($x = q_4$, $y = \bar{q}_7$) or quarks of a different flavour ($x = Q_4$, $y = \bar{Q}_7$). In any of these cases, by permuting $g_{q_1^\perp}^*$ we obtain:
\begin{eqnarray} 0&=& 
C_4\bigl({\color{blue} g^*_{q^\perp_{1}}},\,x^{\oplus},\, \bar{q}_5^{\ominus},\,q_6^{\oplus},\, y^{\ominus},\,g^*_{q^\perp_{5}}\bigr) + 
C_{4}\bigl( x^{\oplus},\,{\color{blue} g^*_{q^\perp_{1}}},\, \bar{q}_5^{\ominus},\,q_6^{\oplus},\, y^{\ominus},\,g^*_{q^\perp_{5}}\bigr)\nonumber\\[1mm] &&+\,  
C_{4}\bigl( x^{\oplus},\,q_5^{\ominus},\,{\color{blue} g^*_{q^\perp_{1}}},\,q_6^{\oplus},\, y^{\ominus},\,g^*_{q^\perp_{5}}\bigr) +
C_{4}\bigl(x^{\oplus},\,q_5^{\ominus},\, q_6^{\oplus},\,{\color{blue} g^*_{q^\perp_{1}}},\, y^{\ominus},\,g^*_{q^\perp_{5}}\bigr)\nonumber\\[1mm] &&+\,
C_{4}\bigl( x^{\oplus},\,q_5^{\ominus},\,q_6^{\oplus},\, y^{\ominus},\, {\color{blue} g^*_{q^\perp_{1}}},\,g^*_{q^\perp_{5}}\bigr) \,. \end{eqnarray}
Note, however, that care should be taken regarding the signs when permuting $g^*_{q_{n-3}^\perp}$ in CEVs (or in PEVs): an additional minus sign should be included whenever an odd permutation of quarks emerges. These are best illustrated using examples. 
For $g^* g^* \rightarrow g \bar{q} q$, permuting $g^*_{q_{4}^\perp}$ gives 
\begin{align}
0=&\ C_{3}\big(g^*_{q_{1}^\perp},\, g_4^\oplus,\, \bar{q}_5^{\ominus},\, q_6^{\oplus},\, g^*_{q_{4}^\perp}\big)
-C_{3}\big( q_6^{\oplus},\, g^*_{q_{1}^\perp},\, g_4^\oplus,\, \bar{q}_5^{\ominus},\,g^*_{q_{4}^\perp}\big) 
+ C_{3}\big(  \bar{q}_5^{\ominus},\, q_6^{\oplus},\,g^*_{q_{1}^\perp},\, g_4^{\oplus},\, g^*_{q_{4}^\perp}\big)\nonumber \\
& \quad
+ C_{3}\big(g_4^\oplus,\, \bar{q}_5^{\ominus},\, q_6^{\oplus},\, g^*_{q_{1}^\perp},\, g^*_{q_{4}^\perp}\big).
\end{align}
Similarly, for the $g^* g^* \to x\,q\,\bar{q}\,y$ example, permuting~$g_{q_5^\perp}^*$, while adhering to our convention where~$g^*_{q^\perp_{5}}$ is the last entry in the colour-ordered CEV, introduces additional signs for odd permutations of fermions:
\begin{align}
0=\ & C_4\bigl(g^*_{q^\perp_{1}},\, x^{\oplus},\, \bar{q}_5^{\ominus},\, q_6^{\oplus},\, y^{\ominus},\, g^*_{q^\perp_{5}}\bigr)  + C_4\bigl(x^{\oplus},\, \bar{q}_5^{\ominus},\, q_6^{\oplus},\, y^{\ominus},\, g^*_{q^\perp_{1}},\, g^*_{q^\perp_{5}}\bigr) 
\nonumber\\
& \pm  C_4\bigl(\bar{q}_5^{\ominus},\, q_6^{\oplus},\, y^{\ominus},\, g^*_{q^\perp_{1}},\, x^{\oplus},\, g^*_{q^\perp_{5}}\bigr) \mp   C_4\bigl(q_6^{\oplus},\, y^{\ominus},\, g^*_{q^\perp_{1}},\, x^{\oplus},\, \bar{q}_5^{\ominus},\, g^*_{q^\perp_{5}}\bigr) 
\nonumber\\
& \pm C_4\bigl(y^{\ominus},\, g^*_{q^\perp_{1}},\, x^{\oplus},\, \bar{q}_5^{\ominus},\, q_6^{\oplus},\, g^*_{q^\perp_{5}}\bigr)\,,
\end{align}
where the `$\pm$' should be read as `$+$' when $x$ and $y$ are gluons and as `$-$' when these are quarks, and vice versa for $\mp$ on the fourth term.

Following the same rules, for the colour-ordered PEVs $g^*g \to g\, q\, \bar{q}\, g$, by permuting $g_{q_4^\perp}^*$, we obtain:
\begin{eqnarray}  
0&=&P_{1\to 4}\left(g_2^\oplus,g_3^\ominus,q_4^\oplus,\bar{q}_5^\ominus,g_6^\ominus,
g_{q_4^\perp}^*\right)
+P_{1\to 4}\left(g_3^\ominus,q_4^\oplus,\bar{q}_5^\ominus,g_6^\ominus,g_2^\oplus,
g_{q_4^\perp}^*\right)\non 
&& +\,P_{1\to 4}\left(q_4^\oplus,\bar{q}_5^\ominus,g_6^\ominus,g_2^\oplus,g_3^\ominus,
g_{q_4^\perp}^*\right)-
P_{1\to 4}\left(\bar{q}_5^\ominus,g_6^\ominus,g_2^\oplus,g_3^\ominus,q_4^\oplus,
g_{q_4^\perp}^*\right)\non 
&& +\,
P_{1\to 4}\left(g_6^\ominus,g_2^\oplus,g_3^\ominus,q_4^\oplus,\bar{q}_5^\ominus,
g_{q_4^\perp}^*\right)\,.\label{uonepev}
\end{eqnarray}
For $ g^*q \to q\, Q\, \bar{Q}\, g$ we have
\begin{eqnarray} 0&=& P_{1\to 4}\bigl(q_2^\oplus,\ \bar{q}_3^\ominus,\ Q_4^\oplus,\ \bar{Q}_5^\ominus,\ g_6^\oplus,\ g_{q_4^\perp}^*\bigr) + P_{1\to 4}\bigl( g_6^\oplus,\ q_2^\oplus,\ \bar{q}_3^\ominus,\ Q_4^\oplus,\ \bar{Q}_5^\ominus,\ g_{q_4^\perp}^*\bigr) \nonumber\\[1mm] &&
-\,
P_{1\to 4}\bigl(\bar{Q}_5^\ominus,\ g_6^\oplus,\ q_2^\oplus,\ \bar{q}_3^\ominus,\ Q_4^\oplus,\ g_{q_4^\perp}^*\bigr) + P_{1\to 4}\bigl(Q_4^\oplus,\ \bar{Q}_5^\ominus,\ g_6^\oplus,q_2^\oplus,\ \bar{q}_3^\ominus,\ g_{q_4^\perp}^*\bigr) \nonumber\\[1mm]  && -\, P_{1\to 4}\bigl( \bar{q}_3^\ominus,\ Q_4^\oplus,\ \bar{Q}_5^\ominus,\ g_6^\oplus,\ q_2^\oplus,\ g_{q_4^\perp}^*\bigr) \,. \end{eqnarray}
All the identities in this section have been explicitly checked using the library. Several other examples can be found in \texttt{Examples.nb}. 

In the next two sections, \ref{kk} and \ref{reversalk}, we review the Kleiss--Kuijf and reversal identities for amplitudes and discuss the resulting relations for MREVs. We note that the off-shell photon decoupling identities discussed above can also be derived from the Kleiss-Kuijf and reversal identities. Nevertheless they provide an alternative  representation of these relations and serve us here for consistency checks for colour-ordered MREVs.

\subsection{Kleiss-Kuijf relation for MREVs}\label{kk}

The Kleiss-Kuijf relation~\cite{Kleiss:1988ne} provides a systematic way of reducing the set of colour-ordered amplitudes of gluons into a minimal basis. This relation has been generalised to amplitudes with quark-antiquark pairs in Refs.~\cite{Melia:2013bta,Johansson:2015oia}. 
In Appendix~\ref{sec:BriefKKrelation}
we briefly summarise how this relation can be derived by comparing the expression for the colour-dressed amplitude between the trace basis and the DDM~\cite{DelDuca:1999rs} one. Here we wish to derive the Kleiss-Kuijf relation for MREVs from the corresponding relation for amplitudes. We first consider pure gluon MREVs, and illustrate the extension to MREVs which include quark-antiquark pairs using explicit examples.

We first prepare the Kleiss-Kuijf relation for pure gluon amplitudes. We fix $g_1^{h_1}$ and~$g_2^{h_2}$ and consider a bipartition of the remaining outgoing gluons, $\{g_3^{h_3}, g_4^{h_4}, \ldots, g_{n-1}^{h_{n-1}}\}$. To simplify the notation, we define $\sigma_i \equiv g_{\sigma_i}^{h_{\sigma_i}}$, with $i \in \{3, 4, \ldots, n\}$. Then, for a \emph{given} permutation of the two subsets, the relation reads:
\begin{align}
        & A_n\left(g_1^{h_1},g_{n}^{h_n}, \sigma_m, \ldots, \sigma_4, g_2^{h_2},g_{3}^{h_3}, \sigma_{m+1}, \ldots, \sigma_{n-1}\right)\non&=(-1)^{m}\sum_\shuffle  A_n(g_1^{h_1}, g_2^{h_2},\{\sigma_4, \ldots, \sigma_m,g_{n}^{h_n}\}\shuffle\{g_{3}^{h_3},\sigma_{m+1}, \ldots, \sigma_{n-1}\}),\label{kkrel}
\end{align}
where the sum here is over all shuffles of the two sets of gluons, $\{\sigma_4, \ldots, \sigma_m, g_{n}^{h_n}\}$ and $\{g_{3}^{h_3}, \sigma_{m+1}, \ldots, \sigma_{n-1}\}$. The shuffle product is defined as usual, constructing  all possible orderings of the union of the two sets, such that the relative ordering among the elements of each of the initial sets remains unchanged throughout.
The factor $(-1)^{m}$ in Eq.~(\ref{kkrel}) corresponds to applying the commutator $m-2$ times in order to flip $\{\sigma_4, \ldots, \sigma_m, g_{n}^{h_n}\}$ across $g_2$ using Eq.~\eqref{fffftoTTTT}.

\begin{figure}
       \centering
\includegraphics[width=1\linewidth]{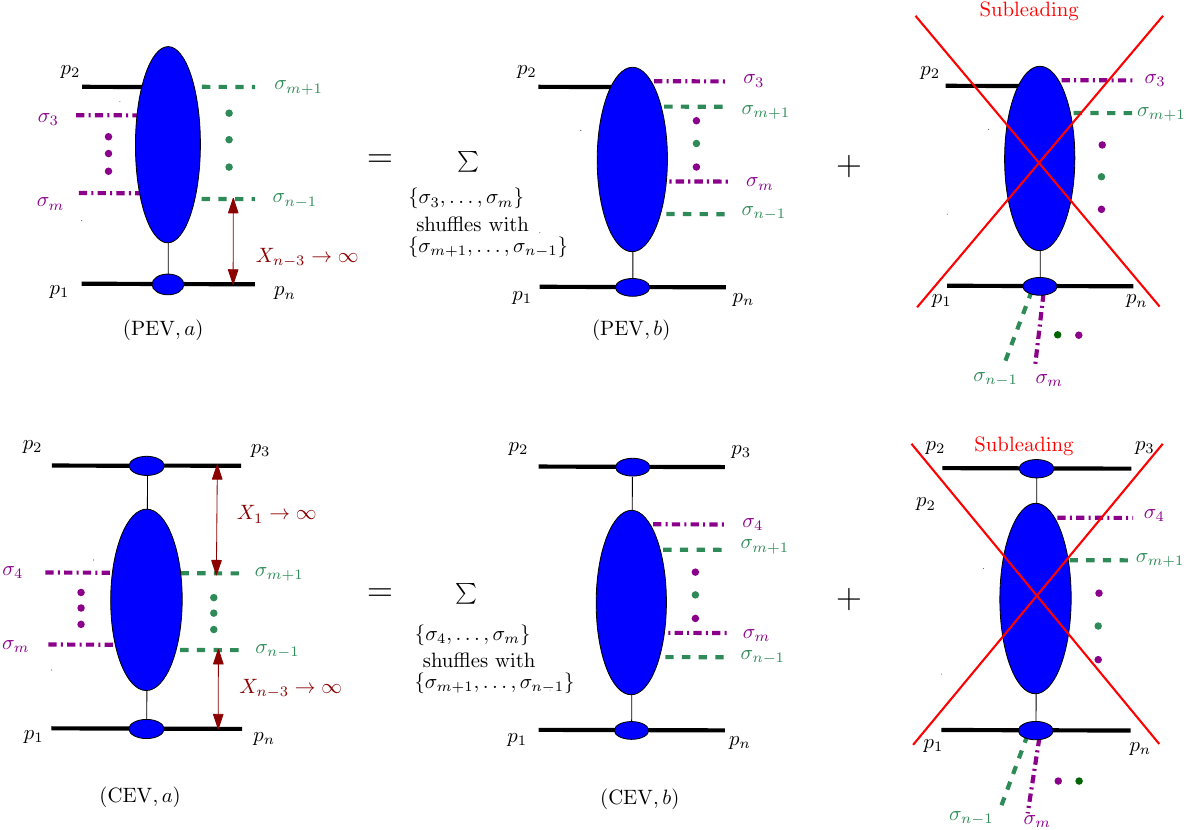}
 \caption{Derivation of Kleiss-Kuijf relation for PEVs (top) and CEV (bottom) based on the corresponding relations for colour-ordered amplitudes. The last colour-ordered configuration shown in each case is power suppressed in the relevant high-energy limit considered, similarly to Figure~\ref{fig:u(1)}.}
    \label{fig:kkpev}
   \end{figure}

To proceed with the derivation of the relation for CEVs we now specialise Eq.~(\ref{kkrel}) to cases where the helicities are conserved for $\{g_2^{h_2},\; g_3^{-h_2}\}$ and $\{g_1^{h_1},\; g_n^{-h_1}\}$.
Now consider the high-energy limit 
\begin{align}
&\lim_{y_{3}\gg \,y_4\,\simeq\,\ldots\,\simeq\, y_{n-1}\gg \,y_{n}} A_n\left(g_1^{h_1},g_{n}^{-h_1}, \sigma_m, \ldots, \sigma_4, g_2^{h_2},g_{3}^{-h_2}, \sigma_{m+1}, \ldots, \sigma_{n-1}\right)
{\to} 
\non 
& =(-1)^{m} \sum_\shuffle   A_n\left( g_{n}^{-h_1},g_1^{h_1}, g_2^{h_2},g_{3}^{-h_2},\left\{\sigma_{4}, \ldots, \sigma_m\right\} \shuffle\left\{ \sigma_{m+1}, \ldots, \sigma_{n-1}\right\}\right) +\text{subleading},\label{powersupressed}
\end{align}
where ``subleading'' refers to terms that are power-suppressed in the high-energy limit considered.   
This derivation is illustrated in 
Figure~\ref{fig:kkpev}. There, the leftmost diagram (a) depict the left-hand side of Eq.~(\ref{powersupressed}), while the other diagrams show contributions to the shuffle sum on the right-hand side of Eq.~(\ref{powersupressed}).
Referring back to the general extraction formula for the CEV in~Eq.~(\ref{cevmulti}) together with Eq.~(\ref{powersupressed}), the figure shows that any colour-ordered CEV in which the 
virtual gluons are not adjacent can be written in terms of CEVs in which they are; similarly, any PEV in which the virtual gluon is not adjacent to the incoming parton in the colour ordering, can be traded for ones in which it is.
   So we obtain
\begin{align}
   C_{n-4}\big(&\sigma_m, \ldots, \sigma_5, \sigma_4,g_{q_{1}^{\perp}}^*, \sigma_{m+1}, \ldots, \sigma_{n-1}, g_{q_{n-3}^{\perp}}^*\big)\nonumber\\[1mm]
   =\; & (-1)^{m} \sum_\shuffle C_{n-4}\big(
   g_{q_{1}^{\perp}}^*, \,
   \{\sigma_4,\sigma_5, \ldots, \sigma_m\} \shuffle \{\sigma_{m+1}, \ldots, \sigma_{n-1}\}, \,
   g_{q_{n-3}^{\perp}}^*\big)\,.
   \label{gluoncc}
\end{align}
Similarly for PEVs:
\begin{eqnarray}
      && \hspace*{-40pt}
      P_{1\to n-3}\big( \sigma_m, \ldots, \sigma_4, \sigma_3,p_2^{h_2},\sigma_{m+1}, \ldots, \sigma_{n-1},g_{q_{n-3}^{\perp}}^*\big) \non 
   &=& (-1)^{m} \sum P_{1\to n-3}\big(p_2^{h_2},\left\{\sigma_3,\sigma_4, \ldots, \sigma_m\right\} \shuffle\left\{\sigma_{m+1}, \ldots, \sigma_{n-1}\right\},g_{q_{n-3}^{\perp}}^* \big)\,.
   \label{gluonpp}
\end{eqnarray}
Just as in the amplitude examples shown in Eqs.~\eqref{eq728}--\eqref{eq730}, the CEV and PEV formulae in Eqs.~\eqref{gluoncc} and \eqref{gluonpp} can be generalized to include quark-antiquark pairs, taking into account the additional signs associated with odd permutations of fermions. This is illustrated in the following two examples:
\begin{align}
     C_3\big(\bar{q}_5^\ominus,\, q_4^\oplus,\, g_{q_1^\perp}^*,\, g_6^\oplus,\, g_{q_4^\perp}^*\big) 
     = &\,
 - C_3\big(g_{q_1^\perp}^*,\, g_6^\oplus,\, q_4^\oplus,\, \bar{q}_5^\ominus,\, g_{q_4^\perp}^*\big)  
  -C_3\big(g_{q_1^\perp}^*,\, q_4^\oplus,\, 
  \bar{q}_5^\ominus,\, g_6^\oplus,\, g_{q_4^\perp}^*\big) \nonumber\\
&\qquad\quad
 -C_3\big(g_{q_1^\perp}^*,\, q_4^\oplus,\, g_6^\oplus,\, \bar{q}_5^\ominus,\, g_{q_4^\perp}^*\big)\,;
\end{align}
\begin{align}
&\hspace*{-20pt}P_{1\to 4}\big(\bar{Q}_4^\ominus ,Q_3^\oplus ,g_2^\ominus,{q}_5^\oplus ,\bar{q}_6^\ominus,g_{q_4^\perp}^* \big) 
\notag \\=&  - \bigg[ P_{1\to 4}\big(g_2^\ominus, Q_3^\oplus, \bar{Q}_4^\ominus, {q}_5^\oplus, \bar{q}_6^\ominus, g_{q_4^\perp}^*\big) - P_{1\to 4}\big(g_2^\ominus, Q_3^\oplus, {q}_5^\oplus, \bar{Q}_4^\ominus, \bar{q}_6^\ominus, g_{q_4^\perp}^*\big) \notag \\
&\hspace{1cm}+ P_{1\to 4}\big(g_2^\ominus, Q_3^\oplus, {q}_5^\oplus, \bar{q}_6^\ominus, \bar{Q}_4^\ominus, g_{q_4^\perp}^*\big) + P_{1\to 4}\big(g_2^\ominus, {q}_5^\oplus, Q_3^\oplus, \bar{Q}_4^\ominus, \bar{q}_6^\ominus, g_{q_4^\perp}^*\big) \notag \\
&\hspace{1cm}- P_{1\to 4}\big(g_2^\ominus, {q}_5^\oplus, Q_3^\oplus, \bar{q}_6^\ominus, \bar{Q}_4^\ominus, g_{q_4^\perp}^*\big) + P_{1\to 4}\big(g_2^\ominus, {q}_5^\oplus, \bar{q}_6^\ominus, Q_3^\oplus, \bar{Q}_4^\ominus, g_{q_4^\perp}^*\big)
\bigg]\,.
\end{align}
In \texttt{Examples.nb}, we have provided checks of  relations of this type for MREVs with up to four parton emissions including, for example, 
\begin{eqnarray}
    \hspace*{-40pt}C_4\left(g_7^{\ominus}, g_4^{\oplus},g_{q_{1}^{\perp}}^*, Q_5^{\ominus}, \bar{Q}_6^{\oplus},g_{q_{5}^{\perp}}^* \right)
    \eqn
    \sum_{\shuffle }  C_4\big(g_{q_{1}^{\perp}}^*,{ \{g_4^{\oplus},g_7^{\ominus} \}\shuffle \{Q_5^{\ominus}, \bar{Q}_6^{\oplus}\}},g_{q_{5}^{\perp}}^*\big)
\end{eqnarray}
and
\begin{eqnarray}
    P_{1\to 4}\big(g_4^{\oplus}, g_3^{\oplus} , g_2^\ominus, Q_5^{\oplus}, \bar{Q}_6^{\ominus},g_{q_{4}^{\perp}}^*\big)
    \eqn
    \sum_{\shuffle }  P_{1\to 4}\big(g_2^\ominus, \{g_3^{\oplus}, g_4^{\oplus} \}\shuffle \{ Q_5^{\oplus}, \bar{Q}_6^{\ominus}\},g_{q_{4}^{\perp}}^*\big).
\end{eqnarray}

\subsection{Reversal identity for MREVs}
\label{reversalk}

\begin{figure}
    \centering
    \includegraphics[width=1\linewidth]{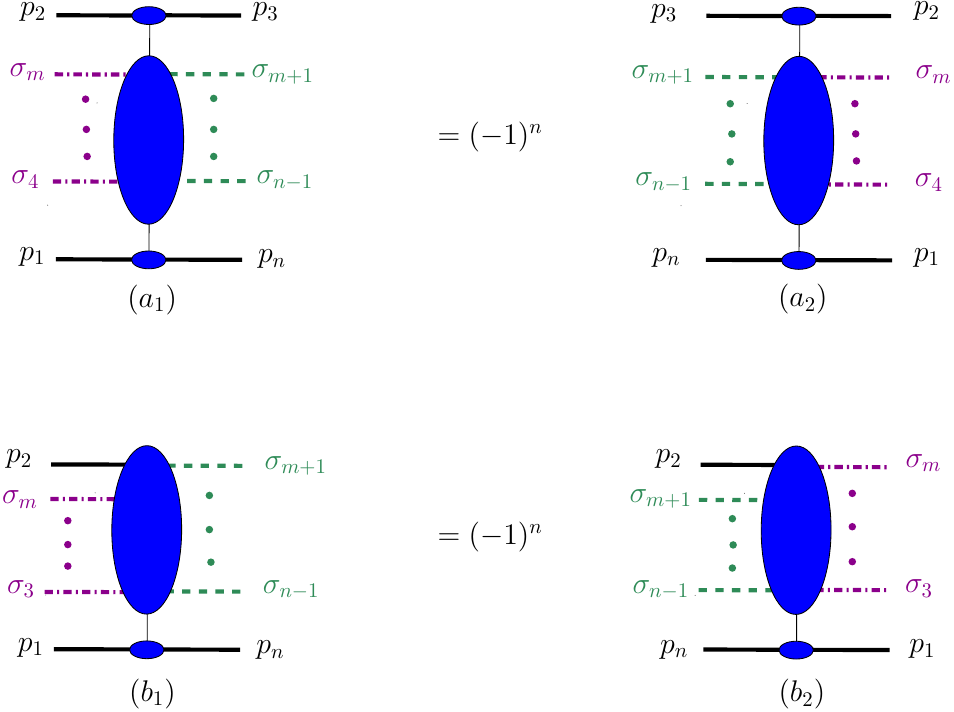}
    \caption{Diagrammatic illustration of the reversal identity, from amplitudes to the CEVs ($a_1$, $a_2$) and PEVs ($b_1$, $b_2$). }
    \label{fig:rev}
\end{figure}

Pure-gluon colour-ordered amplitudes admit the following reversal identity~\cite{Bern:1990ux}: 
\begin{eqnarray}
&& \hspace{-40pt} 
A_n\Big(
p_{n}^{h_n},\, 
p_{1}^{h_1},\, 
g_{\sigma_4}^{h_{\sigma_4}},\, \ldots,\, g_{\sigma_m}^{h_{\sigma_m}},\, 
g_{2}^{h_2},\, 
g_{3}^{h_3},\, 
g_{\sigma_{m+1}}^{h_{\sigma_{m+1}}},\, \ldots,\, g_{\sigma_{n-1}}^{h_{\sigma_{n-1}}}
\Big)
\nonumber \\
&=& (-1)^{n}\,
A_n\Big(
g_{\sigma_{n-1}}^{h_{\sigma_{n-1}}},\, \ldots,\, g_{\sigma_{m+1}}^{h_{\sigma_{m+1}}},\, 
g_{3}^{h_3},\, 
g_{2}^{h_2},\, 
g_{\sigma_m}^{h_{\sigma_m}},\, \ldots,\, g_{\sigma_4}^{h_{\sigma_4}},\, 
g_{1}^{h_1},\, 
g_{n}^{h_n}
\Big).
\label{amp_rev}
\end{eqnarray}
This relation, which is illustrated in Fig.~\ref{fig:rev}, follows directly from the colour-ordered Feynman rules, see e.g.~\cite{Delduca:Lecture_in_amplitudes,Henn:2014yza}. We now use this result to derive the  reversal identity for gluon CEVs.
To this end we consider the high-energy limit on both sides of Eq.~\eqref{amp_rev} as follows:
\begin{eqnarray}
&&
 \lim_{y_{3}\gg \,y_4\,\simeq\,\ldots\,\simeq\, y_{n-1}\gg \,y_{n}}
A_n\left(
p_{n}^{h_n},\, 
p_{1}^{h_1},\, 
g_{\sigma_4}^{h_{\sigma_4}},\, \ldots,\, g_{\sigma_m}^{h_{\sigma_m}},\, 
g_{2}^{h_2},\, 
g_{3}^{h_3},\, 
g_{\sigma_{m+1}}^{h_{\sigma_{m+1}}},\, \ldots,\, g_{\sigma_{n-1}}^{h_{\sigma_{n-1}}}
\right)
\nonumber
\\
&&=\;
 s \,
  P_{1\to 1}\big(  g_2^{h_2},\,g_3^{h_3},\, g_{q_1^\perp}^*\big)\,\frac{1}{t_1}\,
  C_{n-4}\big( g_{\sigma_4}^{h_{\sigma_4}},\,\ldots,\, g_{\sigma_m}^{h_{\sigma_m}},\,g_{q_1^\perp}^*,\, g_{\sigma_{m+1}}^{h_{\sigma_{m+1}}},\,\ldots,\, g_{\sigma_{n-1}}^{h_{\sigma_{n-1}}},\,g_{q_{n-3}^\perp}^*\big)\,
 \notag\\
&&\qquad
 \times\, \frac{1}{t_{n-3}}
  P_{1\to 1}\big(g_1^{h_1},\, g_{q_{n-3}^\perp}^*,\, g_n^{h_n}\big) 
\nonumber
\\
&&=\;
 s \, 
  P_{1\to 1}\big(  g_3^{h_3},\,g_2^{h_2},\, g_{q_1^\perp}^*\big)\,\frac{1}{t_1} \,
(-1)^n  C_{n-4}\big(g_{\sigma_{n-1}}^{h_{\sigma_{n-1}}},\, \ldots,\, g_{\sigma_{m+1}}^{h_{\sigma_{m+1}}},\,g_{q_1^\perp}^*,\, g_{\sigma_m}^{h_{\sigma_m}},\, \ldots,\, g_{\sigma_4}^{h_{\sigma_4}},\, g_{q_{n-3}^\perp}^*\big)\,
 \notag\\
&&\qquad
 \times\, \frac{1}{t_{n-3}}
  P_{1\to 1}\big(g_n^{h_n},\, g_{q_{n-3}^\perp}^*,\, g_1^{h_1}\big).
\end{eqnarray}
\!\!\!Due to the Bose symmetry of the colour-dressed PEV and the antisymmtery of its colour factor (see Eq.~(\ref{positivePEV_g})), the colour-stripped PEV is also antisymmetric. Thus, the reversal of each $P_{1 \to 1}$ picks up a sign. Therefore, we arrive at
\begin{align}
&  C_{n-4}\big( g_{\sigma_4}^{h_{\sigma_4}},\,\ldots,\, g_{\sigma_m}^{h_{\sigma_m}},\,g_{q_1^\perp}^*,\, g_{\sigma_{m+1}}^{h_{\sigma_{m+1}}},\,\ldots,\, g_{\sigma_{n-1}}^{h_{\sigma_{n-1}}},\,g_{q_{n-3}^\perp}^*\big)\nonumber \\[1mm]
&\qquad =\, (-1)^{n}\, C_{n-4}\big(g_{\sigma_{n-1}}^{h_{\sigma_{n-1}}},\, \ldots,\, g_{\sigma_{m+1}}^{h_{\sigma_{m+1}}},\,g_{q_1^\perp}^*,\, g_{\sigma_m}^{h_{\sigma_m}},\, \ldots,\, g_{\sigma_{4}}^{h_{\sigma_{4}}},\, g_{q_{n-3}^{\perp}}^* \big)\,.
\end{align}
A similar reversal identity holds for pure-gluon PEVs:
\begin{align}
&  P_{1\to n-3}\big( g_{\sigma_3}^{h_{\sigma_3}},\,\ldots,\, g_{\sigma_m}^{h_{\sigma_m}},\,g_{2}^{h_{2}},\, g_{\sigma_{m+1}}^{h_{\sigma_{m+1}}},\,\ldots,\, g_{\sigma_{n-1}}^{h_{\sigma_{n-1}}},\,g_{q_{n-3}^\perp}^*\big)\nonumber \\[1mm]
&\qquad =\, (-1)^{n-1}\, P_{1\to n-3}\big(g_{\sigma_{n-1}}^{h_{\sigma_{n-1}}},\, \ldots,\, g_{\sigma_{m+1}}^{h_{\sigma_{m+1}}},\,g_{2}^{h_{2}},\,g_{\sigma_m}^{h_{\sigma_m}},\, \ldots,\, g_{\sigma_{3}}^{h_{\sigma_{3}}},\, g_{q_{n-3}^{\perp}}^* \big)\,.
\end{align}
The way these reversal identities follow from that of the amplitude is illustrated in Fig.~\ref{fig:rev}.

Similar relations apply when fermions are included. In this case an odd permutation of fermions introduces an additional factor of $(-1)$ as we have seen before; for example
\begin{align}
  &C_{4}\big(g_7^\ominus,\, g_5^\oplus,\, g_{q_1^\perp}^*,\, g_4^\oplus,\, g_6^\ominus,\, g_{q_5^\perp}^*\big)
=  C_{4}\big(g_6^\ominus,\, g_4^\oplus,\, g_{q_1^\perp}^*,\, g_5^\oplus,\, g_7^\ominus,\, g_{q_5^\perp}^*\big)
   \nonumber \\
&C_{4}\big(g_7^\ominus,\, g_5^\oplus,\, g_{q_1^\perp}^*,\, q_4^\oplus,\, \bar{q}_6^\ominus,\, g_{q_5^\perp}^*\big)
= - C_{4}\big(\bar{q}_6^\ominus,\, q_4^\oplus,\, g_{q_1^\perp}^*,\, g_5^\oplus,\, g_7^\ominus,\, g_{q_5^\perp}^*\big)
   \nonumber \\
  &
  C_{4}\big(g_{q_1^\perp}^*,\, \bar{q}_4^\ominus,\, q_5^\oplus,\, \bar{Q}_4^\ominus,\, Q_5^\oplus,\, g_{q_5^\perp}^*\big)
= C_{ 4}\big(Q_5^\oplus,\, \bar{Q}_4^\ominus,\, q_5^\oplus,\, \bar{q}_4^\ominus,\, g_{q_1^\perp}^*,\, g_{q_5^\perp}^*\big);
\end{align}
\begin{align}
  &P_{1\to 4}\big(g_6^\oplus,\, g_4^\oplus,\, g_2^\oplus,\, g_3^\ominus,\, g_5^\ominus,\, g_{q_4^\perp}^*\big)
   =  P_{1\to 4}\big(g_5^\ominus,\, g_3^\ominus,\, g_2^\oplus,\, g_4^\oplus,\, g_6^\oplus,\, g_{q_4^\perp}^*\big) 
   \nonumber \\
  &P_{1\to 4}\big(g_6^\oplus,\, g_4^\oplus,\, q_2^\oplus,\, \bar{q}_3^\ominus,\, g_5^\ominus,\, g_{q_4^\perp}^*\big)
   = - P_{1\to 4}\big(g_5^\ominus,\, \bar{q}_3^\ominus,\, q_2^\oplus,\, g_4^\oplus,\, g_6^\oplus,\, g_{q_4^\perp}^*\big)
   \nonumber \\
  &P_{1\to 4}\big(g_3^\ominus,\, q_2^\oplus,\, \bar{q}_4^\ominus,\, Q_5^\ominus,\, \bar{Q}_6^\ominus,\, g_{q_4^\perp}^*\big)
   =\; P_{1\to 4}\big(\bar{Q}_6^\ominus,\, Q_5^\ominus,\, \bar{q}_4^\ominus,\, q_2^\oplus,\, g_3^\ominus,\, g_{q_4^\perp}^*\big).
\end{align}
More examples are provided in \texttt{Examples.nb}. Next, we move on to discuss the supersymmetric Ward identity.
\subsection{Supersymmetric Ward identities among PEVs for quark and gluons}

Let us begin by recalling how supersymmetric Ward identities emerge and how they can be used to constrain tree-level amplitudes in QCD. Then, by considering suitable high-energy limits we will deduce relations between PEVs.
This requires a brief summary of the notation. For more complete reviews we direct the reader to Refs.~\cite{Delduca:Lecture_in_amplitudes,Elvang:2015rqa,Dixon:1996wi,Mangano:1990by}.
 
Scattering amplitudes are defined by the connected part of the vacuum expectation value of a (time-ordered) product of fields:  
\begin{align}
\mathcal A_{0 \to n} \sim \langle \Omega | \, a_{p_1}  \ldots a_{p_n} | \Omega \rangle,
\end{align}  
where $a_{p_i}$, acting on the ``out'' vacuum $ \langle \Omega |$ state annihilates a particle with momentum~$p_i$ and helicity~$h_i$.
Here we will be using the gauge multiplet of ${\cal N}=1$ sYM, where the annihilation operators include $a_{g^\oplus_i}$ ($a_{g^\ominus_i}$)
and $b_{q^\oplus_i}$ ($b_{{q}^\ominus_i}$), which annihilate, respectively, a gluon and a gluino with positive (negative) helicities. The corresponding colour-ordered amplitudes map directly onto the \emph{single-flavour} QCD ones, where the gluino is replaced by a quark (hence the notation for $b_{q}$ above). In the remainder of this subsection we shall restrict ourselves to single-flavour QCD and not distinguish between supersymmetric amplitudes and QCD ones.

The supersymmetric  Ward identity~\cite{Grisaru:1977px} emerges directly from the annihilation of the vacuum state \( | \Omega \rangle \) by the supersymmetric generator: 
\begin{equation}
\label{Qannihilates}
    {Q} | \Omega \rangle = 0, \qquad   {Q}^\dagger  |\Omega\rangle= 0.
\end{equation} 
This implies that the vacuum expectation value of the commutator between the supersymmetric generator and the annihilation operators vanishes:
\begin{equation}
\begin{aligned}
0 &= \bigl\langle \Omega \bigl| \big[ Q^\dagger,\, a_{p_1}  \ldots a_{p_n}\big]\bigr| \Omega \bigr\rangle \\
&= \sum_{i=1}^n \bigl\langle \Omega \bigl| a_{p_1} \cdots \bigl[ Q^\dagger,\, a_{p_i}\bigr] \cdots  a_{p_n}\bigr| \Omega \bigr\rangle,
\end{aligned}
\label{eq:SUSY_Ward}
\end{equation}  
where in the second line we expressed the commutator as a sum of commutators of $Q^\dagger$ with individual annihilation operators\footnote{A similar equation holds upon replacing $Q^\dagger$ by $Q$.}. 
The latter expression is useful because each of these commutators is prescribed by supersymmetry as follows:
\begin{subequations}
\label{RaisingLowering}
\begin{align}
   \text{raising spin:}\qquad &\big[{Q}_{\alpha}, a_{g_i^\ominus}\big]=\big|i\big\rangle _\alpha b_{{q}_i^\ominus},\qquad \big[Q_\alpha, b_{q_i^\oplus}\big]=\big|i\big\rangle _\alpha a_{{g}_i^\oplus},\\ 
   &\big[Q_\alpha, a_{g_i^\oplus}\big]=0,\hspace{1.75cm} \big[Q_\alpha,b _{q_i^\ominus}\big]=0,
   \\
   \text{lowering spin:}\qquad  &\big[{Q}^\dagger_{\dot{\alpha}}, b_{q_i^\ominus}\big]=\big|i\big] _{\dot{\alpha}}a_{{g}_i^\ominus},\qquad \big[{Q}^\dagger_{\dot{\alpha}}, a_{g_i^\oplus}\big]=\big|i\big]_{\dot{\alpha}} b_{{q}_i^\oplus},
   \\
    &\big[ {Q}^\dagger_{\dot{\alpha}}, b_{q_i^\oplus}\big]=0,\hspace{1.75cm}  \big[{Q}^\dagger_{\dot{\alpha}}, a_{g_i^\ominus}\big]=0.
\end{align}
\end{subequations}
As Eq.~(\ref{RaisingLowering}) shows, the supersymmetric generator carries an ${SL}(2,\mathbb{C})$ index similarly to a Weyl spinor, and when they raise or lower spin, an additional spinor is introduced, which compensates for the difference in the little group scaling between a gluino and a gluon.
The commutation relations in Eq.~(\ref{RaisingLowering}) imply that the supersymmetry generator \(Q\) raises the spin by 1/2, while its Hermitian conjugate \(Q^\dagger\) lowers the spin by 1/2.
Upon using them, the identity in Eq.~(\ref{eq:SUSY_Ward}) 
translates into a relation between amplitudes involving different bosonic and fermionic external states. In the context of QCD amplitudes, these would be external states involving a different admixture of gluons and quarks.

To demonstrate Eq.~(\ref{eq:SUSY_Ward}) explicitly in the framework of colour-ordered amplitudes, consider the following example: 
\begin{align}
\label{susy_amp_example1}
\begin{split}
0&=\left\langle \Omega\left|\left[Q^\dagger_{\dot {\alpha}}, a_{g_1^\ominus} a_{g_2^\ominus}a _{g_3^\oplus} b_{\bar {q}_4^\ominus } a_{g_5^\oplus } \right]\right| \Omega\right\rangle\,.
\end{split}
\end{align}
This expectation value must vanish owing to Eq.~(\ref{Qannihilates}).
Using the commutation relations in Eq.~(\ref{RaisingLowering}) and contracting the free spinor index with a reference spinor $[k|^{\dot{\alpha}}$, we get
\begin{align}
\label{susy_amp_example2}
\begin{split}
0& = \left[k 3\right]A_{2\to 3}(g_2^\ominus, {\color{blue}q_{3}^\oplus},\bar q_{4}^\ominus,g_5^\oplus ,g_1^\ominus)+\left[k 4\right]A_{2\to 3}(g_2^\ominus, g_{3}^\oplus,{\color{blue}g_{4}^\ominus },g_5^\oplus,g_1^\ominus)
        \\  &\hspace{1cm}-\left[k 5\right]
 A_{2\to 3}(g_2^\ominus, g_{3}^\oplus,\bar 
q_{4}^\ominus,{\color{blue}q_5^\oplus},g_1^\ominus)\,.
\end{split}
\end{align}
In this way Eq.~(\ref{susy_amp_example1}) amounts to a linear relation between the three terms on the right-hand side of Eq.~(\ref{susy_amp_example2}). The negative sign in the third term arises because \(Q^\dagger\)  crosses a fermionic annihilation operator.

Let us now describe how these supersymmetric relations between on-shell amplitudes translate into relations between PEVs. We begin by considering the example of Eq.~(\ref{susy_amp_example2}).
Taking the high-energy limit where
\(  y_2,\, y_3,\, y_4\gg  y_5,\, y_1 \),
and further choosing $y_k$ to be of the same order of magnitude as $y_5$ and $y_1$,
we find that
the third term in Eq.~\eqref{susy_amp_example2} becomes power suppressed compared to the first two terms. 
This suppression occurs because the relevant $t$-channel propagator in $A_{2\to 3}(g_2^\ominus, g_{3}^\oplus,\bar 
 q_{4}^\ominus,{\color{blue}q_5^\oplus},g_1^\ominus)$ going across the large rapidity span separating between partons 1 and~5 and the remaining three is that of a quark, as opposed to a gluon which arises in the first two amplitudes in Eq.~\eqref{susy_amp_example2}.
In turn, the corresponding spinor products multiplying the three amplitudes admit
$[k3]\simeq [k4]\gg [k5]$.
The two surviving terms 
yield a simple relation between pure-gluon and quark-antiquark PEVs:
\begin{eqnarray}
P_{1\to 2}\left(g_2^\ominus, g_{3}^\oplus, g_{4}^\ominus,g_{q_2^\perp}^*\right)
=
-P_{1\to 2}\left(g_2^\ominus, q_{3}^\oplus, \bar  q_{4}^\ominus,g_{q_2^\perp}^*\right)\times \lim_{ y_{3}\,\simeq \,y_{4}\gg\, y_k}  \frac{\sbb{k}{3}}{ \sbb{k}{4}}\,,\label{SUSY5}
\end{eqnarray}
where
\begin{equation}
    \lim_{ y_{3}\,\simeq \,y_{4}\gg\, y_k}  \frac{\sbb{k}{3}}{ \sbb{k}{4}}=\sqrt{X_1}\,.
\end{equation}
This makes manifest the fact that Eq.~(\ref{SUSY5}) only depends on the degrees of freedom of the PEVs, independently of the amplitude from which it was extracted.

In general, Eq.~(\ref{SUSY5}) can be generalised to $n$-point MHV (anti-MHV) configurations containing exactly one positive (negative) helicity parton, forming a simple relation between pure gluon PEVs and those containing a single
quark-antiquark pair. 
For PEVs originating in NMHV amplitudes, there is no such simple one-to-one relation. Instead, pure gluon PEVs are expressible as linear combinations of two or more PEVs containing a single quark-antiquark pair. For example, one of the supersymmetric Ward identities for $gg^*\to 3$-{parton} PEVs is:
\begin{eqnarray}
		&&\hspace*{-30pt}
        P_{1\to 3}\left(g_2^\oplus, g_{3}^\ominus, g_{4}^\oplus, g_{5}^\ominus,g_{q_3^\perp}^*\right)
  \non \eqn    
 \lim_{y_{3,4,5}\gg y_k }  \bigg[\frac{\ab{k}{5}}{\ab{k}{4}}P_{1\to 3}\left(g_2^\oplus, g_{3}^\ominus, q_{4}^\oplus, \bar{q}_{5}^\ominus,g_{q_3^\perp}^*\right)- \frac{\ab{k}{3}}{\ab{k}{4}}P_{1\to 3}\left(g_2^\oplus, \bar{q}_{3}^\ominus, q_{4}^\oplus, g_{5}^\ominus,g_{q_3^\perp}^*\right)\bigg]\,,\non 
 \label{eq643}
	\end{eqnarray}
    where
    \begin{equation}
        \lim _{y_{3,4,5} \gg y_k}\frac{\langle k 5\rangle}{\langle k 4\rangle}=\frac{1}{\sqrt{X_2}} \qquad   \lim _{y_{3,4,5} \gg y_k}\frac{\langle k 3\rangle}{\langle k 4\rangle}=\sqrt{X_1}.
        \label{eq64423}
    \end{equation}
Supersymmetric Ward identities
for NMHV PEVs with four final-state partons involve more terms. For example:
\begin{align}
 P_{1\to 4}\left(g_2^\oplus, g_{3}^\ominus, g_{4}^\oplus ,g_{5}^\ominus,g_{6}^\oplus,g_{q_4^\perp}^*\right)=  &\!\!
  \lim _{ y_{3,4,5,6}\gg y_k}\bigg[-P_{1\to 4}\left(q_2^\oplus, \bar{q}_{3}^\ominus, g_{4}^\oplus ,g_{5}^\ominus,g_{6}^\oplus,g_{q_4^\perp}^*\right) \frac{\sbb{k}{2}}{\sbb{k}{3}}\non  &\qquad\qquad\,\,+ P_{1\to 4}\left(g_2^\oplus, \bar{q}_{3}^\ominus, q_{4}^\oplus ,g_{5}^\ominus,g_{6}^\oplus,g_{q_4^\perp}^*\right)\frac{\sbb{k}{4}}{\sbb{k}{3}}\non  &\qquad\qquad\,\, + P_{1\to 4}\left(g_2^\oplus, \bar{q}_{3}^\ominus, g_{4}^\oplus ,g_{5}^\ominus,q_{6}^\oplus,g_{q_4^\perp}^*\right)\frac{\sbb{k}{6}}{\sbb{k}{3}}\bigg]\,,
  \label{PEV1to4SusyWard}
\end{align}
where
\begin{equation}
\begin{split}
   & \lim _{ y_{3,4,5,6}\gg y_k}  \frac{[k 2]}{[k 3]}=\frac{i(1+X_3+X_2 X_3+X_1 X_2 X_3)}{\sqrt{X_1X_2X_3}}\quad
   \\
   &\lim _{ y_{3,4,5,6}\gg y_k}  \frac{[k 4]}{[k 3]}=\frac{1}{\sqrt{X_1}}\qquad\lim _{ y_{3,4,5,6}\gg y_k}  \frac{[k 6]}{[k 3]}=\frac{1}{\sqrt{X_1X_2X_3}}.
\end{split}
\end{equation}
Of course, these are just a few of many supersymmetric Ward identities between PEVs. More examples can be found in~\texttt{Examples.nb}.
These identities provide many consistency checks from our results, but more importantly, they capture the highly-constrained and yet rich mathematical structure these objects admit.

A final comment is that in contrast to PEVs, no supersymmetric Ward identities exist among CEVs. To see why, consider for example taking the limit $y_3 \gg y_4 \simeq y_5 \gg y_k$ (i.e., $X_1 \to \infty$) in Eq.~\eqref{eq643}. After factoring out the dependence on $P_{1 \rightarrow 1}\big(g_2^{\oplus}, g_3^{\ominus}, g_{q_1^\perp}^*\big)/|q_1^\perp|^2$, we obtain
\begin{align}
\label{eq6431}
		\hspace*{-20pt}
        C_{2}\left(g_{q_1^\perp}^*, g_{4}^\oplus, g_{5}^\ominus,g_{q_3^\perp}^*\right)
   =&   
\frac{1}{\sqrt{X_2}} C_{2}\left(g_{q_1^\perp}^*, q_{4}^\oplus, \bar{q}_{5}^\ominus,g_{q_3^\perp}^*\right) \\ & - \frac{|q_1^\perp|^2}{P_{1 \rightarrow 1}\big(g_2^{\oplus}, g_3^{\ominus},g_{q_1^\perp}^*\big)}\lim_{X_1 \to \infty} \sqrt{X_1}P_{1\to 3}\left(g_2^\oplus, \bar{q}_{3}^\ominus, q_{4}^\oplus, g_{5}^\ominus,g_{q_3^\perp}^*\right), \nonumber
\end{align}
where we replaced the spinor bracket ratios using Eq.~\eqref{eq64423}. For the last term in Eq.~(\ref{eq6431}), the scaling is $\mathcal{O}(1)$ because $P_{1\to 3}\big(g_2^\oplus,\, \bar{q}_{3}^\ominus,\, q_{4}^\oplus,\, g_{5}^\ominus,\, g_{q_3^\perp}^*\big) \sim 1/\sqrt{X_1}$, which corresponds to quark propagation in the $t$-channel. 

The topic of quark Reggeization has been pursued in Refs.~\cite{Fadin:1976nw,Fadin:1977jr,Bogdan:2002sr,Bogdan:2006af} and more recently in Ref.~\cite{Moult:2017xpp}.
Assuming the last term in Eq.~(\ref{eq6431}) factorises into a PEV and CEV coupled by the $t$-channel exchange of a quark, we would obtain
\begin{align}
\label{sWard_q_Reggon}
C_{2}\left(g_{q_1^\perp}^*, g_{4}^\oplus, g_{5}^\ominus,g_{q_3^\perp}^*\right)
=&   
(X_2)^{-\frac{1}{2}} C_{2}\left(g_{q_1^\perp}^*, q_{4}^\oplus, \bar{q}_{5}^\ominus,g_{q_3^\perp}^*\right) 
\\ & - 
\frac
{P_{1 \rightarrow 1}\big(g_2^{\oplus}, g_3^{\ominus},g_{q_1^\perp}^*\big)}
{P_{1 \rightarrow 1}\big(g_2^{\oplus}, \bar{q}_3^{\ominus},q_{q_1^\perp}^*\big)}
C_{2}\left(q_{q_1^\perp}^*, q_{4}^\oplus, g_{5}^\ominus,g_{q_3^\perp}^*\right), \nonumber
\end{align}
where the ratio of PEVs is a simple phase.
A detailed study of MREVs with $t$-channel quark exchange lies outside the scope of this work. However, we note that identities such as Eq.~(\ref{sWard_q_Reggon}) can be used to determine such objects from the known MREVS with $t$-channel gluon exchange.

\section{Conclusions}
\label{sec:discuss}

In this paper we have provided a systematic determination of all central and peripheral emission vertices in QCD with up to four final-state partons, both quarks and gluons. We have tabulated these in the 
Mathematica library, \texttt{\href{https://github.com/YuyuMo-UoE/Multi-Regge-Emission-Vertices}{Multi-Regge-Emission-Vertices}} (\texttt{MREV})~\cite{yuyu_2017_github} and provided formulae for their colour dressing.

A key to our strategy has been the use of the MSLCVs, Eq.~(\ref{minival:lorentzinvdd}), which avoids the inherent redundancy of spinor products, and allows us to straightforwardly expand the amplitudes in the relevant high-energy limit. These $3n-9$ (real) variables neatly separate between the longitudinal degrees of freedom and the transverse ones, making their invariance with respect to a Lorentz boost along the beam direction manifest. More generally, these variables are Lorentz invariant aside from a phase corresponding to a reference direction in the transverse plane. This additional degree of freedom is necessary, and indeed sufficient, to represent all spinor products, and hence any helicity amplitude. Being free of any redundancy, and directly involving ratios of lightcone momentum components, the MSLCVs are ideally suited to taking high-energy limits.
While in this paper we focus entirely on tree-level MREVs, the strategy of using these variables applies in much the same way at loop level, as has already been demonstrated at one loop in the case of the two-gluon CEV in sYM~\cite{Byrne:2022wzk} and two-parton PEVs in both sYM and QCD~\cite{Byrne:2023nqx}. Closely-related variables have also been used in the context of the two-loop five-point amplitude in multi-Regge kinematics in sYM~\cite{Caron-Huot:2020vlo} and the two-loop single-gluon CEV in QCD~\cite{Buccioni:2024gzo,Abreu:2024xoh}.

We summarised the explicit expressions of all MREVs, classifying all their physical and spurious poles in section~\ref{resultsof CEVPEV}. Upon taking a common denominator form, all spurious singularities cancel (in practice, in order to keep the expressions compact,  we do not express them in this form). Since we use a minimal set of independent variables, the MSLCVs, this cancellation is automatic. 
We note that while the spurious poles cancel outright in tree amplitudes and MREVs, this is not so at loop level, where rational functions appear multiplying transcendental functions. Classifying the various tree-level singularities then pays off when studying loop corrections to MREVs, as demonstrated in~\cite{Byrne:2022wzk,Byrne:2023nqx}.

We discussed the known relations between colour-ordered tree amplitudes, including photon decoupling, Kleiss-Kuijf, reversal and supersymmetric Ward identities.
In each case we determined whether and how these relations translate into relations between PEVs, CEVs, or both (section~\ref{relationsPEVCEV}). We stress that while some of these relations operate within the space of QCD MREVs, some operate in the larger space of colour-ordered objects defined in the supersymmetric context. An example of the latter is the photon decoupling identity for PEVs in Eq.~(\ref{PEVqbargstar}). Another interesting case is the supersymmetric Ward identity between colour-ordered amplitudes, which translates into relations between QCD PEVs, such as Eqs.~(\ref{eq643}) and (\ref{PEV1to4SusyWard}) mixing pure gluon PEVs and ones containing quark-antiquark pairs, but does not directly translate into relations between CEVs. Instead, upon taking the appropriate high-energy limit, these relate power-suppressed contributions associated with quark propagation in the $t$ channel across a large rapidity span, to ordinary (leading-power) QCD CEVs. 
An example is provided in Eq.~(\ref{eq6431}).
To the extent that these relations can be extended to loop level, for the first tower(s) of high-energy logarithms, they provide an interesting setting in which quark Reggeization can be explored.

We have conducted comprehensive checks of the {\tt{MREV}} library~\cite{yuyu_2017_github}, first by taking further high energy, soft and collinear kinematic limits of the MREVs and verifying their factorization properties (Appendix~\ref{Appendix:Fact_in_limits}) and then by verifying that they satisfy the required photon decoupling, Kleiss-Kuijf, reversal and supersymmetric Ward identities.

The quark-antiquark gluon CEV result determined here completes the tree-level information required for the NNLO computation of the Lipatov vertex. We have also provided an independent determination of the three-gluon CEV which was computed before in Refs.~\cite{DelDuca:1999iql,Antonov:2004hh,Duhr:2009uxa}. Figure~\ref{fig:Kernel_NNLO} depicts these ingredients along with loop corrections of lower multiplicity CEVs, some of which have also been determined in recent years, including the single-gluon CEV to two-loop order~\cite{Buccioni:2024gzo,Abreu:2024xoh} and the CEV for the emission of two gluons~\cite{Byrne:2022wzk}, or a $q\bar q$ pair to one-loop order (the two-gluon CEV is only known so far in ${\cal N}=4$ sYM).
The present work represents another step towards completion of the full set of CEVs needed for the NNLO BFKL kernel in QCD.

\acknowledgments
We are grateful to Duncan Miller for collaboration at an early stage of this project. 
We would like to thank Claude Duhr, Giuseppe De Laurentis,  Kanghoon Lee, Jiajie Mei and Ben Page for helpful discussions.
YM is supported by Edinburgh Global Research Scholarship. 
EG is supported by the STFC Consolidated Grant \emph{Particle Physics at the Higgs Centre}. EB is supported by the Royal Society through Grant URF\textbackslash R1\textbackslash201500.
For the purpose of open access, the authors have applied a Creative Commons Attribution (CC BY) licence to any Author Accepted Manuscript version arising from this submission.

\appendix

\section{Review of multi-parton kinematics}
\label{frame}

Here we collect our conventions for momenta and spinor products used throughout the paper.  Following Appendix A in \cite{DelDuca:1999iql}, we consider a $2\to n-2$ process.
We choose a centre-of-mass frame where the two incoming particles, 1 and 2, are directly scattering at each other:
\begin{equation}
\label{wqyug}
\begin{aligned}
			& p_1=\left(\frac{p_1^{-}}{2}, 0,0, \frac{-p_1^{-}}{2}\right) \equiv\left(0, p_1^{-} ; 0,0\right), \\
			& p_2=\left(\frac{p_2^{+}}{2}, 0,0, \frac{p_2^{+}}{2}\right) \equiv\left(p_2^{+}, 0 ; 0,0\right)\,.
\end{aligned}
\end{equation}
The remaining partons have non-zero transverse momenta, for which we will be using the complex momentum representation, $p_{\perp}=p^x+i p^y$. Thus, a four momentum vector is:
\begin{eqnarray}
\label{kijew}
    p_i
    \eqn
\left(\left(p_i^{+}+p_i^{-}\right) / 2, \operatorname{Re}\big[{p_{i}^\perp}\big], \operatorname{Im}\big[{p_{i}^\perp}\big],\left(p_i^{+}-p_i^{-}\right) / 2\right)\non
    &\equiv& \big|{p_{i}^\perp}\big|\left(e^{y_i}, e^{-y_i} ; \cos \phi_i, \sin \phi_i\right)
    \non
    &=&
    \big(p_i^+,p_i^-;{p}_{i}^{\perp}\big),
	\end{eqnarray}
    where each particle has a rapidity defined as 
 \begin{eqnarray}
     y_i=\frac{1}{2} \log \left(\frac{p_i^{+}}{p_i^{-}}\right) ,
     \label{rapidity}
 \end{eqnarray}
 and in the second line of eq.~(\ref{kijew}) the following on-shell condition is used
  \begin{eqnarray}
	    \big|{p_{i}^\perp}\big|^2=p_i^{+} p_i^{-}.\label{onshell}
	\end{eqnarray}
Momentum conservation yields
\begin{equation}
0 =\sum_{i=3}^n {p_{i}^\perp}, \qquad
p_2^{+} =-\sum_{i=3}^{n} p_i^{+} \quad {\rm and} \quad
p_1^{-} =-\sum_{i=3}^{n} p_i^{-}.
\label{momentumconserva}
\end{equation}
With this, we have 
\begin{eqnarray}
    \begin{aligned}
s & =2 p_1 \cdot p_2=p_1^-p_2^+=\sum_{i, j=3}^n p_i^{+} p_j^{-}, \\
s_{2 i} & =2 p_2 \cdot p_i=p_i^-p_2^+=-\sum_{j=3}^n p_i^{-} p_j^{+}, \\
s_{1 i} & =2 p_1 \cdot p_i=p_i^+p_1^-=-\sum_{j=3}^n p_i^{+} p_j^{-}, \\
s_{i k} & =2 p_i \cdot p_k=p_i^{+} p_k^{-}+p_i^{-} p_k^{+}-{p_{i}^\perp} \bar{p}_{k}^\perp-{\bar{p}_{i}^\perp} p_{k }^\perp.
\end{aligned}
\end{eqnarray}
Following Appendix A in \cite{DelDuca:1999iql}, the Weyl spinor products are related to particle momenta by, \begin{eqnarray}
\label{spinoconven}
		\begin{aligned}
			& \left\langle p_i p_j\right\rangle={p_{i}^\perp} \sqrt{\frac{p_j^{+}}{p_i^{+}}}-p_{j }^\perp \sqrt{\frac{p_i^{+}}{p_j^{+}}}, \\
			& \left\langle p_2 p_i\right\rangle=-i \sqrt{\frac{-p_2^{+}}{p_i^{+}}} {p_{i}^\perp} \text {, } \\
			& \left\langle p_i p_1\right\rangle=i \sqrt{-p_1^{-} p_i^{+}} \text {, } \\
			& \left\langle p_2 p_1\right\rangle=-\sqrt{p_2^{+} p_1^{-}}.     \\
			&
		\end{aligned}
	\end{eqnarray}
We finish this section by listing several identities used in this paper
\begin{align}
    [k p] &= \operatorname{sign}\left(k^0 p^0\right)\langle p k\rangle^*,  \label{spinoridentities_1} \\
    \langle k|\slashed{q}|p] &= \langle k q\rangle [q p], \label{spinoridentities_2} \\
    [p|\slashed{q}| k\rangle  &= [p q]\langle q k\rangle, \label{spinoridentities_3} 
\end{align}
and 
\begin{align}
     \sum_{i=1}^n [j i]\langle i k\rangle &= 0 . \label{spinoridentities_4}
\end{align}

  \section{Factorization properties in various kinematic limits}
\label{Appendix:Fact_in_limits}
  \subsection{High-energy limit} 
\label{gqqwqffq}
Within the high-energy factorization framework discussed in section~\ref{defPEVandCEV}, when the rapidity separation between partons becomes large, scattering amplitudes can be decomposed into lower-point high-energy building blocks, such as MREVs~\cite{DelDuca:1998kx,DelDuca:1999iql}. The same methodology can also be applied starting from MREVs~\cite{DelDuca:1998cx,Fadin:1993wh,Fadin:1994fj,Fadin:1996yv,Byrne:2022wzk,Byrne:2023nqx}, and it serves as a crucial cross-check for our results.

In the appropriate large rapidity separation limit, say $y_{m} \gg y_{m+1}$ (corresponding to $p_{m}^{+} \gg p_{m+1}^{+}$ as in Eq.~\eqref{kijew}), the CEV at leading order further factorizes into a product of lower-point CEVs and a $t$-channel (gluon) propagator pole, $1/|q_{m-2}^\perp|^2$. 
 \begin{align}
  \label{qwkl1}
  \begin{split}
&\hspace*{-5pt}
\lim\limits_{y_{m}\gg y_{{m+1}}} C_{n-4}\big( g_{q_1^\perp}^* ,p_{4}^{h_{4}} , p_{5}^{h_{5}} , \ldots, p_{m}^{h_{m}} , p_{{m+1}}^{h_{{m+1}}}, \ldots, p_{{n-1}}^{h_{{n-1}}}, g_{q_{n-3}^\perp}^*\big)
\\
& =
C_{m-3}\big( g_{q_1^\perp}^*, p_{4}^{h_{4}}, \ldots, p_{m}^{h_{m}}, g_{q_{m-2}^\perp}^* \big)\,
\frac{1}{|q_{m-2}^\perp|^2}
\,C_{n-m-1}\big( g_{q_{m-2}^\perp}^*, p_{{m+1}}^{h_{{m+1}}}, \ldots, p_{{n-1}}^{h_{{n-1}}}, g_{q_{n-3}^\perp}^* \big)
\\
&\hspace{1cm}+ O\left( \sqrt{\frac{p_{{m+1}}^{+}}{p_{m}^{+}} }\right).
\end{split}
\end{align}

As an illustrative example, consider the colour-ordered CEV with a single gluon accompanied by a quark-antiquark pair emission. In the limit $y_{4}\gg y_{5}$, the factorization takes the form:
 \begin{eqnarray}
\lim\limits_{y_{4}\gg y_{{5}}}  C_{3}\big(g_{q_1^\perp}^*,g_4^{\oplus}, q_5^{\oplus}, \bar{q}_6^{\ominus},g_{q_4^\perp}^*\big)=C_{1}(g_{q_1^\perp}^*,g_4^{\oplus},g_{q_2^\perp}^*)\frac{1}{t_2} C_{2}( g_{q_2^\perp}^*,q_5^{\oplus}, \bar{q}_6^{\ominus},g_{q_4^\perp}^*).
 \end{eqnarray}
That is, in MSLCV, when taking the limit $X_2 \equiv \frac{p_4^+}{p_5^+}\to \infty$, and according to Eqs.~\eqref{cccppp} to~\eqref{eq444}, the contributions from the second, third, and fifth terms in Eq.~\eqref{gqqb} factorize into Eqs.~\eqref{lipatov} and \eqref{qqb}.
In Appendix \ref{code_ckeching}, we provide our corresponding checking code.

The factorization property holds in an analogous way for PEVs. Under the appropriate large rapidity separation condition, say $y_{m} \gg y_{{m+1}}$, a multi-particle PEV factorizes into a lower-point PEV and a CEV with $t$-channel gluon propagation between the large rapidity gap:
 \begin{eqnarray}
&\lim\limits_{y_{m}\gg y_{{m+1}}}&P_{1\to  n-3}\big( p_{2}^{h_{2}} , p_{3}^{h_{3}} , \ldots p_{m}^{h_{m}} ,p_{{m+1}}^{h_{{m+1}}}, \ldots, p_{n-1}^{h_{n-1}},g_{q_{n-3}^\perp}^*\big)
\non
\eqn
P_{1\to m-2}\big(p_{2}^{h_{2}} , \ldots p_{m}^{h_{m}} ,g_{q_{m-2}^\perp}^* \big)\frac{1}{|q_{{m-2}}^\perp|^2}C_{n-m-1}\big(g_{q_{m-2}^\perp}^* ,p_{{m+1}}^{h_{{m+1}}}, \ldots, p_{n-1}^{h_{n-1}},g_{q_{n-3}^\perp}^*\big).
\non
\label{qwkl2}
  \end{eqnarray}
For example, a 4-parton-emission PEV under ${y_{5} \gg y_{6}}$ factorizes as:
\begin{equation}
 \lim_{y_{5} \gg y_{6}}\; 
   P_{1\to4}\bigl( q_2^\oplus,g_3^\ominus,g_4^\oplus ,\bar q_5^\ominus,g_6^\oplus,\, g_{q_{4}^{\perp}}^*\bigr)
=P_{1\to3}\bigl( q_2^\oplus,g_3^\ominus,g_4^\oplus ,\bar q_5^\ominus,\, g_{q_3^\perp}^* \bigr)
    \frac{1}{\bigl|q_{3}^{\perp}\bigr|^2}  C_{{1}}\bigl( g_{q_3^\perp}^*,\, g_6^\oplus,\, g_{q_{4}^{\perp}}^*\bigr).
    \label{eqb4}
\end{equation}
We used the code given in the Appendix \ref{pevhepfac} to verify that Eq.~\eqref{eqb4} holds with the MREVs in the library.
More high-energy limit factorization checks are provided in the \texttt{Examples.nb} file.
While our checks focus on the $X_{m-2}\equiv  {p_{{m}}^{+}  }/{p_{{m+1}}^{+}}\to \infty$ limit, one may also 
consider other high-energy limits, such as~\(X_{m-2} \to 0\). Detailed considerations for six-particle amplitudes can be found in Appendix~G.1 of~\cite{Byrne:2022wzk}. Notably, the $O\bigl(\sqrt{p_{{m+1}}^{+}/p_{m}^{+}}\bigr)$ corrections in Eq.\,\eqref{qwkl1} may encode a quark propagating across the large rapidity gap~\cite{Fadin:1976nw,Fadin:1977jr}, whose treatment requires a separate analysis beyond the scope of the present work.

 \subsection{Soft limit}
The soft behavior of MREVs in QCD fundamentally originates from Weinberg’s soft theorem for scattering amplitudes~\cite{Weinberg:1965nx}. In the framework of colour-ordered parton amplitudes, when particle~$i$ corresponds to a soft gluon emission, the amplitude universally factorizes at leading order in the soft momentum into a lower-multiplicity amplitude multiplied by a $\mathcal{O}\left(1 / p_{\text {soft }}\right)$ soft factor, which captures the associated infrared singularity:
\begin{align}
&\hspace{-1cm}\lim_{p_i^\mu \rightarrow 0} A_{2\to n-2}(1,2,3,4,\ldots,{\color{blue}g_i^{h_i}},\ldots,n-1,n)\nonumber\\[1ex]
  =& A_{2\to n-3}\bigl(1,2,3,4,\ldots,i-1,i+1,\ldots,n-1,n\bigr)  \operatorname{Soft}\big(p_{i-1}, {\color{blue}g_i^{h_i}}, p_{i+1}\big)\,,
  \label{eqb5}
\end{align}
where $p_{i+1}$ and $p_{i-1}$ can be either gluons or (anti-)quarks while $g_i^{h_i}$ is the soft gluon. The colour-ordered soft factor is given by
\begin{gather}
\operatorname{Soft}\big(p_{i-1}, g_i^{h_i}, p_{i+1}\big)
=\lim _{p_i^\mu \rightarrow 0}\frac{\varepsilon^{h_i}\bigl(p_i\bigr) \cdot p_{i-1}}{p_i \cdot p_{i-1}}
-\frac{\varepsilon^{h_i}\bigl(p_i\bigr) \cdot p_{i+1}}{p_i \cdot p_{i+1}}, \label{eq:soft_helicity}
\end{gather}
or in spinor-helicity variables:
\begin{subequations}
\label{eq:soft_plus_minus}
    \begin{gather}
\operatorname{Soft}\big(p_{i-1}, g_i^{\oplus}, p_{i+1}\big)
=\lim _{p_i^\mu \rightarrow 0}\frac{\langle (i-1)\,\, (i+1) \rangle}{\langle (i-1)\, i \rangle\, \langle i\, (i+1) \rangle}, \label{eq:soft_plus}\\[2ex]
\operatorname{Soft}\big(p_{i-1}, g_i^{\ominus}, p_{i+1}\big)
=-\lim _{p_i^\mu \rightarrow 0}\frac{[\, (i-1)\, \,(i+1)\,]}{[\, (i-1)\, i\,][\, i\, (i+1)\,]}\,. \label{eq:soft_minus}
\end{gather}
\end{subequations}
Aside from the soft gluon itself, the soft factor is independent of both the helicities and the particle species of the colour-adjacent partons \({i-1}\) and~\({i+1}\) appearing in Eqs.~\eqref{eq:soft_helicity} and~\eqref{eq:soft_plus_minus}, and it is also invariant under rescaling of these momenta.

Having reviewed the soft factorization of colour-ordered amplitudes, our analysis will focus exclusively on the single-gluon soft limits of colour-ordered MREVs. In contrast to the case of scattering amplitudes, the position of the soft gluon within the colour ordering of a PEV or CEV induces two distinct factorization patterns:\\

\begin{itemize}
  \item \textbf{Central soft limit:} Occurs when the soft gluon~$i$, together with its colour-adjacent hard partons~$(i\pm1)$, lie within the on-shell emission sequence of the PEV or CEV.
  \vspace{0.3cm}
  \item \textbf{Edge Soft Limit:} Arises when the soft gluon~$i$ is colour-adjacent to an off-shell gluon.\\
\end{itemize}
In the following, we elaborate on these two cases in turn.

\subsubsection*{Central soft limit}
Let us derive the factorization properties of a CEV in the soft limit from Eq.~\eqref{eqb5}.
To this end we consider the two 
relevant limits in different orders: on the one hand the soft limit of the emitted gluon labelled \(i\) (with \(4 < i < n-1\), satisfying centrality), and on the other the high-energy limit needed to extract the CEV from the amplitude, as in Eq.~\eqref{cevmulti}. 
By applying the high-energy limit first we obtain:
\begin{align}
&\hspace*{-40pt}
\lim_{p_i^{\mu} \to 0}\left(
\lim_{y_3 \gg \,y_4 \,\simeq\, \cdots \,\simeq\, y_{n-1} \gg \,y_n}\;
A_{2\to n-2}\big(1,\, 2,\, 3,\, 4,\, \ldots,\, {\color{blue}g_i^{h_i}},\, \ldots,\, n-1,\, n \big)\right)
\nonumber
\\
&=
s\, P_{1\to 1}\big(2,\, 3,\, g_{q_1^\perp}^* \big)
\,\frac{1}{t_1}\,\frac{1}{t_{n-3}}\,
P_{1\to 1}\big(g_{q_{n-3}^\perp}^*,\, n,\, 1\big)
\nonumber\\
&\quad\times
\lim_{p_i^{\mu} \to 0}C_{n-4}\big(g_{q_1^\perp}^*,\, 4,\, \ldots,\, i-1,\,{\color{blue}g_i^{h_i}},\, i+1,\, \ldots,\, n-1,\, g_{q_{n-3}^\perp}^*\big),
\label{eq:first_fourth}
\end{align}
and by applying the soft limit first we get:
\begin{align}
&\hspace*{-40pt}
\lim_{y_3 \gg \,y_4 \,\simeq\, \cdots \,\simeq\, y_{n-1} \gg \,y_n}\;\left(
\lim_{p_i^{\mu} \to 0}
A_{2\to n-2}\big(1,\, 2,\, 3,\, 4,\, \ldots,\, {\color{blue}g_i^{h_i}},\, \ldots,\, n-1,\, n \big)\right)
\nonumber
\\
&=
\lim_{y_3 \gg y_4 \,\simeq\, \cdots \,\simeq\, y_{n-1} \gg \,y_n}\;
A_{2\to n-3}\big(2,\, 3,\, 4,\, \ldots,\, i-1,\, i+1,\, \ldots,\, n-1,\, n,\, 1 \big)
\nonumber\\
&\quad\hspace*{90pt}
\times\operatorname{Soft}\big(p_{i-1},\, {\color{blue}g_i^{h_i}},\, p_{i+1}\big)
\nonumber\\
&= 
s\, P_{1\to 1}\big(2,\, 3,\, g_{q_1^\perp}^* \big)
\,\frac{1}{t_1}\,\frac{1}{t_{n-3}}\,
P_{1\to 1}\big(g_{q_{n-3}^\perp}^*,\, n,\, 1\big)
\nonumber\\
&\quad\times
C_{n-5}\big(g_{q_1^\perp}^*,\, 4,\, \ldots,\,i-1,\,i+1,\, \ldots,\, n-1,\, g_{q_{n-3}^\perp}^*\big)
\operatorname{Soft}\big( p_{i-1},\, {\color{blue}g_i^{h_i}},\,  p_{i+1}\big)
\label{eq:fourth_fifth}
\end{align}
Comparing Eqs. \eqref{eq:first_fourth} and \eqref{eq:fourth_fifth} and stripping off the common factors 
\[
s\, P_{1\to 1}\big(2,\, 3,\, g_{q_1^\perp}^* \big)
\,\frac{1}{t_1}\,\frac{1}{t_{n-3}}\,
P_{1\to 1}\big(g_{q_{n-3}^\perp}^*,\, n,\, 1\big)
\]
we are left with
\begin{align}
&\lim_{p_i^{\mu} \to 0}
C_{n-4}\big(g_{q_1^\perp}^*,\, 4,\, \ldots,\, i-1,\, {\color{blue}g_i^{h_i}},\, i+1,\, \ldots,\, n-1,\, g_{q_{n-3}^\perp}^*\big)
\nonumber\\
&\hspace{2cm}= 
C_{n-5}\big(g_{q_1^\perp}^*,\, 4,\, \ldots,\, i-1,\, i+1,\, \ldots,\, n-1,\, g_{q_{n-3}^\perp}^*\big)
\, \operatorname{Soft}\big(p_{i-1},\, {\color{blue}g_i^{h_i}},\, p_{i+1}\big).
\label{softC}
\end{align}
Therefore, as long as colour-adjacent neighbours of the soft gluon remain within the emission sequence, the CEV factorizes in the same way as the amplitude, and similarly for the~PEV:
\begin{align}
\lim_{p_{i} \rightarrow 0} &\, P_{1\to n-3}\big( 2,\, \ldots,\, i-1,\,{\color{blue}g_i^{h_i}},\,  i+1,\,\ldots,\, n-1,\, g_{q_{n-3}^\perp}^* \big) \nonumber\\
&\hspace{2cm}= \, P_{1\to n-4}\big( 2,\, \ldots,\, i-1,\, i+1,\, \ldots,\, n-1,\, g_{q_{n-3}^\perp}^* \big)
\, \operatorname{Soft}\big(p_{i-1},\, {\color{blue}g_i^{h_i}},\, p_{i+1}\big).
\end{align}

As a concrete demonstration, consider the soft gluon with  
\begin{equation}
    p_5^\mu \rightarrow \lambda\,p_5^\mu= (\lambda  p_5^+,\,\lambda   p_5^-;\,\lambda  p_5^\perp)
\end{equation}    
where we scale \(\lambda \to 0\). In this limit, the soft factor \( \operatorname{Soft}\big(p_4, g_5^{\oplus}, p_6\big)\sim \frac{1}{\lambda}\), as shown in Eq.~\eqref{eq:soft_helicity}. Implementing the MSLCV parameterization from Eq.~\eqref{minival:nonlorentzinvdd}, this corresponds to the following scaling laws:
 \begin{eqnarray}
     X_2\to \frac{X_2}{\lambda},\   z_2\to \frac{z_2}{\lambda},\  X_3\to {\lambda X_3}.
     \label{softparamenter}
 \end{eqnarray}
Considering for example the colour-ordered three-gluon CEV~$C_{3}\big(g_{q_1^\perp}^*,g_4^{\oplus}, g_5^{\oplus}, g_6^{\ominus},g_{q_4^\perp}^*\big)$ of Eq.~\eqref{CEVgggPPM}, using the parameterization of Eq.~\eqref{softparamenter}, one can verify that the CEV factorizes into a two-gluon CEV given in Eq.~\eqref{C2mixed_hel},
 multiplied by a soft factor as follows:
 \begin{eqnarray}
\lim\limits_{p_5^\mu\to \lambda p_5^\mu }C_{3}\left(g_{q_1^\perp}^*,g_4^{\oplus}, g_5^{\oplus}, g_6^{\ominus},g_{q_4^\perp}^*\right)=\operatorname{Soft}\left(p_4, g_5^{\oplus}, p_6\right)C_{2}\left(g_{q_1^\perp}^*,g_4^{\oplus},g_6^{\ominus},g_{q_4^\perp}^*\right),\non \label{amp_soft}
 \end{eqnarray}
 where in MSLCV
 \begin{eqnarray}
 \label{SoftFactorExample}
 \operatorname{Soft}\left(p_4, g_5^{\oplus}, p_6\right)=\lim\limits_{{p_5^\mu\to \lambda p_5^\mu }}\frac{\langle 46\rangle}{\langle 45\rangle\langle 56\rangle}=\frac{1}{   \lambda }\frac{\left(z_1-1\right) z_2^2 \left(X_2 X_3 z_1-z_3+1\right)}{z_1
 \left(X_2 z_1-z_2\right) \left(X_3 z_2-z_3+1\right)
  q_1^\perp }\,.
 \end{eqnarray}
Note the homogeneous scaling of Eq.~(\ref{SoftFactorExample}) as~$1/\lambda$ subject to the scaling law, Eq.~(\ref{softparamenter}).
The corresponding code that checks this factorization property is given in Appendix~\ref{cevsopft}.

\subsubsection*{Edge soft limit}

There are situations which the central soft limit does not cover. These correspond to the edge soft limit. Still working with Eq.~\eqref{eqb5}, we can consider the case where gluon 4 becomes soft, i.e.\ \(p_4^\mu \to 0\). Then, under the rapidity ordering \(y_3 \gg y_4 \,\simeq\, \ldots \,\simeq\, y_{n-1} \gg y_n\), we can derive the edge soft limit factorization for the CEV by applying the soft and high-energy limit in different orders.
To begin with, consider applying first the high-energy limit which defines the CEV, and then the soft limit:
\begin{align}
&\hspace*{-40pt}
\lim_{p_4^\mu \to 0}
\left(
\lim_{y_3 \gg y_4\,\simeq\, \ldots \,\simeq\,\, y_{n-1} \gg y_n}
A_{2\to n-2}\big( 2,\, 3,\, {\color{blue}g_4^{h_4}},\, 5,\, \ldots,\, n,\, 1 \big)\right)
\nonumber \\
&=
s\, P_{1\to 1}\big(2,\, 3,\, g_{q_1^\perp}^*\big)
\, \frac{1}{t_1}\, \frac{1}{t_{n-3}}\,
P_{1\to 1}\big(g_{q_{n-3}^\perp}^*,\, n,\, 1\big)\nonumber\\
&\quad \times 
\lim_{p_4^\mu \to 0}
C_{n-4}\big(g_{q_1^\perp}^*,\, {\color{blue}g_4^{h_4}},\, 5,\, \ldots,\, n-1,\, g_{q_{n-3}^\perp}^*\big)\,.
\label{cevmultc2}
\end{align}
Next, consider taking the soft limit first, and then the high-energy limit:
\begin{align}
&\hspace*{-40pt}
\lim_{y_3 \gg y_4\,\simeq\, \ldots \,\simeq\, y_{n-1} \gg y_n}\lim_{p_4^\mu \to 0}
A_{2\to n-2}\big( 2,\, 3,\, {\color{blue}g_4^{h_4}},\, 5,\, \ldots,\, n,\, 1 \big) \nonumber\\
&=
\lim_{y_3 \gg y_4\,\simeq\, \ldots \,\simeq\, y_{n-1} \gg y_n}
A_{2\to n-3}\big( 2,\, 3,\, 5,\, \ldots,\, n-1,\, n,\, 1 \big)
\, \operatorname{Soft}\big(3,\, {\color{blue}g_4^{h_4}},\, 5\big)
\nonumber
\\
&=
s\, P_{1\to 1}\big(2,\, 3,\, g_{q_1^\perp}^*\big)
\, \frac{1}{t_1}\, \frac{1}{t_{n-3}}\,
P_{1\to 1}\big(g_{q_{n-3}^\perp}^*,\, n,\, 1\big)
\nonumber
\\
&\quad \times 
C_{n-5}\big(g_{q_1^\perp}^*,\, 5,\, \ldots,\, n-1,\, g_{q_{n-3}^\perp}^*\big)\,
\operatorname{Soft}\big(+,\, {\color{blue}g_4^{h_4}},\, 5\big),
\label{cevmultc1}
\end{align}
where we define
\begin{equation}
\operatorname{Soft}\left(+,g_4^{h_4}, p_{{5}}\right)\equiv \lim_{y_3 \gg \, y_4\,\simeq\,  y_{5}}\operatorname{Soft}\left(p_{{3}},g_4^{h_4},p_{{5}}\right)\,.
\end{equation}
Here we used the inherent rescaling invariance of the soft factor to manifest its independence on $p_3$.
Combining Eqs.~\eqref{cevmultc1} and~\eqref{cevmultc2}, we find that
\begin{align}
    &\lim_{p_4^\mu \to 0}
    C_{n-4}\big(g_{q_1^\perp}^*,\, {\color{blue}g_4^{h_4}},\, 5,\, \ldots,\, n-1,\, g_{q_{n-3}^\perp}^*\big)
    \nonumber\\
&\hspace{2cm}=  
    C_{n-5}\big(g_{q_1^\perp}^*,\, 5,\, \ldots,\, n-1,\, g_{q_{n-3}^\perp}^*\big)\,
    \operatorname{Soft}\big(+,\, {\color{blue}g_4^{h_4}},\, 5\big).
\label{edgesofttheoremo}
\end{align}

Let us now take a closer look at \(\operatorname{Soft}\big(+,\, {\color{blue}g_4^{h_4}},\, 5\big)\).
Following Eq.~\eqref{spinoconven} we find that, in general, for the edge soft factor with \(p_{x}^{+} \gg p_i^{+} \sim\, p_j^{+}\), 
\begin{equation}
\operatorname{Soft}\left(+, g_i^{\oplus}, p_j\right) = \lim_{p_{x}^{+} \gg\, p_i^{+} \sim\, p_j^{+}} \operatorname{Soft}\left(p_x, g_i^{\oplus}, p_j\right)=\lim_{p_{x}^{+} \gg \,p_i^{+} \sim \,p_j^{+}} \frac{\langle x j\rangle}{\langle x i\rangle\langle i j\rangle}
    = \frac{p_{j}^\perp}{p_{i}^\perp} \sqrt{\frac{p_i^{+}}{p_j^{+}}} \frac{1}{\langle ij \rangle}.
    \label{sf}
\end{equation}
Following similar analysis, 
\begin{equation}
\operatorname{Soft}\left(+, g_i^{\ominus}, p_j\right)=\frac{\bar p_j^{\perp}}{\bar p_i^{\perp}} \sqrt{\frac{p_i^{+}}{p_j^{+}}} \frac{1}{[ i j]}.
\end{equation}
We have thus seen that despite the soft gluon being adjacent to the virtual gluon, the factorization of the CEV in this limit is consistent: the kinematic dependence on the right-hand side of Eq.~(\ref{edgesofttheoremo}) involves only the variables the CEV depends upon.

A similar conclusion holds of course when the soft gluon is colour-adjacent to the off-shell gluon that connects to negative rapidities:
\begin{align}
    &\lim_{p_{n-1}^\mu \to 0}
    C_{n-4}\big(
       g_{q_1^\perp}^*,\,
        4,\, 5,\, \ldots,\, n-2,\,
        {\color{blue}g_{n-1}^{h_{n-1}}},\,
        g_{q_{n-3}^\perp}^*
    \big)
    \nonumber\\
    &\hspace{2cm}=
    C_{n-5}\big(
        g_{q_1^\perp}^*,\,
        4,\, 5,\, \ldots,\, n-2,\,
        g_{q_{n-3}^\perp}^*
    \big)
    \;\operatorname{Soft}\big(p_{n-2},\, {\color{blue}g_{n-1}^{h_{n-1}}},\, -\big).
\end{align}
Here, the edge soft factor $\operatorname{Soft}(p_{n-2},\, g_{n-1}^{h_{n-1}},\, -)$ is defined analogously to Eq.~\eqref{sf}. Specifically, from Eq.~\eqref{spinoconven}, when $p_i^{+} \sim p_j^{+} \gg p_{y}^{+}$, we have
\begin{equation}
      \operatorname{Soft}\left(p_j,\, g_i^{\oplus},\, -\right)
      \equiv \lim_{p_i^{+} \sim p_j^{+} \gg \,p_{y}^{+}} \operatorname{Soft}\left(p_j,\, g_i^{\oplus},\, p_{y}\right)
      = \lim_{p_i^{+} \sim p_j^{+} \gg\, p_{y}^{+}}
      \frac{\left\langle j\, {y} \right\rangle}
      {\left\langle j\, i \right\rangle \left\langle i\, {y} \right\rangle}
      = \sqrt{\frac{p_j^+}{p_i^+}}\, \frac{1}{\langle j\,i\rangle}.
      \label{qwqw}
\end{equation}
Following similar analysis, \begin{equation}
        \operatorname{Soft}\left(p_j,\, g_i^{\ominus},\, -\right)=\sqrt{\frac{p_j^{+}}{p_i^{+}}} \frac{1}{ [j i]} .
\end{equation}

This completes our discussion of the CEV edge soft limit. We find that the PEV likewise admits the same factorization in the edge soft limit.
\begin{align}
    &\lim_{p_{n-1}^\mu \to 0}
    P_{1 \to n-3}\Big(
        2,\, 3,\, \ldots,\, n-2,\, g_{n-1}^{h_{n-1}},\, g_{q_{n-3}^\perp}^*
    \Big)
    \nonumber\\
    &\hspace{2cm}=
    P_{1 \to n-4}\Big(
        2,\, 3,\, \ldots,\, n-2,\, g_{q_{n-3}^\perp}^*
    \Big)
    \;\operatorname{Soft}\big(p_{n-2},\, g_{n-1}^{h_{n-1}},\, -\big).
\end{align}
The soft limit factorization examples above and several others can be found in the MREV package file \texttt{Examples.nb}.
To close the soft limit section, we note that the soft limit of real parton emission has been systematically extended beyond single-gluon to two-parton~\cite{Catani:1999ss}, three-parton~\cite{DelDuca:2022noh,Catani:2022hkb} and most recently four-parton~\cite{Chen:2024hvp} soft emission currents. Such currents can also be obtained from the CEVs and PEVs, however, we leave this for future work.

\subsection{Collinear Limit}
\label{wohac_collinear}%
The collinear behaviour of MREVs in QCD originates from the collinear factorization of scattering amplitudes. For colour-ordered parton amplitudes, collinear singularities emerge exclusively when \emph{colour-adjacent} partons become collinear, as described by the factorization formula:
\begin{align}
&\hspace*{-30pt}A_{2\to n-2}\bigl(
  2, 3, 4,\ldots,i,i+1,\ldots,n-1,n,1
\bigr)
\nonumber\\[1ex]
\xrightarrow{i \| i+1}&\sum_{h_c=\pm} A_{2\to n-3}\bigl(  2, 3, 4,\ldots,P^{h_c},\ldots,n-1,n,1
\bigr) \times\, \text{Sp}_{P^{-h_c}}\bigl(
    x_{i}P^{h_{i}},\, x_{{i+1}}P^{h_{{i+1}}}
\bigr),
\label{collinear_mother}
\end{align}
where the collinear momentum \( P = p_i + p_{i+1} \) is parameterized as $p_i=x_i P, p_{i+1}=x_{i+1} P$ , with momentum fractions satisfying $x_i+x_{i+1}=1$. 

Considering two colour-adjacent partons $i$ and $j$, the collinear splitting amplitudes, which exhibit $1/\sqrt{s_{ij}}$ behaviour as $s_{ij}=(p_i+p_j)^2 \to 0$, are given by \cite{DelDuca:1999iql,Bern:1994zx,Mangano:1990by}:
\begin{align}
\label{collinears}
\begin{array}{ll}
\operatorname{Sp}_{g^\ominus}\left(g_i^{\oplus}, g_j^{\oplus}\right)=\frac{1}{\sqrt{x_i x_j}\langle ij\rangle}
&\hspace*{30pt}
\operatorname{Sp}_{g^\oplus}
\left(g_i^{\oplus}, g_j^{\ominus}\right)=\frac{x_j^2}{\sqrt{x_i x_j}\langle ij\rangle}
\\
\operatorname{Sp}_{g^\oplus}\left(\bar q_i^{\ominus}, q_j^{\oplus}\right)=\frac{-x_i}{\langle ij\rangle}
&\hspace*{30pt}
\operatorname{Sp}_{g^\oplus}\left(\bar{q}_i^{\oplus},  q_j^{\ominus}\right)=\frac{x_j}{\langle ij\rangle}
\\
\operatorname{Sp}_{\bar q^\ominus}\left(q_i^{\oplus}, g_j^{\oplus}\right)=\frac{1}{\sqrt{ x_j}\langle ij\rangle}
&\hspace*{30pt} \operatorname{Sp}_{q^\oplus}\left(\bar q_i^{\ominus}, g_j^{\oplus}\right)=\frac{x_i}{\sqrt{ x_j}\langle ij\rangle}\\
\operatorname{Sp}_{\bar q^\ominus}\left(g_i^{\oplus}, q_j^{\oplus}\right)=\frac{1}{\sqrt{ x_i}\langle ij\rangle}
&\hspace*{30pt}
\operatorname{Sp}_{q^\oplus}\left(g_i^{\oplus}, \bar q_j^{\ominus}\right)=\frac{x_j}{\sqrt{ x_i}\langle ij\rangle}.
\end{array}
\end{align}
Other helicity configurations can be obtained through complex conjugation of these expressions in Eq. \eqref{collinears}.
 
Equipped with this, let us now consider the collinear limit of MREVs. Consider first the CEV case. Taking the high-energy limit of Eq.~\eqref{collinear_mother} as in Eq.~\eqref{cevmulti}, we have:
\begin{align}
\label{hjqwqeg} 
&\hspace*{-25pt}\lim_{y_3 \gg\, y_4\,\simeq\, \ldots \,\simeq\, y_{n-1} \gg y_n} A_{2\to n-2}\bigl(
  2, 3, 4,\ldots,i,i+1,\ldots,n-1,n,1
\bigr) \nonumber \\
&\hspace*{25pt}=\, s\, P_{1\to 1}\left(2, 3, g_{q_1^\perp}^*\right) \frac{1}{t_1} \frac{1}{t_{n-3}} P_{1\to 1}\left(g_{q_{n-3}^\perp}^*, n, 1\right) \nonumber \\[1ex]
&\hspace*{65pt}\times C_{n-4}\left(g_{q_1^\perp}^*, 4, \ldots, i,i+1, \ldots , n-1, g_{q_{n-3}^\perp}^*\right)
\,.
\end{align}
Taking first the collinear limit $i||(i+1)$ and then the high-energy limit we find:
\begin{align}
\label{hjqwqeg11}
\begin{split}
&\hspace*{1pt}\lim_{y_3 \gg\, y_4\,\simeq\, \ldots \,\simeq\, y_{n-1} \gg y_n} A_{2\to n-2}\bigl(
  2, 3, 4,\ldots,i,i+1,\ldots,n-1,n,1
\bigr) 
\\
&\xrightarrow{i \| i+1} s\, P_{1\to 1}\left(2, 3, g_{q_1^\perp}^*\right) \frac{1}{t_1} \frac{1}{t_{n-3}} P_{1\to 1}\left(g_{q_{n-3}^\perp}^*, n, 1\right)  \\
& \hspace*{40pt}\times \sum_{h_c=\pm} C_{n-5}\left(g_{q_1^\perp}^*, 4, \ldots, P^{h_c}, \ldots , n-1, g_{q_{n-3}^\perp}^*\right)
\, \text{Sp}_{P^{-h_c}}\bigl(
 x_{i}P^{h_{i}},\, x_{{i+1}}P^{h_{{i+1}}}
\bigr).
\end{split}
\end{align}
Thus, from Eqs.~\eqref{hjqwqeg} and \eqref{hjqwqeg11}, we have
\begin{align}
\label{ccpev}
\begin{split}
& C_{n-4}\left(g_{q_1^\perp}^*, 4, \ldots, i,i+1, \ldots , n-1, g_{q_{n-3}^\perp}^*\right)
\\
&\hspace*{40pt}
\xrightarrow{i \| i+1}\sum_{h_c=\pm} C_{n-5}\left(g_{q_1^\perp}^*, 4, \ldots, P^{h_c}, \ldots , n-1, g_{q_{n-3}^\perp}^*\right)
\, \text{Sp}_{P^{-h_c}}\bigl(
x_{i}P^{h_{i}},\, x_{{i+1}}P^{h_{{i+1}}}
\bigr).
\end{split}
\end{align}
The collinear behaviour for PEVs follows similarly
\begin{align}
&\hspace*{-40pt}
P_{1\to n-3}\bigl(
2,\,, \ldots, i,\, {i+1}, \ldots, {n-1},\, g_{q_{n-3}^{\perp}}^*
\bigr) \nonumber\\[1ex]
\hspace*{20pt}
\xrightarrow{i \| i+1}&\sum_{h_c=\pm}\, P_{1\to n-4}\bigl(
2, \ldots,  P^{h_c},\ldots, {n-1},\, g_{q_{n-3}^{\perp}}^*
\bigr)\,\, \text{Sp}_{P^{-h_c}}\bigl(
x_{i}P^{h_{i}},\, x_{{i+1}}P^{h_{{i+1}}}
\bigr).
\end{align}
Next, we present an explicit example illustrating the collinear limit of a four-parton CEV.
 \subsubsection*{Example: collinear limit of a four-parton CEV}

We analyze the collinear limit $p_5 \parallel p_6$ in the CEV $C_{4}\big(g_{q_1^\perp}^*,g_4^{\oplus}, q_5^{\oplus}, \bar{q}_6^{\ominus}, g_7^{\ominus},g_{q_5^\perp}^*\big)$. 
Our first goal is to derive the MSLCV behaviour in the limit. The momenta $p_5$ and $p_6$ merge along a shared direction $P$, parameterized as 
\begin{equation}
\label{collSplit}
p_5\to x_5P,\qquad p_6\to x_6P,\qquad\text{ with } x_5+x_6=1.
\end{equation}
We should manifest the splitting directly in the MSLCVs. Firstly, the ratio of momenta amongst the collinear pair in Eq.~(\ref{collSplit}) directly fixes the MSLCV $X_3$, defined in Eq.~\eqref{minival:nonlorentzinvdd} as the ratio of plus lightcone momenta:  
\begin{figure}
        \centering
        \includegraphics[width=1\linewidth]{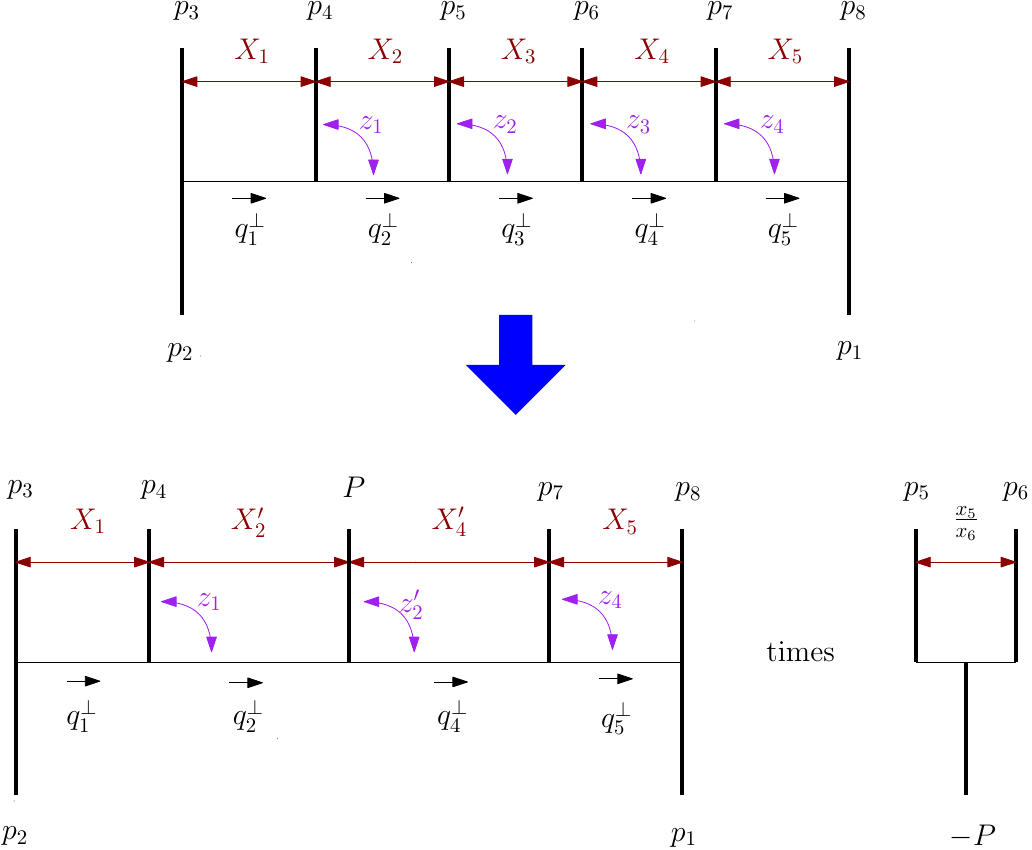}
 \caption{
The MSLCVs for 8-point kinematics and their reduction via collinear factorization to 7-point kinematics in the collinear limit $p_5 \parallel p_6$, with $p_5 \to x_5 P$ and $p_6 \to x_6 P$. In terms of MSLCV, this corresponds in the 8-point case to $z_3 \to 1+z_2
{x_5}/{x_6}$ and $X_3 \to {x_5}/{x_6}$. The resulting 7-point kinematics, in terms of the new MSLCV, are related to those of the original 8-point configuration as \hbox{$\{z_2' = ({z_2 + x_6})/{x_5},\, X_2' = X_2 x_5,\, 
{X_4' = {X_4}/{x_6}}\}$}.
}
        \label{fig:acs}
    \end{figure}
    \begin{equation}
         X_3=\frac{p_5^+}{p_6^+}\to \frac{x_5}{x_6}.
 \end{equation}
Secondly, from the MSLCV definition given in Eq.~\eqref{minival:nonlorentzinvdd}, one can first derive \begin{equation}
\frac{q_{3}^\perp}{q_{4}^\perp} = -\frac{z_3}{1-z_3}
    \label{hjfqwe}
\end{equation} 
using momentum conservation. Then from Eq.~\eqref{hjfqwe}, using the definitions of \(z_2\) and \(z_3\), we obtain:
\begin{equation}
\left.
\begin{aligned}
& z_2=-\frac{q_3^\perp }{p_5^\perp} = -\frac{q_3^\perp }{x_5P^\perp }  \\
& z_3=-\frac{q_4^\perp }{p_6^\perp} = -\frac{q_4^\perp }{x_6P^\perp }
\end{aligned}\right\}
\qquad\Longrightarrow \qquad
\frac{z_2}{z_3}=-\frac{x_6 (1-z_3)}{x_5z_3}.\label{perpRela}
 \end{equation}
Solving Eq.~\eqref{perpRela} yields $z_3\to 1+z_2 {x_5}/{x_6}$, which also holds for the conjugate variable $\bar z_3$. Consequently, in the MSLCVs, the collinear limit  corresponds to
 \begin{eqnarray}
  \left\{z_3\to \frac{x_5
   z_2+x_6}{x_6},\bar z_3\to \frac{x_5
   \bar z_2+x_6}{x_6},X_3\to \frac{x_5}{x_6}\right\}.
\label{collinearsub}
 \end{eqnarray}

Consider now the spinor products corresponding to the collinear pair (which can be determined in terms of the MSLCVs using Eqs.~(\ref{spinoconven}) and (\ref{minival:nonlorentzinvdd})):
\begin{equation}
\langle 56\rangle=\frac{{\color{purple}(1+X_{3}z_{2}-z_{3})}}{\sqrt{X_{3}}(z_{2}-1)(z_{3}-1)}\qquad [ 56]=\frac{{\color{purple}(1+X_{3}\bar z_{2}-\bar z_{3})}}{\sqrt{X_{3}}(\bar z_{2}-1)(\bar z_{3}-1)}\,.
\end{equation}
We observe that 
in the collinear limit, Eq.~(\ref{collinearsub}),
both spinor products $\langle 56 \rangle$ and $[56]$ simultaneously vanish. 
Thus, in the CEV $C_{4}\big(g_{q_1^\perp}^*,g_4^{\oplus}, q_5^{\oplus}, \bar{q}_6^{\ominus}, g_7^{\ominus},g_{q_5^\perp}^*\big)$ terms containing the $1/\langle 56 \rangle$ and $1/[56]$ factors emerge as the dominant divergences in the collinear limit $p_5 \parallel p_6$. 
Upon substituting Eq.~\eqref{collinearsub} into the MSLCV representation of this CEV, one indeed obtains the expected factorization of Eq.~\eqref{ccpev}.
Our next task is to determine how the lower-point CEV depends on the collinear parameters $x_4$, $x_5$, and the original lightcone variables MSLCVs, i.e. $X_2$, $X_4$, and $z_2$. 
Figure~\ref{fig:acs}, illustrates the transition from the original four-parton CEV~$C_{4}\big(g_{q_1^\perp}^*,g_4^{\oplus}, q_5^{\oplus}, \bar{q}_6^{\ominus}, g_7^{\ominus},g_{q_5^\perp}^*\big)$ to the 3-parton CEV~ $C_{3}\big(g_{q_1^\perp}^*,g_4^{\oplus}, g_P^h, g_7^{\ominus},g_{q_5^\perp}^*\big).$ 
This corresponds to the mapping: 
\begin{equation}
\label{collMap}
\{X_2, X_3, X_4, q_{1}^\perp,z_1, z_2,  z_3,  z_4\} \quad \longrightarrow\quad 
\{X_2^{\prime}, X_4^{\prime}, q_{1}^\perp, z_1,  z_2^{\prime},  z_4\},
\end{equation}
where the primed variables corresponding to $C_3$. Their definition according to~\eqref{minival:nonlorentzinvdd} is:
\begin{align*}
X_2^\prime &= \frac{p_4^{+}}{P^{+}}, &
X_4^\prime &= \frac{P^{+}}{p_7^{+}}, &
z_2^\prime &= -\frac{q_4^\perp}{P^\perp}.
\end{align*}
The relation defining the collinear map (\ref{collMap}) are:
\begin{subequations}
 \begin{align}
&\text{Transverse:} \quad 
    z_2^\prime=-\frac{q_4^\perp}{P^\perp} = -x_5\frac{q_3^\perp-p_6^\perp }{p_5^\perp} = x_5\left(z_2+\frac{x_6}{x_5}\right); \\
    &\text{Longitudinal:} \quad 
X_2^\prime =  \frac{p_4^+}{P^+}  =  \frac{x_5 p_4^+}{p_5^+} =X_2x_5, \quad \quad 
 X_4^\prime=  \frac{P^+}{p_7^+} = \frac{p_6^+}{x_6 p_7^+}=  \frac{X_4}{x_6}\,  .
\end{align}
\end{subequations}
This leads to the complete transformation rules:
\begin{equation}
\left\{
z_2^\prime = {z_2x_5 + x_6},\ 
X_2^\prime = X_2x_5,\ 
X_4^\prime = \frac{X_4}{x_6}
\right\}. \label{eq:transform}
\end{equation}
In the collinear limit $p_5 \parallel p_6$,~\eqref{collinearsub}, the function
$
C_{4}(g_{q_1^\perp}^*,\,g_4^{\oplus},\, q_5^{\oplus},\, \bar{q}_6^{\ominus},\, g_7^{\ominus},\, g_{q_5^\perp}^*)
$
has simple poles at $z_3 = 1 + \frac{x_5}{x_6} z_2$ and $\bar{z}_3 = 1 + \frac{x_5}{x_6} \bar{z}_2$, whose residues are, respectively,
 \begin{eqnarray}
    && \hspace*{-30pt}\underset{z_3= 1+\frac{x_5}{x_6}z_2}{\text{Res}}C_{4}(g_4^\oplus,q_5^\oplus,\bar{q}_6^{\ominus},g_7^\ominus)(X_2,\frac{x_5}{x_6},X_4,q_1^\perp,\bar q_1^{\perp},z_1,\bar{z}_1,z_2,\bar{z}_2,z_3,\bar{z}_3,z_4,\bar{z}_4)\non \eqn C_{3}(g_4^\oplus,g_P^\ominus,g_7^\ominus)\big(X_2^\prime,X_4^\prime,q_1^\perp,\bar{q}_1^{\perp},z_1,\bar{z}_1,z_2^\prime,\bar{z}_2^\prime,z_4,\bar{z}_4\big)\underset{z_3= 1+\frac{x_5}{x_6} z_2}{\text{Res}}\operatorname{Sp}_{g_{-P}^\oplus}\left(x_{5} q_P^{\oplus}, x_{6} q_P^{\ominus}\right),\non 
 \end{eqnarray} 
 and
 \begin{eqnarray}
    && \hspace*{-30pt}\underset{\bar z_3= 1+\frac{x_5}{x_6}\bar z_2}{\text{Res}}C_{4}(g_4^\ominus,q_5^\oplus,\bar{q}_6^{\ominus},g_7^\oplus)(X_2,\frac{x_5}{x_6},X_4,q_1^\perp,\bar q_1^{\perp},z_1,\bar{z}_1,z_2,\bar{z}_2,z_3,\bar{z}_3,z_4,\bar{z}_4)\non \eqn C_{3}(g_4^\ominus,g_P^\oplus,g_7^\oplus)\left(X_2^\prime,X_4^\prime,q_1^\perp,\bar q_1^{\perp},z_1,\bar{z}_1,z_2^\prime,\bar{z}_2^\prime,z_4,\bar{z}_4\right)\underset{\bar z_3= 1+\frac{x_5}{x_6} \bar z_2}{\text{Res}}\operatorname{Sp}_{g_{-P}^\ominus}\left(x_{5} q_P^{\oplus}, x_{6} q_P^{\ominus}\right),\non 
 \end{eqnarray} 
 with the primed variable substitution according to~\eqref{eq:transform}.
 These correspond to the two terms with helicities plus and minus in (\ref{ccpev}).
The code implementing this check is given in the Mathematica file \texttt{Examples.nb}.

\section{Brief introduction to the Kleiss-Kuijf relation}
\label{sec:BriefKKrelation}

In this appendix we will illustrate how the Kleiss-Kuijf relation can be derived for pure gluon amplitudes. We then comment on its generalization to amplitudes containing one or two quark-antiquark pairs. In section~\ref{kk} we then use these relations to derive corresponding ones for MREVs.
For a comprehensive discussion and justification of the general-multiplicity case, we refer the reader to Refs.~\cite{DelDuca:1999rs,Weinzierl:2016bus}.

One way to obtain the Kleiss-Kuijf relation between colour-ordered amplitudes is to compare between the trace basis and the DDM~\cite{DelDuca:1999rs} basis for a colour-dressed amplitude. Let us illustrate this using the six-point colour-dressed pure gluon amplitude $\mathcal{A}(1,2,3,4,5,6)$. This amplitude can be expressed in either the trace basis or the DDM basis, respectively, as follows:
\begin{align}
\mathcal	A(1,2,3,4,5,6)=&\sum_{\sigma\in S_5}\tr(T^{a_1}T^{a_{\sigma_2}}T^{a_{\sigma_3}}T^{a_{\sigma_4}}T^{a_{\sigma_5}}T^{a_{\sigma_6}}) A(1,\sigma_2,\sigma_3,\sigma_4,\sigma_5,\sigma_6)
	\label{tbb}
	\\
	=&\sum_{\sigma\in S_4}F^{a_1a_{\sigma_2}x_1}F^{x_1a_{\sigma_3}x_2}F^{x_2a_{\sigma_4}x_3}F^{x_3a_{\sigma_5}a_6}A(1,\sigma_2,\sigma_3,\sigma_4,\sigma_5,6),\label{fbb}
\end{align}
where for brevity we denote each particle by $i\equiv g_i^{h_i}$. 
Using the six-point version of the relation \eqref{fffftoTTTT}, e.g. 
\begin{eqnarray}
	F^{a_1 a_2 x_1} F^{x_1 a_3 x_2} F^{x_2 a_4 x_3}F^{x_3 a_5 a_6}  = \operatorname{tr}\left(T^{a_1}\left[T^{a_2},\left[T^{a_3},\left[T^{a_4},\left[T^{a_{5}}, T^{a_6}\right] \right]\right]\right]\right) \,,
    \label{liudianver}
\end{eqnarray}
we find that when going from the DDM basis expression in \eqref{fbb} to the trace basis in \eqref{tbb}, the DDM basis elements contribute to the specific trace $ \operatorname{tr}\left(T^{a_1} T^{a_2} T^{a_3} T^{a_6} T^{a_5} T^{a_4}\right)$ as
\begin{align}
    & (F^{a_2}F^{a_3}F^{a_4}F^{a_5})_{a_1a_6} A(1,2,3,4,5,6)
    + (F^{a_2}F^{a_4}F^{a_3}F^{a_5})_{a_1a_6} A(1,2,4,3,5,6)
 \nonumber  \\
    &  + (F^{a_2}F^{a_4}F^{a_5}F^{a_3})_{a_1a_6} A(1,2,4,5,3,6)
    + (F^{a_4}F^{a_2}F^{a_3}F^{a_5})_{a_1a_6} A(1,4,2,3,5,6)
   \nonumber\\
    &  + (F^{a_4}F^{a_2}F^{a_5}F^{a_3})_{a_1a_6} A(1,4,2,5,3,6)
    + (F^{a_4}F^{a_5}F^{a_2}F^{a_3})_{a_1a_6} A(1,4,5,2,3,6)
   \nonumber \\
    =&\ \operatorname{tr}\left(T^{a_1} T^{a_2} T^{a_3} T^{a_6} T^{a_5} T^{a_4}\right)
    \Big( 
        A(1,2,3,4,5,6)
        + A(1,2,4,3,5,6)
        + A(1,2,4,5,3,6)
      \nonumber  \\
        &\hspace{6em}
        + A(1,4,2,3,5,6)
        + A(1,4,2,5,3,6)
        + A(1,4,5,2,3,6)
    \Big)
    + \ldots
\label{eq723}
    \\
    =&\ \operatorname{tr}\left(T^{a_1} T^{a_2} T^{a_3} T^{a_6} T^{a_5} T^{a_4}\right)
    A(1,2,3,6,5,4)
    + \ldots
\label{eq724}
\end{align}
where the ellipsis stand for other elements in the trace basis according to Eq.~\eqref{liudianver}.
Since the trace basis elements are linearly independent, it follows from 
Eqs.~\eqref{eq723} and \eqref{eq724}, that
\begin{align}
A(1,2,3,6,5,4)=\sum_{\shuffle}A(1,\{2,3\}\shuffle \{4,5\},6),
\end{align}
where $\shuffle$ means the shuffle product.
From this perspective, the 
Kleiss-Kuijf relation is a 
consistency condition between the DDM basis decomposition in Eq.~\eqref{fbb} and the trace basis one in Eq.~\eqref{tbb}. 
One similarly has 
\begin{align}
A(1,2,3,7,6,5,4)=-\sum_{\shuffle}A(1,\{2,3\}\shuffle \{4,5,6\},7)
\end{align}
where the additional minus sign originates from the extra layers of nested commutators in Eq.~\eqref{liudianver}.

Working with fermions in the adjoint representation, e.g. in the context of ${\cal N}=4$ sYM, 
the result easily generalises to fermions, where one simply needs to account for the extra sign arising from odd permutations of fermions. 
Then, since the colour-ordered amplitudes do not depend on the colour representation, these relations carry over to QCD.
For example,
\begin{align}
	A(1,q_2,{\bar{q}}_3,6,5,4) &= \sum_{\shuffle} A\big(1,\{q_2,{\bar{q}}_3\} \shuffle \{4,5\},6\big), 
    \label{eq728}\\
	A(1,2,3,6,{\bar{q}}_5,q_4) &= -\sum_{\shuffle} A\big(1,\{2,3\} \shuffle \{q_4,{\bar{q}}_5\},6\big),
    \label{wqcwe}
\end{align}
and a more interesting one:
\begin{align}
	&A(1,Q_2,{\bar{Q}}_3,6,{\bar{q}}_5,q_4) \nonumber 
	\\=& -\sum_{\shuffle} (-1)^{w(\shuffle)} A\big(1,\{Q_2,{\bar{Q}}_3\} \shuffle \{q_4,{\bar{q}}_5\},6\big),
   \nonumber   \\
    =& - \Big(
    A(1,Q_2,{\bar{Q}}_3,q_4,{\bar{q}}_5,6)
    - A(1,Q_2,q_4,{\bar{Q}}_3,{\bar{q}}_5,6)
    + A(1,Q_2,q_4,{\bar{q}}_5,{\bar{Q}}_3,6)
    \nonumber \\
    &\qquad
    + A(1,q_4,Q_2,{\bar{Q}}_3,{\bar{q}}_5,6)
    - A(1,q_4,Q_2,{\bar{q}}_5,{\bar{Q}}_3,6)
    + A(1,q_4,{\bar{q}}_5,Q_2,{\bar{Q}}_3,6)
    \Big),
    \label{eq730}
\end{align}
where the sign of each term is fixed by $w(\shuffle)$, which equals $1$ and $0$ for fermion permutations that are odd and even, respectively, relative to the reference ordering $\{Q_2, \bar{Q}_3, q_4, \bar{q}_5\}$.

\section{Colour-dressed PEVs with three emissions}
\label{colourdressingExample}
 
 \subsection{PEV \texorpdfstring{$g^*g \rightarrow g \bar{q} q$}{g* g → g qbar q}}
The PEV for $g g^* \rightarrow g \bar{q} q$ was first computed in Ref.~\cite{DelDuca:1999iql}. Here, we review the colour dressing and present the full PEV for~$\mathcal{P}_{1\to3}^{g \to g \bar{q} q}\big(g_2^{\oplus},\, g_3^{\oplus},\, \bar{q}_4^{\ominus},\, q_5^{\oplus},\, g_{q_{3}^\perp}^*\big)$ in MSLCV form.
 \begin{equation}
     \begin{aligned}
& \hspace*{-30pt}\mathcal{P}_{1\to3}^{g \to g \bar{q} q}\big(g_2^{\oplus} , g_3^{\oplus}, \bar{q}_4^{\ominus}, q_5^{\oplus},g_{q_{3}^\perp}^*\big)= \\
& g^3\Big[\big(T^{c_3} T^{d_2} T^{d_3}\big)_{i_5 \bar{\imath}_4} P_{1\to3}\big(g_2^{\oplus},g_3^{\oplus},\bar{q}_4^{\ominus},q_5^{\oplus},g_{q_{3}^\perp}^*\big) \\
&+\big(T^{d_2} T^{c_3} T^{d_3}\big)_{i_5 \bar{\imath}_4} P_{1\to3}\big(g_3^{\oplus},\bar{q}_4^{\ominus},q_5^{\oplus},g_2^{\oplus},g_{q_{3}^\perp}^*\big) \\
& +\big(T^{d_3} T^{c_3} T^{d_2}\big)_{i_5 \bar{\imath}_4}P_{1\to3}\big(g_2^{\oplus},\bar{q}_4^{\ominus},q_5^{\oplus},g_3^{\oplus},g_{q_{3}^\perp}^*\big) \\
&+\big(T^{d_3} T^{d_2} T^{c_3}\big)_{i_5 \bar{\imath}_4} P_{1\to3}\big(\bar{q}_4^{\ominus},q_5^{\oplus},g_3^{\oplus},g_2^{\oplus},g_{q_{3}^\perp}^*\big) \\
& +\big(T^{d_2} T^{d_3} T^{c_3}\big)_{i_5 \bar{\imath}_4} P_{1\to3}\big(\bar{q}_4^{\ominus}, q_5^{\oplus}, g_2^{\oplus} , g_3^{\oplus},g_{q_{3}^\perp}^*\big) \\
&+\big(T^{c_3} T^{d_3} T^{d_2}\big)_{i_5 \bar{\imath}_4} P_{1\to3}\big(g_3^{\oplus}, g_2^{\oplus} , \bar{q}_4^{\ominus}, q_5^{\oplus},g_{q_{3}^\perp}^*\big)\Big]\,,
\end{aligned}\label{wjke}
 \end{equation}
where the  colour-ordered PEVs entering~\eqref{wjke} are:
 \begin{align}
     P_{1\to3}\big(g_2^{\oplus} , g_3^{\oplus}, \bar{q}_4^{\ominus}, q_5^{\oplus},g_{q_{3}^\perp}^*\big)=&-\frac{X_1 X_2^{5/2} \big(z_1-1\big) z_1 z_2}{\big(X_1 X_2+X_2+1\big){}^2 \big(X_1-z_1+1\big) \big(X_2 z_1-z_2+1\big)
   {q_1^\perp}^2}\non P_{1\to3}\big( g_3^{\oplus}, \bar{q}_4^{\ominus}, q_5^{\oplus},g_2^{\oplus} ,g_{q_{3}^\perp}^*\big)=&\frac{X_2^{3/2} \big(z_1-1\big){}^2 \big(z_2-1\big) z_2}{\big(X_1 X_2+X_2+1\big){}^2 \big(X_1-z_1+1\big) \big(X_2 z_1-z_2+1\big)
   {q_1^\perp}^2}\non P_{1\to3}\big(g_2^{\oplus} , \bar{q}_4^{\ominus}, q_5^{\oplus}, g_3^{\oplus},g_{q_{3}^\perp}^*\big)=&   \nonumber
   \\& \hspace{-2cm}\frac{X_2^{5/2} \big(z_1-1\big){}^2 z_1 \big(z_2-1\big) z_2}{\big(X_1 X_2+X_2+1\big){}^2 \big(X_2 z_1-z_2+1\big) \big(X_1 X_2 z_1-(1-z_1)(1-z_2)\big) {q_1^\perp}^2}\non P_{1\to3}\big( \bar{q}_4^{\ominus}, q_5^{\oplus}, g_3^{\oplus},g_2^{\oplus} ,g_{q_{3}^\perp}^*\big)=& \nonumber
   \\& \hspace{-2cm}-\frac{X_1 X_2^{5/2} \big(z_1-1\big) z_1 \big(z_2-1\big) z_2}{\big(X_1 X_2+X_2+1\big){}^2 \big(X_2 z_1-z_2+1\big) \big(X_1 X_2
   z_1-(1-z_1)(1-z_2)\big) {q_1^\perp}^2}\non P_{1\to3}\big( \bar{q}_4^{\ominus}, q_5^{\oplus} ,g_2^{\oplus}, g_3^{\oplus},g_{q_{3}^\perp}^*\big)=&\frac{X_2^{3/2} \big(z_1-1\big) \big(z_2-1\big) z_2}{\big(X_1 X_2+X_2+1\big){}^2 \big(X_2 z_1-z_2+1\big) {q_1^\perp}^2}\non P_{1\to3}\big(g_3^{\oplus}, g_2^{\oplus} , \bar{q}_4^{\ominus}, q_5^{\oplus},g_{q_{3}^\perp}^*\big)=&\frac{X_2^{5/2} \big(z_1-1\big) z_1 z_2}{\big(X_1 X_2+X_2+1\big){}^2 \big(X_2 z_1-z_2+1\big) {q_1^\perp}^2}\,.
   \label{eqc32}
 \end{align}
The pole $\big(X_1 X_2 + X_2 + 1\big)$ is always positive in the physical region, as it corresponds to ratios of plus light-cone components, as shown in Eq.~\eqref{minival:nonlorentzinvdd}.
The other poles in Eq.~\eqref{eqc32} are collinear poles. They correspond to the following spinor brackets:
\begin{subequations}
\begin{align}
    \ab{3}{4} &= \frac{ {\left(X_1 - z_1 + 1\right)}\, q_1^\perp }{ \sqrt{X_1}\, (z_1 - 1) } \\
    \ab{3}{5} &= \frac{ {\left(X_1 X_2 z_1 - (1-z_1)(1-z_2)\right)}\, q_1^\perp }
    { \sqrt{X_1}\, \sqrt{X_2}\, (z_1 - 1)\, (z_2 - 1) } \\
    \ab{4}{5} &= \frac{ { \left(X_2 z_1 - z_2 + 1\right)}\, q_1^\perp }
    { \sqrt{X_2}\, (z_1 - 1)\, (z_2 - 1) }\,. \label{collinearpoles}
\end{align}
\end{subequations}
As in Eq.~\eqref{minival:nonlorentzinvdd} (see also section \ref{wohac_collinear}) these poles in Eq.~\eqref{collinearpoles} can be reached within the physical region by varying the transverse momenta to~$z_i=1$.
 
Next we show how the individual colour-ordered PEVs in Eq.~\eqref{wjke} are obtained from \texttt{MREV}~\cite{yuyu_2017_github}. One first
loads the data in terms of MSLCVs
\begin{lstlisting}
Import["PEV_1to3.m"],
\end{lstlisting}
and initialises the momentum representation via
 \begin{align*}
&\mathtt{pevtopevp[\#] \& /@} \nonumber\\[1mm]
&\begin{array}{lll}
\big\{\mathtt{pev[P, P, qM, qP]}, \quad& \mathtt{pev[P, qPc, qMc, Pc]}, \quad  &\mathtt{pev[P, qP, qM, P]}\\[1mm]
\hspace{0.2cm}\mathtt{pev[P, Pc, qPc, qMc]}, \quad& \mathtt{pev[P, P, qPc, qMc]},
\quad& \mathtt{pev[P, Pc, qM, qP]}\big\},
\end{array}\,.
\end{align*}
The code conventions are introduced in section~\ref{sec:4partonCEVsPEVs}. The momentum representations for the colour-ordered PEVs in Eq.~\eqref{wjke} can then be obtained using the following command:
\begin{align}
P_{1\to 3}\left(g_2^{\oplus} ,g_3^{\oplus}, \bar{q}_4^{\ominus}, q_5^{\oplus}, g_{q_{3}^\perp}^* \right)
&= \mathtt{pevp[P, P, qM, qP][3,4,5]}\\
P_{1\to 3}\left(g_3^{\oplus}, \bar{q}_4^{\ominus}, q_5^{\oplus}, g_2^{\oplus}, g_{q_{3}^\perp}^* \right)
&= \mathtt{pevp[P, qPc, qMc, Pc][5, 4, 3]}\\
P_{1\to 3}\left( g_2^{\oplus}, \bar{q}_4^{\ominus}, q_5^{\oplus}, g_3^{\oplus}, g_{q_{3}^\perp}^* \right)
&= \mathtt{pevp[P, qP,qM,  P][4, 5, 3]}\\
P_{1\to 3}\left( \bar{q}_4^{\ominus}, q_5^{\oplus}, g_3^{\oplus},g_2^{\oplus}, g_{q_{3}^\perp}^* \right)
&= \mathtt{pevp[P, Pc, qPc, qMc][3, 5, 4]}\\
P_{1\to 3}\left(\bar{q}_4^{\ominus}, q_5^{\oplus},g_2^{\oplus}, g_3^{\oplus}, g_{q_{3}^\perp}^* \right)
&= \mathtt{pevp[P, P, qPc, qMc][3, 5, 4]}\\
P_{1\to 3}\left(g_3^{\oplus},g_2^{\oplus}, \bar{q}_4^{\ominus}, q_5^{\oplus}, g_{q_{3}^\perp}^* \right)
&= \mathtt{pevp[P, qM, qP, Pc][ 4, 5,3]}.
\end{align}

 \subsection{PEV \texorpdfstring{$\bar{q} g^*\to q\bar{Q}Q$}{qbar g* → q Qbar Q}}
Here we consider PEVs with two quark-antiquark pairs. The momentum representation for the full PEV for $\mathcal{P}_{1\to 3}^{\bar{q} \to q \bar{Q} Q}\big(\bar{q}^\ominus_{2},\, q^\oplus_{3},\, \bar{Q}^\ominus_{4},\, Q^\oplus_{5},\, g_{q_3^\perp}^*\big)$ was presented in~\cite{DelDuca:1999iql}. We show how to obtain this result using our \texttt{MREV}~\cite{yuyu_2017_github} library. 

The colour dressing of the PEV $\mathcal{P}_{1\to 3}^{\bar{q} \to q \bar{Q} Q}\big(\bar{q}^\ominus_{2},\, q^\oplus_{3},\, \bar{Q}^\ominus_{4},\, {Q}^\oplus_{5},\, g_{q_3^\perp}^*\big)$ is given by

\begin{eqnarray}
     \begin{aligned}
& \hspace*{-20pt}\mathcal{P}_{1\to 3}^{\bar{q} \to q \bar{Q} Q}\left(\bar{q}^\ominus_{2} , {q}^\oplus_{3}, \bar{Q}^\ominus_{4}, {Q}^\oplus_{5},g_{q_3^\perp}^*\right)= \\
& g^3\left[T_{j_5 \bar{\imath}_2}^c \delta_{i_3 \bar{\jmath}_4} P\left(\bar{q}^\ominus_{2},{q}^\oplus_{3},\bar{Q}^\ominus_{4},{Q}^\oplus_{5},g_{q_3^\perp}^*\right)-\frac{1}{N_c} T_{i_3\bar{\imath}_2}^c \delta_{j_5 \bar{\jmath}_4} P\left(\bar{q}^\ominus_{2},{Q}^\oplus_{5},\bar{Q}^\ominus_{4},{q}^\oplus_{3},g_{q_3^\perp}^*\right)\right. \\
& \left.+T_{i_3 \bar{\jmath}_4}^c \delta_{j_5 \bar{\imath}_2} P\left(\bar{Q}^\ominus_{4},{Q}^\oplus_{5},\bar{q}^\ominus_{2},{q}^\oplus_{3},g_{q_3^\perp}^*\right)-\frac{1}{N_c} T_{j_5 \bar{\jmath}_4}^c \delta_{i_3 \bar{\imath}_2}P\left(\bar{Q}^\ominus_{4},{q}^\oplus_{3},\bar{q}^\ominus_{2},{Q}^\oplus_{5},g_{q_3^\perp}^*\right)\right] \\
& -\delta_{q Q}(1 \leftrightarrow 3) \text {. } \\
&
\end{aligned}
\label{labedcaz}
 \end{eqnarray}
 We first load the PEVs, written in MSLCVs, using:
\begin{lstlisting}
Import["PEV_1to3.m"],
\end{lstlisting}
and then initialise the momentum representation as
 \begin{align*}
&\mathtt{pevtopevp[\#] \& /@} \nonumber\\[1mm]
&\begin{array}{ll}
\big\{\mathtt{ pev[\{qM,q\},\{qP,q\},\{qM,Q\},\{qP,Q\}]}, \quad& \mathtt{ pev[\{qM,q\},\{qP,Q\},\{qM,Q\},\{qP,q\}]},\\[1mm]
\hspace{0.2cm}\mathtt{ pev[\{qM,q\},\{qP,q\},\{qPc,Q\},\{qMc,Q\}]}, \quad&\mathtt{ pev[\{qM,q\},\{qP,Q\},\{qPc,q\},\{qMc,Q\}]}\big\}\,.
\end{array}
\end{align*}
Then, the colour-ordered PEV appearing in \eqref{labedcaz} in the momentum representation are given by
 \begin{eqnarray}
       P\left(\bar{q}^\ominus_{2},{q}^\oplus_{3},\bar{Q}^\ominus_{4},{Q}^\oplus_{5},g_{q_3^\perp}^*\right)\eqn\mathtt{pevp[\{qM,q\},\{qP,q\},\{qM,Q\},\{qP,Q\}][3,4,5]} \non P\left(\bar{q}^\ominus_{2},{Q}^\oplus_{5},\bar{Q}^\ominus_{4},{q}^\oplus_{3},g_{q_3^\perp}^*\right)\eqn \mathtt{pevp[\{qM,q\},\{qP,Q\},\{qM,Q\},\{qP,q\}][5,4,3]}\non P\left(\bar{Q}^\ominus_{4},{Q}^\oplus_{5},\bar{q}^\ominus_{2},{q}^\oplus_{3},g_{q_3^\perp}^*\right)\eqn \mathtt{pevp[\{qM,q\},\{qP,q\},\{qPc,Q\},\{qMc,Q\}][3,5,4]}\non P\left(\bar{Q}^\ominus_{4},{q}^\oplus_{3},\bar{q}^\ominus_{2},{Q}^\oplus_{5},g_{q_3^\perp}^*\right)\eqn \mathtt{pevp[\{qM,q\},\{qP,Q\},\{qPc,q\},\{qMc,Q\}][5,3,4]}.
 \end{eqnarray}

\section{Examples of \texttt{MREV} library of PEVs and CEVs}
\label{sec:mrev}

In this section we provide additional information regarding the use of our \texttt{MREV} library~\cite{yuyu_2017_github}, supplementing its introduction in section~\ref{sec:4partonCEVsPEVs}. An important goal of this appendix is to demonstrate   some of the checks we performed. We begin with a brief explanation of the initialization  process of the package and a summary of  frequently-used functions. We then provide examples verifying the factorization properties of the obtained MREVs in high-energy and soft limits. We also provide examples of the photon decoupling identity to demonstrate some of the checks we performed for the colour-ordered MREVs results collected in the library.  For brevity, we do not display here examples for collinear factorization, the Kleiss-Kuijf identity, the reversal identity, or the supersymmetric Ward identity. For interested readers, we provide several examples of each of these relations in the accompanying notebook \texttt{Examples.nb} in \texttt{MREV} library.

\subsection{Initialising}
One can use the following command to initialise the package:
\begin{lstlisting}
In[1]:= SetDirectory["path/to/MREVinit.m"];
%set directory, please modify the path,
In[2]:= Import["MREVinit.m"];  %initialise the package.
\end{lstlisting}
\subsection{List of useful function}
The initialisation process defines the following frequently-used functions.
\begin{lstlisting}
 List of functions:

1.  getMinimalgn[#, number of points]:
    Convert momentum representation to minimal set of lightcone variables(MSLCVs) representation. The "number of points" should be an integer with a value larger than the minimal multiplicity of the amplitude required to obtain the corresponding PEV or CEV.

2.  cevtocevp[#]: (appliable for 2-4 emissions)
     Convert CEV in MSLCV to CEV in momentum representation.
   pevtopevp[#]:
     Convert PEV in MSLCV to PEV in momentum representation.

3. Variable: "nptxX1/nptx"
   This variable contains a randomly chosen numerical point for the MSCLV, and it can be used for numerical tests of PEVs and CEVs of up to 4 emissions.
   The user may redefine the random point.

4. getargumentcev[#]:
   Automatically fills in the MSLCV dependence for a CEV with up to 4 emissions.
   getargumentpev[#]:
   Automatically fills in the MSLCV dependence for a PEV with up to 4 emissions.
\end{lstlisting}

Up to now, no CEV or PEV data (previously determined MREVs stored in the library) has been loaded; the data is contained in the \texttt{.m} files listed in table \ref{tab:PEVandCEVfiles}. To proceed with our examples, let us load the relevant CEV and PEV data in MSLCV needed for this section.
\begin{lstlisting}
Import["CEV_1_and_2.m"]              
Import["CEV_3.m"]          
Import["CEV_4_qqbgg.m"]
Import["PEV 1to3.m"]
Import["PEV_1to4_qqbggg.m"]
Import["PEV_1to4_qqbqqbg.m"].
\end{lstlisting}

\subsection{Checking factorization}
In this section, we present explicit examples for checking the high-energy and soft factorization of both types of MREVs.

\label{code_ckeching}
\subsubsection*{High-energy limit  factorization of a three-emission CEV }
In this section, we check the following factorization property of the quark-antiquark gluon CEV
\begin{align}
\lim\limits_{X_2\to \infty}  C_{3}\left(g_{q_1^\perp}^*,g_4^{\oplus}, q_5^{\oplus}, \bar{q}_6^{\ominus},g_{q_4^\perp}^*\right)=C_{1}(g_{q_1^\perp}^*,g_4^{\oplus},g_{q_2^\perp}^*)\frac{1}{t_2} C_{2}( g_{q_2^\perp}^*,q_5^{\oplus}, \bar{q}_6^{\ominus},g_{q_4^\perp}^*).
\end{align}
\begin{lstlisting}
In[1]:= 
getmrk$analytic$c[pl_, X2_] := Series[pl, {X2, (*@$\infty$@*), 0}]//Normal 
(*define function for taking the high-energy limit of interest*)

In[2]:= 
LHS = cev[P, qP, qM][(*@$
  \texttt{X}_2, \texttt{X}_3, 
  \texttt{q}[1,``\perp"], \texttt{q}[1,``\perp*"], 
  \texttt{z}_1, \texttt{zb}_1, \texttt{z}_2, \texttt{zb}_2, 
  \texttt{z}_3, \texttt{zb}_3
$@*)] // 
{getmrk$analytic$c[#, (*@$\texttt{X}_2$@*)]} &;     
 (*taking the high-energy limit analytically*).
 \end{lstlisting}
 \begin{lstlisting}
In[3]:= 
exp2 = {{cev[P][(*@$
  \texttt{q}[1,``\perp"], \texttt{q}[1,``\perp*"], 
  \texttt{z}_1, \texttt{zb}_1
$@*)], 
cev[qP, qM][(*@$
  \texttt{X}_3, \texttt{q}[2,``\perp"], \texttt{q}[2,``\perp*"], 
  \texttt{z}_2, \texttt{zb}_2, \texttt{z}_3, \texttt{zb}_3
$@*)], (*@$1/(\texttt{q}[2,``\perp"] \texttt{q}[2,``\perp*"])$@*)}};
 (*anticipated factorization form*)
     
RHS = (Times @@@ (exp2 /. cevc -> cev)) // getMinimalgn[#, 7] &;
 \end{lstlisting}

 \begin{lstlisting}
In[4]:= diff = LHS - RHS // Simplify
(*comparing*)

Out[4]= {0}.
\end{lstlisting}

\subsubsection*{High-energy limit  factorization of a four-emission CEV }
Here we consider the factorization of a multiplicity four CEV in the high-energy limit. While feasible, this check is computationally intensive to perform analytically. For practical purposes, a numerical check is perfectly adequate, and we demonstrate it in what follows.

We first define the numerical high-energy limit target function, \texttt{NMRKcheck8N}, in which \texttt{X2\_} represents an arbitrary longitudinal momentum ratio. It is important to note that the tree-level propagator \(t\) has already been factored out within the definition of this function.

\begin{lstlisting}
In[5]:= 
NMRKcheck8N[x_, n_, X2_, t_] :=   x *(t // getMinimalgn[#, n] &) // ReplaceAll[X2 -> 2^32].
\end{lstlisting}
Next we check the factorization of $C_4$ in the high-energy limit 
\begin{align}
    \lim_{X_3\to \infty} C_4(g_{q_1^\perp}^*,\bar{q}_4^\ominus,q_5^\oplus,q_6^\oplus,\bar{q}_7^\ominus,g_{q_5^\perp}^*)=C_2(g_{q_1^\perp}^*,\bar{q}_4^\ominus,q_5^\oplus,g_{q_3^\perp}^*) \frac{1}{t_3} C_2(g_{q_3^\perp}^*,q_6^\oplus,\bar{q}_7^\ominus,g_{q_5^\perp}^*).
\end{align}
To this end we write the left-hand side and right-hand side as follows:
\begin{lstlisting}
In[6]:= 
LHS = cev[qM, qP, qP, qM][(*@$
  \texttt{X}_2, \texttt{X}_3, \texttt{X}_4, 
  \texttt{q}[1,``\perp"], \texttt{q}[1,``\perp*"], 
  \texttt{z}_1, \texttt{zb}_1, \texttt{z}_2, \texttt{zb}_2, 
  \texttt{z}_3, \texttt{zb}_3, \texttt{z}_4, \texttt{zb}_4
$@*)] //       NMRKcheck8N[#, 8, (*@$\texttt{X}_3$@*), (*@$\texttt{q}[3,``\perp"] \texttt{q}[3,``\perp*"]$@*)] & // Normal;
       
RHS = cev[qM, qP][(*@$\texttt{X}_2, \texttt{q}[1,``\perp"], \texttt{q}[1,``\perp*"], \texttt{z}_1, \texttt{zb}_1, \texttt{z}_2, \texttt{zb}_2$@*)] 
       cev[qP, qM][(*@$\texttt{X}_4, \texttt{q}[3,``\perp"], \texttt{q}[3,``\perp*"], \texttt{z}_3, \texttt{zb}_3, \texttt{z}_4, \texttt{zb}_4$@*)] //      getMinimalgn[#, 8] &;.
\end{lstlisting}
Using a random point,
\begin{align}
\text{nptx} = \left\{
\begin{aligned}
& x_2 = 1.2, \quad x_3 = 1.32, \quad x_4 = 1.4, \\
& q_{1}^\perp = 1.5 + 1.42619\,\mathrm{i}, \quad \bar q_{1}^\perp= 1.5 - 1.42619\,\mathrm{i}, \\
& z_1 = 1.386 + 1.16829\,\mathrm{i}, \quad \bar{z}_1 = 1.386 - 1.16829\,\mathrm{i}, \\
& z_2 = 1.21346 + 1.11293\,\mathrm{i}, \quad \bar{z}_2 = 1.21346 - 1.11293\,\mathrm{i}, \\
& z_3 = 1.87891 + 1.76891\,\mathrm{i}, \quad \bar{z}_3 = 1.87891 - 1.76891\,\mathrm{i}, \\
& z_4 = 1.12346 + 1.49334\,\mathrm{i}, \quad \bar{z}_4 = 1.12346 - 1.49334\,\mathrm{i}
\end{aligned}
\right\},
\end{align}
we then compare the results:
\begin{lstlisting}
In[7]:=LHS /. (*@$\texttt{nptx}$@*)
In[8]:=RHS /. (*@$\texttt{nptx}$@*)
In[9]:=%-%%
--------------------------------------------------
Out[7]= 2.06044 - 0.139504 I
Out[8]= 2.06044 - 0.139505 I
Out[9]= -4.07642*10^-7 - 3.05558*10^-7 I,
\end{lstlisting}
demonstrating that the difference indeed vanishes at the given numerical accuracy.

\subsubsection*{High-energy limit  factorization of a four-emission PEV }
\label{pevhepfac}
In this section, we check 
\begin{align}
    \lim_{X_3\to \infty} P_{1\to 4}(q_2^\oplus,g_3^\ominus,g_4^\oplus,\bar{q}_5^\ominus,g_6^\oplus,g_{q_4^\perp}^*)=P_{1\to 3}(q_2^\oplus,g_3^\ominus,g_4^\oplus,\bar{q}_5^\ominus,g_{q_3^\perp}^*)\frac{1}{t_3} C_2(g_{q_3^\perp}^*,g^\oplus_6,g_{q_5^\perp}^*).
    \label{cad4}
\end{align}
We prepare the PEV that will be checked:
\begin{lstlisting}
checkfn = pev[qP, M, P, qM, P] /. pev -> pevc
nc = 3;
expression = checkfn;.
\end{lstlisting}
We then fill in the PEV MSLCV arguments
\begin{lstlisting}
In[10]:= exp1 = expression @@ (expression // getargumentpev)

Out[10]= pevc[qP, M, P, qM, P][(*@$
  \texttt{X}_1, \texttt{X}_2, \texttt{X}_3, 
  \texttt{q}[1,``\perp"], \texttt{q}[1,``\perp*"], 
  \texttt{z}_1, \texttt{zb}_1, 
  \texttt{z}_2, \texttt{zb}_2, 
  \texttt{z}_3, \texttt{zb}_3
$@*)].
\end{lstlisting}
We can define functions to take the MRK limit numerically first and then evaluate the function at the numerical point in Eq. \eqref{npt2}:
\begin{lstlisting}
getmrkX1[pl_, X2_] :=  ReplaceAll[nptxX1][ReplaceAll[{X2 -> 2^64}][Tofn[pl]]].
\end{lstlisting}
\begin{equation}
\begin{aligned}
\mathtt{nptxX1} = \big\{\, 
& X_1=1.5,\;\; X_2=1.2,\;\; X_3=1.32,\;\; X_4=1.4, \\
& q_{1}^\perp = 1.5 + 1.42619\,\mathrm{i}, \;\; \bar q_{1}^\perp = 1.5 - 1.42619\,\mathrm{i}, \\
& z_1 = 1.386 + 1.16829\,\mathrm{i},\;\;  \bar{z}_1 = 1.386 - 1.16829\,\mathrm{i}, \\
& z_2 = 1.21346 + 1.11293\,\mathrm{i},\;\;  \bar{z}_2 = 1.21346 - 1.11293\,\mathrm{i}, \\
& z_3 = 1.87891 + 1.76891\,\mathrm{i},\;\;  \bar{z}_3 = 1.87891 - 1.76891\,\mathrm{i}, \\
& z_4 = 1.12346 + 1.49334\,\mathrm{i},\;\;  \bar{z}_4 = 1.12346 - 1.49334\,\mathrm{i}
\, \big\}.
\end{aligned}
\label{npt2}
\end{equation}
We then initialize the LHS of Eq. \eqref{cad4} numerically
\begin{lstlisting}
In[11]:= LHS = exp1 // {getmrkX1[#, (*@$\texttt{X}_3$@*)]} &

Out[11]= {-0.0855234 - 0.0537592 I},

\end{lstlisting}
and initialize the RHS of Eq. \eqref{cad4} based on factorization:
\begin{lstlisting}
 exp2 = {{pevc[qP, M, P, qM][(*@$
  \texttt{X}_1, \texttt{X}_2, 
  \texttt{q}[1,"\perp"], \texttt{q}[1,"\perp*"], 
  \texttt{z}_1, \texttt{zb}_1, \texttt{z}_2, \texttt{zb}_2
$@*)],        cevc[P][(*@$\texttt{q}[3,"\perp"], \texttt{q}[3,"\perp*"], \texttt{z}_3, \texttt{zb}_3$@*)],       (*@$1/(\texttt{q}[3,"\perp"] \texttt{q}[3,"\perp*"])$@*)}};

In[13]:= RHS = (Times @@@ (exp2 /. {pevc -> pev, cevc -> cev})) //   getMinimalgn[#, 7] & // ReplaceAll[nptxX1]

Out[13]= {-0.0855234 - 0.0537592 I}
\end{lstlisting}
Then, for the randomly chosen numerical point given in Eq.~\eqref{npt2},
the difference vanishes at the required numerical accuracy:
\begin{lstlisting}
In[14]:= diff = LHS - RHS 

Out[14]= {-1.38778*10^{-17} + 3.46945*10^{-17} I}.
\end{lstlisting}

\subsubsection*{Soft Limit of a CEV}
\label{cevsopft}

To verify the soft factorization described in Eq.~\eqref{amp_soft}, we first prepare the function to be checked
\begin{lstlisting}
 checkfn = cev[P, P, M];
(*input the CEV needed to be checked*),
\end{lstlisting}
and the soft factor corresponding to the positive-helicity gluon,
\begin{lstlisting}
qcw = ab[4, 6]/(ab[4, 5] ab[5, 6]) // getMinimalgn[#, 7] & // Factor;
soft = qcw /. (*@$\texttt{X}_4 -> \texttt{X}_1$@*) // 
    Map[Series[#, {(*@$\texttt{X}_1$@*), \[Infinity], 0}] &] // Normal;
(*input the soft factor*).
\end{lstlisting}
We prepare the left-hand side (LHS) by taking the soft limit of the original CEV, \\
\(
\lim_{k_5^\mu \to \lambda k_5^\mu} C_3\big(g_{q_1^{\perp}}^*,\, q_4^{\oplus},\, g_5^{\oplus},\, \bar{q}_6^{\ominus},\, g_{q_4^{\perp}}^*\big)
\), 
and dividing it by the corresponding soft factor,
\\ \(
 \operatorname{Soft}(p_4,\, g_5^{\oplus},\, p_6)
\):
\begin{lstlisting}
exp1s = cevc[P, P, M][(*@$
  \texttt{X}_2, \texttt{X}_3, 
  \texttt{q}[1,"\perp"], \texttt{q}[1,"\perp*"], 
  \texttt{z}_1, \texttt{zb}_1, \texttt{z}_2, \texttt{zb}_2, 
  \texttt{z}_3, \texttt{zb}_3
$@*)] /. cevc -> cev;
LHS = (Divide[#, soft] & /@ exp1s) /. {(*@$       \texttt{X}_2 -> \texttt{X}_2/\lambda,       \texttt{z}_2 -> \texttt{z}_2/\lambda,       \texttt{zb}_2 -> \texttt{zb}_2/\lambda,        \texttt{X}_3 -> \lambda \texttt{X}_3$@*)} //     Map[Series[#, {(*@$\lambda$@*), 0, 0}] &] // Normal // PowerExpand;
(*LHS: high point soft limit divide the soft factor*).
\end{lstlisting}
We prepare the RHS 2-point CEV of Eq.~\eqref{amp_soft} with the light-cone variable expressed in terms of the original ones as \(X_2' = \frac{p_4^+}{p_6^+} = X_2 X_3\):
\\
\begin{lstlisting}
RHS = cevc[P, M][(*@$
  \texttt{X}_2 \texttt{X}_3, 
  \texttt{q}[1,"\perp"], \texttt{q}[1,"\perp*"], 
  \texttt{z}_1, \texttt{zb}_1, \texttt{z}_3, \texttt{zb}_3
$@*)] /. cevc -> cev// PowerExpand;
(*RHS: lower point*)
\end{lstlisting}
Finally, we make a comparison 
\begin{lstlisting}
In[15]:= df = LHS - RHS // PowerExpand // Simplify

Out[15]= 0
\end{lstlisting}
confirming the soft limit of interest.
Examples of the PEV soft limit and collinear limit checks for MREVs are provided in the~\texttt{Examples.nb} file.

\subsection{Numerical check of relations between colour-ordered MREVs }
\label{u1u1checkcehc}

\textbf{Check of the photon decoupling identity~\eqref{qjfoqfwq}}:
\begin{lstlisting}
cevtocevp[#] & /@{
    cev[P, qP, qM], cev[qP, P, qM], cev[qP, qM, P], cev[qP, qM, Pc]};
   (*Initialising Momentum representation.*)
   \end{lstlisting}
   \begin{lstlisting}
In[16]:= (cev[P,qP,qM][1,4,5,6]+cev[qP,P,qM][1,5,4,6]+cev[qP,qM,P][1,5,6,4]
    +cev[qP,qM,Pc][1,5,6,4] /.     cev -> CEVp)     // Map[getMinimalgn[#, 8] &] // Map[ReplaceAll[nptxX1]]
   (*Convert back to MSLCV and check numerically*)
    -------------------------------------
Out[16]:= 3.88578*10^-16 - 1.27676*10^-15 I,
\end{lstlisting}
where the numerical point \texttt{nptxX1} is given in Eq. \eqref{npt2}.\\

\noindent
\textbf{Check of the photon decoupling identity~\eqref{qwjoingqg}}:
\begin{lstlisting}
    pevtopevp[#]&/@(
    {
    pev[qP,qM,P,qP,qM], pev[qP,qM,qP,P,qM],
    pev[qP,qM,qP,qM,P], pev[qP,qM,qP,qM,Pc], 
    pev[qP,P,qM,qP,qM]
    };
   (*Initialising Momentum representation.*)
   \end{lstlisting}
   \begin{lstlisting}
    In[17]:= {
    pev[qP,qM,P,qP,qM][3,4,5,6], pev[qP,qM,qP,P,qM][3,5,4,6], 
    pev[qP,qM,qP,qM,P][3,5,6,4], pev[qP,qM,qP,qM,Pc][3,5,6,4], 
    pev[qP,P,qM,qP,qM][4,3,5,6]
     }/. PEVu -> PEVup // Total //     getMinimalgn[#, 8] & // ReplaceAll[nptxX1] // AbsoluteTiming
   (*Convert back to MSLCV and check numerically*)
    -------------------------------------
    Out[17]:= {4.86426, 1.47451*10^-16 + 2.81025*10^-16 I}.
\end{lstlisting}

\bibliographystyle{JHEP} 
\bibliography{main}

\end{document}